\documentclass[%
 reprint, 
 twocolumn,
 amsmath,amssymb,amsfonts
 aps,
 pre,
floatfix,
showpacs,
showkeys 
]{revtex4-1}
\usepackage{color}
\usepackage{graphicx}
\usepackage{dcolumn}
\usepackage{bm}
\usepackage[colorlinks,linkcolor=red,citecolor=blue,urlcolor=blue]{hyperref}
\usepackage{tikz}
\usepackage{pifont} 
\usepackage{afterpage}
\usepackage{stmaryrd} 
\usetikzlibrary{arrows}
\usepackage{tabularx}
\usepackage{diagbox}
\newcolumntype{C}{>{\centering\arraybackslash}p{2.5em}}

\allowdisplaybreaks

\usepackage{notes2bib}

\begin{document}



\title{Comprehensive comparison of collision models in the lattice Boltzmann framework: Theoretical investigations}

\author{Christophe Coreixas}
\email{Corresponding author: christophe.coreixas@unige.ch}
\author{Bastien Chopard}
\author{Jonas Latt}
\affiliation{University of Geneva, Geneva, Switzerland.}

\date{\today}

                 
\begin{abstract}
\textcolor{black}{Over the last decades, several types of collision models have been proposed to extend the validity domain of the lattice Boltzmann method (LBM)\textcolor{black}{, each of them being introduced in its own formalism}. The present article proposes a formalism that describes all these methods within a common mathematical framework, and in this way allows us to draw direct links between them. 
Here, the focus is put on single and multirelaxation time collision models in either their raw moment, central moment, cumulant or regularized form. In parallel with that, several bases (non orthogonal, orthogonal, Hermite) are considered for the polynomial expansion of populations. General relationships between moments are first derived to understand how moment spaces are related to each other. 
\textcolor{black}{In addition, a review of collision models further sheds light on collision models that can be rewritten in a linear matrix form.} More quantitative mathematical studies are then carried out by comparing explicit expressions for the post collision populations. Thanks to this, it is possible to deduce the impact of both the polynomial basis (raw, Hermite, central, central Hermite, cumulant) and the inclusion of regularization steps on isothermal LBMs. \textcolor{black}{Extensive results are provided for the D1Q3, D2Q9, and D3Q27 lattices, the latter being further extended to the D3Q19 velocity discretization.}
Links with the most common two and multirelaxation time collision models are also provided for the sake of completeness. 
\textcolor{black}{The present work ends by emphasizing the importance of an accurate representation of the equilibrium state, independently of the choice of moment space.
As an addition to the theoretical purpose of the present article, general instructions are provided to help the reader with the implementation of the most complicated collision models.}}
\end{abstract}                 


\maketitle

\twocolumngrid

\section{\label{sec:intro}INTRODUCTION}

During the past three decades, the lattice Boltzmann method (LBM) was proven to be of particular interest to the field of Computational Fluid Dynamics. This started showing its efficiency for the simulation of isothermal and weakly compressible flows, where it is now considered to be a particularly tough challenger for more conventional fluid solvers based on the \textcolor{black}{solving} of Navier-Stokes-Fourier equations~\cite{MANOHA_AIAA_2846_2015, RUMSEY_AIAA_1258_2018}. In the mean time, the validity domain of LBMs was extended to a very large set of phenomena which are not restricted anymore to Fluid Mechanics~\cite{SUCCI_Book_2018}. Despite these successes, severe stability issues were also encountered when simulating high Reynolds number flows with the BGK-LBM~\cite{RICOT_JCP_228_2009,MARIE_JCP_333_2017}. A great number of collision models have been proposed to circumvent this issue. These models are either based on static or dynamic single (multiple) relaxation time(s). 

Most common LBMs based on dynamic relaxation times can be divided into two categories: (1) Large Eddy simulation (LES) based LBMs, and (2) Entropic LBMs (ELBMs). LES-LBMs rely on the well known fact that underresolved turbulent scales must be accounted for through the use of a subgrid scale model~\cite{SAGAUT_Book_2006,SAGAUT_CMA_59_2010}. This model mainly consists in mimicking the behavior of small scales through the dissipation of structures at the grid cut off size. This can be done modifying the relaxation time to take into account the additional eddy viscosity which usually scales as the strain rate tensor.

Regarding ELBMs, these models ensure the H-theorem to be valid after the velocity discretization of the Boltzmann equation~\cite{KARLIN_PRL_81_1998,BOGHOSIAN_PRSLA_457_2001a}. This is done solving a minimization problem at each grid point and time step. This leads to a variable relaxation time that locally self-adjusts to the flow. Hence a non constant dynamic viscosity is obtained with ELBMs, especially when underresolved mesh grids are used for the simulation of high Reynolds (turbulent) flows. In that sense, ELBMs seem very similar to LES-LBMs. In fact, Malaspinas \emph{et al.} proved the behavior of the ELBM was sharing similarities with the standard Smagorinsky subgrid scale model~\cite{SAGAUT_Book_2006} since its dynamic viscosity also scales as the strain rate tensor, and its value tends towards zero when the resolution of the mesh grid is increased~\cite{MALASPINAS_PRE_78_2008}. Nevertheless, solving a minimization problem at each grid node and for each time step induces a nonnegligible extra CPU cost, which is higher than for subgrid scale models, while sharing an equivalent accuracy~\cite{HAMDI_InBook_2018}.

Hence, one may prefer to rely on KBC models to improve the numerical stability of LBMs at a low CPU cost. These collision models are based on an approximation to the minimization problem, and use an analytic formula for the computation of the dynamic relaxation time~\cite{KARLIN_PRE_90_2014a,BOSCH_ESAIM_52_2015,BOSCH_PRE_92_2015,DORSCHNER_PRE_94_2016,DORSCHNER_PRE_97_2018,FLINT_PRE_97_2018,BOSCH_EPL_122_2018}.
They can further decouple the relaxation of shear modes from acoustic and ghost modes, hence, freeing themselves from the generation of spurious vortices induced by the use of a varying shear viscosity in underresolved conditions. Nevertheless, it is worth noting that some theoretical work remains to be done in order to properly define the validity range of the approximated minimization problem~\cite{MATTILA_PRE_91_2015}. Eventually, one can further find in the literature stabilization techniques based on other kinds of dissipation control~\cite{BROWNLEE_PRE_74_2006,BROWNLEE_PRE_75_2007,BROWNLEE_PA_387_2008,GORBAN_PA_414_2014}.

All the aforementioned collision models, despite showing a great stability gain as compared to the BGK operator, cannot be easily studied from a mathematical viewpoint due to the dynamic computation of the relaxation time.

Regarding LBMs based on static relaxation times, d'Humi\`{e}res originally proposed to apply the collision step within the moment space to improve the numerical stability of the resulting LBM through the increase of the number of free parameters. This ended up in a multirelaxation time (MRT) collision model for the D3Q19 velocity discretization~\cite{DHUMIERE_PAA_159_1992}. With this approach, several relaxation times controlled the relaxation process of moments toward their respective equilibrium state. This model was further popularized by the 2D formulation (D2Q9) of Lallemand and Luo~\cite{LALLEMAND_PRE_61_2000}. In their work, stability gains were obtained through fine tuning of these free parameters via a linear stability analysis. Nevertheless, finding optimal values for all relaxation times rapidly becomes a tedious task, especially in the 3D case. 
This is why two relaxation time (TRT) models were then proposed by Ginzburg \emph{et al.}~\cite{GINZBURG_CCP_3_2008a,GINZBURG_CCP_3_2008b}. In their most simple and naive form, these models can be related to both Lallemand's and Luo's work in 2D, and d'Humi\`{e}re \emph{et al.} 3D extension~\cite{DHUMIERES_TRS_360_2002}, where a particular distinction is made between the relaxation times of even ($\tau^+$) and odd moments ($\tau^-$). While the former is related to the dynamic viscosity, the second one is a free parameter. Using very thorough mathematical derivations, general relationships between $\tau^+$ and $\tau^-$ were proposed through the so called `magic number' $\Lambda$ to improve both the accuracy and the stability of these TRT-LBMs~\cite{DHUMIERE_CMA_58_2009,GINZBURG_JSP_139_2010,KUZMIN_CMWA_61_2011}. Both types of collision model were shown to be of particular interest for the simulation of various types of phenomena~\cite{HAO_JPS_186_2009,XU_JCP_230_2011,WANG_CMA_65_2013,SERVANCAMAS_AWR_32_2009,KUZMIN_CMWA_61_2011,SILVA_PRE_96_2017}.

To further improve the numerical stability of MRT-LBMs, several authors proposed to perform the collision step in the moment space associated to the comoving reference frame~\cite{GEIER_PRE_73_2006,GEIER_EPJST_171_2009,GELLER_CMA_65_2013,LYCETTBROWN_CMWA_67_2014,DUBOIS_CCP_17_2015,NING_IJNMF_82_2016,FEI_PRE_96_2017,DEROSIS_PRE_95_2017,FEI_PRE_97_2018,DEROSIS_PRE_99_2019,DEROSIS_PRE_Submitted_2019}. These models were originally named as cascaded LBM by Geier \emph{et al.}~\cite{GEIER_PRE_73_2006} due to the construction of high order moments from lower order ones in a cascade like way. Despite an improved stability for high Reynolds flow simulations, this model was still suffering from Galilean invariance defects that were partially corrected later by the same authors~\cite{GEIER_EPJST_171_2009}. Nonetheless, the complex implementation of the cascaded process made it difficult to extend this collision model to both a broader range of physics, and to other kinds of lattices. This deficiency was overcome, for example, by Lycett-Brown and Luo who simplified the derivation of the cascaded mechanism, and applied it to the simulation of multiphase flows~\cite{LYCETTBROWN_CMWA_67_2014,LYCETT_BROWN_PRE_94_2016}. In the mean time, Dubois \emph{et al.} properly reformulated the cascaded model in terms of collisions performed in the central moment (CM) space~\cite{DUBOIS_CCP_17_2015}. The latter being well known from the point of view of Statistics, more solid mathematical foundations were then provided to these models. 
The same authors also quantified the stability gain obtained with this kind of collision models through a linear stability analysis~\cite{DUBOIS_CRM_343_2015}. In their work, they further confirmed the superiority, in terms of linear stability, of the tensor product basis as compared to the more standard ones used in common MRT models. All of these led to the extension of CM-LBMs to very different fields of research such as shallow water and magnetohydrodynamics among others~\cite{DEROSIS_CMAME_319_2017,DEROSIS_PRE_95_2017,DEROSIS_JT_19_2018}.

The last version of these MRT-LBMs is based on a collision step occurring this time in the cumulant space. By definition, cumulants are another kind of statistical quantities, such as raw  and central moments, that allow the description of both continuous and discrete probability distribution functions~\cite{COOK_Bio_38_1951}. They were originally used for the simulation of gas dynamics by Seeger \emph{et al.}~\cite{SEEGER_CMT_12_2000,SEEGER_CMT_14_2002,SEEGER_PhD_2003}. Their method was based on solving equations obtained by taking cumulants of the Boltzmann equation.
In the LBM context, Geier \emph{et al.} did confirm the stability improvement induced by the collision model based on cumulants instead of raw moments~\cite{GEIER_CMA_70_2015}. Nonetheless, it was done in an extremely reduced configuration where the diffusive scaling was adopted, thus reducing the validity of the comparison to flows governed by the incompressible Navier-Stokes equations~\cite{JUNK_JCP_210_2005,KRUGER_Book_2017}. Hence, no information regarding the generation and propagation of acoustic waves can be deduced from either their work or from those that followed~\cite{GEIER_JCP_348_2017a,GEIER_JCP_348_2017b,GEIER_CF_166_2018}. More surprisingly, no comparison with central moments was provided for the most complicated case, i.e, the study of the flow around a sphere. As a consequence, it is still not clear if cumulant methods are (or not) always more stable than CM-LBMs at the time of writing. 

In parallel to MRT-like collision operators, regularization steps were proposed to increase the stability of the BGK-LBM in the low viscosity limit. This kind of LBMs was originally proposed by Ladd \emph{et al.} to reduce the memory consumption~\cite{LADD_JSP_104_2001}. Nevertheless, it was rapidly shown to be able to improve the numerical stability by filtering out nonhydrodynamic modes (Latt and Chopard~\cite{LATT_ARXIV_2005,LATT_MCS_72_2006a}), and ensuring the rotational invariance of the numerical scheme (Chen \emph{et al.}~\cite{CHEN_PA_362_2006}). This model was further extended to high order velocity discretizations by Zhang \emph{et al.}~\cite{ZHANG_PRE_74_2006}. A recursive version of these models was recently derived for standard~\cite{MALASPINAS_ARXIV_2015}, and high order LBMs~\cite{COREIXAS_PRE_96_2017}. This recursive approach was proven to increase the numerical stability of LBMs as compared to the original regularized collision model~\cite{COREIXAS_PhD_2018}. It was further shown to be competitive against state of the art Navier-Stokes and LBM solvers for the simulation of jet noise~\cite{BROGI_JASA_142_2017}. This confirmed its capability for high fidelity simulations, including computational aeroacoustics of moderate  to high Reynods number flows. In the mean time, Mattila \emph{et al.} worked on a regularization step operating in the comoving reference frame~\cite{MATTILA_PF_29_2017}. They noticed that by neglecting particular central Hermite coefficients, they were able to recover the recursive formulation of Malaspinas. These coefficients being related to diffusive phenoma, they were supposed to be negligible for high Reynolds number flows. In the end, all the aforementioned regularized models were also employed to simulate a large panel of distinct phenomena~\cite{NIU_PRE_76_2007,VIGGEN_PRE_87_2013,WANG_PRE_92_2015,COREIXAS_PRE_96_2017,BROGI_JASA_142_2017,BA_PRE_97_2018,MONTESSORI_CF_167_2018}.

Despite a wide variety of collision models, only very few comprehensive and consistent comparative studies have been conducted in the past. Most of them are restricted to less than four collision models, or are only based on numerical test cases, and contain only very few sound comments on their respective theoretical discrepancies. The former issue most likely comes from the fact that each type of collision model is usually introduced in its own framework. Consequently, it is complicated to express all the collision models within the very same framework in order to eventually compare them to each other from a general viewpoint. Regarding the lack of theoretical comparisons, it can be explained by the nonnegligible mathematical background required to properly derive links between collision models. Nonetheless, rigorous comparisons relying on different tools are also available in the literature. As an example, several authors used linear stability analysis to quantify the numerical stability of collision models~\cite{LALLEMAND_PRE_61_2000, SIEBERT_PRE_77_2008,ADHIKARI_PRE_78_2008,MARIE_JCP_228_2009,RICOT_JCP_228_2009,KUZMIN_CMWA_61_2011,XU_JCP_230_2011,DUBOIS_CRM_343_2015,HOSSEINI_IJMPC_28_2017,CHAVEZMODENA_CF_172_2018,COREIXAS_PhD_2018,WISSOCQ_JCP_380_2019}. In addition, it is worth noting that studies dedicated to the derivation of links between several kinds of LBMs from a (more or less) theoretical viewpoint can also be found in the literature. Among them, one may refer to Refs.~\cite{ASINARI_PRE_78_2008,GINZBURG_CCP_3_2008a,GORBAN_PA_414_2014,GEIER_CMA_70_2015,DUBOIS_CCP_17_2015,COREIXAS_PRE_96_2017} and therein references.
     
\textcolor{black}{Knowing all of this, the present work proposes a formalism that describes all static methods within a common mathematical framework, and in this way allows us to draw direct links between them. \textcolor{black}{This is done considering the LBM in its general form, i.e, with no restriction to the incompressible regime.} 
In addition, it is intended to answer several questions that arise from the literature. For instance, one can wonder if all collision models can be rewritten in a linear matrix form as originally proposed by Higuera \emph{et al.}~\cite{HIGUERA_EPL_9_1989}, and further discussed in the context of LBMs by several authors~\cite{BENZI_PR_222_1992,DHUMIERE_PAA_159_1992}. If this assumption were to be true, then it would be possible to easily switch from one collision model to another through (1) matrix products and (2) adjustments of relaxation frequencies. One can also wonder if the equilibrium state does or not depend on the Galilean invariance properties of the collision model, and consequently, if the resulting macroscopic behavior of the LBM also depends on the collision model~\cite{GEIER_PRE_73_2006,GEIER_EPJST_171_2009, GEIER_CMA_70_2015,GEIER_JCP_348_2017a}. \textcolor{black}{As a consequence, the present work can be considered as the first (theoretical) stage of a large project that intends to compare collisions models from both a theoretical and a numerical viewpoint.}}

The rest of the paper is organized as follows. Recalls on Statistics are first provided in Sec.~\ref{sec:stat}. Thanks to them, the way all kinds of statistical quantities are linked is determined by relating their moment generating function to each other. All collision models are then thoroughly reviewed, and potential links between the models that have been previously pointed out in the literature are \textcolor{black}{recalled} (Sec.~\ref{sec:LBM}). In this way, a first picture of all the families of collision models is drawn. In the following sections~(\ref{sec:CompStudy1},~\ref{sec:CompStudy2} and~\ref{sec:CompStudy3}), relationships derived between moments are used to carry out, in a straightforward way, the comparative study of post collision populations for lattices of increasing complexity (D1Q3, D2Q9, D3Q27 and D3Q19). All algebraic manipulations are detailed for the most simple lattices, hence allowing the reader to readily extend the present approach to any kind of LBMs. This comparative study ends with a reflection on macroscopic equations resulting from the LBM. General conclusions regarding the main discrepancies and similarities between collision models are finally made in Sec.~\ref{sec:Conclusion}.

For the sake of completeness, several appendices are also provided to help the interested reader properly understand all concepts addressed in the present work. Lattice structures and partial Bell polynomials are described in Apps.~\ref{sec:Lattices} and~\ref{sec:Bell}. Relationships between one-, two- and three-dimensional statistical quantities are gathered in Apps.~\ref{sec:RelationshipsUnivariate},~\ref{sec:RelationshipsBivariate} and~\ref{sec:RelationshipsTrivariate} respectively. Linear transformations leading to the matrix form of collision models are compiled in App.~\ref{sec:LinearTransMatrix} for the D2Q9 lattice. Explanations regarding the way to build them are also provided, which allows a straightforward calculation of these matrices for any type of lattice. The D3Q27 formulations of all types of populations are recalled in App.~\ref{sec:3Dextension}. It is also explained how to easily derive their D3Q19 counterparts. Eventually, the most general form of equilibrium states is presented in App.~\ref{sec:EqFunction}.

\textcolor{black}{To help the reader with the implementation of the most complicated collision models, general instructions are finally provided as Supplemental material~\bibnote[SupMat]{See Supplemental Material at [\protect \url{COREIXAS_PRE_2019_Theo_SupMat.pdf}] for instructions regarding the implementation of most collision models considered in the present work.}. More specifically, pseudo codes, describing the computation of each type of moment, as well as, equilibrium and post collision populations, are supplied for both the D2Q9~\bibnote[SupMatQ9]{See Supplemental Material for instructions regarding the implementation of D2Q9 collision models at [\protect \url{Comprehensive_Collision_Models_D2Q9.txt}].} and the D3Q27~\bibnote[SupMatQ27]{See Supplemental Material for instructions regarding the implementation of D3Q27 collision models at [\protect \url{Comprehensive_Collision_Models_D3Q27.txt}].} formulations.} 

\section{Theoretical background on Statistics\label{sec:stat}}

\subsection{Motivations}

\textcolor{black}{As it will be shown in Sec.~\ref{sec:derivationLBMs}, the choice of the moment space directly impacts the derivation of LBMs through their populations
$$f_i = \sum_n c_{\textcolor{black}{i,}n} P_{\textcolor{black}{i,}n},$$ 
where $c_{\textcolor{black}{i,}n}$ is a coefficient related to the moment $P_{\textcolor{black}{i,}n}$, both yet to be defined.
To properly quantify differences between LBMs, it is then essential to understand how all kinds of moments are derived, and how they are linked to each other. In the present work, it is proposed to relate families of moments through their sole common feature, namely, their generating function. This method further simplifies their calculation in any $D$-dimension, $D$ being the number of physical dimensions.}
 
\subsection{Moment generating function}
To properly understand how relationships between raw, Hermite, central, central Hermite moments and cumulants are derived, it is necessary to introduce the notion of (raw) moment generating function (MGF) $\mathcal{M}$ of a probability distribution function $f$,
\begin{equation}
\mathcal{M}(\textcolor{black}{X}) = \displaystyle{\int} \exp(\xi \textcolor{black}{X})f(\xi)\:\mathrm{d}\xi,
\end{equation}
By expanding the generating function of monomials in power series, 
$$\Phi_M=\exp(\xi \textcolor{black}{X}) = \displaystyle{\sum_n} \dfrac{(\xi \textcolor{black}{X})^n}{n!},$$
the MGF can be directly linked to raw moments $M_n$ of $f$ as follows,
\begin{equation}
\mathcal{M}(\textcolor{black}{X}) = \displaystyle{\sum_n}\left(\displaystyle{\int} \xi^n f(\xi)\:\mathrm{d}\xi\right)\dfrac{\textcolor{black}{X}^n}{n!} = \displaystyle{\sum_n}M_n\dfrac{\textcolor{black}{X}^n}{n!},
\label{eq:MGF}
\end{equation}
where $n! = n\times (n-1)\times ... \times 2 \times 1$. Hence, (raw) moments of order $n$ are obtained through the partial derivatives of the MGF~(\ref{eq:MGF}) evaluated at $\textcolor{black}{X}=0$,
\begin{equation}
M_n = \mathcal{M}^{(n)}(0),
\end{equation}
$\mathcal{M}^{(n)}$ being the $n$th derivative of $\mathcal{M}$ with respect to $\textcolor{black}{X}$. In the rest of the paper, $M_0=1$ will be assumed. This corresponds to the assumption of normalized moments, as already used by several authors in the lattice Boltzmann community~\cite{KARLIN_PA_389_2010,LYCETTBROWN_CMWA_67_2014,BOSCH_ESAIM_52_2015}.

\subsection{Hermite moment generating function}

Hermite moments are defined through their generating function,
\begin{equation}
\mathcal{H}(\textcolor{black}{X}) = \displaystyle{\int} \exp(\xi \textcolor{black}{X} - c_s^2 \textcolor{black}{X}^2/2)f(\xi)\:\mathrm{d}\xi,
\end{equation}
where $\Phi_H=\exp(\xi \textcolor{black}{X} - c_s^2 \textcolor{black}{X}^2/2)$ is the generating function of Hermite polynomials as they are usually defined in the lattice Boltzmann framework, i.e, taking into account the lattice constant $c_s$~\cite{COREIXAS_PRE_96_2017}. As a reminder, Hermite polynomials up to $n=6$ are
\begin{align}
H_{0}(\xi) &= 1, \notag\\
H_{1}(\xi) &= \xi,  \notag\\
H_{2}(\xi) &= \xi^2-c_s^2, \notag\\
H_{3}(\xi) &= \xi^3-3 c_s^2\xi,   \label{eq:HermitePoly}\\
H_{4}(\xi) &= \xi^4-6 c_s^2 \xi^2+3 c_s^4, \notag\\
H_{5}(\xi) &= \xi^5-10 c_s^2 \xi^3+15 c_s^4 \xi,  \notag\\
H_{6}(\xi) &= \xi^6-15 c_s^2 \xi^4+45 c_s^4 \xi ^2-15 c_s^6. \notag
\end{align}
Since (Hermite) polynomials are linear combinations of monomials, then relationships between Hermite  and raw moments will be the same as those between their polynomial counterparts. To derive these relationships, let us start noticing that both polynomial generating functions are linked as follows,
$$\Phi_H=\exp(\xi \textcolor{black}{X} -c_s^2 \textcolor{black}{X}^2/2) = \Phi_M\exp(-c_s^2 \textcolor{black}{X}^2/2).$$
Since these polynomials are obtained by deriving $n$ times $\Phi_H$ ($\Phi_M$) about $\textcolor{black}{X}=0$, the only expression that needs to be evaluated is the $n$th derivative of $\exp(-c_s^2 \textcolor{black}{X}^2/2)$. While its general formulation is quite complicated, it can easily be shown by induction that
\begin{equation}
\hspace*{-0.275cm} \dfrac{\mathrm{d}^{(k)}}{\mathrm{d}\textcolor{black}{X}^k}e^{-c_s^2 \textcolor{black}{X}^2/2}\vert_{\textcolor{black}{X}=0}= \left\{
 \begin{array}{c l}
 1 & \text{if } k=0\\
 (-c_s^2)^{k/2}(k-1)!! & \text{if } k \text{ is even}\\
 0 & \text{otherwise}
 \end{array}\right.
 \label{eq:formula}
\end{equation}
with \textcolor{black}{`!!' standing for the double factorial, i.e., $$(k-1)!! = (k-1)\times (k-3)\times ... \times 3 \times 1.$$} 
Hence,
\begin{equation}
\mathcal{H}^{(n)}(\textcolor{black}{X}) = \sum_{k=0}^n \binom{n}{k} \bigg(\dfrac{\mathrm{d}^{(k)}}{\mathrm{d}\textcolor{black}{X}^k}e^{-c_s^2 \textcolor{black}{X}^2/2}\bigg)\mathcal{M}^{(n-k)}(\textcolor{black}{X}),
\label{eq:HermiteRaw}
\end{equation}
with $\binom{n}{k}=n!/k!(n-k)!$ being the binomial coefficient, and where Hermite moments are defined as
$$A_n = \mathcal{H}^{(n)}(0).$$
The inversion formula is simply obtained changing $-c_s^2$ to $+c_s^2$ in Eq.~(\ref{eq:HermiteRaw}),
$$\mathcal{M}^{(n)}(\textcolor{black}{X}) = \sum_{k=0}^n \binom{n}{k} \bigg(\dfrac{\mathrm{d}^{(k)}}{\mathrm{d}\textcolor{black}{X}^k}e^{c_s^2 \textcolor{black}{X}^2/2}\bigg)\mathcal{H}^{(n-k)}(\textcolor{black}{X}),$$
where now,
\begin{equation}
\hspace*{-0.275cm} \dfrac{\mathrm{d}^{(k)}}{\mathrm{d}\textcolor{black}{X}^k}e^{c_s^2 \textcolor{black}{X}^2/2}\vert_{\textcolor{black}{X}=0}= \left\{
 \begin{array}{c l}
 1 & \text{if } k=0\\
 (c_s^2)^{k/2}(k-1)!! & \text{if } k \text{ is even}\\
 0 & \text{otherwise}
 \end{array}\right.
 \label{eq:formula2}
\end{equation} 
 
\subsection{Central moment generating functions}

Regarding central moments $\widetilde{M}_n$, their generating function reads as
\begin{align}
\widetilde{\mathcal{M}}(\textcolor{black}{X}) &= \displaystyle{\int} \exp[(\xi-M_1)\textcolor{black}{X}]f(\xi)\:\mathrm{d}\xi \notag\\[0.1cm]
					&= \displaystyle{\sum_n}\left[\displaystyle{\int} (\xi-M_1)^n f(\xi)\:\mathrm{d}\xi\right]\dfrac{\textcolor{black}{X}^n}{n!}\notag\\[0.1cm]
&= \displaystyle{\sum_n}\widetilde{M}_n\dfrac{\textcolor{black}{X}^n}{n!}.
\end{align}
A simple way to link raw  and central moments is to express the central moment generating function as
\begin{equation*}
\widetilde{\mathcal{M}}(\textcolor{black}{X}) = \exp(-M_1 \textcolor{black}{X})\mathcal{M}(\textcolor{black}{X}).
\end{equation*}
Using the binomial formula, the $n$th derivative of $\widetilde{\mathcal{M}}(\textcolor{black}{X})$ then becomes
\begin{equation*}
\widetilde{\mathcal{M}}^{(n)}(\textcolor{black}{X})= \sum_{k=0}^n \binom{n}{k} (-M_1)^{n-k}\mathcal{M}^{(k)}(\textcolor{black}{X}).
\end{equation*}
Eventually, evaluating the $n$th derivative about $\textcolor{black}{X}=0$ leads to
\begin{equation}
\widetilde{M}_n = \sum_{k=0}^n \binom{n}{k} (-M_1)^{n-k} M_k.
\label{eq:recCentralRaw}
 \end{equation} 
The inversion formula is simply obtained noticing that
\begin{equation*}
\mathcal{M}(\textcolor{black}{X}) = \exp(M_1 \textcolor{black}{X})\widetilde{\mathcal{M}}(\textcolor{black}{X}).
\end{equation*}
Following the very same steps as before, one ends up with
\begin{equation}
M_n = \sum_{k=0}^n \binom{n}{k} (M_1)^{n-k} \widetilde{M}_k.
\label{eq:recRawCentral}
 \end{equation}

Since Hermite polynomials are linear combinations of monomials, the above derivation can be applied to link raw  and central moments in the Hermite expansion framework,
\begin{equation}
\widetilde{A}_n = \sum_{k=0}^n \binom{n}{k} (-A_1)^{n-k} A_k,
\label{eq:recHermiteCentralRaw}
 \end{equation}
where $\widetilde{A}_n$ are central Hermite moments, and $A_1 = M_1$ since $H_1(\xi) = \xi$. 
Finally, the inversion formula is
\begin{equation}
A_n = \sum_{k=0}^n \binom{n}{k} (A_1)^{n-k} \widetilde{A}_k.
\label{eq:recHermiteRawCentral}
 \end{equation}
 
\subsection{Cumulant generating function}

 When it comes to cumulants, their generating function $\mathcal{K}$ is defined as~\cite{COOK_Bio_38_1951}
\begin{equation}
\mathcal{K}(\textcolor{black}{X}) = \ln[\mathcal{M}(\textcolor{black}{X})].
\label{eq:cumulMomentGenFunc}
\end{equation} 
Its power series expansion allows us to link cumulants and raw moments as follows, 
\begin{align}
\displaystyle{\sum_n} K_n \dfrac{\textcolor{black}{X}^n}{n!}
 	&= \ln\bigg[M_0\bigg(\displaystyle{\sum_{p}} \dfrac{M_p}{M_0} \dfrac{\textcolor{black}{X}^p}{p!}\bigg)\bigg] \notag\\[0.1cm]
	&= \ln(M_0) + \ln\bigg( 1 + \displaystyle{\sum_{p\geq 1}} \dfrac{M_p}{M_0} \dfrac{\textcolor{black}{X}^p}{p!}\bigg) \notag\\[0.1cm]
	&= \ln(M_0) + \displaystyle{\sum_{q\geq 1}}\dfrac{(-1)^{q-1}}{q}\bigg(\displaystyle{\sum_{p\geq 1}} \dfrac{M_p}{M_0} \dfrac{\textcolor{black}{X}^p}{p!}\bigg)^q \notag\\[0.1cm]
	&= \displaystyle{\sum_{q\geq 1}}\dfrac{(-1)^{q-1}}{q}\bigg(\displaystyle{\sum_{p\geq 1}} M_p \dfrac{\textcolor{black}{X}^p}{p!}\bigg)^q,
\label{eq:expansionCumulMomentGenFunc}
\end{align}
where the series expansion of the logarithmic part was computed using
$$\ln(1+ \textcolor{black}{Y}) = \displaystyle{\sum_{q\geq 1}}  \dfrac{(-1)^{q-1}}{q}\textcolor{black}{Y}^q,$$
with $\textcolor{black}{Y} = \sum_{p\geq 1} M_p \textcolor{black}{X}^p/p!.$

To determine the inversion formula, the starting point is
\begin{equation}
\mathcal{M}(\textcolor{black}{X}) = \exp[\mathcal{K}(\textcolor{black}{X})].
\label{eq:momentCumulGenFunc}
\end{equation}
Taking its power series expansion, the following expression is derived,
\begin{align}
\displaystyle{\sum_n} M_n \dfrac{\textcolor{black}{X}^n}{n!}
 	&= \exp\bigg(\displaystyle{\sum_p} K_p \dfrac{\textcolor{black}{X}^p}{p!}\bigg) \notag\\[0.1cm]
	&= \displaystyle{\sum_{q\geq 1}}\dfrac{1}{q!}\bigg(\displaystyle{\sum_{p\geq 1}} K_p \dfrac{\textcolor{black}{X}^p}{p!}\bigg)^q.  
\label{eq:expansionMomentCumulGenFunc}
\end{align}


 
In their current formulation, it is quite difficult to use Eqs.~(\ref{eq:expansionCumulMomentGenFunc}) and~(\ref{eq:expansionMomentCumulGenFunc}). This is why two ways of simplifying these formulas, as originally proposed by Kendall~\cite{KENDALL_AE_10_1940}, are presented hereafter:

\emph{`Handmade' expansion of the double sum}: It consists on expanding the right hand side of these formulas assuming truncated sums. Relationships are then obtained identifying polynomials coefficients of $t^n$ with those of $t^{p+q}$, where $p+q=n$ (uniqueness of the power series expansion). In other words, by considering $p\leq n$ and $q\leq n$, one just needs to collect all coefficients of $t^n$. Even if this method is rather general, it rapidly leads to lengthy calculations even for small values of $n$.

\emph{Partial Bell polynomials}: They are used in Combinatorial Mathematics to study set partitions~\cite{BELL_AM_1934,WITHERS_1994}. 
These polynomials are defined as follows,
\begin{equation}
\displaystyle{\sum_{k=\textcolor{black}{q}}^{n} B_{n,k}(x_1,...,x_{n-k+1}})\dfrac{\textcolor{black}{X}^n}{n!}=\dfrac{1}{q!}\bigg(\displaystyle{\sum_{p\geq 1}} x_p \dfrac{\textcolor{black}{X}^p}{p!}\bigg)^q,
\label{eq:BellPoly}
\end{equation}
where $B_{n,k}$ are partial Bell polynomials. They can be individually computed thanks to the recursive formula
\begin{equation*}
B_{n,k}(x_1,...,x_{n-k+1})= \displaystyle{\sum_{l=1}^{n-k+1}\binom{n-1}{l-1}x_l B_{n-l,k-1}}, 
\end{equation*}
with
\begin{equation*}
B_{0,0}=1,\: B_{n,0}=0\:(k\geq 1),\: B_{0,k}=0 \:(n\geq 1).   
\end{equation*}
Expressions up to $n=6$, and further details concerning these polynomials, are given in App.~\ref{sec:Bell}. 
Injecting Eq.~(\ref{eq:BellPoly}) into Eq.~(\ref{eq:expansionMomentCumulGenFunc}), the $n$th raw moment is simply expressed as
\begin{equation}
M_n = \displaystyle{\sum_{k=1}^n} B_{n,k}(K_1,...,K_{n-k+1}). 
\end{equation} 
Likewise, the $n$th cumulant is obtained via the inversion formula of partial Bell polynomials~\cite{KENDALL_AE_10_1940},
\begin{equation}
K_n = \displaystyle{\sum_{k=1}^n} (-1)^{k-1}(k-1)! B_{n,k}(M_1,...,M_{n-k+1}). 
\end{equation} 

To obtain formulas between cumulants and other statistical quantities, one simply needs to switch between the quantity of interest to its raw moment counterpart, and then use the above formulas. Relationships between all kinds of moments, up to $n=6$, are detailed in App.~\ref{sec:RelationshipsUnivariate}.  

\subsection{Multivariate extensions\label{sec:MultiExt}}

In the $D$-dimensional case, the multivariate generating moment function is now defined as
\begin{equation}
\mathcal{M}(\bm{\textcolor{black}{X}}) = \displaystyle{\int} \exp\bigg(\sum_{k=1}^D \xi_k \textcolor{black}{X}_k\bigg)f(\bm{\xi})\:\mathrm{d}\bm\xi,
\label{eq:MultivariateMomentGenFunc}
\end{equation}
where $\bm{\textcolor{black}{X}}=(\textcolor{black}{X}_1,...,\textcolor{black}{X}_D)$ and $\bm{\xi}=(\xi_1,...,\xi_D)$ are now vectors. As in the univariate case, moments are obtained through successive differentiations of $\mathcal{M}(\bm{\textcolor{black}{X}})$ about $\bm{\textcolor{black}{X}}=\bm{0}$. For the most complicated case ($D=3$), this leads to
\begin{equation}
M_{pqr} = \mathcal{M}^{(p,q,r)}(0,0,0).
\end{equation}
$M_{pqr}$ is a raw moment of order $n=p+q+r$. $\mathcal{M}^{(p,q,r)}$ is the $n$th derivative of $\mathcal{M}$, where $p$, $q$, $r$ are the number of differentiations with respect to $\xi_x$, $\xi_y$ and $\xi_z$. 

While rederiving all relationships between statistical quantities using Eq.~(\ref{eq:MultivariateMomentGenFunc}) is relatively straightforward, the computation of multivariate cumulants rapidly becomes lengthy. It is proposed hereafter to rely on a rather elegant way to extend the derivation of cumulant formulas to the $D$-dimensional case. 
The most general way relies on advanced knowledge of set partitioning. Here, it is preferred to use a simpler, yet rigorous, method based on a particular kind of differential operator. This approach was originally proposed by Kendall~\cite{KENDALL_AE_10_1940}, and explained in more details by Cook~\cite{COOK_Bio_38_1951}. Let us start from
\begin{equation}
M_3 = K_3 + 3 K_2 K_1 + K_1^3.
\label{eq:Moment2CentralUni3} 
\end{equation}
Each component of $M_{n}$ or $K_{n}$ depends on $\xi_x^{n}$ in such a way that Eq.~(\ref{eq:Moment2CentralUni3}) can be rewritten as
\begin{equation}
M(\xi_x^3) = K(\xi_x^3) + 3 K(\xi_x^2) K(\xi_x) + K(\xi_x)^3.
\label{eq:Moment2CentralUni3Kendall}
\end{equation}
Applying the operator $\Delta_{y}^x \equiv \xi_y \partial/\partial \xi_x$ to the variable of $M(\xi_x^3)$ leads to~\cite{KENDALL_AE_10_1940,COOK_Bio_38_1951},
\begin{align}
\Delta_{y}^x[M(\xi_x^3)] &= \xi_y \frac{\partial}{\partial \xi_x}M(\xi_x^3) \notag\\
							 &= M(3\xi_x^2\xi_y) \notag\\
							 &= 3M(\xi_x^2\xi_y) \notag\\
							 &= 3M_{21}
\label{eq:Kendall1}
\end{align}
When it comes to the RHS terms of Eq.~(\ref{eq:Moment2CentralUni3Kendall}), the following formulas are obtained,
\begin{align}
\Delta_{y}^x[K(\xi_x^3)] &= 3K_{21}, \\
\Delta_{y}^x[3K(\xi_x^2) K(\xi_x)] &= 6K_{11}K_{10} + 3K_{20}K_{01},\\
\Delta_{y}^x[K(\xi_x)^3] &= 3 K_{10}^2 K_{01}.
\label{eq:Kendall2}
\end{align}
Using Eqs.~(\ref{eq:Kendall1})-(\ref{eq:Kendall2}), $M_{21}$ is eventually linked to cumulants through
\begin{equation}
M_{21} = K_{21} + K_{20}K_{01} + 2 K_{11}K_{10} + K_{10}^2 K_{01}.
\end{equation}
Applying $\Delta_y^x$ a second time leads to
\begin{equation}
M_{12} = K_{12} + 2 K_{11}K_{01} + K_{02}K_{10} + K_{10} K_{01}^2,
\end{equation}
while applying it a third time further gives the expression
\begin{equation}
M_{03} = K_{03} + 3 K_{02}K_{01} + K_{01}^3.
\end{equation}
The formula for $M_{30}$ is simply obtained noticing that $\xi_x^3 = \xi_x^3\xi_y^0$, and then taking $0$ as second subscript for all terms in Eq.~(\ref{eq:Moment2CentralUni3}),
\begin{equation}
M_{30} = K_{30} + 3 K_{20}K_{10} + K_{10}^3.
\end{equation}
As a way to confirm the validity of the above third order bivariate moments, one just has to sum all subscripts for each moment. If the bivariate formula is correct, then one should obtain the univariate formula for $M_3$ as a result.

To derive trivariate ($D=3$) formulas, two operators of the same kind are required, namely, $\Delta_{z}^x$ and $\Delta_{z}^y$. Using them, it is straightforward to obtain trivariate expressions, such as,
\begin{align}
M_{201} &= \Delta_{z}^x[M_{30}]/3 \notag\\
			 &= K_{201} + K_{200}K_{001} + 2 K_{101}K_{100} + K_{100}^2 K_{001},
\end{align}
and    
\begin{align}
M_{111} &= \Delta_{z}^y[M_{12}]/2 \notag\\
			 &= K_{111} + K_{110}K_{001} + K_{101}K_{010} \notag\\
			 & + K_{011}K_{100} + K_{100}K_{010}K_{001}.
\end{align}
Once again, one can check the validity of the above formulas summing all subscripts for each moment, and comparing the result with the univariate case, i.e, $M_3$.
While Kendall's diffential operators might appear rather empirical, it is important to know that it relies on rigorous mathematical derivations originating from the multivariate version of the Taylor series expansion of generating functions~\cite{KENDALL_AE_10_1940,COOK_Bio_38_1951}. Using this method, only univariate formulas up to $n=4$ ($n=6$) are needed to derive bivariate (trivariate) statistical quantities of interest for the D2Q9 (\textcolor{black}{D3Q19,} D3Q27) lattice. Univariate formulas are summarized in App.~\ref{sec:RelationshipsUnivariate}, whereas expressions needed for the implementation of the D2Q9-, \textcolor{black}{D3Q19-} and D3Q27-LBMs are compiled in Apps.~\ref{sec:RelationshipsBivariate} and~\ref{sec:RelationshipsTrivariate}.

\subsection{\textcolor{black}{Application to the normal distribution}\label{sec:StatGaussian}}

In Kinetic Theory, the normal (or Gaussian) distribution plays a major role in the modeling of the collision term. Indeed, the relaxation process is usually based on the BGK assumption which says that collisions make distribution functions tend towards their equilibrium state~\cite{BHATNAGAR_PR_94_1954}. The latter follows a normal distribution function and is usually named as the Maxwell-Boltzmann equilibrium (distribution) function. It reads as
\begin{equation}
f^{eq}(\rho, \bm{u}, \theta)=\dfrac{\rho}{(2\pi c_s^2 \theta)^{D/2}}\exp\bigg[-\dfrac{(\bm{\xi}-\bm{u})^2}{2 c_s^2 \theta}\bigg],
\label{eq:MaxwellBoltzmannEq}
\end{equation}
where $D$ is the number of physical dimensions, $\rho$ being the density, $\bm{u}$ the macroscopic velocity, $\bm\xi$ the mesoscopic velocity, $\theta=T/T_0$ the reduced temperature, and $c_s$ the isothermal \textcolor{black}{(or Newtonian)} sound speed. In the present work, only isothermal LBMs are considered, which implies that $T=T_0$ so that $\theta=1$. 

Injecting Eq.~(\ref{eq:MaxwellBoltzmannEq}) in the definition of raw, central, Hermite, central Hermite moments and cumulants, one can derive their equilibrium counterparts. Another way to obtain all equilibrium quantities is to start with equilibrium raw moments in either the univariate or the multivariate case, and then use corresponding formulas (Apps.~\ref{sec:RelationshipsUnivariate}-\ref{sec:RelationshipsTrivariate}) to compute other families of moments.

Starting with \textcolor{black}{one-dimensional} equilibrium raw moments, one obtains up to $n=4$
\begin{equation*}
\begin{array}{l}
M_0^{eq} = 1,\:
M_1^{eq} = u_x,\:
M_2^{eq} = u_x^2 + c_s^2,\\[0.1cm]
M_3^{eq} = u_x^3 + 3u_x c_s^2,\:
M_4^{eq} = u_x^4 + 6u_x^2 c_s^2 + 3c_s^4.
\end{array}
\end{equation*}
Related central moments are then
\begin{equation*}
\begin{array}{c}
\widetilde{M}_0^{eq} = 1,
\widetilde{M}_1^{eq} = 0,
\widetilde{M}_2^{eq} = c_s^2,
\widetilde{M}_3^{eq} = 0, 
\widetilde{M}_4^{eq} = 3c_s^4,
\end{array}
\end{equation*}
while equilibrium Hermite moments read as,
\begin{equation*}
\begin{array}{c}
A_0^{eq} = 1,
A_1^{eq} = u_x,
A_2^{eq} = u_x^2,
A_3^{eq} = u_x^3, 
A_4^{eq} = u_x^4,
\end{array}
\end{equation*}
and equilibrium central Hermite moments are expressed as,
\begin{equation*}
\begin{array}{c}
\widetilde{A}_0^{eq} = 1,
\widetilde{A}_1^{eq} = \widetilde{A}_2^{eq} = \widetilde{A}_3^{eq} = \widetilde{A}_4^{eq} = 0.
\end{array}
\end{equation*}
Eventually, equilibrium cumulants are
$$
K_0^{eq} = 0,
K_1^{eq} = u_x,
K_2^{eq} = c_s^2,
K_3^{eq} = 0, 
K_4^{eq} = 0.
$$

Central moments are usually preferred to their counterparts in the reference frame at rest because, for the former, the collision is applied in a moment space that is not impacted by Galilean invariance issues~\cite{GEIER_PRE_73_2006,GEIER_EPJST_171_2009,NING_IJNMF_82_2016}. Nonetheless, lattice dependent defects still remain in $\widetilde{M}^{eq}_{2p}$ ($p > 0$) through terms proportional to $c_s$ \textcolor{black}{that are nonnegligible in the weakly compressible limit, since $c_s\gg u$ for the latter}. Thus one may wonder if, in the \emph{isothermal} case, it would not be wiser to apply the collision in the central Hermite moment space instead. In addition, one of the reason behind the use of cumulants is that they allow us to easily quantify deviations from a normal distribution since $K^{eq}_n=0$ for $n\geq 3$~\cite{MCCULLAGH_Book_2018}. But once again, it seems more logical to rely on central Hermite moments since $\widetilde{A}^{eq}_n=0$ for $n\geq 1$ in the \emph{isothermal} case. Nevertheless, one should keep in mind that $\widetilde{A}^{eq}_n\neq 0$ for $n\geq 2$ in the \emph{thermal} case. Hence, cumulants may be a preferable choice for the latter case.

Multivariate formulations of all kinds of equilibrium moments are simply obtained thanks to the isotropy of the multivariate normal distribution. Hence, multivariate equilibrium moments are built through products of univariate ones. 

Henceforth, only equilibrium moments computed via the Maxwell-Boltzmann distribution function will be considered. \textcolor{black}{But it is also worth noting that depending of the physics of interest, other equilibrium distribution functions might be used, e.g., those derived from the principle of maximum entropy~\cite{KOGAN_JAMM_29_1965,PRESSE_RMP_85_2013}. Interestingly, the latter methodology was used in the lattice Boltzmann community for the derivation of exact~\cite{ANSUMALI_EPL_63_2003,ASINARI_PRE_79_2009} and approximated~\cite{FRAPOLLI_PRE_92_2015,ATIF_PRE_98_2018} equilibrium distribution functions.}

\section{Review of collision models\label{sec:LBM}}

The LBM is a very specific numerical scheme employed to solve the lattice Boltzmann equation -- a set of Boltzmann equations resulting from the discretization of the velocity space.

\subsection{The lattice Boltzmann method}

This method describes the space and time evolution of the velocity distribution function (VDF) $f_i(\bm{x},\bm{\xi}_i,t)$, also referred to as `population' in the present work. Roughly speaking, this quantity can be assimilated to the number of fictive particles at a point $(\bm{x},t)$ and characterized by a given velocity $\bm{\xi}_i$.  Its space and time evolution is obtained following two (dimensionless) successive steps
\begin{align}
f_i(\bm{x} +\bm{\xi}_i, t+1) =f_i^*(\bm{x}, t), \label{eq:StreamingStep}\\
f^*_i(\bm{x}, t) = f^{eq}_i(\bm{x}, t) + (1 - \Omega) f^{neq}_i(\bm{x}, t),\label{eq:CollisionStep}
\end{align}
where $f^*_i$ are post collision VDFs.
Eq.~(\ref{eq:StreamingStep}) corresponds to the streaming step that propagates $f_i$ to their neighboring nodes following the direction $\bm{\xi}_i$.  Eq.~(\ref{eq:CollisionStep}) is the collision step that locally takes into account the rate of change of $f_i$ due to collisions, where $\Omega$ is the collision model in its general form, $f_i^{eq}$ is the equilibrium VDF, and $f_i^{neq}$ is the deviation from the equilibrium state. Altogether, they form the famous `Collide and Stream' algorithm which shows both CPU time efficiency and accuracy~\cite{KRUGER_Book_2017}. 

Hereafter, the discussion will be restricted to the D2Q9 lattice. This is done for the sake of simplicity, and because it is sufficient to explain the differences between collision models of interest in the present work.

\subsection{BGK collision model\label{sec:BGK}}

In the particular case of the BGK collision operator,
$$\Omega = \omega_{\nu},\: f^{neq}_i = (f_i-f^{eq,2}_i),$$
with $\omega_{\nu} = 1/(\tau + 1/2)$ the collision frequency, $\tau$ being the relaxation time, $\nu$ the kinematic viscosity, and $f^{eq,2}_i$ is the standard second order equilibrium VDF~\cite{QIAN_EPL_17_1992,SHAN_JFM_550_2006}. Using this collision model, severe stability issues have been reported in the zero viscosity limit~\cite{SUCCI_Book_2018}. While one can improve the numerical stability of the LBM changing either the space or the time discretizations of the streaming and collision steps~\cite{LEE_JCP_215_2006,SURMAS_EPJST_171_2009,GUO_TRS_369_2011,GUO_PRE_88_2013}, it is usually preferred to derive new collision models introducing new forms for either $\Omega$~\cite{DHUMIERE_PAA_159_1992,LALLEMAND_PRE_61_2000,GINZBURG_CCP_3_2008a,GEIER_PRE_73_2006,DUBOIS_CCP_17_2015,GEIER_CMA_70_2015} or $f_i^{neq}$~\cite{LATT_ARXIV_2005,LATT_MCS_72_2006a,MALASPINAS_ARXIV_2015,COREIXAS_PRE_96_2017}. Doing so, both the efficiency and the accuracy can be preserved in most of the cases, and the stability of the resulting LBM is drastically improved. 

\subsection{Raw moment space\label{sec:Rev_MRT_TRT}}

To circumvent the stability issue, it has first been proposed to apply the collision step in the moment space through a MRT collision model~\cite{LALLEMAND_PRE_61_2000},
$$\bm{\Omega_{\mathrm{MRT}}} = \bm{M^{-1}SM},\: f^{neq}_i = (f_i-f^{eq,2}_i),$$ 
where $\bm S$ is the collision matrix which may have a diagonal form depending on the choice of moments. $\bm M$ and $\bm M^{-1}$ are commonly defined as \emph{orthogonal} matrices allowing us to move from the 9-dimensional velocity space $$(f_0,f_1,f_2,f_3,f_4,f_5,f_6,f_7,f_8)$$ to the 9-dimensional \emph{orthogonal} moment space 
$$({\rho},{j_x},{j_y},{e},{p_{xx}},{p_{xy}},{q_x},{q_y},{\varepsilon}),$$
and vice versa. The polynomials basis related to this moment space is
\begin{equation}
\mathcal{B}_{\mathrm{MRT}} = (M_{\rho},M_{j_x},M_{j_y},M_{e},M_{p_{xx}},M_{p_{xy}},M_{q_x},M_{q_y},M_{\varepsilon}),
\label{eq:MRTBasis}
\end{equation}
and it is built using the Gram-Schmidt orthogonalization procedure based on a unweighted scalar product.  

In the present case, the collision matrix is defined as~\cite{LALLEMAND_PRE_61_2000} $$\bm{S}_{\mathrm{MRT}} =\mathrm{diag}(0,0,0,\omega_e,\omega_{\nu},\omega_{\nu},\omega_q,\omega_q,\omega_{\varepsilon}).$$ Hence, the linear matrix form of $\bm \Omega$ allows the proper control of the relaxation process of each moment. This approach can drastically increase the numerical stability of the MRT-LBM, but this will depend on the choice of both the moment space, $\bm M$ and $\bm M^{-1}$~\cite{DUBOIS_CRM_343_2015}, and the relaxation frequencies composing the collision matrix $\bm S$~\cite{LALLEMAND_PRE_61_2000}. This last point becomes a huge drawback in either the 3D case or when dealing with high order LBMs, since the number of parameters that need to be fine tuned rapidly becomes very large. 

To reduce the number of free parameters in a general way, the TRT-LBM was proposed~\cite{GINZBURG_CCP_3_2008a,GINZBURG_CCP_3_2008b}. In the context of the orthogonal basis $\mathcal{B}_{\mathrm{MRT}}$~(\ref{eq:MRTBasis}), it relies on the following decomposition into a symmetric (even moments), and an antisymmetric (odd moments) sub-bases~\cite{KRUGER_Book_2017}
$$\mathcal{B}_{\mathrm{TRT}} =\mathcal{B}_{\mathrm{MRT}}^+ \cup \mathcal{B}_{\mathrm{MRT}}^-,$$
with 
$$\mathcal{B}_{\mathrm{MRT}}^+ = (M_{\rho},M_{e},M_{p_{xx}},M_{p_{xy}},M_{\varepsilon}),$$
and $$\mathcal{B}_{\mathrm{MRT}}^- = (M_{j_x},M_{j_y},M_{q_x},M_{q_y}).$$
Hence, the relaxation of even  and odd moments is decoupled using two relaxation frequencies. \textcolor{black}{In the above context,} this leads to
$$\bm{S}_{\mathrm{TRT}} =\mathrm{diag}(0,0,0,\omega^{+},\omega^{+},\omega^{+},\omega^{-},\omega^{-},\omega^{+}),$$
with $\omega^{+}=\omega_{\nu}$, $(1/\omega^{-}-1/2)=\Lambda/(1/\omega^{+}-1/2)$, and $\Lambda$ the `magic parameter' that controls both accuracy and stability of the TRT-LBM~\cite{DHUMIERE_CMA_58_2009,GINZBURG_JSP_139_2010,KUZMIN_CMWA_61_2011,SILVA_PRE_96_2017}.

\textcolor{black}{It is worth noting that, by relying on its original set of relaxation frequencies, the orthogonal MRT model is not well suited for aeroacoustic simulations. Indeed, the standard value of the relaxation frequency related to acoustic modes ($\omega_e$) is always far from $2$, hence leading to huge values of the bulk viscosity~\cite{LALLEMAND_PRE_61_2000,DHUMIERES_TRS_360_2002}. This further increases the stability of the corresponding LBM by overdissipating acoustic waves~\cite{MARIE_JCP_228_2009,XU_JCP_230_2011,GENDRE_PRE_96_2017}.} But from the theoretical viewpoint, the major defect of the above models is that they \textcolor{black}{originally
relied} on polynomial bases including high order terms that are not compliant with the velocity discretization. Taking the example of $q_x$, its associated vector is
$$M_{q_x} = [-5 + 3(\xi_{i,x}^2+\xi_{i,y}^2)]\xi_{i,x},$$
and it reduces to
$$M_{q_x} = (-2+\xi_{i,y}^2)\xi_{i,x},$$
due to the aliasing defects $\xi_{i,x}^3=\xi_{i,x}$ and $\xi_{i,y}^3=\xi_{i,y}$ of the D2Q9 lattice, where $\xi_{i,x},\xi_{i,y}\in\{0,\pm1\}$. The same issue is also encountered with $M_{q_y}$ and $M_{\varepsilon}$. Unfortunately, this aliasing problem remains present for D3Q19 and D3Q27 lattices where it might be even more dominant. \textcolor{black}{Nevertheless, both defects can easily be circumvented by using: (1) the correct value of the bulk viscosity to compute $\omega_e$, and (2) a polynomial basis that is not affected by aliasing issues (see for example Eq.~(\ref{eq:TPBasis})).}

It is also worth noting that these models were shown to recover the behavior of the BGK collision operator based on $f^{eq,2}_i$ if one neglects higher order velocity dependent terms. Knowing that a fourth order equilibrium state $f^{eq,4}_i$ increases the linear stability of the LBM~\cite{COREIXAS_PhD_2018,WISSOCQ_JCP_380_2019}, it would be interesting to determine if these orthogonal MRT and TRT models recover the behavior of the corresponding BGK-LBM. A few words regarding this possible link can be found in Sec.~\ref{sec:D2Q9_MRT_LL}.

 \textcolor{black}{In the rest of the paper, LBMs based on the orthogonal basis~(\ref{eq:MRTBasis}), and its 3D extensions~\cite{DHUMIERES_TRS_360_2002,SUGA_CMA_69_2015,FAKHARI_JCP_341_2017}, will be denoted as orthogonal MRT and TRT models.}

To find polynomial bases compliant with the order of accuracy of the velocity discretization, it was proposed to build $D$-dimensional bases following tensor product rules of 1D bases~\cite{KARLIN_PA_389_2010}. Since the D2Q9 lattice is a second order velocity discretization~\cite{SHAN_JFM_550_2006}, these rules lead to the construction of a basis only composed of polynomials of the form $\xi_{i,x}^n\xi_{i,y}^m$ with $n\leq 2$ and $m \leq 2$. The most natural basis compliant with the D2Q9 lattice is then  
\begin{equation}
\begin{array}{r}
\mathcal{B}_{\mathrm{TP}} = (1,\xi_{i,x},\xi_{i,y},\xi_{i,x}^2,\xi_{i,y}^2,\xi_{i,x}\xi_{i,y},\xi_{i,x}^2\xi_{i,y},\:\:\: \\[0.2cm]
\xi_{i,x}\xi_{i,y}^2,\xi_{i,x}^2\xi_{i,y}^2),
\end{array}
\label{eq:TPBasis}
\end{equation}
where the subscript TP stands for the tensor product formalism. In fact, any basis composed of linear combinations of $\mathcal{B}_{\mathrm{TP}}$ elements would also be correct. Hence, its Hermite counterpart (HTP) could also be used~\cite{MALASPINAS_ARXIV_2015,COREIXAS_PRE_96_2017},
\begin{equation}
\begin{array}{r}
\mathcal{B}_{\mathrm{HTP}} = (H_{i,00},H_{i,10},H_{i,01},H_{i,20},H_{i,02},H_{i,11},\:\:\: \\[0.2cm]
H_{i,21},H_{i,12},H_{i,22}).
\end{array}
\label{eq:HTPBasis}
\end{equation}
If one wants to further decouple shear and bulk viscosities, one simply needs to change $\xi_{i,x}^2$ and $\xi_{i,y}^2$ ($H_{i,20}$ and $H_{i,02}$) to $\xi_{i,x}^2+\xi_{i,y}^2$ and $\xi_{i,x}^2-\xi_{i,y}^2$ ($H_{i,20}+H_{i,02}$ and $H_{i,20}-H_{i,02}$) as proposed in Refs.~\cite{KARLIN_ARXIV_2011,KARLIN_PRE_90_2014a,LYCETTBROWN_CMWA_67_2014}.

This TP approach paved the way for a systematic derivation of new MRT-LBMs as introduced hereafter. It is finally worth noting that other kinds of TRT models were also proposed in different contexts~\cite{ANSUMALI_EPJB_56_2007,ADHIKARI_PRE_78_2008}, but they will not be studied in the present work since they do not introduce new concepts as compared to those already contained within the orthogonal TRT model.

\subsection{Central moment space\label{sec:Rev_CM}}

Despite a nonnegligible increase in stability of both the MRT- and TRT-LBMs, issues were still encountered when simulating high Reynolds number flows. It was then suggested by several authors to improve the numerical stability of MRT-LBMs by applying the collision step in the comoving reference frame~\cite{GEIER_PRE_73_2006,GEIER_EPJST_171_2009,GELLER_CMA_65_2013,LYCETTBROWN_CMWA_67_2014,DUBOIS_CCP_17_2015,NING_IJNMF_82_2016,FEI_PRE_96_2017,DEROSIS_PRE_95_2017,FEI_PRE_97_2018,DEROSIS_PRE_99_2019,DEROSIS_PRE_Submitted_2019}. 

These models rely on two modifications of the previous MRT-LBMs. First, the collision is applied in the central moment (CM) space based on either the TP~\cite{GEIER_PRE_73_2006,GEIER_EPJST_171_2009,LYCETTBROWN_CMWA_67_2014,FEI_PRE_96_2017,DEROSIS_PRE_95_2017,FEI_PRE_97_2018,DEROSIS_PRE_99_2019,DEROSIS_PRE_Submitted_2019} or the orthogonal MRT formalism~\cite{DUBOIS_CCP_17_2015,DUBOIS_CRM_343_2015}, i.e,
$$\mathcal{B}_{\mathrm{CM}} = \widetilde{\mathcal{B}}_{\mathrm{TP}}\:\: \mathrm{or}\:\: \widetilde{\mathcal{B}}_{\mathrm{MRT}},$$
where the tilde stands for the velocity shift of each vector of the basis: $\widetilde{\xi}_{i,x} = \xi_{i,x}-u_x$ and $\widetilde{\xi}_{i,y} = \xi_{i,y}-u_y$. Second, the equilibrium state is also expanded following the TP formalism, leading to the inclusion of third and fourth order terms in the definition of $f_i^{eq}$. It is important to note that while these terms were added in an \emph{a posteriori} manner in the first (cascaded) model~\cite{GEIER_PRE_73_2006,GEIER_EPJST_171_2009}, the TP formalism offers an \emph{a priori} and \emph{systematic} way to include high order terms without exceeding the order of accuracy of the lattice. \textcolor{black}{By extending the previous reasoning on equilibrium moments to the equilibrium population itself, one can derive new equilibrium states that includes high-order Hermite polynomials. This improves the Galilean invariance of the resulting LBM in a \emph{systematic} and \emph{a priori} way, using either the CM formalism~\cite{HUANG_PRE_97_2018,DEROSIS_PRE_99_2019,DEROSIS_PRE_Submitted_2019} or any other one. This point will further be investigated in Sec.~\ref{sec:MacroEq}.}

The CM-LBM can be summarized by
$$\bm{\Omega_{\mathrm{CM}}} = \bm{M^{-1}N^{-1}SNM},\: f^{neq}_i = (f_i-f^{eq,\mathrm{ext}}_i),$$ 
where $\bm{N}$ and $\bm{N^{-1}}$ are the velocity dependent matrices used to move from the raw to the central moment space, and vice versa~\cite{DUBOIS_CCP_17_2015,FEI_PRE_96_2017}. In the particular case of a flow at rest ($u_x=u_y=0$), these matrices reduce to the identity matrix, and the \textcolor{black}{orthogonal MRT (or TP) collision model is} recovered. $f^{eq,\mathrm{ext}}$ is the extended equilibrium state that includes up to fourth order (six order) terms for the D2Q9 (D3Q27) velocity discretization~\cite{HUANG_PRE_97_2018,DEROSIS_PRE_99_2019,DEROSIS_PRE_Submitted_2019}. Both of these modifications lead to a drastic stability gain, and especially when the TP formalism is adopted for the moment space~\cite{DUBOIS_CRM_343_2015}. This is why only this formalism will be considered in the present work.

Eventually, it is important to note that despite a further stability increase, these models were mostly validated imposing acoustically related moments (those including the trace of second order moments) to their equilibrium value in the single phase and isothermal case~\cite{GEIER_PRE_73_2006,GEIER_EPJST_171_2009,GELLER_CMA_65_2013,NING_IJNMF_82_2016}. This translates in a severe overdissipation of acoustic waves. As a consequence, \textcolor{black}{it is of paramount importance to properly decouple the stability increase induced by the CM-LBM itself, from the one induced by the bulk viscosity. A few results about this problem can be found in Ref.~\cite{DUBOIS_CRM_343_2015}, but more in depth investigations are still required to properly decouple these two stabilization mechanisms. For example, it would be interesting to quantify the impact of the moment space, and of relaxation frequencies, on both academic and realistic configurations that would include acoustically related phenomena.}

\subsection{Regularization steps and Hermite moments}

In parallel of the derivation of CM-LBMs, another type of collision models have been proposed to increase the numerical stability of the BGK-LBM without accounting for collisions in the moment space. Instead, these regularized collision models aim at filtering out nonhydrodynamic contributions that appear during the streaming step~\cite{LATT_ARXIV_2005,LATT_MCS_72_2006a,ZHANG_PRE_74_2006,NIU_PRE_76_2007,COREIXAS_PRE_96_2017,COREIXAS_PhD_2018}. This is done projecting the nonequilibrium VDF on Hermite polynomials up to a given order $N$. Hence, this regularized collision model can be summarized as
$$\Omega_{\mathrm{PRN}} = \omega_{\nu},\: f^{neq}_i = w_i\sum_{n=0}^N \dfrac{1}{n! c_s^{2n}}\bm{a}^{neq}_n : \bm{H}_{i,n},$$
where `:' is the tensor index contraction, $w_i$ are the quadrature weights, $c_s$ is the lattice constant, $\bm{a}_n$ are Hermite coefficients, and the subscript PRN stands for the projection based regularization at order $N$, with $N=2$ in the original model~\cite{LATT_ARXIV_2005,LATT_MCS_72_2006a}, whereas $N\geq 3$ for high order LBMs~\cite{ZHANG_PRE_74_2006,NIU_PRE_76_2007,COREIXAS_PRE_96_2017,COREIXAS_PhD_2018}. In the original PR framework, Hermite nonequilibrium coefficients read as
\begin{equation}
\bm{a}^{neq}_n =  \sum_{i}(f_i-f^{eq,2}_i)\bm{H}_{i,n}.
\label{eq:HermiteCoeffPR}
\end{equation}
The above description corresponds to the way the regularization step was first introduced. In fact, it can be reinterpreted in terms of a MRT collision model based on $\mathcal{B}_{\mathrm{HTP}}$~(\ref{eq:HTPBasis}) and the corresponding extended equilibrium state~\cite{COREIXAS_PhD_2018}
$$\bm{\Omega_{\mathrm{PR2}}} = \bm{M^{-1}SM},\: f^{neq}_i = (f_i-f^{eq,4}_i),$$ 
and where,
$\bm{S_{\mathrm{PR2}}}=\mathrm{diag}(0,0,0,\omega_{\nu},\omega_{\nu},\omega_{\nu},\omega_3,\omega_3,\omega_4)$,
with $\omega_3=\omega_4=1$. This is the sole set of relaxation frequencies that will be considered for this collision model in the present work. It is interesting to note that a more general MRT-LBM based on the Hermite moment space was also introduced in Refs.~\cite{SHAN_IJMPC_18_2007,CHEN_IJMPC_25_2014}.

Despite its great success in various fields~\cite{NIU_PRE_76_2007,WANG_PRE_92_2015,BA_PRE_97_2018,MONTESSORI_CF_167_2018}
, it was recently proven that this model does not filter out all nonhydrodynamic contributions~\cite{MALASPINAS_ARXIV_2015,COREIXAS_PRE_96_2017}. The reason lies in the fact that part of them are still present in $\bm{a}^{neq}_n$ via $f_i-f^{eq,4}_i$. Indeed, the latter only reduces to the first order (Navier-Stokes) nonequilibrium VDF $f^{(1)}_i$ in the continuum limit, which might no be valid anymore when the mesh becomes very coarse, i.e, nonnegligible Knudsen number based on the grid cell size. To counter this defect, its was proposed to impose the correct nonequilibrium part through a recursive computation of $\bm{a}^{(1)}_n$ using formulas derived from the Chapman-Enskog expansion~\cite{MALASPINAS_ARXIV_2015,COREIXAS_PRE_96_2017,BROGI_JASA_142_2017,COREIXAS_PhD_2018}. This recursive regularized (RR) collision model is defined as
$$\bm{\Omega_{\mathrm{RRN}}} = \bm{M^{-1}SM},\: f^{neq}_i = w_i \sum_{n=0}^N \dfrac{1}{n! c_s^{2n}}\bm{a}^{(1)}_n : \bm{H}_{i,n},$$
with $\bm{S}=\omega_{\nu} \bm{I}$ in the original model, $\bm{I}$ being the identity matrix. The recursive formulas for the computation of $\bm{a}^{(1)}_n$ were derived in Refs.~\cite{MALASPINAS_ARXIV_2015,COREIXAS_PRE_96_2017}. They are recalled in the case of the D2Q9 lattice in Eq.~(\ref{eq:RRformulas_D2Q9}), whereas formulas for the D3Q27 are summarized in Eq.~(\ref{eq:RRformulas_D3Q27}). The single relaxation time (SRT) PR collision model was shown to recover the behavior of the BGK operator when it is applied to the \emph{complete} Hermite polynomial basis of the D2Q9 lattice, i.e, $\omega_4=\omega_3=\omega_{\nu}$~\cite{COREIXAS_PRE_96_2017}. When it comes to the RR procedure, no direct link with the BGK operator is known at the time of writing.

Before moving to LBMs based on cumulants, it is worth noting that extensions of the PR model to other moment spaces have been proposed in the literature. 
A first extension was proposed in the TRT context~\cite{SILVA_PRE_96_2017}. Imposing $\Lambda_{q}^- \Lambda_{\varepsilon}^+=1/4$, it was noticed that nonhydrodynamic contributions could be filtered out in the same spirit as for the original PR model. Unfortunately, this model suffers the same problems as the orthogonal TRT, since it also relies on the same polynomial basis $\mathcal{B}_{\mathrm{TRT}}$. 
Another extension was proposed in the framework of collisions occurring in the comoving reference frame. Originally, the purpose of such an extension was to improve the Galilean invariance of LBMs relying on the central Hermite polynomial expansion of VDFs~\cite{CHEN_PATENT_Collision_2015}. This formalism was further used to increase the stability of the PR model for the simulation of high Reynolds number flows~\cite{MATTILA_PF_29_2017}. Discarding several \emph{diffusive} terms in an adhoc manner, the authors recovered recursive formulas obtained from the Chapman-Enskog expansion in the context of the isothermal RR collision model. This suggests that there might be an underlying link between the CHM and the RR frameworks. More information will be given in Sec.~\ref{sec:D2Q9_Reg} regarding this last point.
  
\subsection{Cumulant space}

The most recent improvement of the MRT-LBM is based on a collision step occurring in the cumulant space~\cite{GEIER_CMA_70_2015,GEIER_JCP_348_2017a,GEIER_JCP_348_2017b}. The main advantage of cumulants, as compared to standard raw and central moments, is their ability to quantify the deviation of a probability density function with respect to a Gaussian distribution~\cite{MCCULLAGH_Book_2018}, in our case, the normalized equilibrium VDF $f^{eq}/\rho$. This was confirmed in Sec.~\ref{sec:StatGaussian} where the only nonzero equilibrium cumulants were the first, and the second ones: $$K^{eq}_1=u_x,\:K^{eq}_2=c_s^2,\: K^{eq}_n=0\: (n\geq 3).$$
By definition, cumulants belong to a type of statistical quantities quite different from raw and central moments. Nevertheless, they directly flow from (the logarithm of) the moment generating function as explained in Sec.~\ref{sec:stat}. Besides, they also share some similarities with central moments in the sense that their second and third order terms are strictly equivalent, as already precised in Refs.~\cite{KENDALL_AE_10_1940,COOK_Bio_38_1951}, and further illustrated in Apps.~\ref{sec:RelationshipsUnivariate}-\ref{sec:RelationshipsTrivariate}. This means that without doing any calculations, it is known for sure that CM-LBM and cumulant-LBM (K-LBM) are identical for the D1Q3 lattice.

Due to the nonlinear relationships between cumulants and the other moment like quantities for $n\geq 4$, it is not possible to express the collision model in a linear matrix form. Nonetheless, one obtains a very simple algorithm using formulas derived to \textcolor{black}{link} central moments with respect to the cumulant one, and vice versa (Apps.~\ref{sec:RelationshipsUnivariate}-\ref{sec:RelationshipsTrivariate}). According to Refs.~\cite{GEIER_CMA_70_2015,GEIER_JCP_348_2017a,GEIER_JCP_348_2017b}, the K-LBM cannot recover the behavior of the BGK-LBM. This point will be further studied in Sec.~\ref{sec:D2Q9_Cumul}.

\subsection{Partial conclusions}

Almost all collision models can be rewritten in a linear matrix form,
\begin{equation}
\bm{\Omega} = \bm{M^{-1}SM},
\label{eq:MatrixFormColl}
\end{equation}
with the exception of the K-LBM. Hence, it is already known for sure that this LBM belongs to a completely different family of LBMs. For collision models that satisfy Eq.~(\ref{eq:MatrixFormColl}), another distinction can be made between those relying on the approximation of the continuum limit $f^{neq}_i=f_i-f^{eq}_i$, and the RR collision model which imposes $f^{neq}_i=f^{(1)}_i$ instead.  Regarding models operating in the comoving reference frame, their linear matrix form depends on the local velocity. Hence it seems complicated, at first sight, to find relationships with other collision models expressed in the reference frame at rest. Consequently, four different groups seem to emerge from the above review: (1) RM, HM and PR, (2) CM and CHM, (3) RR and (4) cumulant based collision models. Nevertheless, the possible relationship between RR- and CHM-LBMs remains to be addressed.

The rest of the paper will mainly be devoted to the validation of the above assumptions. To do so, it will be confirmed if the form of the equilibrium state is independent of the moment space or not. In the meantime, possible sets of relaxation frequencies allowing to link different types of collision model will be sought.  

For the sake of completeness, the derivation of all linear transformations matrices $\bm{M}$ and $\bm{M^{-1}}$ are detailed for the D2Q9 lattice in App.~\ref{sec:LinearTransMatrix}. General explanations regarding their construction are also provided, which makes possible their derivation for both the D3Q19 and D3Q27 lattices in a straightforward manner.

\section{Theoretical comparison using the D1Q3 lattice \label{sec:CompStudy1}}

The purpose of this section is to compare, in the one-dimensional case, the most common collision models encountered in the LBM framework through their equilibrium, pre and post collision populations. To do so, the derivation of raw, Hermite, central, central Hermite LBMs is first recalled before moving to the comparative study.

\subsection{Discrete moments and populations \label{sec:derivationLBMs}}
Let us consider the one-dimensional discretization of the velocity space using the following three discrete velocities $(\xi_0,\xi_{\pm 1})=(0,\pm 1)$, namely, the D1Q3 lattice~\cite{KRUGER_Book_2017}. This model has three degrees of freedom meaning it allows, at best, to preserve the first three moments of the VDF during the velocity discretization, i.e, $\forall n\in\{0,1,2\}$,
\begin{equation}
\rho P_n = \displaystyle\int{f \mathcal{P}_n \mathrm{d}\xi_x} = \sum_i f_i \mathcal{P}_{i,n}, 
\label{eq:MomentConservation}
\end{equation}
where $P_n$ is a normalized moment of order $n$, yet to be defined. $\mathcal{P}_n$, and $\mathcal{P}_{i,n}$ (its discrete counterpart), also depend on the framework on which the collision model relies. Four different moment spaces will be considered hereafter. Hence, $\mathcal{P}_{i,n}$ can be defined using the
\begin{enumerate}
\item Raw moment (RM) space
\begin{equation}
\xi_{i,x}^n,
\end{equation}
\item Hermite moment (HM) space
\begin{equation}
H_{i,n}(\xi_{i,x}),
\end{equation}
\item Central moment (CM) space
\begin{equation}
\widetilde{\xi}_{i,x}^n=(\xi_{i,x}-u_x)^n,
\end{equation}
\item Central Hermite moment (CHM) space
\begin{equation}
\widetilde{H}_{i,n}(\xi_{i,x})=H_{i,n}(\xi_{i,x}-u_x),
\end{equation}  
\end{enumerate}
where $H_{i,n}$ is the discrete version of the Hermite polynomial of degree $n$, as introduced in Eq.~(\ref{eq:HermitePoly}). The tilde symbol stands for the central moment version of the polynomial basis considered for the collision process. 

Once the moment space has been chosen, Eq.~(\ref{eq:MomentConservation}) is enforced adopting the correct form of $f_{+1}$, $f_0$ and $f_{-1}$. Assuming the collision takes place in the RM space, the following system needs to be solved
\begin{equation}
\left\{
\begin{array}{r @{\: = \:} l}
\rho M_0 & f^{\mathrm{RM}}_{+1} + f^{\mathrm{RM}}_0 + f^{\mathrm{RM}}_{-1}\\[0.1cm]
\rho M_1 & f^{\mathrm{RM}}_{+1} - f^{\mathrm{RM}}_{-1}\\[0.1cm]
\rho M_2 & f^{\mathrm{RM}}_{+1} + f^{\mathrm{RM}}_{-1}
\end{array}
\right.
\label{eq:fEq1DmomentMatching}
\end{equation} 
which gives
$$f_0^{\mathrm{RM}} = \rho \left(M_0-M_2\right) \:\:\text{and}\:\:
f_{\sigma}^{\mathrm{RM}} = \dfrac{\rho}{2}\left(\sigma M_1 + M_2\right),$$
with $\sigma = \pm 1$.
Corresponding equilibrium VDFs are then obtained computing equilibrium moments $M_n^{eq}$ \textcolor{black}{that satisfy the moment conservation rule~(\ref{eq:MomentConservation})}. By using the \emph{isothermal} moments of the continuous Maxwell equilibrium VDF~(\ref{eq:MaxwellBoltzmannEq}), one obtains
\begin{equation}
\left\{
\begin{array}{r @{\: = \:} l}
M_0^{eq} & 1\\[0.1cm]
M_1^{eq} & u_x\\[0.1cm]
M_2^{eq} & u_x^2 + c_s^2
\end{array}
\right.
\label{eq:EqMoments_RM}
\end{equation}    
which leads to 
$$f_0^{eq,\mathrm{RM}} = \rho [1-(u_x^2+c_s^2)],\:
f^{eq,\mathrm{RM}}_{\sigma} = \dfrac{\rho}{2}[\sigma u_x + (u_x^2+c_s^2)].$$
$c_s$ is the isothermal \textcolor{black}{(or Newtonian)} speed of sound, and it corresponds to the lattice constant when the lattice Boltzmann unit system is adopted~\cite{VIGGEN_PhD_2014}.

Instead of performing collisions in the moment space using the linear matrix form of the collision operator, it is preferred here to work directly with VDFs. Post collision VDFs then read as
\begin{equation}
f_0^{*,\mathrm{RM}} = \rho (M_0^*-M_2^*),\:
f_{\sigma}^{*,\mathrm{RM}} = \dfrac{\rho}{2}(\sigma M_1^* + M_2^*),
\label{eq:postCollVDF}
\end{equation}
with $M_n^*$ the post collision moment defined as
\begin{equation}
M_n^* = (1-\omega_n)M_n + \omega_n M_n^{eq},
\label{eq:postCollMoments}
\end{equation}
where $\omega_n=1/(\tau_n+1/2)$ is the relaxation frequency associated to $M_n$, and $\tau_n$ the corresponding relaxation time. In the case of the D1Q3 lattice, only mass and momentum are collision invariants, thus one can freely choose the value of their corresponding relaxation frequency. Most of the time, one imposes $\omega_0=\omega_1=0$ in the absence of external forces~\cite{GINZBURG_CCP_3_2008a}. 

Injecting equilibrium moments~(\ref{eq:EqMoments_RM}) into the definition of post collision moments~(\ref{eq:postCollMoments}), post collision VDFs~(\ref{eq:postCollVDF}) now read as
\begin{align}
f_0^{*,\mathrm{RM}} &= \rho \left[1-\left(1-\omega_{\nu}\right) M_2 - \omega_{\nu}\left(u_x^2+c_s^2\right)\right],\\[0.1cm]
f_{\sigma}^{*,\mathrm{RM}} &= \dfrac{\rho}{2} \left[\sigma u_x+\left(1-\omega_{\nu}\right) M_2 + \omega_{\nu}\left(u_x^2+c_s^2\right)\right],
\end{align}
where $\omega_{\nu}=\omega_2$ is the relaxation time related to the kinematic viscosity $\nu$ through \textcolor{black}{$1/\omega_{\nu}=\nu/c_s^2+1/2$}.

\subsection{Impact of the moment space (D1Q3) \label{sec:ImpactOfMomentSpaceD1Q3}}

By solving Eq.~(\ref{eq:MomentConservation}) in different moment frameworks, various forms of post collision VDFs are obtained. In the HM framework, 
\begin{equation}
\begin{array}{l}
f_0^{*,\mathrm{HM}} = \rho \big[\big(1-c_s^2\big)-\big(1-\omega_{\nu}\big) A_2 - \omega_{\nu} u_x^2\big],\\[0.2cm]
f_{\sigma}^{*,\mathrm{HM}} = \dfrac{\rho}{2} \big[c_s^2 + \sigma u_x+\big(1-\omega_{\nu}\big) A_2 + \omega_{\nu} u_x^2\big],
\end{array}
\label{eq:postCollVDF_HM}
\end{equation}
since $A_0^*=A_0^{eq}=1$, $A_1^*=A_1^{eq}=u_x$, $A_2^{eq}=u^2_x$. Using central moments, these VDFs are then
\begin{equation}
\begin{array}{l}
f_0^{*,\mathrm{CM}} = \rho \big[\big(1-u_x^2\big)- \big(1-\omega_{\nu}\big) \widetilde{M}_2 - \omega_{\nu} c_s^2\big],\\[0.2cm]
f_{\sigma}^{*,\mathrm{CM}} = \dfrac{\rho}{2} \big[\sigma u_x + u_x^2 +\big(1-\omega_{\nu}\big) \widetilde{M}_2 + \omega_{\nu} c_s^2\big],
\end{array}
\label{eq:postCollVDF_CM}
\end{equation}
where $\widetilde{M}_0^*=\widetilde{M}_0^{eq}=1$, $\widetilde{M}_1^*=\widetilde{M}_1^{eq}=0$, $\widetilde{M}_2^{eq}=c_s^2$. Eventually, the CHM framework leads to
\begin{equation}
\begin{array}{l}
f_0^{*,\mathrm{CHM}} = \rho \big[1-c_s^2-u_x^2 - \big(1-\omega_{\nu}\big) \widetilde{A}_2\big],\\[0.2cm]
f_{\sigma}^{*,\mathrm{CHM}} = \dfrac{\rho}{2} \big[c_s^2 + \sigma u_x + u_x^2 +\big(1-\omega_{\nu}\big) \widetilde{A}_2\big],
\end{array}
\label{eq:postCollVDF_CHM}
\end{equation}
using $\widetilde{A}_0^*=\widetilde{A}_0^{eq}=1$, $\widetilde{A}_1^*=\widetilde{A}_1^{eq}=0$, $\widetilde{A}_2^{eq}=0$. 

Regarding equilibrium moments, it is interesting to note that one can simply derive them applying two simple rules to Eq.~(\ref{eq:EqMoments_RM}). First, one can switch from equilibrium raw or central moments to their Hermite counterparts by neglecting $c_s$-dependent terms. The reason behind this is that Hermite polynomials reduce to monomials when terms proportional to $c_s$ are discarded. Second, the change from equilibrium raw (or Hermite) moments to their central (Hermite) versions is done discarding velocity dependent terms. This is explained by the fact that in the comoving reference frame, equilibrium moments are velocity independent. Consequently, all \emph{isothermal} equilibrium moments, but the zeroth, are \textcolor{black}{null} in the CHM framework.

At first sight, all moment spaces seem to lead to different expressions for VDFs. Nonetheless, links can be drawn between all the above populations. Starting from post collision VDFs obtained within the RM framework~(\ref{eq:postCollVDF}),
\begin{equation}
\begin{array}{r@{\: = \:}l}
f_0^{*,\mathrm{RM}} & \rho \Big(1-M_2^*\Big)\\[0.2cm]
	  & \rho \Big(1-(1-\omega_{\nu})M_2 - \omega_{\nu} M^{eq}_2\Big)\\[0.2cm]
	  & \rho \Big(1-(1-\omega_{\nu})(A_2+c_s^2) - \omega_{\nu} (A^{eq}_2+c_s^2)\Big)\\[0.2cm]
	  & \rho \Big(1-c_s^2 -(1-\omega_{\nu})A_2 - \omega_{\nu} A^{eq}_2\Big) \\[0.2cm]
	  & \rho \Big(1-c_s^2 -A_2^*\Big)\\[0.2cm]
	  & f_0^{*,\mathrm{HM}},
\end{array}
\end{equation}
and
\begin{equation}
\begin{array}{r@{\: = \:}l}
f_{\sigma}^{*,\mathrm{RM}} & \dfrac{\rho}{2} \Big(\sigma u_x + M_2^*\Big)\\[0.2cm]
& \dfrac{\rho}{2} \Big(\sigma u_x + (1-\omega_{\nu})M_2 + \omega_{\nu} M^{eq}_2\Big)\\[0.2cm]
& \dfrac{\rho}{2} \Big(\sigma u_x + (1-\omega_{\nu})(A_2+c_s^2) + \omega_{\nu} (A^{eq}_2+c_s^2)\Big)\\[0.2cm]
& \dfrac{\rho}{2} \Big(c_s^2+ \sigma u_x + (1-\omega_{\nu})A_2 + \omega_{\nu} A^{eq}_2\Big) \\[0.2cm]
& \dfrac{\rho}{2} \Big(c_s^2+ \sigma u_x + A_2^*\Big)\\[0.2cm]
& f_{\sigma}^{*,\mathrm{HM}}.
\end{array}
\end{equation}
Thus, the definition of post collision VDFs originating from the HM framework~(\ref{eq:postCollVDF_HM}) are recovered from those of the RM approach. Moving now to the CM framework~(\ref{eq:postCollVDF_CM}),
\begin{equation}
\begin{array}{r@{\: = \:}l}
f_0^{*,\mathrm{CM}} & \rho \Big(1 - u_x^2 - \widetilde{M}_2^*\Big)\\[0.2cm]
	  & \rho \Big(1 - u_x^2 - (1-\omega_{\nu})\widetilde{M}_2 - \omega_{\nu} \widetilde{M}^{eq}_2\Big)\\[0.2cm]
	  & \rho \Big(1-c_s^2- u_x^2 -(1-\omega_{\nu})\widetilde{A}_2 - \omega_{\nu} \widetilde{A}^{eq}_2\Big)\\[0.2cm]
	  & \rho \Big(1-c_s^2- u_x^2 -\widetilde{A}_2^*\Big)\\[0.2cm]
	  & f_0^{*,\mathrm{CHM}},
\end{array}
\end{equation}
and
\begin{equation}
\hspace*{-0.25cm}
\begin{array}{r@{\: = \:}l}
f_{\sigma}^{*,\mathrm{CM}} & \dfrac{\rho}{2} \Big(\sigma u_x + u_x^2 + \widetilde{M}_2^*\Big)\\[0.2cm]
& \dfrac{\rho}{2} \Big(\sigma u_x + u_x^2 + (1-\omega_{\nu})\widetilde{M}_2 + \omega_{\nu} \widetilde{M}^{eq}_2\Big)\\[0.2cm]
& \dfrac{\rho}{2} \Big(c_s^2+ \sigma u_x + u_x^2 + (1-\omega_{\nu})\widetilde{A}_2 + \omega_{\nu} \widetilde{A}^{eq}_2\Big) \\[0.2cm]
& \dfrac{\rho}{2} \Big(c_s^2+ \sigma u_x + u_x^2 + \widetilde{A}_2^*\Big)\\[0.2cm]
& f_{\sigma}^{*,\mathrm{CHM}},
\end{array}
\end{equation}
which proves that post collision VDFs originating from the CHM framework~(\ref{eq:postCollVDF_CHM}) are also recovered. 

The question remains regarding the validity of this result when the reference frame is changed. Starting with,
\begin{equation}
\begin{array}{r@{\: = \:}l}
f_0^{*,\mathrm{RM}} & \rho \Big(1-M_2^*\Big),\\[0.2cm]
f_{\sigma}^{*,\mathrm{RM}} & \dfrac{\rho}{2} \Big(\sigma u_x + u_x^2 + \widetilde{M}_2^*\Big),
\end{array}
\end{equation}
and using Eq.~(\ref{eq:recRawCentral}), one obtains $M_1^* = \widetilde{M}_1^* + u_x$ and $M_2^* = \widetilde{M}_2^* + u_x^2$. This leads to
\begin{equation}
\begin{array}{r@{\: = \:}l}
f_0^{*,\mathrm{RM}} & \rho \Big[1-\big(\widetilde{M}_2^*+u_x^2\big)\Big]=f_0^{*,\mathrm{CM}},\\[0.2cm]
f_{\sigma}^{*,\mathrm{RM}} & \dfrac{\rho}{2} \Big[\sigma u_x + \big(\widetilde{M}_2^*+u_x^2\big)\Big]=f_{\sigma}^{*,\mathrm{CM}},
\end{array}
\end{equation}
and it results in the equivalence between raw and central moment frameworks for the D1Q3 lattice. 

In summary, raw, Hermite, central and central Hermite frameworks recovers the very same behavior (BGK) when the D1Q3 lattice is employed. Imposing $\omega_{\nu}=1$, one can further confirm that all models share the very same equilibrium state. 


It is important to note that the K-LBM reduces to the CM-LBM as far as the D1Q3 lattice is employed since only moments up to the second order are considered for this velocity discretization. To quantify the possible benefit of the cumulant collision model in 1D, one must consider LBMs derived from high order velocity discretizations, such as the (zero-one-three) D1Q5 or the D1Q7 lattices~\cite{CHIKATAMARLA_PRE_79_2009,KARLIN_PA_389_2010,FRAPOLLI_PhD_2017}.

The last point that need to be checked is the way regularized collision operators are related to the above collision models. To answer this question, it is necessary to link populations derived through the Gauss-Hermite quadrature~\cite{GRADb_CPAM_2_1949,SHAN_PRL_80_1998,SHAN_JFM_550_2006} with those obtained via the conservation of Hermite moments~(\ref{eq:postCollVDF_HM}).

\subsection{Gauss-Hermite quadrature and regularized collision models \label{sec:WeightedScalarProduct}}

Another way to derive populations associated to a given velocity discretization is based on the Gauss-Hermite (GH) quadrature~\cite{GRADb_CPAM_2_1949,SHAN_PRL_80_1998,SHAN_JFM_550_2006}. This mathematical tool allows the exact preservation of certain properties (moments up to a certain order $N$) during the discretization of the velocity space. This method defines, in a unique and \emph{systematic} way, the discrete populations $f_i$ as
\begin{equation}
f_i = w_i\sum_{n=0}^N \dfrac{1}{n! c_s^{2n}}\bm{a}_n : \bm{H}_{i,n},
\label{eq:VDF_GH_Quad}
\end{equation}
where the quadrature weights $w_i$ and the lattice constant $c_s$ can be obtained, for example, ensuring the preservation of Hermite polynomial orthogonality properties up to the order $N$~\cite{PHILIPPI_PRE_73_2006}. 
For the D1Q3 lattice, the above expression reduces to,
\begin{equation}
f_i = w_i\bigg[a^{eq}_0 H_{i,0}+\dfrac{1}{c_s^2}a^{eq}_1 H_{i,1}+\dfrac{1}{2c_s^4}(a^{eq}_2+a^{neq}_2) H_{i,2}\bigg],
\end{equation}
with $w_0=2/3$, $w_{\sigma}=1/6$, $c_s=1/\sqrt{3}$ and $a^{eq}_n=\rho u_x^n$ in the isothermal case~\cite{MALASPINAS_ARXIV_2015}. Thus,
\begin{equation}
\begin{array}{r@{\: = \:}l}
f_0^{*,\mathrm{GH}} & w_0\big(\rho-a^*_2/2 c_s^2 \big)\\[0.1cm]
& \rho\big(2/3-A^*_2\big)\\[0.1cm]
& \rho\big(1-c_s^2-A^*_2\big)\\[0.1cm]
& f_0^{*,\mathrm{HM}}.
\end{array}
\end{equation}
using $a^*_2 = \sum_i f^*_i H_{i,2} =\rho A^*_2$. Furthermore,
\begin{equation}
\begin{array}{r@{\: = \:}l}
f_{\sigma}^{*,\mathrm{GH}} & w_{\sigma}\big(\rho+\sigma \rho u_x+a^*_2(1-c_s^2)/2c_s^4 \big)\\[0.1cm]
& \dfrac{\rho}{2}\big(c_s^2+\sigma u_x + A^*_2\big)\\[0.1cm]
& f_{\sigma}^{*,\mathrm{HM}}.
\end{array}
\end{equation}

Since both PR and RR collision models rely on this Gauss-Hermite quadrature, and that RR and PR approaches are equivalent for moments of order $n\leq 2$~\cite{COREIXAS_PRE_96_2017}, it can be concluded that both regularization steps recover the behavior of all aforementioned collision models. 

All in all, every collision model considered in the present work reduces to the BGK collision model, as far as the D1Q3 lattice is employed. The question is now to determine if the present conclusion remains valid in both 2D and 3D cases. To answer it, the $D$-dimensional extension based on the tensor product formulation will be adopted for both monomial~\cite{KARLIN_PA_389_2010}, and Hermite polynomial bases~\cite{COREIXAS_PhD_2018}. This step will allow us to properly understand which moments should be included in the derivation of populations, in the case of tensor product based LBMs such as the D2Q9 or the D3Q27 lattices (Secs.~\ref{sec:CompStudy2} and ~\ref{sec:CompStudy3} respectively).

\section{Theoretical comparisons using the D2Q9 lattice \label{sec:CompStudy2}}

\subsection{\label{sec:2DLBM} $D$-dimensional extension}
The $D$-dimensional extension is done following the tensor product rules proposed in Refs.~\cite{KARLIN_PA_389_2010,KARLIN_PTRSL_369_2011}. In the most general (3D) case, they read as
\begin{enumerate}
\item \emph{Lattice}
\begin{equation}
\xi_{(i,j,k)} = (\xi_{i},\xi_{j},\xi_{k}),
\label{eq:TPvel}
\end{equation}
\item \emph{Population}
\begin{equation}
f_{(i,j,k)} = \rho \phi_{i}\phi_{j}\phi_{k},
\label{eq:TPpop}
\end{equation}
with $\phi_{i}=f_{i}/\rho$, $\phi_{j}=f_{j}/\rho$, $\phi_{k}=f_{k}/\rho$,
\item \emph{Weight}
\begin{equation}
w_{(i,j,k)} = w_{i}w_{j}w_{k}.
\label{eq:TPweight}
\end{equation}
\end{enumerate}
If not otherwise stated, the $3$-tuple $(i,j,k)$ (or $2$-tuple $(i,j)$) will be used instead of the conventional single index $i$ to describe a discrete velocity and its associated population.

Before moving to the derivation of post collision VDFs in the 2D case, it is important to note that by following the above rules, it can be shown (App.~\ref{sec:EqFunction}) that all equilibrium states derived following these rules recover the exact same form whatever the polynomial basis used, and in any $D$-dimension. This important result flows from the fact that they already share the same equilibrium state in the 1D case, as demonstrated in Sec.~\ref{sec:CompStudy1}. 
  
Thanks to tensor product rules~(\ref{eq:TPvel}) and~(\ref{eq:TPpop}), populations evolving in the RM space now read as
\begin{align*}
f^{\mathrm{RM}}_{(0,0)} &= \rho \left(M_{00}-M_{20}\right)\left(M_{00}-M_{02}\right), \\
f^{\mathrm{RM}}_{(\sigma,0)} &= \dfrac{\rho}{2}\left(\sigma M_{10} + M_{20}\right) \left(M_{00}-M_{02}\right), \\
f^{\mathrm{RM}}_{(0,\lambda)} &= \dfrac{\rho}{2}\left(M_{00}-M_{20}\right)\left(\lambda M_{01} + M_{02}\right), \\
f^{\mathrm{RM}}_{(\sigma,\lambda)} &= \dfrac{\rho}{4}\left(\sigma M_{10} + M_{20}\right)\left(\lambda M_{01} + M_{02}\right),
\end{align*}
with $(\sigma,\lambda)\in\{-1,+1\}^2$. This leads to
\begin{subequations}
\begin{align}
f^{\mathrm{RM}}_{(0,0)} &= \rho \left(M_{00}-(M_{20}+M_{02}) + M_{22}\right), \\
f^{\mathrm{RM}}_{(\sigma,0)} &= \dfrac{\rho}{2}\left(\sigma M_{10} + M_{20} - \sigma M_{12} - M_{22}\right), \\
f^{\mathrm{RM}}_{(0,\lambda)} &= \dfrac{\rho}{2}\left(\lambda M_{01} + M_{02} - \lambda M_{21} - M_{22}\right), \\
f^{\mathrm{RM}}_{(\sigma,\lambda)} &= \dfrac{\rho}{4}\left(\sigma\lambda M_{11} + \sigma M_{12} + \lambda M_{21} + M_{22}\right),
\end{align}
\end{subequations}
where the isotropy of the VDF \textcolor{black}{itself} is enforced up to the second order (in each direction) imposing~\cite{KARLIN_PA_389_2010} $$M_{p}^{\mathrm{1D}}M_{q}^{\mathrm{1D}}=M_{p0}^{\mathrm{2D}}M_{0q}^{\mathrm{2D}}=M_{pq},$$ 
with $(p,q)\in\{0,1,2\}^2$.

\noindent Hence, post collision populations for the RM framework read as 
\begin{subequations}
\begin{align}
f^{*,\mathrm{RM}}_{(0,0)} &= \rho \left(M_{00}^*-(M_{20}^*+M_{02}^*) + M_{22}^*\right), \\
f^{*,\mathrm{RM}}_{(\sigma,0)} &= \dfrac{\rho}{2}\left(\sigma M_{10}^* + M_{20}^* - \sigma M_{12}^* - M_{22}^*\right), \\
f^{*,\mathrm{RM}}_{(0,\lambda)} &= \dfrac{\rho}{2}\left(\lambda M_{01}^* + M_{02}^* - \lambda M_{21}^* - M_{22}^*\right), \\
f^{*,\mathrm{RM}}_{(\sigma,\lambda)} &= \dfrac{\rho}{4}\left(\sigma\lambda M_{11}^* + \sigma M_{12}^* + \lambda M_{21}^* + M_{22}^*\right),
\end{align}
\label{eq:Q9postCollVDF_RM}
\end{subequations}
where post collision raw moments are defined as
\begin{align*}
M_{20}^*+M_{02}^*&=(1-\omega_{\nu_b})(M_{20}+M_{02}) + \omega_{\nu_b} (M_{20}^{eq}+M_{02}^{eq}),\\
M_{20}^*-M_{02}^*&=(1-\omega_{\nu})(M_{20}-M_{02}) + \omega_{\nu} (M_{20}^{eq}-M_{02}^{eq}),\\
M_{11}^*&=(1-\omega_{\nu})M_{11} + \omega_{\nu} M_{11}^{eq},\\
M_{21}^*&=(1-\omega_3)M_{21} + \omega_3 M_{21}^{eq},\\
M_{12}^*&=(1-\omega_3)M_{12} + \omega_3 M_{12}^{eq},\\
M_{22}^*&=(1-\omega_4)M_{22} + \omega_4 M_{22}^{eq},
\end{align*}
with $\nu_b$ being the bulk viscosity, and $\nu$ the kinematic viscosity. $\omega_3$ and $\omega_4$ are the relaxation frequencies associated to third and fourth order moments respectively. Furthermore, corresponding equilibrium moments are
\begin{subequations}
\begin{align}
M_{20}^{eq}+M_{02}^{eq} &= u_x^2+u_y^2+2 c_s^2,\\
M_{20}^{eq}-M_{02}^{eq} &= u_x^2-u_y^2,\\
M_{11}^{eq} &= u_x u_y,\\
M_{21}^{eq} &= (u_x^2+c_s^2) u_y,\\
M_{12}^{eq} &= u_x(u_y^2+c_s^2),\\
M_{22}^{eq} &= (u_x^2+c_s^2)(u_y^2+c_s^2).
\end{align}
\label{eq:EqMoments_RM_2D}
\end{subequations}
From this, the collision matrix cannot be diagonal if one requires $\nu_b\neq \nu$~\cite{KRUGER_Book_2017}. A simple remedy consists in replacing $(\xi_{i,x}^2,\xi_{i,y}^2)$ in $\mathcal{B}_{\mathrm{TP}}$ by $(\xi_{i,x}^2 + \xi_{i,y}^2,\xi_{i,x}^2 - \xi_{i,y}^2)$ as proposed in Ref.~\cite{BOSCH_ESAIM_52_2015}. 

\textcolor{black}{From a more general viewpoint, attention must be paid to the polynomial basis that is chosen since it may introduce extra computations as compared to the original one, and this overhead can become nonnegligible in the 3D case. As an example, this was investigated in Ref.~\cite{FEI_PRE_97_2018} by comparing the CPU times required by several CM-LBMs to simulate the very same configuration.}

From now on, let us determine if previous conclusions regarding the impact of the polynomial basis in 1D are still valid in the 2D case.

\subsection{\label{sec:ImpactOfMomentSpaceD2Q9} Impact of the moment space (D2Q9)}
Regarding Hermite, central and central Hermite moment spaces, their post collision populations are obtained following the very same tensor product rules as before (Eqs.~(\ref{eq:TPvel}) and~(\ref{eq:TPpop})). In the HM framework, one obtains
\begin{widetext}
\begin{subequations}
\begin{align}
\begin{split}
\label{eq:Q9postCollVDF_HM}
f^{*,\mathrm{HM}}_{(0, 0)} &= \rho[C^2- C (A^*_{20}+ A^*_{02})+ A^*_{22}],\\
f^{*,\mathrm{HM}}_{(\sigma, 0)}&= \frac{\rho}{2}[C (c_s^2+\sigma  u_x) +C A^*_{20}-c_s^2 A^*_{02}-\sigma  A^*_{12}-A^*_{22}],\\
f^{*,\mathrm{HM}}_{(0, \lambda)}&= \frac{\rho}{2}[C (c_s^2+\lambda  u_y)-c_s^2 A^*_{20}+C A^*_{02}-\lambda  A^*_{21}-A^*_{22}],\\
f^{*,\mathrm{HM}}_{(\sigma, \lambda)}&=\frac{\rho}{4}[c_s^2 (c_s^2+\sigma  u_x+\lambda  u_y)+\sigma \lambda A^*_{11}+\textcolor{black}{c_s^2( A^*_{20}+A^*_{02})}+\lambda  A^*_{21}+\sigma  A^*_{12}+A^*_{22}],\end{split}\\\notag\\
\intertext{while for the CM framework}
\begin{split}
\label{eq:Q9postCollVDF_CM}
f^{*,\mathrm{CM}}_{(0, 0)} &= \rho[U_x U_y+ 4 u_x u_y \widetilde{M}^*_{11} -U_y \widetilde{M}^*_{20}- U_x \widetilde{M}^*_{02}+ 2 u_y \widetilde{M}^*_{21}+ 2 u_x \widetilde{M}^*_{12}+ \widetilde{M}^*_{22}],\\
f^{*,\mathrm{CM}}_{(\sigma, 0)} &= \frac{\rho}{2}[u_x \sigma_x U_y-2 \sigma_{2x} u_y \widetilde{M}^*_{11}+U_y \widetilde{M}^*_{20}-u_x \sigma_x  \widetilde{M}^*_{02}-2 u_y \widetilde{M}^*_{21}-\sigma_{2x} \widetilde{M}^*_{12}-\widetilde{M}^*_{22}],\\
f^{*,\mathrm{CM}}_{(0, \lambda)} &= \frac{\rho}{2}[U_x u_y \lambda_y-2 u_x \lambda_{2y} \widetilde{M}^*_{11}-u_y \lambda_y \widetilde{M}^*_{20}+U_x \widetilde{M}^*_{02}-\lambda_{2y} \widetilde{M}^*_{21}-2 u_x \widetilde{M}^*_{12}-\widetilde{M}^*_{22}],\\
f^{*,\mathrm{CM}}_{(\sigma, \lambda)} &=\frac{\rho}{4}[u_x \sigma_x u_y \lambda_y+ \sigma_{2x} \lambda_{2y} \widetilde{M}^*_{11}+ u_y \lambda_y \widetilde{M}^*_{20}+ u_x \sigma_x  \widetilde{M}^*_{02}+ \lambda_{2y} \widetilde{M}^*_{21} + \sigma_{2x} \widetilde{M}^*_{12}+ \widetilde{M}^*_{22}],\end{split}\\\notag\\
\intertext{and for the CHM framework}
\begin{split}
\label{eq:Q9postCollVDF_CHM}
f^{*,\mathrm{CHM}}_{(0, 0)} &= \rho[C_x C_y+ 4 u_x u_y \widetilde{A}^*_{11}- C_y \widetilde{A}^*_{20}- C_x \widetilde{A}^*_{02}+ 2 u_y \widetilde{A}^*_{21}+ 2 u_x \widetilde{A}^*_{12}+ \widetilde{A}^*_{22}],\\
f^{*,\mathrm{CHM}}_{(\sigma, 0)} &= \frac{\rho}{2}[C_{\sigma}C_y-2 \sigma_{2x} u_y \widetilde{A}^*_{11}+C_y \widetilde{A}^*_{20}-C_{\sigma} \widetilde{A}^*_{02}-2 u_y \widetilde{A}^*_{21} -\sigma_{2x} \widetilde{A}^*_{12}-\widetilde{A}^*_{22}],\\
f^{*,\mathrm{CHM}}_{(0, \lambda)} &= \frac{\rho}{2}[C_x C_{\lambda}-2 u_x \lambda_{2y} \widetilde{A}^*_{11}-C_{\lambda} \widetilde{A}^*_{20}+C_x \widetilde{A}^*_{02}-\lambda_{2y} \widetilde{A}^*_{21}-2 u_x \widetilde{A}^*_{12}-\widetilde{A}^*_{22}],\\
f^{*,\mathrm{CHM}}_{(\sigma, \lambda)} &=\frac{\rho}{4}[C_{\sigma} C_{\lambda}+ \sigma_{2x} \lambda_{2y} \widetilde{A}^*_{11}+ C_{\lambda} \widetilde{A}^*_{20}+ C_{\sigma} \widetilde{A}^*_{02}+ \lambda_{2y} \widetilde{A}^*_{21}+ \sigma_{2x} \widetilde{A}^*_{12}+ \widetilde{A}^*_{22}],
\end{split}
\end{align}
\label{eq:}
\end{subequations}
\end{widetext}
with 
\begin{equation*}
\left\{
\begin{array}{l}
U_x = 1-u_x^2\\[0.1cm]
U_y = 1-u_y^2\\[0.1cm]
\sigma_x=\sigma + u_x\\[0.1cm]
\sigma_{2x}=\sigma + 2u_x\\[0.1cm]
\lambda_y=\lambda + u_y\\[0.1cm]
\lambda_{2y}=\lambda + 2u_y
\end{array}
\right. 
\quad\text{and} \quad
\left\{
\begin{array}{l}
C=1-c_s^2\\[0.1cm]
C_x=1-c_s^2-u_x^2\\[0.1cm]
C_y=1-c_s^2-u_y^2\\[0.1cm]
C_{\sigma}=c_s^2+u_x (\sigma +u_x)\\[0.1cm]
C_{\lambda}=c_s^2+u_y (\lambda +u_y)
\end{array}
\right..
\end{equation*}
While formulas expressed in RM and HM frameworks are rather simple, their counterparts in the comoving reference frame rapidly become lengthy. It is even worse in the 3D case, as detailed in App.~\ref{sec:3Dextension}. Populations based on the Gauss-Hermite quadrature do recover the very same expression as those expressed within the HM framework. Hence they will only be considered in Sec.~\ref{sec:D2Q9_Reg} to further study possible links between RR- and CHM-LBMs.

For the D2Q9 lattice, HM equilibrium moments necessary for the collision step are defined as,
\begin{equation*}
\begin{array}{l}
A_{20}^{eq}= u_x^2,A_{02}^{eq}= u_y^2,A_{11}^{eq} = u_x u_y,\\[0.1cm]
A_{21}^{eq}= u_x^2 u_y,A_{12}^{eq} = u_x u_y^2,A_{22}^{eq} = u_x^2 u_y^2,
\end{array}
\label{eq:EqMoments_HM_2D}
\end{equation*}
while CM formulas read as
\begin{equation*}
\widetilde{M}_{20}^{eq}=\widetilde{M}_{02}^{eq}= c_s^2,\: \widetilde{M}_{11}^{eq} = \widetilde{M}_{21}^{eq}= \widetilde{M}_{12}^{eq}= 0,\:\widetilde{M}_{22}^{eq} = c_s^4.
\label{eq:EqMoments_CM_2D}
\end{equation*}
Finally, expressions for the CHM framework are
\begin{equation*}
\widetilde{A}_{20}^{eq}=\widetilde{A}_{02}^{eq}= \widetilde{A}_{11}^{eq} = \widetilde{A}_{21}^{eq}= \widetilde{A}_{12}^{eq}= \widetilde{A}_{22}^{eq} = 0.
\label{eq:EqMoments_CHM_2D}
\end{equation*}

Now that all required formulas have been provided, let us start with the comparison between post collision VDFs obtained with both the RM and the HM frameworks. Considering the population at rest in the RM framework, 
\begin{align*}
f_{(0,0)}^{*,\mathrm{RM}} &= \rho [1-(M_{20}^*+M_{02}^*) + M_{22}^*]\\
         &= \rho [(1 -c_s^2)^2  - (1 - c_s^2)(A_{20}^*+A_{02}^*) + A_{22}^*]\\ 
         &\quad + \rho(\omega_{\nu_b}-\omega_4)(A_{20}^{neq}+A_{02}^{neq}),
\end{align*}
so that
\begin{subequations}
\label{eq:Comp_RM_HM}
\begin{align}
f_{(0,0)}^{*,\mathrm{RM}} &= f_{(0,0)}^{*,\mathrm{HM}} + \Delta f^{*,\mathrm{HM}}_{22},\\
\intertext{where $\Delta f^{*,\mathrm{HM}}_{22}$ corresponds to the deviation encountered during the relaxation of $M_{22}^*$, which eventually impacts the relaxation of both $A^*_{20}$ and $A^*_{02}$ in the HM framework. Similar results are also obtained with other populations,}
f_{(\sigma,0)}^{*,\mathrm{RM}} &= f_{(\sigma,0)}^{*,\mathrm{HM}} - \Delta f^{*,\mathrm{HM}}_{22}/2, \\
f_{(0,\lambda)}^{*,\mathrm{RM}} &= f_{(0,\lambda)}^{*,\mathrm{HM}} - \Delta f^{*,\mathrm{HM}}_{22}/2, \\
f_{(\sigma,\lambda)}^{*,\mathrm{RM}} &= f_{(\sigma,\lambda)}^{*,\mathrm{HM}} + \Delta f^{*,\mathrm{HM}}_{22}/4.
\end{align}
\end{subequations}
It flows from these comparisons that it is not possible anymore to freely switch between raw and Hermite moments. This is explained by the fact that $M^*_{22}$ is not only \textcolor{black}{linked} to $A^*_{22}$ but also to $A^*_{20}+A^*_{02}$, while these two kinds of moments are relaxed using two different frequencies, namely, $\omega_4$ and $\omega_{\nu_b}$. The equivalency between these two approaches is then lost when $\omega_4\neq\omega_{\nu_b}$. \textcolor{black}{One can further suppose that discrepancies between these approaches will tend to zero when phenomena linked to compressibility effects are negligible, i.e, $A^{neq}_{20}+A^{neq}_{02}\approx 0$. These deviations} cannot be observed through the comparison of pre collision or equilibrium VDFs, hence the use of post collision VDFs for the comparison of different frameworks. 

Since relationships between raw and Hermite moments are the same as those between central and central Hermite moments, it is known for sure that CM and CHM frameworks only merge \textcolor{black}{for the same reasons as before}.

Let us continue with the comparison between RM and CM frameworks. A similar computation shows that 
\begin{subequations}
\begin{equation}
f_{(0,0)}^{*,\mathrm{RM}} = f_{(0,0)}^{*,\mathrm{CM}} + \Delta f^{*,\mathrm{CM}}_{22},
\end{equation}
where now,
\begin{align*}
\Delta f^{*,\mathrm{CM}}_{22} &= \rho(\omega_3-\omega_4) (2u_y \widetilde{M}_{21}^{neq} + 2u_x \widetilde{M}_{12}^{neq})\\
&+ \rho(\omega_{\nu_b}-\omega_4) (u_x^2+u_y^2)(\widetilde{M}_{20}^{neq}+\widetilde{M}_{02}^{neq})/2\\
&+ \rho(\omega_{\nu}-\omega_4) (u_y^2-u_x^2)(\widetilde{M}_{20}^{neq}-\widetilde{M}_{02}^{neq})/2\\
&+ \rho(\omega_{\nu}-\omega_4) (4 u_x u_y)\widetilde{M}_{11}^{neq},
\end{align*}
since $M_{22}$ is linked to \textcolor{black}{all central moments} $\widetilde{M}_{22}$, $\widetilde{M}_{21}$, $\widetilde{M}_{12}$, $\widetilde{M}_{20}$ and $\widetilde{M}_{02}$ (see Eq.~(\ref{eq:CM2RM_Q9})). 
The relationships obtained for other populations are
\begin{align}
f_{(\sigma,0)}^{*,\mathrm{RM}} &= f_{(\sigma,0)}^{*,\mathrm{CM}} - \sigma\Delta f^{*,\mathrm{CM}}_{12}/4 - \Delta f^{*,\mathrm{CM}}_{22}/2,\\
f_{(0,\lambda)}^{*,\mathrm{RM}} &= f_{(0,\lambda)}^{*,\mathrm{CM}} -\lambda\Delta f^{*,\mathrm{CM}}_{21}/4 - \Delta f^{*,\mathrm{CM}}_{22}/2,\\
f_{(\sigma,\lambda)}^{*,\mathrm{RM}} &= f_{(\sigma,\lambda)}^{*,\mathrm{CM}}+( \sigma\Delta f^{*,\mathrm{CM}}_{12} + \lambda\Delta f^{*,\mathrm{CM}}_{21})/8 \notag\\
&\qquad\qquad\qquad\qquad\qquad\:\:+ \Delta f^{*,\mathrm{CM}}_{22}/4
\end{align}
\end{subequations}
with
\begin{widetext}
\begin{align*}
\Delta f^{*,\mathrm{CM}}_{12} &= \rho\big[u_x (\omega_3-\omega_{\nu_b})(\widetilde{M}_{20}^{neq}+\widetilde{M}_{02}^{neq})- u_x (\omega_3-\omega_{\nu})(\widetilde{M}_{20}^{neq}-\widetilde{M}_{02}^{neq})+4 u_y (\omega_3-\omega_{\nu})\widetilde{M}_{11}^{neq}\big],\\
\Delta f^{*,\mathrm{CM}}_{21} &= \rho\big[u_y (\omega_3-\omega_{\nu_b})(\widetilde{M}_{20}^{neq}+\widetilde{M}_{02}^{neq})+ u_y (\omega_3-\omega_{\nu})(\widetilde{M}_{20}^{neq}-\widetilde{M}_{02}^{neq}) +4 u_x (\omega_3-\omega_{\nu})\widetilde{M}_{11}^{neq}\big].
\end{align*}
\end{widetext}
As expected, even more deviations are present for other populations since they contain third order RM terms ($M_{21}$ and $M_{12}$) that will impact the relaxation of second order CM terms ($\widetilde{M}_{20}$ and $\widetilde{M}_{02}$).
It is then clear that RM and CM frameworks \emph{only} share the same behavior when $\omega_4=\omega_3=\omega_{\nu_b}=\omega_{\nu}$, and a fortiori, when the behavior of the BGK collision is recovered by both models. On the contrary, if a  MRT approach is adopted then these deviations can only be neglected in the case of a flow at rest ($u_x=u_y=0$), for which both frameworks merge as already mentioned in Sec.~\ref{sec:Rev_CM}. \textcolor{black}{Eventually, the case where $\widetilde{M}_{20}^{neq}+\widetilde{M}_{02}^{neq}$, $\widetilde{M}_{20}^{neq}-\widetilde{M}_{02}^{neq}$ and $\widetilde{M}_{11}^{neq}$ are simultaneously negligible is very restrictive, since it would imply that both compressibility and shear effects are also negligible. To the best of the authors' knowledge, this is very unlikely to happen. As a consequence, this possibility will not be considered in the rest of the paper.}

Once again, since relationships between Hermite and central Hermite moments are the same as those between raw and central moments, one can affirm that above conclusions can be extended to HM and CHM frameworks:
\begin{equation}
f_{i}^{*,\mathrm{HM}} = f_{i}^{*,\mathrm{CHM}}
\Longleftrightarrow \left\{
\begin{array}{l}
\omega_4=\omega_3=\omega_{\nu_b}=\omega_{\nu},\\
u_x=u_y=0\quad\mathrm{otherwise.}
\end{array}
\right.
\end{equation}

In summary, the choice of the moment space used for the D2Q9-LBM does have an impact on the resulting post collision VDFs, when either a MRT approach is considered, or the flow is not at rest. More precisely, the deviation between collision models expressed in either the reference frame at rest or the comoving reference frame are not constant, and they depend on the local flow velocity for MRT approaches. These discrepancies between models naturally emerge from the 2D representation, while it was not present with the D1Q3 lattice.

\subsection{Orthogonal MRT and TRT models \label{sec:D2Q9_MRT_LL}}

These two collision models are based on the expansion of populations over a polynomial basis built through the Gram-Schmidt orthogonalization procedure~(\ref{eq:MRTBasis}). The use of this basis leads to the following formulas for VDFs
\begin{subequations}
\label{eq:VDF_MRT_LL}
\begin{align}
f^{\mathrm{LL}}_{(0,0)} &=  \dfrac{1}{9}\left({\rho}-{e} + {\varepsilon}\right), \\
f^{\mathrm{LL}}_{(\sigma,0)} &= \dfrac{1}{36}\left(4 {\rho} + 6 \sigma {j_x} -{e} + 9 {p_{xx}}- 6 \sigma {q_x}-2 {\varepsilon}\right), \\
f^{\mathrm{LL}}_{(0,\lambda)} &= \dfrac{1}{36}\left(4 {\rho} + 6 \lambda {j_y} -{e} - 9 {p_{xx}}- 6 \lambda {q_y}-2 {\varepsilon}\right), \\
f^{\mathrm{LL}}_{(\sigma,\lambda)} &= \dfrac{1}{36}(4 {\rho} + 6 \sigma {j_x} + 6 \lambda {j_y} +2{e} + 9\sigma\lambda {p_{xy}}\notag\\
&\qquad\qquad+3 \sigma {q_x} +3 \lambda {q_y}+ {\varepsilon}),
\end{align}
\end{subequations}
where the superscript LL stands for Lallemand's and Luo's MRT model introduced in Ref.~\cite{LALLEMAND_PRE_61_2000}. Interestingly, it is possible to show that the equilibrium form of Eq.~(\ref{eq:VDF_MRT_LL}) equals $f_i^{eq,4}$ if one does not neglect high order velocity terms in the equilibrium moments
\begin{align*}
\rho^{eq} &= \rho,\: j_x^{eq} = \rho u_x,\: j_y^{eq} = \rho u_y,\: e^{eq} = \rho[3(u_x^2+u_y^2)-2],\\
p^{eq}_{xx} &= \rho(u_x^2-u_y^2), \: p^{eq}_{xy} = \rho u_x u_y,\: q_x^{eq} = \rho u_x(u_y^2-1),\\
q_y^{eq} &= \rho (u_x^2-1)u_y,\: \varepsilon^{eq} = \rho [1 -3(u_x^2+u_y^2) + 9 u_x^2 u_y^2].
\end{align*}  
By fixing all relaxation frequencies to $\omega_{\nu}$, the orthogonal LL-MRT recovers the behavior of the BGK-LBM based on $f_i^{eq,4}$. By extension, the orthogonal TRT-LBM also recovers its behavior fixing $\omega^+=\omega^-=\omega_{\nu}$. Thus, it is quite surprising that both models usually neglect $\mathcal{O}(u^3)$ and $\mathcal{O}(u^4)$ terms while the use of $f_i^{eq,4}$ might improve the linear stability of the resulting LBM, as already proven for the BGK operator and for both regularization steps in Ref.~\cite{COREIXAS_PhD_2018}.

The question is now to determine if these collision models are further related to other models based on different moment spaces. Rewriting LL moments within the HM framework,
\begin{align}
{\rho} &= \rho A_{00}, \:\:{j_x} = \rho A_{10}, \:\:{j_y} = \rho A_{01}, \nonumber\\
{p_{xx}} &= \rho A_{20}-\rho A_{02}, \:\:{p_{xy}} = \rho A_{11}, \nonumber\\
{e} &= 3(\rho A_{20}+\rho A_{02}) -2 \rho A_{00},\label{eq:HM2LL} \\
{q_x} &= 3 \rho A_{12} - \rho A_{10}, \:\:{q_y} = 3 \rho A_{21} - \rho A_{01}, \nonumber\\
{\varepsilon}&= 9 \rho A_{22} -3(\rho A_{20}+\rho A_{02}) + \rho A_{00}.\nonumber
\end{align}   
Thus the only issue that will be encountered with post collision VDFs will originates once again from the relaxation of the fourth order term ${\varepsilon}^*$. Indeed,
\begin{align*}
\varepsilon^* &= (1-\omega_{\varepsilon})\varepsilon + \omega_{\varepsilon}\varepsilon^{eq}\\
&= 9 \rho A_{22}^* -3(\rho A_{20}^*+\rho A_{02}^*) + \rho -\Delta f^{*,\mathrm{HM}}_{\varepsilon}
\end{align*}
with $\Delta f^{*,\mathrm{HM}}_{\varepsilon}=3 (\omega_{\varepsilon}-\omega_{\nu_b}) \rho(A_{20}^{neq}+ A_{02}^{neq})$ and $\omega_{\varepsilon}=\omega_4$ so that,
\begin{subequations}
\begin{align}
f_{(0,0)}^{*,\mathrm{LL}} &= f_{(0,0)}^{*,\mathrm{HM}} - \Delta f^{*,\mathrm{HM}}_{\varepsilon}/9, \\
f_{(\sigma,0)}^{*,\mathrm{LL}} &= f_{(\sigma,0)}^{*,\mathrm{HM}} + \Delta f^{*,\mathrm{HM}}_{\varepsilon}/18, \\
f_{(0,\lambda)}^{*,\mathrm{LL}} &= f_{(0,\lambda)}^{*,\mathrm{HM}} + \Delta f^{*,\mathrm{HM}}_{\varepsilon}/18, \\
f_{(\sigma,\lambda)}^{*,\mathrm{LL}} &= f_{(\sigma,\lambda)}^{*,\mathrm{HM}} - \Delta f^{*,\mathrm{HM}}_{\varepsilon}/36.
\end{align}
\end{subequations}
As for RM and HM frameworks, the LL-MRT-LBM reduces to the HM-MRT-LBM if the collision step of the fourth order and the bulk moments follow the very same relaxation rate, i.e, $\omega_{\varepsilon}=\omega_{\nu_b}$. \textcolor{black}{Once again, the same result is obtained when compressibility effects are negligible ($A_{20}^{neq}+ A_{02}^{neq}\approx 0$)}. One can derive the very same constraints for the LL-MRT-LBM to recover the behavior of the RM-LBM. In the same spirit, one can further come to the conclusion that LL and CM frameworks merge if a SRT collision model is employed, since it reduces both approaches to the BGK-LBM. In the case of a MRT collision model, the LL-LBM reduces to the CM-LBM in the zero velocity limit, as it was also shown in the context of the RM collision model~(\ref{eq:Comp_RM_HM}). In the particular case of the orthogonal TRT approach, the relaxation of all odd (or even) moments are linked together. Hence, the condition $\omega_{\varepsilon}=\omega_{\nu_b}$ is always satisfied for this collision model. In addition, the behavior of the BGK collision model is also recovered when $\omega^+=\omega^-=\omega_{\nu}$.
 
\subsection{Regularized collision models\label{sec:D2Q9_Reg}}

The interesting thing about the Gauss-Hermite quadrature is that it remains valid whatever the number of dimensions or the lattice considered~\cite{GRADb_CPAM_2_1949,SHAN_PRL_80_1998,SHAN_JFM_550_2006}. Nevertheless, it was not used (in its original form) to define which high order components should be included in the definition of populations. In the context of the D2Q9 lattice, the original basis used for the polynomial expansion was
$$\mathcal{B}_{\mathrm{Q9}}=(H_{i,00},H_{i,10},H_{i,01},H_{i,11},H_{i,20},H_{i,02}).$$ 
One can notice that $\mathcal{B}_{\mathrm{Q9}}$ is not a `basis' from the mathematical viewpoint since it does not contain nine elements. In order to improve the description of populations, it was proposed by several authors to include three more elements~\cite{DELLAR_PRE_65_2002,DELLAR_JCP_259_2014,MALASPINAS_ARXIV_2015,COREIXAS_PRE_96_2017}. While the first approaches added these elements through a Gram-Schmidt orthogonalization procedure, the last two relied on Hermite polynomials that were orthogonal with respect to those belonging to $\mathcal{B}_{\mathrm{Q9}}$. \textcolor{black}{Nonetheless, they all end up with the same \emph{complete} (and true) basis}
$$\mathcal{B}_{\mathrm{Q9}}^{\mathrm{complete}}=\mathcal{B}_{\mathrm{Q9}} \cup (H_{i,21},H_{i,12},H_{i,22}),$$
that corresponds to $\mathcal{B}_{\mathrm{HTP}}$~(\ref{eq:HTPBasis})\textcolor{black}{, as explained in Ref.~\cite{ADHIKARI_PRE_78_2008}.}
This basis is the starting point to compare all kinds of regularization steps against each other. Let us first consider projection based regularized (PR) models in both the reference frame at rest~\cite{LATT_ARXIV_2005,LATT_MCS_72_2006a,CHEN_PA_362_2006} 
\begin{align*}
f_i = w_i &\bigg[a_{00} H_{i,00}+\dfrac{1}{c_s^2}(a_{10} H_{i,10}+a_{01} H_{i,01})\\
&+\dfrac{1}{2c_s^4} (a_{20} H_{i,20} + a_{02} H_{i,02} +2 a_{11} H_{i,11})\\
&+\dfrac{1}{6c_s^6} (a_{21} H_{i,21} + a_{12} H_{i,12})+\dfrac{1}{24c_s^8} a_{22} H_{i,22}\bigg],
\end{align*}
and in the comoving reference frame where central Hermite coefficients are defined as~\cite{CHEN_PATENT_Collision_2015,MATTILA_PF_29_2017} 
\begin{align}
\widetilde{a}_{nm}=\sum_i f_i \widetilde{H}_{i,nm},
\end{align}
so that Hermite coefficients are now computed using relationships between Hermite and central Hermite moments (App.~\ref{sec:RelationshipsBivariate}). This leads to
\begin{align}
a_{00} &= \rho,\: a_{10} = \rho u_x,\: a_{01} = \rho u_y,\notag\\
a_{20} &= \widetilde{a}_{20} + \rho u_x^2,\: a_{02} = \widetilde{a}_{02} + \rho u_y^2,\: a_{11} = \widetilde{a}_{11} + \rho u_x u_y,\notag\\
a_{21} &= \widetilde{a}_{21} +u_y \widetilde{a}_{20} +2 u_x \widetilde{a}_{11} + \rho u_x^2 u_y,\notag\\
a_{12} &= \widetilde{a}_{12} +u_x \widetilde{a}_{02} +2 u_y \widetilde{a}_{11} + \rho u_x u_y^2,\label{eq:CHM_PR}\\
a_{22} &= \widetilde{a}_{22} +2 u_y \widetilde{a}_{21}+2 u_x \widetilde{a}_{12} +  u_y^2 \widetilde{a}_{20} +  u_x^2 \widetilde{a}_{02}\notag\\
&\quad + 4 u_x u_y \widetilde{a}_{11} +  \rho u_x^2 u_y^2.\notag
\end{align} 
where $ \widetilde{a}_{00}=\rho$ and $ \widetilde{a}_{10}= \widetilde{a}_{01}=0$ have been assumed. Restricting ourselves to the isothermal case, equilibrium coefficients are then computed as
\begin{align*}
a^{eq}_{00} &= \rho,\quad a^{eq}_{10} = \rho u_x,\quad a^{eq}_{01} = \rho u_y,\\
a^{eq}_{20} &= \rho u_x^2,\quad a^{eq}_{02} = \rho u_y^2, \quad a^{eq}_{11} = \rho u_x u_y,\\
a^{eq}_{21} &= \rho u_x^2 u_y,\quad a^{eq}_{12} = \rho u_x u_y^2,\quad a^{eq}_{22} = \rho u_x^2 u_y^2,
\end{align*} 
while nonequilibrium coefficients are
\begin{align*}
a^{neq}_{00} &= 0,\:\:a^{neq}_{10} = 0,\:\:a^{neq}_{01} = 0,\\
a^{neq}_{20} &= \widetilde{a}^{neq}_{20},\:\:a^{neq}_{02} = \widetilde{a}^{neq}_{02},\:\:a^{neq}_{11} = \widetilde{a}^{neq}_{11},\\
a^{neq}_{21} &= \widetilde{a}^{neq}_{21} +u_y \widetilde{a}^{neq}_{20} +2 u_x \widetilde{a}^{neq}_{11},\\
a^{neq}_{12} &= \widetilde{a}^{neq}_{12} +u_x \widetilde{a}^{neq}_{02} +2 u_y \widetilde{a}^{neq}_{11},\\
a^{neq}_{22} &= \widetilde{a}^{neq}_{22} +2 u_y \widetilde{a}^{neq}_{21}+2 u_x \widetilde{a}^{neq}_{12} +  u_y^2 \widetilde{a}^{neq}_{20} +  u_x^2 \widetilde{a}^{neq}_{02}\\
&\quad + 4 u_x u_y \widetilde{a}^{neq}_{11}.
\end{align*}
In Ref.~\cite{MATTILA_PF_29_2017}, the authors suggested to discard part of nonequilibrium (diffusive) terms $\widetilde{a}^{neq}$ based on the fact that these terms should be negligible for high Reynolds number flows. Doing so, they noticed that by imposing $\widetilde{a}^{neq}_{22}=\widetilde{a}^{neq}_{21}=\widetilde{a}^{neq}_{12}=0$, they were able to recover the recursive formulas of the RR approach~\cite{MALASPINAS_ARXIV_2015}, 
\begin{align}
a^{(1)}_{21} &= u_y a^{neq}_{20}+ 2 u_x a^{neq}_{11},\quad a^{(1)}_{12} = u_x a^{neq}_{02}+ 2 u_y a^{neq}_{11},\notag\\
a^{(1)}_{22} &= 2 (u_y a^{(1)}_{21}+  u_x a^{(1)}_{12}) -u_y^2 a^{neq}_{20}-u_x^2 a^{neq}_{02} - 4 u_x u_y a^{neq}_{11}\notag\\
&= u_y^2 a^{neq}_{20} + u_x^2 a^{neq}_{02} + 4 u_x u_y a^{neq}_{11}.
\label{eq:RRformulas_D2Q9}
\end{align}
where $a^{neq}_{20}$, $a^{neq}_{02}$ and $a^{neq}_{11}$ are computed using Eq.~(\ref{eq:HermiteCoeffPR}).
While this result was obtained in an adhoc manner, it can be shown that it originates from particular values of relaxation times in the CHM framework. Indeed, post collision central Hermite coefficients are expressed as
\begin{equation}
\widetilde{a}^*_{pq}=\widetilde{a}^{eq}_{pq}+\bigg(1-\dfrac{1}{\overline{\tau}_{pq}}\bigg)\widetilde{a}^{neq}_{pq},
\label{eq:CHM_PR_Coll}
\end{equation}
where \textcolor{black}{$\overline{\tau}_{pq}=1/2+\tau_{pq}$ is the discrete} relaxation time, with $(p,q)\in\{0,1,2\}^2$ for the D2Q9 lattice. Considering $\overline{\tau}_{21}=\overline{\tau}_{12}=\overline{\tau}_{22}=1$, corresponding coefficients are fixed to their equilibrium values. In the CHM framework, all equilibrium moments are zero with the exception of $\widetilde{a}^{eq}_{00}$ (see Sec.~\ref{sec:StatGaussian}). Hence, RR formulas~(\ref{eq:RRformulas_D2Q9}) are indeed recovered when Eqs.~(\ref{eq:CHM_PR}) and~(\ref{eq:CHM_PR_Coll}) are employed to compute $a^*_{pq}$ with this particular set of relaxation times. The SRT-RR-LBM then corresponds to a PR approach in the comoving reference frame where high order ($p+q\geq 3$) nonequilibrium contributions are filtered out.

Furthermore, \textcolor{black}{both the MRT-RR-LBM (with $\overline{\tau}_{\nu_b}^{\mathrm{RR}}=1$) and the cascaded based LBM share the same behavior when $\overline{\tau}_{\nu_b}^{\mathrm{CM}}=\overline{\tau}_{21}^{\mathrm{CM}}=\overline{\tau}_{12}^{\mathrm{CM}}=\overline{\tau}_{22}^{\mathrm{CM}}=1$. This result is pertinent if and only if the reader has no interest in acoustically related phenomena.} 

Eventually, since it is mandatory to use several relaxation times in the CHM framework to recover $f_i^{*,\mathrm{RR}}$, then the RR collision model cannot recover the behavior of the BGK collision operator. Hence the MRT-RR procedure does belong to \emph{another category} of collision models.

\subsection{Cumulant space \label{sec:D2Q9_Cumul}}

Here, the discrepancies between collision steps occurring in the CM and the cumulant spaces are investigated. As demonstrated in Apps.~\ref{sec:RelationshipsBivariate} and~\ref{sec:RelationshipsTrivariate}, cumulants are not equivalent anymore to central moments when high order terms ($n\geq 4$) have to be taken into account in the expansion of populations. In the context of the D2Q9 lattice, it has been shown that one should include the fourth order term $\widetilde{M}_{22}$. If one follows the definition of post collision cumulants, as they are defined in Refs~\cite{GEIER_CMA_70_2015,GEIER_JCP_348_2017a,GEIER_JCP_348_2017b}, one obtains
$$\widetilde{M}_{22}^*  =  K_{22}^* + K_{20}^*K_{02}^* + 2(K_{11}^*)^2.$$
Post collision populations then read as
\begin{widetext}
\begin{align}
\begin{split}
f^{*,\mathrm{K}}_{(0, 0)} &= \rho[U_x U_y+ 4 u_x u_y K^*_{11} -U_y K^*_{20}- U_x K^*_{02}+ 2 u_y K^*_{21}+ 2 u_x K^*_{12}+ 2(K^*_{11})^2+ K^*_{20}K^*_{02} + K^*_{22}],\\
f^{*,\mathrm{K}}_{(\sigma, 0)} &= \frac{\rho}{2}[u_x \sigma_xU_y-2 \sigma_{2x} u_y K^*_{11}+U_y K^*_{20}-u_x \sigma_x K^*_{02}-2 u_y K^*_{21}-\sigma_{2x} K^*_{12}-2(K^*_{11})^2 - K^*_{20}K^*_{02} - K^*_{22}],\\
f^{*,\mathrm{K}}_{(0, \lambda)} &= \frac{\rho}{2}[U_x u_y \lambda_y-2 u_x \lambda_{2y} K^*_{11}-u_y \lambda_y K^*_{20}+U_x K^*_{02}-\lambda_{2y} K^*_{21}-2 u_x K^*_{12}-2(K^*_{11})^2- K^*_{20}K^*_{02} - K^*_{22}],\\
f^{*,\mathrm{K}}_{(\sigma, \lambda)} &=\frac{\rho}{4}[u_x u_y \sigma_x \lambda_y+ \sigma_{2x} \lambda_{2y} K^*_{11}+ u_y \lambda_y K^*_{20}+ u_x \sigma_x K^*_{02}+ \lambda_{2y} K^*_{21}+ \sigma_{2x} K^*_{12} + 2(K^*_{11})^2+ K^*_{20}K^*_{02} + K^*_{22}].
\end{split}
\end{align}
\end{widetext}
with $U_x = 1-u_x^2$,
$\sigma_x=\sigma + u_x$,
$\sigma_{2x}=\sigma + 2u_x$,
$U_y = 1-u_y^2$,
$\lambda_y=\lambda + u_y$ and
$\lambda_{2y}=\lambda + 2u_y$.
From this, one can see that deviations from their CM counterparts come from the fact that
\begin{align*}
\widetilde{M}_{22}^* &= (1-\omega_4)\widetilde{M}_{22} + \omega_4 \widetilde{M}_{22}^{eq}\\
							  &= (1-\omega_4)(K_{22} + K_{20}K_{02} + 2K_{11}^2)\\
							 &\quad + \omega_4 (K_{22}^{eq} + K_{20}^{eq}K_{02}^{eq} + 2(K_{11}^{eq})^2)
\end{align*}
while
\begin{align*}
K_{22}^* &= (1-\omega_4) K_{22}+\omega_4  K_{22}^{eq},\\
(K_{11}^*)^2 &= (1-\omega_{\nu})^2 K_{11}^2 + \omega_{\nu}(1-\omega_{\nu}) K_{11}K_{11}^{eq}+\omega_{\nu}^2  (K_{11}^{eq})^2,\\
K_{20}^*K_{02}^* &= (1-\omega_{\nu})^2 K_{20}K_{02}+\omega_{\nu}^2  K_{20}^{eq}K_{02}^{eq} \\
&\quad+ \omega_{\nu}(1-\omega_{\nu}) ( K_{20}K_{02}^{eq}+ K_{20}^{eq}K_{02}),
\end{align*}
where for the sake of simplicity the relaxation frequency of $K_{11}$, $K_{20}$ and $K_{02}$ was taken as $\omega_{\nu}$. The latter assumption amounts to impose $\omega_{\nu_b}=\omega_{\nu}$~\cite{KRUGER_Book_2017}. 

Eventually, post collision populations can be rewritten as
\begin{subequations}
\begin{align}
f_{(0,0)}^{*,\mathrm{CM}} &= f_{(0,0)}^{*,\mathrm{K}} + \Delta f^{*,\mathrm{K}}_{22}, \\
f_{(\sigma,0)}^{*,\mathrm{CM}} &= f_{(\sigma,0)}^{*,\mathrm{K}} - \Delta f^{*,\mathrm{K}}_{22}/2, \\
f_{(0,\lambda)}^{*,\mathrm{CM}} &= f_{(0,\lambda)}^{*,\mathrm{K}} - \Delta f^{*,\mathrm{K}}_{22}/2, \\
f_{(\sigma,\lambda)}^{*,\mathrm{CM}} &= f_{(\sigma,\lambda)}^{*,\mathrm{K}} + \Delta f^{*,\mathrm{K}}_{22}/4,
\end{align}
\end{subequations}
with
\begin{align*}
\Delta f^{*,\mathrm{K}}_{22} &= \rho[(1-\omega_4) - (1-\omega_{\nu})^2](K_{20}K_{02} + 2K_{11}^2) \\
&\quad - \rho\omega_{\nu}(1-\omega_{\nu})(K_{20}K_{02}^{eq} +K_{20}^{eq}K_{02} + 2K_{11}^{eq}K_{11}) \\
&\quad + \rho(\omega_4-\omega_{\nu})[K_{20}^{eq}K_{02}^{eq} + 2(K_{11}^{eq})^2].
\end{align*}
From this, one needs to impose $\omega_4=\omega_{\nu}=1$ to make the deviation $\Delta f^{*,\mathrm{K}}_{22}$ disappear. In that particular case, the kinematic viscosity cannot be chosen freely. Hence, it is \emph{not possible} to recover the behavior of the BGK collision operator using cumulants, as already assumed in Refs.~\cite{GEIER_CMA_70_2015,GEIER_JCP_348_2017a,GEIER_JCP_348_2017b}. As a consequence, the K-LBM belongs to \emph{another} category of LBMs, \emph{different} from those presented in Secs.~\ref{sec:ImpactOfMomentSpaceD2Q9}-\ref{sec:D2Q9_Reg}. 

\textcolor{black}{Before moving to the investigation of 3D models, the interested reader may refer to the Supplemental material~\bibnotemark[SupMatQ9] for more information regarding the way to implement D2Q9-LBMs in either their RM, HM, CM, CHM, K or RR formulation. Thanks to them, one can further derive instructions for orthogonal models (LL and TRT) by switching from HM to LL moments using Eq.~(\ref{eq:HM2LL}).}

\section{Further investigations \label{sec:CompStudy3}}

This section is dedicated to several studies. First, collision models will be compared in the 3D case considering both the D3Q27 and the D3Q19 lattices. Second, the impact of collision models on the resulting macroscopic equations will be investigated.

\subsection{D3Q27-LBMs}

The most straightforward way to build LBMs based on the D3Q27 lattice is to consider it as a tensor product of three D1Q3 lattices (one for each spatial direction). Doing so, it is possible to rely on rules defined in Eqs.~(\ref{eq:TPvel})-(\ref{eq:TPweight}). This approach then leads to populations that are expanded over tensor product like polynomial bases. For raw moments,
\begin{align*}
\mathcal{B}_{\mathrm{TP}}^{\mathrm{Q27}} = &(1,\xi_{100},\xi_{010},\xi_{001},\xi_{200},\xi_{020},\xi_{002},\xi_{110},\xi_{101},\xi_{011},\\
&\:\: \xi_{210},\xi_{120},\xi_{201},\xi_{102},\xi_{021},\xi_{012},\xi_{111},\xi_{220},\xi_{202},\\
&\:\: \xi_{022},\xi_{211},\xi_{121},\xi_{112},\xi_{221},\xi_{212},\xi_{122},\xi_{222}),
\label{eq:TPBasis_Q27}
\end{align*}
using the shorthand notation $\xi_{pqr}=\xi_{i,x}^p\xi_{i,y}^q\xi_{i,z}^r$. For other moment spaces, one just needs to change from monomials ($\xi_{pqr}$) to either Hermite polynomials ($H_{i,pqr}$), central monomials ($\widetilde{\xi}_{pqr}$) or central Hermite polynomials ($\widetilde{H}_{i,pqr}$). One can notice that by using the tensor product rules a true basis is obtained, i.e, it contains the same number of elements as the number of velocities in the lattice. In addition, one can prove that once again population obtained through the Gauss-hermite quadrature do share the same expression as those obtained via the tensor product rules.

Knowing the basis corresponding to each framework, one can easily derive post collision populations for the D3Q27 lattice (App.~\ref{sec:3Dextension}). The comparison of collision models follow the same steps as those presented in both the 1D and the 2D cases, with the exception that now calculations are more complex. Taking the example of the RM framework, post collision populations are
\begin{widetext}
\begin{subequations}
\begin{align}
f^{*,\mathrm{RM}}_{(0, 0, 0)}&=\rho [1-M^*_{200}-M^*_{020}-M^*_{002}+M^*_{220}+M^*_{202}+M^*_{022}-M^*_{222}],\\
f^{*,\mathrm{RM}}_{(\sigma, 0, 0)}&= \dfrac{\rho}{2} [ \sigma  u_x +M^*_{200} -\sigma  M^*_{120} -\sigma  M^*_{102} -M^*_{220} -M^*_{202} +\sigma  M^*_{122} +M^*_{222}],\\
f^{*,\mathrm{RM}}_{(0, \lambda, 0)}&= \dfrac{\rho}{2} [\lambda  u_y +M^*_{020} -\lambda  M^*_{210} -\lambda  M^*_{012} -M^*_{220} -M^*_{022} +\lambda  M^*_{212} +M^*_{222}],\\
f^{*,\mathrm{RM}}_{(0, 0, \delta)}&= \dfrac{\rho}{2} [  \delta  u_z +M^*_{002} -\delta  M^*_{201} -\delta  M^*_{021} -M^*_{202} -M^*_{022} +\delta  M^*_{221} +M^*_{222} ],\\
f^{*,\mathrm{RM}}_{(\sigma, \lambda, 0)}&= \dfrac{\rho}{4} [ \sigma\lambda  M^*_{110}+\lambda  M^*_{210}+\sigma  M^*_{120}-\sigma\lambda    M^*_{112}+M^*_{220}-\lambda  M^*_{212}-\sigma  M^*_{122}-M^*_{222}  ],\\
f^{*,\mathrm{RM}}_{(\sigma, 0, \delta)}&= \dfrac{\rho}{4} [\sigma\delta  M^*_{101} +\delta  M^*_{201} +\sigma  M^*_{102} -\sigma \delta M^*_{121} +M^*_{202} -\delta  M^*_{221} -\sigma  M^*_{122} -M^*_{222}],\\
f^{*,\mathrm{RM}}_{(0, \lambda, \delta)}&= \dfrac{\rho}{4} [\lambda\delta  M^*_{011} +\delta  M^*_{021} +\lambda  M^*_{012} -\lambda\delta  M^*_{211} +M^*_{022} -\delta  M^*_{221} -\lambda  M^*_{212} -M^*_{222}],\\
f^{*,\mathrm{RM}}_{(\sigma, \lambda, \delta)}&= \dfrac{\rho}{8} [\sigma\lambda\delta M^*_{111}+ \lambda\delta  M^*_{211}+\sigma\delta  M^*_{121}+\sigma\lambda  M^*_{112}+\delta  M^*_{221}+\lambda  M^*_{212}+\sigma  M^*_{122}+M^*_{222}],
\end{align}
\label{eq:VDFs_Q27_RM}
\end{subequations}
\end{widetext}
with $(\sigma,\lambda,\delta)\in\{-1,+1\}^3$. Before moving to the comparison itself, one must define the number of relaxation frequencies required for the collision step. To do so, one simply needs to categorize all types of moments employed in Eq.~(\ref{eq:VDFs_Q27_RM}). As an example, \textcolor{black}{one can suppose} that $M_{210}$ and all its cyclic permutations ($M_{201},M_{120},M_{102},M_{021}$ and $M_{012}$) should be relaxed to their equilibrium value via the very same relaxation frequency, \textcolor{black}{the latter being different for the relaxation of $M_{111}$}. Consequently, one possible set of relaxation frequencies \textcolor{black}{would be}

\begin{equation}
 \begin{array}{c @{\quad \longrightarrow \quad} c}
 M^*_{200},M^*_{020},M^*_{002} &  \omega_{1},\\
 M^*_{110},M^*_{101},M^*_{011} &  \omega_{2},\\
 M^*_{210},M^*_{201},M^*_{120},M^*_{102},M^*_{021},M^*_{012} &  \omega_{3},\\
 M^*_{111} &  \omega_{4},\\
 M^*_{220}, M^*_{202}, M^*_{022} &  \omega_{5},\\
 M^*_{211}, M^*_{121}, M^*_{112} &  \omega_{6},\\
 M^*_{221}, M^*_{212}, M^*_{122} &  \omega_{7},\\
 M^*_{222} &  \omega_{8},
 \end{array}
 \label{eq:omegaD3Q27}
 \end{equation} 
\textcolor{black}{which is a particular case of the collision matrix adopted in Ref.~\cite{FEI_PRE_97_2018}. Here, $\omega_{\nu_b}=\omega_{\nu}$ is assumed, and this corresponds to $s_+=s_2$ and $s_-=0$ in their framework. This is done for the sake of simplicity, and more importantly, without significantly impacting the conclusions that are drawn below.}

Hereafter, the comparative study will be simplified by starting from $f^{*,\mathrm{RM}}_i$~(\ref{eq:VDFs_Q27_RM}), and using relationships between each kind of statistical quantities (App.~\ref{sec:RelationshipsTrivariate}) to derive deviations between each type of collision models. If one considers the HM framework, deviations with the RM approach come from fourth and higher order moments. Considering the following moments
\begin{subequations}
\begin{align}
M_{220} &= A_{220} + c_s^2 (A_{200} + A_{020}) + c_s^4, \label{eq:subEq1}\\
M_{211} &= A_{211} + c_s^2 A_{011}, \label{eq:subEq2}\\
M_{221} &= A_{221} + c_s^2(A_{201}+A_{021}) + c_s^4 u_z,  \label{eq:subEq3}\\
M_{222} &= A_{222} + c_s^2(A_{220} + A_{202} + A_{022}) \notag\\
			&\quad+ c_s^4(A_{200}+A_{020}+A_{002}) + c_s^6,  \label{eq:subEq4}
\end{align}
\end{subequations}
is then sufficient to determine all discrepancies between both families of moments. \textcolor{black}{As previously discussed for the D2Q9 lattice, if one discards flows where compressibility and shear phenomena are negligible, then} formulas on fourth order moments~(\ref{eq:subEq1} and~\ref{eq:subEq2}) lead to the constraints $\omega_{5}=\omega_{1}$ and $\omega_{6}=\omega_{2}$ to enforce the equivalence between RM and HM frameworks. Expressions for fifth and six order moments~(\ref{eq:subEq3} and~\ref{eq:subEq4}) further implies that $\omega_{7}=\omega_{3}$ and $\omega_{8}=\omega_{5}=\omega_{1}$, \textcolor{black}{since it is quite difficult to derive further constraints based on the physical meaning of these moments, at least in a straightforward manner. Considering that} $\omega_{1}$ and $\omega_{2}$ are related to the dissipation of acoustic and shear waves, only $\omega_{3}$ and $\omega_{4}$ are free parameters. As for the D2Q9 lattice, both RM and HM frameworks do recover the behavior of the BGK collision operator when only one relaxation frequency \textcolor{black}{is used}. 
The very same conclusions can be drawn regarding CM and CHM frameworks.

For RM and CM frameworks, the relaxation of each raw moment $M_{pqr}$ impacts the relaxation of all lower order moments $\widetilde{M}_{p'q'r'}$ such as $p'\leq p$, $q'\leq q$ and $r'\leq r$. Hence, the configuration for which both approaches merge corresponds to the use of a single relaxation frequency (BGK). One can further show that if a MRT approach is considered, then both frameworks recover the same behavior if and only if the flow is at rest. Once again, one comes to the same conclusions for HM and CHM frameworks. 

Orthogonal MRT and TRT collision models are based on the Gram-Schmidt orthogonalization procedure~\cite{SUGA_CMA_69_2015,FAKHARI_JCP_341_2017}. The latter does not strongly impact the relationships between families of moments since it relies on linear transformations. Hence, it can be shown that above conclusions remain valid for orthogonal MRT collision models. Once again, orthogonal TRT-LBMs recover the behavior of the HM-LBMs since the relaxations of even moments are controlled by the same relaxation frequency. If one further imposes $\omega^+=\omega^-=\omega_{\nu}$ then the orthogonal TRT approach mimics the behavior of the BGK operator, as it was already the case for the D2Q9 lattice.

When it comes to the PR collision model, it leads to different results when different moment spaces are considered, since it relies on the equilibriation of higher than second order moments ($\omega_3=\omega_4=\omega_5=\omega_6=\omega_7=\omega_8=1$). The question is then to determine if it is still linked to the RR approach when central Hermite moments are employed for the relaxation process. Using relationships between central Hermite and Hermite moments (Sec.~\ref{sec:RelationshipsTrivariate}), one can show that the PR collision model applied in the CHM framework recovers the behavior of the SRT-RR-LBM. Indeed, by following the same steps as in Sec.~\ref{sec:D2Q9_Reg}, one obtains
\begin{subequations}
\begin{align}
a^{(1)}_{210} &= u_y a^{neq}_{200}+ 2 u_x a^{neq}_{110},\\
a^{(1)}_{111} &= u_z a^{neq}_{110}+ u_y a^{neq}_{101}+ u_x a^{neq}_{011},\\
a^{(1)}_{220} &= u_y^2 a^{neq}_{200} + u_x^2 a^{neq}_{020} + 4 u_x u_y a^{neq}_{110},\\
a^{(1)}_{211} &= u_y u_z a^{neq}_{200} + 2 u_x u_z a^{neq}_{110} + 2 u_x u_y a^{neq}_{101} + u_x^2 a^{neq}_{011},\\
a^{(1)}_{221} &= u_y^2 u_z a^{neq}_{200} + u_x^2 u_z a^{neq}_{020} + 4 u_x u_y u_z a^{neq}_{110} + 2 u_x u_y^2 a^{neq}_{101}\notag\\
&\quad + 2 u_x^2 u_y a^{neq}_{011},\\
a^{(1)}_{222} &= u_y^2 u_z^2 a^{neq}_{200} + u_x^2 u_z^2 a^{neq}_{020} + u_x^2 u_y^2 a^{neq}_{002}+ 4 u_x u_y u_z^2 a^{neq}_{110}\notag\\
&\quad  + 4 u_x u_y^2 u_z a^{neq}_{101} + 4 u_x^2 u_y u_z a^{neq}_{011}
\end{align}
\label{eq:RRformulas_D3Q27}
\end{subequations}
where other formulas are derived through cyclic permutations. The above relationships do match recursive formulas flowing from the Chapman-Enskog expansion at the Navier-Stokes-Fourier level~\cite{MALASPINAS_ARXIV_2015,COREIXAS_PRE_96_2017}. Nevertheless, the equivalence is once again lost when several relaxation times are employed for the RR collision model.

The last comparison is related to cumulants and central moments. In the 2D case, it was shown that the deviation was only impacting the sole fourth order term,  while now fourth, fifth and six order terms are impacted. In addition, the deviation was directly proportional to the square of the relaxation frequency, whereas it is now proportional to up to the cube of the relaxation frequency. Due to these \emph{nonlinear} deviations, it is \emph{not possible} to link CM- and K-LBMs. 

In summary, no further links can be derived between collision models for the D3Q27 lattice as compared to the 2D case. This was to be expected from the way the D3Q27 lattice is built using tensor product rules~(\ref{eq:TPvel} and~\ref{eq:TPpop}). Nevertheless, these rules do not hold anymore in the case of the D3Q19 velocity discretization. Consequently, one may wonder if the aforementioned conclusions are still valid for this particular LBM.

\subsection{D3Q19-LBMs \label{sec:D3Q19}}

\begin{table*}[hbtp]
\renewcommand{\arraystretch}{1.25}
\renewcommand{\tabcolsep}{1mm}
\begin{tabular}{| C | C | C C | C C |  C C | C C C | C |}
\hline
\diagbox[width=3.1em]{X}{Y} & BGK & MRT & TRT & RM & HM & CM & CHM & PR & RR & $\text{RR}^*$ & K  \\
\hline
BGK &  -- &   $\bm{\sim}$ &  $\bm{\sim}$ &  $\bm{\sim}$ &  $\bm{\sim}$ &  $\bm{\sim}$ &  $\bm{\sim}$ & \ding{111} & \ding{111} &  \ding{111} &  \ding{111} \\ \hline
MRT &  \ding{51} &   -- &  \ding{51} & $\bm{\sim}$ & $\bm{\sim}$ & $\bm{\sim}$ & $\bm{\sim}$ & \ding{111} & \ding{111} &  \ding{111} &  \ding{111} \\
TRT &  \ding{51} &   $\bm{\sim}$ &  -- & $\bm{\sim}$ & $\bm{\sim}$ &  $\bm{\sim}$ &  $\bm{\sim}$ & \ding{111} & \ding{111} &  \ding{111} &  \ding{111} \\  \hline 
RM &  \ding{51} &   $\bm{\sim}$ & \ding{51} &  -- & $\bm{\sim}$ &  $\bm{\sim}$ & $\bm{\sim}$ & \ding{111} & \ding{111} &  \ding{111} &  \ding{111} \\ 
HM &  \ding{51} & $\bm{\sim}$ &  \ding{51} &  $\bm{\sim}$ & -- &  $\bm{\sim}$ & $\bm{\sim}$ & \ding{51} & \ding{111} &  \ding{111} &  \ding{111} \\  \hline 
CM &  \ding{51} &   $\bm{\sim}$ &  $\bm{\sim}$ & $\bm{\sim}$ &   $\bm{\sim}$ & -- & $\bm{\sim}$ & \ding{111} &  \ding{111} &  \ding{111} &  \ding{111} \\ 
CHM &  \ding{51} &  $\bm{\sim}$ & $\bm{\sim}$ &  $\bm{\sim}$ &   $\bm{\sim}$ & $\bm{\sim}$ & -- & \ding{111} & \ding{51} &  \ding{111} &  \ding{111}  \\ \hline 
PR &  \ding{111} &  \ding{111} & \ding{111}  & \ding{111} &  $\bm{\sim}$ & \ding{111} & \ding{111} & -- & \ding{111} &  \ding{111} &  \ding{111} \\
RR &  \ding{111} &  \ding{111} & \ding{111}  & \ding{111} &  \ding{111} & \ding{111} & $\bm{\sim}$ & \ding{111} & -- &  $\bm{\sim}$ &  \ding{111} \\ \hline
$\text{RR}^*$ &  \ding{111} &  \ding{111} & \ding{111}  & \ding{111} &  \ding{111} & \ding{111} & \ding{111} & \ding{111} & \ding{51} & --  &  \ding{111} \\ \hline
K &  \ding{111} &  \ding{111} & \ding{111}  & \ding{111} &  \ding{111} & \ding{111} & \ding{111} & \ding{111} &  \ding{111} & \ding{111}  &  -- \\
\hline
\end{tabular}
\caption{Can X recover the behavior of Y by using a particular set of relaxation frequencies? Yes (\ding{51}), Partially ($\bm{\sim}$), No (\ding{111}). Collision models considered in this summary are based on: raw moments (RM), Hermite moments (HM), central moments (CM), central Hermite moments (CHM), both projection based (PR) and recursive regularization (RR) steps, cumulants (K). MRT and TRT models corresponds to collision models expressed within an orthogonal basis derived from the Gram-Schmidt orthogonalization procedure. $\text{RR}^*$ is the multirelaxation time version of the RR collision model. Only links that do not impact the resulting physics ($\omega_{\nu}\neq 1$ and $\omega_{\nu_b}\neq 1$) are reported here. Furthermore, a collision model X will be considered to \emph{partially} recover the behavior of another one Y, if the condition to enforce $\Delta f^*=0$  implies that at least \textcolor{black}{of one of} $\omega_n^Y$ cannot be chosen freely anymore, with the most constraining case being the reduction to a single relaxation time approach. Eventually, all reported results are valid for both the D2Q9 and the D3Q27 lattices. They further remain valid for the D3Q19 lattice if and only if one compares collision models relying on the same equilibrium state.}
\label{tab:Links}
\end{table*}

To derive populations for the D3Q19 lattice, it is proposed to start from the D3Q27-LBM and then to discard particular terms that are not compliant with the velocity discretization using a `pruning method'~\cite{KARLIN_PA_389_2010,KARLIN_PTRSL_369_2011}. The latter is of particular interest since it also allows us to determine which moments should be included in the expansion of populations. Hereafter, another way to choose the number of moments that are necessary for the D3Q19 lattice is presented. 

It has been known for a long time that the projection of either the D3Q15 or the D3Q19 lattices, onto the 2D velocity space, leads to the D2Q9 lattice~\cite{KRUGER_Book_2017}. Nonetheless, only the D3Q19 velocity discretization contains all the discrete velocities of the D2Q9 lattice in each of its planes $(x,y)$, $(x,z)$ and $(y,z)$. Hence all moments encountered in the definition of populations in the 2D case should also be accounted for in the present case. In other words, $\mathcal{B}^{\mathrm{Q19}}$ should, at least, contain all the 3D versions of monomials encountered in Eq.~(\ref{eq:TPBasis}). By 3D version, it is meant that, for example, $\xi_{120}$ should be included since $\xi_{12}$ belongs to the 2D basis. 
Starting from Eq.~(\ref{eq:TPBasis}) and considering all cyclic permutations, one ends up with the following polynomial basis
\begin{align*}
\mathcal{B}_{\mathrm{Q19}} = &(1,\xi_{100},\xi_{010},\xi_{001},\xi_{200},\xi_{020},\xi_{002},\xi_{110},\xi_{101},\xi_{011},\\
&\:\: \xi_{210},\xi_{120},\xi_{201},\xi_{102},\xi_{021},\xi_{012},\xi_{220},\xi_{202},\xi_{022}),
\end{align*} 
expressed in the RM framework, and where the shorthand notation $\xi_{pqr}=\xi_{i,x}^p\xi_{i,y}^q\xi_{i,z}^r$ was used for the sake of clarity. This simple yet rigorous reasoning leads to a polynomial basis composed of nineteen elements. 
Thus one can simply move from D3Q27-LBMs to their D3Q19-LBMs discarding $\xi_{111}$, $\xi_{211}$, $\xi_{121}$, $\xi_{112}$, $\xi_{221}$, $\xi_{212}$, $\xi_{122}$, $\xi_{222}$ and then using relationships detailed in App.~\ref{sec:RelationshipsTrivariate}. Doing so, it is possible to extend several collision models to the D3Q19-LBM even if they were originally developed within the framework of the D3Q27 velocity discretization\textcolor{black}{, such as both RR- and K-LBMs (see App.~\ref{sec:3Dextension}). In addition}, this basis was recently used to derive a more efficient CM-LBM based on the D3Q19 lattice~\cite{FEI_PRE_97_2018}. \textcolor{black}{Nonetheless, it should be noted that, at the time of writing, the accuracy of this D3Q19 formulation as not yet been compared to its D3Q27 counterpart.}

Using the above methodology, one can see that $\mathcal{B}^{\mathrm{Q19}}$ still contains fourth order terms. Consequently, all links that have been drawn for the D3Q27 lattice remain valid for the D3Q19 velocity discretization. One just needs to be careful and only compare collision models based on the same equilibrium state (see App.~\ref{sec:EqFunction} for both choices obtained with the D3Q19 lattice).

\textcolor{black}{General instructions for the coding of 3D models, as well as pseudo codes of the D3Q27 formulations, are provided in Supplemental materials~\bibnotemark[SupMat] and~\bibnotemark[SupMatQ27] respectively.}

\subsection{Partial conclusions}
Tab.~\ref{tab:Links} summarizes all derived links between collision models. More precisely, it details all collision models whose behavior can be entirely or partially recovered using another collision model with a particular set of collision frequencies. This is done considering only cases where the resulting physics is not impacted by the choice of the relaxation frequencies. Taking the example of the D2Q9 lattice, if one applies the collision step within the CM framework then one can \emph{entirely} recover the results of the BGK collision model using only one relaxation frequency $\omega^{\mathrm{CM}}_{\nu}$. Besides, the CM-LBM can \emph{partially} recover the behavior of the CHM-LBM if one imposes that $\omega^{\mathrm{CM}}_{\nu_b}=\omega^{\mathrm{CM}}_{4}=\omega^{\mathrm{CHM}}_{4}=\omega^{\mathrm{CHM}}_{\nu_b}$. In other words, this implies that the bulk viscosity cannot be chosen independently of the relaxation coefficient of fourth order moments. The number of free parameters is then reduced, hence the \emph{partial} recovery of the CHM-LBM behavior by the CM-LBM. The most constraining case is the reduction to a single relaxation time approach, where the behaviors of RM, HM, orthogonal MRT and TRT models are also recovered. Eventually, it would be possible to get the same results as those obtained with the (SRT) RR approach by imposing $\omega^{\mathrm{CM}}_{\nu_b}=\omega^{\mathrm{RR}}_{\nu_b}=1$, but this would lead to an extreme overdissipation of acoustic waves~\cite{MARIE_JCP_228_2009,XU_JCP_230_2011,GENDRE_PRE_96_2017}. Thus, this link is not considered in Tab.~\ref{tab:Links}.

\subsection{Macroscopic behavior\label{sec:MacroEq}}
To conclude this comparative study, let us have a look at macroscopic equations flowing from all the different collision models.

\subsubsection{Motivations\label{sec:Motiv}}
\textcolor{black}{Previous investigations suggest that the choice of moment space has a major impact on the macroscopic behavior of LBMs, and more specifically on their Galilean invariance properties (e.g., Refs.~\cite{GEIER_PRE_73_2006,GEIER_EPJST_171_2009,PREMNATH_JSP_143_2011,GEIER_CMA_70_2015,NING_IJNMF_82_2016,GEIER_JCP_348_2017a,GEIER_CF_166_2018} among others). Knowing that these collision models naturally rely on an extended equilibrium state, one might wonder if the improved macroscopic behavior of these models come from either (1) high-order velocity terms of their equilibrium state, (2) the moment space, or (3) both of them. With this idea in mind, the hydrodynamic limit of LBMs will be studied in a general manner, meaning that the asymptotic study will be conducted before the (space-time) numerical discretization of the collision model. This is explained by the fact that one would have to properly distinguish errors resulting from every possible discretization technique (finite difference, finite volume, finite element, discontinuous Galerkin, etc) in order to \emph{objectively} quantify the impact of each collision model on the resulting macroscopic behavior, and this is out of the scope of the present work.
\\ 
\indent In the following, a brief review on the origin of errors encountered in the context of LBMs is first proposed, whereas the investigation of the macroscopic behavior of LBMs is conducted in the second part of this section.}

\subsubsection{Velocity and space-time discretization errors\label{sec:MacroNum}}
\textcolor{black}{The LBM relies on two types of discretization, namely, the velocity and the space-time discretizations. 
To correctly make the distinction between errors that emerge from both of them, let us recall the two main steps that are required for the design of any numerical scheme, and a fortiori for the derivation of LBMs.
\\
\indent The first step consists in selecting the mathematical model corresponding to the desired level of approximation to reality~\cite{HIRSCH_Book_2007}. In the following, and for the sake of simplicity, let us start from the force-free Boltzmann equation (BE),
\begin{equation}
\partial_t f + \bm{\nabla}\cdot(f \bm{\xi}) =\mathcal{C},
\label{eq:BE}
\end{equation}
where $\mathcal{C}$ is the general form of the collision term. The physics governed by Eq.~(\ref{eq:BE}) goes far beyond the validity of standard macroscopic equations of interest, namely, the compressible Navier-Stokes-Fourier equations~\cite{HUANG_Book_2nd_1987}. As a consequence, this mathematical model is too detailed for the level of approximation to reality that is required in the present framework. To simplify it, a physical discretization is applied to the BE. It consists in drastically reducing the number of possible velocities $\xi$ for the propagation of populations $f$. This is done in such a way that the physics of interest is not lost during the overall process. One then ends up with the discrete velocity Boltzmann equation, also known as lattice Boltzmann equation (LBE),
\begin{equation}
\forall i\in\llbracket 1,V\rrbracket,\quad \partial_t f_i + \bm{\nabla}\cdot(f_i \bm{\xi}_i) =\mathcal{C}_i,
\label{eq:LBE}
\end{equation}
with $V$ being the number of discrete velocities required to recover the macroscopic behavior of interest~\cite{SHAN_JFM_550_2006}.
\\
\indent Once this set of partial differential equations has been chosen, the second step consists in choosing a type of space and time discretization in order to be able to numerically solve it~\cite{HIRSCH_Book_2007}. Among the wide panel of numerical discretizations available in the literature, the most commonly used approach is the `Collide and Stream' algorithm (Eqs.~(\ref{eq:CollisionStep}) and~(\ref{eq:StreamingStep})).
\\
\indent To determine the hydrodynamic limit of the LBM, one can either start from the LBE or from its numerical discretization -- the LBM itself. Hence at least two types of asymptotic study are possible. The first methodology is the most commonly used in the lattice Boltzmann community, as it naturally flows from Statistical Physics, and it is based on the Chapman-Enskog expansion of the LBE~\cite{CHAPMAN_Book_3rd_1970,HUANG_Book_2nd_1987}. The second approach is more generally used for the evaluation of numerical errors introduced during the space and time discretization of a given set of equations. It is based on the Taylor expansion of these discretized equations, and it allows the user to rigorously evaluate the order of accuracy of the numerical scheme of interest~\cite{HIRSCH_Book_2007}. In the lattice Boltzmann framework, the Taylor expansion is applied to the LBM itself by considering either a diffusive~\cite{JUNK_JCP_210_2005} or an acoustic~\cite{DUBOIS_CMA_55_2008} scaling. While the former leads to the derivation of error terms with respect to the incompressible Navier-Stokes equations, the latter derives them in the context of their weakly compressible formulation. 
\\
\indent Since the Chapman-Enskog expansion is conducted before the space-time discretization of the LBE, results obtained from this asymptotic study remain valid whatever the numerical discretization considered (finite difference, finite volume, finite element, discontinuous Galerkin, etc). On the contrary, the Taylor expansion leads to the derivation of error terms that flow from both velocity and space-time discretizations. Hence, it is not possible to determine in a straightforward manner which discretization is related to error terms obtained with this expansion. For all of these reasons, the Chapman-Enskog expansion of the LBE seems to be the best approach to draw general conclusions about the hydrodynamic limit of LBMs. Results obtained hereafter will then flow from this asymptotic study. For the sake of completeness, it is also worth noting that the Chapman-Enskog could also be applied to the LBM itself, in order to further derive numerical errors introduced by the space-time discretization~\cite{OTOMO_PA_486_2017,GENDRE_PRE_96_2017}.
}

\subsubsection{Macroscopic equations}

In the present context, only second order velocity discretizations (D1Q3, D2Q9, D3Q19 and D3Q27) are considered.
Let us start with their second order equilibrium state~\cite{QIAN_EPL_17_1992,SHAN_JFM_550_2006},
$$f_i^{eq,2} = \rho w_i \bigg[1+ \dfrac{\bm{\xi}_i\cdot \bm{u}}{c_s^2} + \dfrac{(\bm{\xi}_i\cdot \bm{u})^2}{2 c_s^4} - \dfrac{u^2}{2 c_s^2}\bigg].$$
The corresponding (isothermal) macroscopic equations recovered through the Chapman-Enskog expansion~\cite{CHAPMAN_Book_3rd_1970} are
\begin{equation}
\begin{array}{c}
\partial_t(\rho) + \bm{\nabla}\cdot(\rho\bm{u})=0, \vspace{0.1cm}\\
\partial_t(\rho\bm{u}) + \bm{\nabla}\cdot(\rho\bm{u}^2)= - \bm{\nabla}p + \bm{\nabla}\cdot(\bm{\Pi}'),
\end{array}
\label{Eq:NavierStokes}
\end{equation}
where $ \bm{\Pi}' =\mu \left[\bm{\nabla}\bm{u}+(\bm{\nabla}\bm{u})^T\right] +\mathcal{O}(\mathrm{Ma}^3)$ is the viscous stress tensor with the \textcolor{black}{isothermal and} weakly compressible limitation. The superscript $T$ stands for the transpose operator and $\mu = \tau p$ is the dynamic viscosity. For the D1Q3 lattice, the compressibility error reads as
$$\Delta\Pi_{xx}=\Pi_{xx}-\Pi'_{xx} = \tau \partial_x u_x^3.$$
In the case of the D2Q9 lattice, each component of $\bm{\Pi}$ is affected as follows
\begin{align*}
\Delta\Pi_{xx} &= \tau [\partial_x (u_x^3)+\partial_y (u_x u_y^2)],\\
\Delta\Pi_{xy} &= \Delta\Pi_{yx} = \tau [\partial_x (u_x^2 u_y)+\partial_y (u_x u_y^2)],\\
\Delta\Pi_{yy} &= \tau [\partial_y (u_y^3)+\partial_x (u_x^2 u_y)].
\end{align*}
When it comes to both the D3Q19 and the D3Q27 lattice, even more error terms are obtained
\begin{align*}
\Delta\Pi_{xx} &= \tau [\partial_x (u_x^3)+\partial_y (u_x u_y^2)+\partial_z (u_x u_z^2)],\\
\Delta\Pi_{xy} &= \Delta\Pi_{yx} = \tau [\partial_x (u_x^2 u_y)+\partial_y (u_x u_y^2)+\partial_z (u_x u_y u_z)],\\
\Delta\Pi_{xz} &= \Delta\Pi_{zx} = \tau [\partial_x (u_x^2 u_z)+\partial_y (u_x u_y u_z)+\partial_z (u_x u_z^2)],\\
\Delta\Pi_{yy} &= \tau [\partial_y (u_y^3)+\partial_x (u_x^2 u_y)+\partial_z (u_y u_z^2)],\\
\Delta\Pi_{yz} &= \Delta\Pi_{zy} = \tau [\partial_x (u_x u_y u_z)+\partial_y (u_y^2 u_z)+\partial_z (u_y u_z^2)],\\
\Delta\Pi_{zz} &= \tau [\partial_z (u_z^3)+\partial_x (u_x^2 u_z)+\partial_y (u_y^2 u_z)].
\end{align*}
It is important to note that the general form of the above error terms is \emph{independent} of the collision model framework. Only the truncation order of the equilibrium state $f_i^{eq}$ does have an impact on the compressibility error terms, \textcolor{black}{since~\cite{KRUGER_Book_2017,COREIXAS_PhD_2018}
$$
\dfrac{\Pi_{\alpha\beta}}{\tau} = \partial_t \bigg(\sum_i \xi_{i,\alpha}\xi_{i,\beta} f_i^{eq}\bigg) + \partial_{\gamma}\bigg(\sum_i \xi_{i,\alpha}\xi_{i,\beta}\xi_{\gamma} f_i^{eq}\bigg).
$$}

\noindent It is not possible to get rid of third order terms proportional to $u_{\alpha}^3$ ($\alpha=x,y,z$), due to the aliasing defect $\xi_{i,\alpha}^3=\xi_{i,\alpha}$ that is specific to second order velocity discretizations. Nonetheless, by taking into account third order moments compliant with these velocity discretizations, one can discard the influence of all non diagonal error terms as already pointed out in several studies~\cite{DELLAR_PRE_65_2002,HAZI_JPA_39_2006,DELLAR_JCP_259_2014,MALASPINAS_ARXIV_2015,COREIXAS_PRE_96_2017,HUANG_PRE_97_2018,DEROSIS_PRE_99_2019,DEROSIS_PRE_Submitted_2019}.
This result is valid for both D2Q9 and D3Q27 \textcolor{black}{lattices}, as long as, third order velocity dependent terms are not discarded in the definition of their equilibrium state. For the D3Q19 lattice, error terms proportional to $\partial_z (u_x u_y u_z)$ are still present since $\xi_{i,111}$ is not taken into account in $\mathcal{B}_{\mathrm{Q19}}$. The interested reader may refer to App.~\ref{sec:EqFunction} where full forms of equilibrium states are compiled.

\textcolor{black}{Regarding the impact of the moment space used for the collision process, it is important to understand that all moment spaces, even the cumulant one when fourth- and higher-order cumulants are discarded, can be related to each other through linear transformation matrices (see App.~\ref{sec:LinearTransMatrix}). In this context, the only difference between the macroscopic behavior of LBMs comes from their \emph{original} equilibrium state. As compared to standard collision models (BGK, MRT, PR, etc) which \emph{originally} relied on a second-order equilibrium state, the CM-, CHM-, K- and RR-LBMs were derived using extended ones that include high-order velocity terms. It is then only natural that the latter models reduce the number of velocity dependent error terms present in the viscous stress tensor. Nonetheless, the present reasoning also suggests that by keeping high-order velocity terms, even for standard collision models, then one would also improve the macroscopic behavior of the resulting LBM. This was discussed by several authors~\cite{DELLAR_PRE_65_2002,HAZI_JPA_39_2006,DELLAR_JCP_259_2014,MALASPINAS_ARXIV_2015,COREIXAS_PRE_96_2017,HUANG_PRE_97_2018,DEROSIS_PRE_99_2019,DEROSIS_PRE_Submitted_2019}, and further confirmed through the linear stability analysis of several LBMs~\cite{COREIXAS_PhD_2018}.}

Consequently, \emph{only} the equilibrium state, which depends on the lattice (and not on the collision model), does impact the resulting macroscopic equations recovered by the isothermal and weakly compressible LBMs. \textcolor{black}{For the sake of fairness, however, it is also important to understand that by changing the collision model then one also modifies the \emph{numerical} properties of the LBM. This is explained by the fact that the only error introduced by the numerical discretization comes from the collision term~\cite{DELLAR_CMA_65_2013,KRUGER_Book_2017}. Hence, one can show that the \emph{numerical} behavior of LBMs is drastically impacted by the choice of both the moment space and the relaxation parameters~\cite{DHUMIERE_CMA_58_2009,GEIER_CMA_70_2015,GEIER_JCP_348_2017a,GEIER_CF_166_2018}. Nevertheless, these errors should not be attributed to a physical problem, such as the Galilean invariance issue, but rather to a purely numerical defect, in order not to mislead the reader on this particularly complex topic.}
%
%

\section{\label{sec:Conclusion}CONCLUSION}

The BGK-LBM has been really successful during the last decades due to both its efficiency and accuracy. Nevertheless, this single relaxation time (SRT) collision model is known to suffer from severe stability issues during the simulation of high Reynolds number flows. To circumvent this deficiency numerous collision models has been developed. They can be classified according to the number of relaxation times they rely on, and via the moment space used for the derivation of populations. Most of the time, all these collision models are presented in their own framework with only very few suggestions regarding their possible link with already known models. While authors meanly concentrate on the comparison of collision models through a list of numerical test cases, they seldomly explain observed discrepancies from a theoretical viewpoint.

In this context, the present work focused on the understanding of fundamental differences between the most common collision models. To do so, an extensive search of links between these collision operators was performed in the case of standard LBMs of increasing complexity (D1Q3, D2Q9, D3Q19 and D3Q27 lattices). After drawing relationships between all moment spaces (raw (RM), central (CM),  Hermite (HM), central Hermite (CHM), cumulants (K)) through the use of their corresponding moment generating function, a thorough review of collision models was conducted. The latter drew a first picture of known links between collision models and further showed that all these collision models can be rewritten in a linear matrix form with the exception of the K-LBM. 

In a general way, it was also demonstrated that all collision models recover the very same second, fourth and six order equilibrium states when the D1Q3, the D2Q9 and the D3Q27 lattices are respectively employed for the velocity discretization of the Boltzmann equation. Nonetheless, the use of the D3Q19 lattice led to two different equilibrium states because of its non compliance with tensor product rules. Interestingly, the use of these extended equilibrium states might improve the linear stability of the corresponding LBMs, as recently demonstrated for both BGK and regularized D2Q9-LBMs~\cite{COREIXAS_PhD_2018}. 

\textcolor{black}{Regarding the mathematical comparison of collision models,} while it was not possible to find discrepancies between them in the one-dimensional case, deviations started appearing in the two-dimensional case. Using the D2Q9 lattice and with the assumption of a SRT operator, all models recovered the behavior of the BGK-LBM but both RR- and K-LBMs. Using several relaxation times, a partial equivalency between raw and Hermite frameworks was obtained in both the reference frame at rest (RM and HM) and in the comoving reference frame (CM and CHM), when the collision of bulk and fourth order moments were sharing the same relaxation frequency ($\omega_{\nu_b}=\omega_4$). Still using a multirelaxation time (MRT) approach, it was confirmed that raw and central frameworks can only share the same behavior if the simulated flow is at rest, i.e, imposing $u_x=u_y=0$. 

In addition, the projection based regularization (PR) step was recasted in a collision step occurring in the HM framework, where nonhydrodynamic contributions are actually filtered out imposing $\omega^{\mathrm{HM}}_3=\omega^{\mathrm{HM}}_4=1$. Most importantly, the SRT-RR-LBM was reinterpreted as an extension of the PR approach to the comoving reference frame (CHM space) where third and fourth order contributions were also discarded ($\omega^{\mathrm{CHM}}_3=\omega^{\mathrm{CHM}}_4=1$). Imposing $\omega_{\nu_b}^{\mathrm{CM}}=\omega_{3}^{\mathrm{CM}}=\omega_{4}^{\mathrm{CM}}=1$, the cascaded-LBM was further demonstrated to correspond to the MRT-RR-LBM where $\omega_{\nu_b}^{\mathrm{RR}}=1$. The latter condition is viable if and only if one is not interested in the simulation of acoustically related phenomena. This might explain why the cascaded collision model was usually employed with these collision frequencies in order to simulate incompressible flows. In conclusion, the most general formulation of the MRT-RR-LBM was shown to belong to a completely different kind of collision models. 

Discrepancies between orthogonal (MRT and TRT) and nonorthogonal approaches (RM and HM frameworks) were also highlighted. For the orthogonal MRT, they were shown to originate from the orthogonalization procedure, which is based on the construction of an orthogonal basis through linear combinations of monomials, eventually leading to spurious entanglements between moments of different orders. From this perspective, the orthogonalization procedure seemed to induce more issues than it solved. \textcolor{black}{The orthogonal version of the TRT-LBM was shown to be able to cancel out these spurious entanglements.}

A deviation between the CM- and the K-LBMs appeared due to the inclusion of the fourth order central moment in the expansion of post collision populations. This discrepancy originates from the nonlinear relaxation of cumulants related to this central moment, eventually leading to a radically different behavior as compared to all aforementioned collision models.

The above results were further confirmed for the D3Q27 lattice where the number of deviations increased due to the presence of more high order moments. Guidelines concerning the extension of all collision models to the D3Q19 lattice were also provided. In particular, these explanations make possible the derivation of both RR- and K-LBMs in a straightforward manner (App.~\ref{sec:3Dextension}). The comparative study then led to the very same conclusions as before due to the presence of fourth order contributions in the expansion of populations. Nevertheless, equilibrium states obtained through Gauss-Hermite quadrature and tensor product rules were shown to not be equivalent anymore, meaning that only collision models based on the same equilibrium state should be compared to each other. 

It was finally shown that the reduction of error terms in the macroscopic equations recovered by the LBM -- those originating from the limited accuracy of standard lattices and not from the numerical discretization -- was only due to the form of the equilibrium state and not to moment space employed for the collision process.

\textcolor{black}{For the sake of completeness, three Supplemental materials are also provided to help the reader with the coding of collision models considered in the present work. While the general methodology is discussed in~\bibnotemark[SupMat], pseudo codes dedicated to the coding of the D2Q9 and the D3Q27 formulations are provided in~\bibnotemark[SupMatQ9] and~\bibnotemark[SupMatQ27] respectively.} 

Regarding future works, it is planned to further compare collision models through the evaluation of their linear stability domain. This study will include both results on eigenvalues~\cite{COREIXAS_PhD_2018} and eigenvectors~\cite{WISSOCQ_JCP_380_2019} of corresponding linearized LBMs. Not only will it allow the proper classification of collision models based on their stability domain, but it will also lead to a better understanding of the stabilization mechanism specific to each model. These linear stability analysis should also help to: \textcolor{black}{(1) better quantify errors introduced by the numerical discretization of the collision term, and (2) find optimal values of collision frequencies to further improve the stability of LBMs in the low-viscosity regime}. Collision models offering the best trade off between stability and accuracy will then be compared to dynamic models through both academic and realistic configurations.

In parallel to this, it is also planned to continue the work initiated in Ref.~\cite{COREIXAS_PhD_2018}. This should lead to further extensions of all aforementioned collision models to high order LBMs, for which the simulation of compressible flows induces even more severe stability issues.

\begin{acknowledgments}
Fruitful discussions with G. Wissocq, T. Astoul, F. Renard, D. Ricot, F. Gendre and O. Malaspinas are gratefully acknowledged. The authors would also like to thank the reviewers for their pertinent remarks and insightful questions. This work was supported by the Swiss National Science
Fund SNF through the Sinergia project 170930, `A 3D Cell-Based Simulation Framework for
Morphogenetic Problems'. All mathematical derivations were conducted using `Mathematica' software. 
\end{acknowledgments}  
  
\appendix

\section{\textcolor{black}{Lattices} \label{sec:Lattices}}

Henceforth, the main characteristics of standard velocity discretizations of interest are compiled. Fig.~\ref{fig:Lattices} contains the representation of each velocity discretization, whereas Tab.~\ref{tab:Lattices} gives further information regarding their velocity sets $\bm{\xi}_i$, their associated Gauss-Hermite weights $w_i$ and their lattice constant $c_s$.  
\begin{figure}[h!]
\centering
\begin{tikzpicture}[scale=1,baseline={(current bounding box.center)}]
\draw (2,0,0)--(2,2,0)--(0,2,0); 
\draw [gray,dashed] (0,0,0)--(2,0,0);
\draw [gray,dashed] (0,0,0)--(0,2,0);
\draw [gray,dashed] (0,0,0)--(0,0,2);
\draw (2,0,0) -- (2,0,2);
\draw (0,2,0) -- (0,2,2);
\draw [dotted,fill=gray, opacity=0.2] (1,0,0)--(1,2,0)--(1,2,2)--(1,0,2)--cycle; 
\draw [dotted,fill=red, opacity=0.2] (0,0,1)--(0,2,1)--(2,2,1)--(2,0,1)--cycle; 
\draw [dotted,fill=gray, opacity=0.2] (0,1,0)--(0,1,2)--(2,1,2)--(2,1,0)--cycle; 
\begin{scope}[shift={(1,1,1)}]
\draw (0.003,-0.01) node {$\bullet$};
\draw [line width=0.4mm,>=stealth, ->] (0,0,0) -- (1,1,-1);
\draw [line width=0.4mm,>=stealth, ->] (0,0,0) -- (-1,1,-1);
\draw [line width=0.4mm,>=stealth, ->] (0,0,0) -- (-1,-1,-1);
\draw [line width=0.4mm,>=stealth, ->] (0,0,0) -- (1,-1,-1);
\draw [line width=0.4mm,>=stealth, ->] (0,0,0) -- (1,1,1);
\draw [line width=0.4mm,>=stealth, ->] (0,0,0) -- (-1,1,1);
\draw [line width=0.4mm,>=stealth, ->] (0,0,0) -- (-1,-1,1);
\draw [line width=0.4mm,>=stealth, ->] (0,0,0) -- (1,-1,1);
\draw [red,line width=0.4mm,>=stealth, ->] (0,0,0) -- (1,1,0);
\draw [red,line width=0.4mm,>=stealth, ->] (0,0,0) -- (-1,1,0);
\draw [red,line width=0.4mm,>=stealth, ->] (0,0,0) -- (-1,-1,0);
\draw [red,line width=0.4mm,>=stealth, ->] (0,0,0) -- (1,-1,0);
\draw [line width=0.4mm,,>=stealth, ->] (0,0,0) -- (1,0,1);
\draw [line width=0.4mm,>=stealth, ->] (0,0,0) -- (0,1,1);
\draw [line width=0.4mm,>=stealth, ->] (0,0,0) -- (-1,0,1);
\draw [line width=0.4mm,>=stealth, ->] (0,0,0) -- (0,-1,1);
\draw [line width=0.4mm,,>=stealth, ->] (0,0,0) -- (1,0,-1);
\draw [line width=0.4mm,>=stealth, ->] (0,0,0) -- (0,1,-1);
\draw [line width=0.4mm,>=stealth, ->] (0,0,0) -- (-1,0,-1);
\draw [line width=0.4mm,>=stealth, ->] (0,0,0) -- (0,-1,-1);
\draw [red,line width=0.4mm,,>=stealth, ->] (0,0,0) -- (1,0,0);
\draw [red,line width=0.4mm,>=stealth, ->] (0,0,0) -- (0,1,0);
\draw [line width=0.4mm,>=stealth, ->] (0,0,0) -- (0,0,1);
\draw [red,line width=0.4mm,>=stealth, ->] (0,0,0) -- (-1,0,0);
\draw [red,line width=0.4mm,>=stealth, ->] (0,0,0) -- (0,-1,0);
\draw [line width=0.4mm,>=stealth, ->] (0,0,0) -- (0,0,-1);
\end{scope}
\draw (0,0,2)--(2,0,2)--(2,2,2)--(0,2,2)--cycle; 
\draw (2,2,0) -- (2,2,2);
\draw [red,dashed] (0,0,1)--(0,-1,1);
\draw [red,dashed] (2,0,1)--(2,-1,1);

\begin{scope}[shift={(1,-2,1)}]
\draw [thin,dashed] (-1,-1) grid (1,1);
\draw [line width=0.4mm,>=stealth, ->] (0,0,0) -- (1,1,0);
\draw [line width=0.4mm,>=stealth, ->] (0,0,0) -- (-1,1,0);
\draw [line width=0.4mm,>=stealth, ->] (0,0,0) -- (-1,-1,0);
\draw [line width=0.4mm,>=stealth, ->] (0,0,0) -- (1,-1,0);
\draw [red,line width=0.4mm,,>=stealth, ->] (0,0,0) -- (1,0,0);
\draw [line width=0.4mm,>=stealth, ->] (0,0,0) -- (0,1,0);
\draw [red,line width=0.4mm,>=stealth, ->] (0,0,0) -- (-1,0,0);
\draw [line width=0.4mm,>=stealth, ->] (0,0,0) -- (0,-1,0);
\draw [red](0.003,-0.01) node {$\bullet$};

\end{scope}
\end{tikzpicture}
\qquad
\begin{tikzpicture}[scale=1,baseline={(current bounding box.center)}]
\begin{scope}[shift={(0,1,0)}]
\draw (2,0,0)--(2,2,0)--(0,2,0); 
\draw [gray,dashed] (0,0,0)--(2,0,0);
\draw [gray,dashed] (0,0,0)--(0,2,0);
\draw [gray,dashed] (0,0,0)--(0,0,2);
\draw (2,0,0) -- (2,0,2);
\draw (0,2,0) -- (0,2,2);
\draw [dotted,fill=gray, opacity=0.2] (1,0,0)--(1,2,0)--(1,2,2)--(1,0,2)--cycle; 
\draw [dotted,fill=red, opacity=0.2] (0,0,1)--(0,2,1)--(2,2,1)--(2,0,1)--cycle; 
\draw [dotted,fill=gray, opacity=0.2] (0,1,0)--(0,1,2)--(2,1,2)--(2,1,0)--cycle; 
\begin{scope}[shift={(1,1,1)}]
\draw (0.003,-0.01) node {$\bullet$};
\draw [red,line width=0.4mm,>=stealth, ->] (0,0,0) -- (1,1,0);
\draw [red,line width=0.4mm,>=stealth, ->] (0,0,0) -- (-1,1,0);
\draw [red,line width=0.4mm,>=stealth, ->] (0,0,0) -- (-1,-1,0);
\draw [red,line width=0.4mm,>=stealth, ->] (0,0,0) -- (1,-1,0);
\draw [line width=0.4mm,,>=stealth, ->] (0,0,0) -- (1,0,1);
\draw [line width=0.4mm,>=stealth, ->] (0,0,0) -- (0,1,1);
\draw [line width=0.4mm,>=stealth, ->] (0,0,0) -- (-1,0,1);
\draw [line width=0.4mm,>=stealth, ->] (0,0,0) -- (0,-1,1);
\draw [line width=0.4mm,,>=stealth, ->] (0,0,0) -- (1,0,-1);
\draw [line width=0.4mm,>=stealth, ->] (0,0,0) -- (0,1,-1);
\draw [line width=0.4mm,>=stealth, ->] (0,0,0) -- (-1,0,-1);
\draw [line width=0.4mm,>=stealth, ->] (0,0,0) -- (0,-1,-1);
\draw [red,line width=0.4mm,,>=stealth, ->] (0,0,0) -- (1,0,0);
\draw [red,line width=0.4mm,>=stealth, ->] (0,0,0) -- (0,1,0);
\draw [line width=0.4mm,>=stealth, ->] (0,0,0) -- (0,0,1);
\draw [red,line width=0.4mm,>=stealth, ->] (0,0,0) -- (-1,0,0);
\draw [red,line width=0.4mm,>=stealth, ->] (0,0,0) -- (0,-1,0);
\draw [line width=0.4mm,>=stealth, ->] (0,0,0) -- (0,0,-1);
\end{scope}
\draw (0,0,2)--(2,0,2)--(2,2,2)--(0,2,2)--cycle; 
\draw (2,2,0) -- (2,2,2);

\begin{scope}[shift={(1,-2,1)}]
\draw [line width=0.4mm,,>=stealth, ->] (0,0,0) -- (1,0,0);
\draw [line width=0.4mm,>=stealth, ->] (0,0,0) -- (-1,0,0);
\draw [white,line width=0.4mm,>=stealth, ->] (0,0,0) -- (0,-1,0);
\draw (0.003,-0.01) node {$\bullet$};
\draw (0,-0.75,0) node[above]{$\xi_0$};
\draw (1.05,-0.75,0) node[above]{$\xi_{+1}$};
\draw (-1,-0.75,0) node[above]{$\xi_{-1}$};
\end{scope}
\end{scope}

\end{tikzpicture}
\caption{(Color online) Illustration of the D3Q27, D3Q19, D2Q9 and D1Q3 lattices (from top left to bottom right). Relationships between lattices are highlighted in red. From them, it flows that the structure of the D1Q3 lattice is contained in the D2Q9 lattice, while both the D3Q27 and the D3Q19 lattices rely on the fundamental structure of the D2Q9 lattice.}
\label{fig:Lattices}
\end{figure}
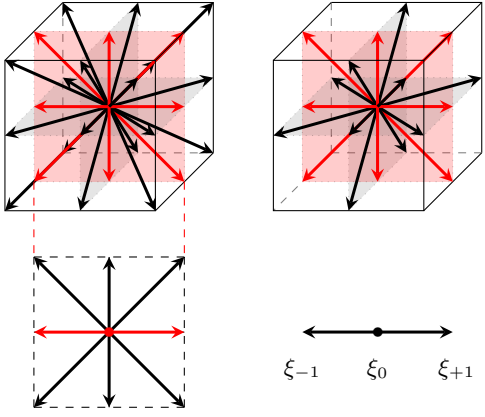
\begin{table}[h!]
\renewcommand{\arraystretch}{1.5}
\renewcommand{\tabcolsep}{2mm}
\begin{tabular}{c c c c c c}
\hline
$\text{Group}$ & $\boldsymbol\xi_i$ & D1Q3 & D2Q9 & D3Q19  & D3Q27 \\\hline
$1$ & $(0,0,0)$ & $2/3$ & $4/9$ & $1/3$ & $8/27$  \\

$2$ & $(1,0,0)$ & $1/6$ & $1/9$ & $1/18$ & $2/27$ \\
$3$ & $(1,1,0)$ &  & $1/36$ & $1/36$ & $1/54$  \\
$4$ & $(1,1,1)$ &  &  &  & $1/216$  \\
\hline
\end{tabular}
\caption{Description of standard lattice structures of interest. Weights $w_i$ related to velocity groups compose the right part of the table. Here, the cyclic permutation is implied, which means for example that $(1,0,0)$ stands for all six possibilities $(\pm1,0,0)$, $(0,\pm1,0)$ and $(0,0,\pm1)$ in the 3D case, while only $(\pm1,0)$, $(0,\pm1)$ and $(\pm1)$ are implied in the 2D and 1D cases respectively. All velocity discretizations share the same lattice constant $c_s=1/\sqrt{3}$.}
\label{tab:Lattices}
\end{table}

\section{Bell polynomials \label{sec:Bell}}

These polynomials are used in combinatorial mathematics to study set partitions~\cite{BELL_AM_1934,WITHERS_1994}. The Bell polynomial of degree $n$ is defined as
\begin{equation}
B_{n}(x_1,...,x_{n})=\displaystyle{\sum_{k=1}^{n} B_{n,k}(x_1,...,x_{n-k+1})},
\end{equation}
where $B_{n,k}$ are partial Bell polynomials, and $x_1,...,x_n$ are $n$ variables. Each $B_{n,k}$ corresponds to the partitioning of a set composed of $n$ elements into $k$ non-empty subsets. 

As an example, let us consider the set $\mathcal{S}=\{a,b,c\}$, and let us find all its possible partitionings (described via $B_{3}$). They are three ways to partition $\mathcal{S}$: 
\begin{enumerate}
\item \emph{One} subset composed of three elements
$$\mathcal{S}_1=\{\{a,b,c\}\},$$
\item \emph{Two} subsets composed of one and two elements
$$\mathcal{S}_2=\{\{a\},\{b,c\}\} \cup \{\{b\},\{a,c\}\} \cup \{\{c\},\{a,b\}\},$$
\item \emph{Three} subsets composed of one element
$$\mathcal{S}_3=\{\{a\},\{b\},\{c\}\}.$$
\end{enumerate}
Through $x_1$, $x_2$ and $x_3$, partial Bell polynomials $B_{3,1}$, $B_{3,2}$ and $B_{3,3}$ allow the mathematical description partitionings $\mathcal{S}_1$, $\mathcal{S}_2$ and $\mathcal{S}_3$ respectively. Assuming $x_p$ indicates the presence of a subset composed of $p$ elements, then $$B_{3,1}(x_1,x_2,x_3)=x_3$$ 
since only \emph{one} subset, composed of three elements, is included in $\mathcal{S}_1$. In addition, $$B_{3,2}(x_1,x_2)=3 x_1x_2$$ because $\mathcal{S}_2$ is composed of (three) subsets divided into \emph{two} blocks of length two and one respectively. Eventually, $$B_{3,3}(x_1)=x_1^3$$ since $\mathcal{S}_3$ contains \emph{three} singletons. Hence, the corresponding Bell polynomial is
$$B_{3}(x_1,x_2,x_3)=x_3 + 3 x_1 x_2 + x_1^3.$$

Applying the very same reasoning for $1 \leq n\leq 6$, Bell polynomials read as 
\begin{align}
B_{1}(x_1) &= x_1,\notag\\
B_{2}(x_1,x_2) &= x_2 + x_1^2,\notag\\
B_{3}(x_1,x_2,x_3) &= x_3 + 3 x_2 x_1 + x_1^3,\notag\\
B_{4}(x_1,...,x_4) &= x_4 + 4 x_1 x_3 + 3 x_2^2 + 6 x_1^2 x_2 + x_1^4,\notag\\
B_{5}(x_1,...,x_5) &= x_5 + 5 x_1 x_4 + 10 x_2x_3 + 10 x_1^2x_3\notag \\
										 &\quad+ 15 x_1 x_2^2 + 10 x_1^3 x_2 + x_1^5,\notag\\
B_{6}(x_1,...,x_6) &= x_6 + 6 x_1 x_5 + 15 x_2x_4 + 15 x_1^2 x_5+ 10 x_3^2  \notag\\
										 &\quad+ 60 x_1x_2x_3 + 20 x_1^3 x_3+ 15 x_2^3 + 45 x_1^2 x_2^2  \notag\\
										 &\quad+ 15 x_1^4 x_2 + x_1^6.
\end{align}
Defining
$y_n={\sum_{k=1}^{n} B_{n,k}(x_1,...,x_{n-k+1})},$
the inversion formula, allowing to express $x_n$ with respect to $y_1,...,y_n$, is~\cite{BELL_AM_1934,WITHERS_1994}
\begin{equation}
x_n=\displaystyle{\sum_{k=1}^{n} (-1)^{k-1}(k-1)! B_{n,k}(y_1,...,y_{n-k+1})}.
\end{equation}

Finally, one obtains relationships between raw moments and cumulants (Eqs.~\ref{eq:Univar_RM_K} and ~\ref{eq:Univar_K_RM}) replacing $(x_n,y_n)$ by $(M_n,K_n)$ in the above formulas. 

\section{Univariate formulas\label{sec:RelationshipsUnivariate}}

In this appendix, formulas used to link all statistical quantities of interest between each other are compiled in the univariate case. 

Relationships between raw and Hermite moments flows from formulas expressing Hermite polynomials with respect to monomials~(\ref{eq:HermitePoly}). Up to $n=6$, they read as 
\begin{equation}
\begin{array}{l}
M_{0} = A_0 = 1, \\[0.1cm]
M_{1} = A_1 = u_x, \\[0.1cm]
M_{2} = A_2+c_s^2,  \\[0.1cm]
M_{3} = A_3+3 c_s^2 u_x, \\[0.1cm]
M_{4} = A_4+6 c_s^2 A_2+3 c_s^4,  \\[0.1cm]
M_{5} = A_5+10 c_s^2 A_3+15 u_x c_s^4,  \\[0.1cm]
M_{6} = A_6+15 c_s^2 A_4+45 c_s^4 A_2+15 c_s^6,   
\end{array}
\end{equation}
and
\begin{equation}
\begin{array}{l}
A_{0} = M_0 = 1, \\[0.1cm]
A_{1} = M_1 = u_x,\\[0.1cm]
A_{2} = M_2-c_s^2,  \\[0.1cm]
A_{3} = M_3-3 c_s^2 u_x, \\[0.1cm]
A_{4} = M_4-6 c_s^2 M_2+3 c_s^4,  \\[0.1cm]
A_{5} = M_5-10 c_s^2 M_3+15 u_x c_s^4,  \\[0.1cm]
A_{6} = M_6-15 c_s^2 M_4+45 c_s^4 M_2-15 c_s^6.
\end{array}
\end{equation}


Continuing with relationships between raw  and central moments, they are simply obtained using recursive formulas introduced in Eq.~(\ref{eq:recRawCentral}),
\begin{equation}
\begin{array}{l @{\:} l}
M_0= & \widetilde{M}_0 = 1, \\
M_1= & \widetilde{M}_1 + u_x=u_x, \\
M_2= & \widetilde{M}_2+u_x^2, \\
M_3= & \widetilde{M}_3 + 3 u_x \widetilde{M}_2  +  u_x^3, \\
M_4= & \widetilde{M}_4 + 4 u_x \widetilde{M}_3  + 6 u_x^2 \widetilde{M}_2 +  u_x^4, \\
M_5= & \widetilde{M}_5 + 5 u_x \widetilde{M}_4  +10 u_x^2 \widetilde{M}_3 + 10 u_x^3 \widetilde{M}_2 +  u_x^5,   \\
M_6= & \widetilde{M}_6 + 6 u_x \widetilde{M}_5  +15 u_x^2 \widetilde{M}_4 +20 u_x^3 \widetilde{M}_3 \\
         &+15 u_x^4 \widetilde{M}_2  + u_x^6,
\end{array}
\end{equation}
and Eq.~(\ref{eq:recCentralRaw}),
\begin{equation}
\begin{array}{l @{\:} l}
\widetilde{M}_0= & M_0 = 1, \\
\widetilde{M}_1= & M_1 -  u_x = 0, \\
\widetilde{M}_2= & M_2-u_x^2, \\
\widetilde{M}_3= & M_3 - 3 u_x M_2  + 2 u_x^3, \\
\widetilde{M}_4= & M_4 - 4 u_x M_3  + 6 u_x^2 M_2 - 3 u_x^4, \\
\widetilde{M}_5= & M_5 - 5 u_x M_4  +10 u_x^2 M_3 -10 u_x^3 M_2 + 4 u_x^5,   \\
\widetilde{M}_6= & M_6 - 6 u_x M_5  +15 u_x^2 M_4 -20 u_x^3 M_3 \\
                         & +15 u_x^4 M_2 -5 u_x^6.
\end{array}
\end{equation}

It is important to note that replacing $M$ and $\widetilde{M}$ by their counterparts in the Hermite framework ($A$ and $\widetilde{A}$) the above formulas remain valid. This is explained by the fact that when the change from the reference frame at rest to the comoving one is done then a simple shift of $\exp(-M_1 t)$ is applied to the raw (or Hermite) moment generating function.

Considering now relationships between raw moments and cumulants, formulas derived thanks to Bell polynomials (App.~\ref{sec:Bell}) lead to
\begin{equation}
\begin{array}{l @{\:} l}
 M_0 = & \exp(K_0)=1 \\[0.1cm]
 M_1 = & K_1 = u_x,\\[0.1cm]
 M_2 = & K_2+u_x^2, \\[0.1cm]
 M_3 = & K_3+3 u_x K_2+u_x^3, \\[0.1cm]
 M_4 = & K_4+4 u_x K_3+3 K_2^2+6 u_x^2 K_2+u_x^4,\\[0.1cm]
 M_5 = & K_5+5 u_x K_4+10 K_3 K_2+10 u_x^2 K_3+15 u_x K_2^2\\[0.1cm]
       &+10 u_x^3 K_2+u_x^5, \\[0.1cm]
 M_6 = & K_6+6 K_5 u_x+15 K_4 K_2+15 u_x^2 K_4\\[0.1cm]
        & +10 K_3^2+60 u_x K_3 K_2+20 u_x^3 K_3+15 K_2^3\\[0.1cm]
        & +45 u_x^2 K_2^2+15 u_x^4 K_2 +u_x^6,
\end{array}
\label{eq:Univar_RM_K}
\end{equation}
and
\begin{equation}
\begin{array}{l @{\:} l}
 K_0 = & \ln(M_0)=0, \\[0.1cm]
 K_1 = & M_1 = u_x, \\[0.1cm]
 K_2 = & M_2-u_x^2, \\[0.1cm]
 K_3 = & M_3 -3 u_x M_2 + 2 u_x^3, \\[0.1cm]
 K_4 = & M_4 -3 M_2^2 -4 u_x M_3+12 u_x^2 M_2 -6 u_x^4,\\[0.1cm]
 K_5 = & M_5 -5 u_x M_4+20 u_x^2 M_3-10 M_3 M_2 \\[0.1cm]
        &+30 u_x M_2^2-60 u_x^3 M_2 + 24 u_x^5, \\[0.1cm]
 K_6 = & M_6-6 u_x M_5-15 M_4 M_2+30 u_x^2 M_4-10 M_3^2 \\[0.1cm]
        &+120 u_x M_3M_2-120 u_x^3 M_3+30 M_2^3 \\[0.1cm]
        &-270 u_x^2 M_2^2+360 u_x^4 M_2-120 u_x^6.
\end{array}
\label{eq:Univar_K_RM}
\end{equation}

Regarding formulas between central moments and cumulants, one just need to move to the comoving reference frame. In fact, this amounts to neglecting terms proportional to $u_x$ in the above formulas~\cite{KENDALL_AE_10_1940,COOK_Bio_38_1951}. Hence 
\begin{equation}
\begin{array}{l @{\:} l}
 \widetilde{M}_0= & \exp(K_0) = 1, \\
 \widetilde{M}_1= & K_1-u_x=0, \\
 \widetilde{M}_2= & K_2, \\
 \widetilde{M}_3= & K_3, \\
 \widetilde{M}_4= & K_4+3 K_2^2,\\
 \widetilde{M}_5= & K_5+10 K_3 K_2,\\
 \widetilde{M}_6= & K_6+15 K_4 K_2+10 K_3^2+15 K_2^3,
\end{array}
\end{equation}
and
\begin{equation}
\begin{array}{l @{\:} l}
 K_0 = & \ln(\widetilde{M}_0) = 0,\\
 K_1 = &  \widetilde{M}_1+u_x = u_x,\\
 K_2 = &  \widetilde{M}_2, \\
 K_3 = & \widetilde{M}_3, \\
 K_4 = &  \widetilde{M}_4 -3 \widetilde{M}_2^2, \\
 K_5 = &  \widetilde{M}_5 -10  \widetilde{M}_3 \widetilde{M}_2, \\
 K_6 = &  \widetilde{M}_6-15  \widetilde{M}_4  \widetilde{M}_2-10  \widetilde{M}_3^2+30  \widetilde{M}_2^3.
\end{array}
\end{equation}


\section{Bivariate formulas\label{sec:RelationshipsBivariate}}

Henceforth, bivariate formulations of relationships between statistical quantities of interest are presented. 

Starting with raw and Hermite moments, the most straightforward way to compute their bivariate relationships is to take advantage of the orthogonality properties of Hermite tensors. Hence, they are simply computed as~\cite{COREIXAS_PhD_2018},
\begin{equation}
\begin{array}{r @{\:} l}
M_{00} =& A_{00}=1,\\[0.1cm]
M_{10} =& A_{10}=u_x,\\[0.1cm]
M_{01} =& A_{01}=u_y,\\[0.1cm]
M_{11} =& M_{10}M_{01} = A_{10}A_{01} = A_{11},\\[0.1cm]
M_{20} =& A_{20}+c_s^2,\\[0.1cm]
M_{02} =& A_{02}+c_s^2,\\[0.1cm]
M_{21} =& M_{20}M_{01} = (A_{20}+c_s^2)A_{01} = A_{21}+c_s^2 u_y,\\[0.1cm]
M_{12} =& M_{10}M_{02} = A_{10}(A_{02}+c_s^2) = A_{12}+c_s^2 u_x,\\[0.1cm]
M_{22} =& M_{20}M_{02} = (A_{20}+c_s^2) (A_{02}+c_s^2)\\[0.1cm]
            =& A_{22}+c_s^2(A_{20}+A_{02}) +c_s^4,
\end{array}
\label{eq:HM2RM_Q9}
\end{equation}
and
\begin{equation}
\begin{array}{r @{\:} l}
A_{00} =& M_{00}=1,\\[0.1cm]
A_{10} =& M_{10}=u_x,\\[0.1cm]
A_{01} =& M_{01}=u_y,\\[0.1cm]
A_{11} =& A_{10}A_{01} = M_{10}M_{01} = M_{11},\\[0.1cm]
A_{20} =& M_{20}-c_s^2,\\[0.1cm]
A_{02} =& M_{02}-c_s^2,\\[0.1cm]
A_{21} =& A_{20}A_{01} = (M_{20}-c_s^2)M_{01} = M_{21}-c_s^2 u_y,\\[0.1cm]
A_{12} =& A_{10}A_{02} = M_{10}(M_{02}-c_s^2) = M_{12}-c_s^2 u_x,\\[0.1cm]
A_{22} =& A_{20}A_{02} = (M_{20}-c_s^2) (M_{02}-c_s^2)\\[0.1cm]
            =& M_{22}-c_s^2(M_{20}+M_{02}) +c_s^4.
\end{array}
\label{eq:RM2HM_Q9}
\end{equation}
One must be careful regarding the above method. Indeed, one must wait until the very end before replacing zeroth and first order moments by their values. As an example, $M_{11}\neq u_x u_y(=M_{10}M_{01})$ since $M_{11}^{neq}=M_{11}-M^{eq}_{11}$ is related to viscous phenomena and is usually non zero.

%
%
The simplest way to extend relationships between raw and central moments to the bivariate case is to rely on the corresponding binomial formula,
\begin{equation*}
\widetilde{M}_{pq} = \sum_{k_x=0}^{p}\sum_{k_y=0}^{q} \binom{p}{k_x}\binom{q}{k_y} (-u_x)^{p-k_x}(-u_y)^{q-k_y} M_{pq}.
 \end{equation*}
Inversion formulas are then obtained discarding the minus sign, 
\begin{equation*}
M_{pq} = \sum_{k_x=0}^{p}\sum_{k_y=0}^{q} \binom{p}{k_x}\binom{q}{k_y} u_x^{p-k_x}u_y^{q-k_y} \widetilde{M}_{pq}.
 \end{equation*}
In the case of the D2Q9 lattice ($p\leq2$ and $q\leq 2$), this leads to
\begin{equation}
\begin{array}{r @{\:} l}
M_{00}= & \widetilde{M}_{00}=1,\\
M_{10}= & \widetilde{M}_{10}+u_x,\\
M_{01}= & \widetilde{M}_{01}+u_y,\\
M_{20}= & \widetilde{M}_{20}+u_x^2,\\
M_{02}= & \widetilde{M}_{02}+u_y^2,\\
M_{11}= & \widetilde{M}_{11}+u_xu_y ,\\
M_{21}= & \widetilde{M}_{21}+u_y \widetilde{M}_{20}+2 u_x \widetilde{M}_{11}+u_x^2 u_y,\\
M_{12}= & \widetilde{M}_{12}+u_x \widetilde{M}_{02}+2 u_y \widetilde{M}_{11}+u_x u_y^2,\\
M_{22}= & \widetilde{M}_{22}+2 u_y \widetilde{M}_{21}+2 u_x \widetilde{M}_{12}\\
        &+u_y^2 \widetilde{M}_{20} +u_x^2 \widetilde{M}_{02}+4 u_x u_y \widetilde{M}_{11}+u_x^2 u_y^2,
\end{array}
\label{eq:CM2RM_Q9}
\end{equation}
and
\begin{equation}
\begin{array}{r @{\:} l}
\widetilde{M}_{00}= & M_{00}=1,\\
\widetilde{M}_{10}= & M_{10}-u_x,\\
\widetilde{M}_{01}= & M_{01}-u_y,\\
\widetilde{M}_{20}= & M_{20}-u_x^2,\\
\widetilde{M}_{02}= & M_{02}-u_y^2,\\
\widetilde{M}_{11}= & M_{11}-u_xu_y ,\\
\widetilde{M}_{21}= & M_{21}-u_y M_{20}-2 u_x M_{11}+2 u_x^2 u_y,\\
\widetilde{M}_{12}= & M_{12}-u_x M_{02}-2 u_y M_{11}+2 u_x u_y^2,\\
\widetilde{M}_{22}= & M_{22}-2 u_y M_{21}-2 u_x M_{12}+u_y^2 M_{20}\\[0.075cm]
                    & +u_x^2 M_{02}+4 u_x u_y M_{11}-3 u_x^2 u_y^2.
\end{array}
\label{eq:RM2CM_Q9}
\end{equation}
\noindent Due to the fact that there is a linear transformation allowing us to move from monomials to Hermite polynomials, these relationships remain valid for Hermite and central Hermite moments. Hence, one simply needs to replace $M$ and $\widetilde{M}$ by $A$ and $\widetilde{A}$ in the above formulas to recover relationships between Hermite and central Hermite moments.

To link raw moments and cumulants in the bivariate case, differential operators introduced in Sec.~\ref{sec:MultiExt}, and proposed by Kendall~\cite{KENDALL_AE_10_1940,COOK_Bio_38_1951}, are employed. Corresponding relationships are then
\begin{equation}
\begin{array}{l @{\:} l}
M_{00} = & \exp(K_{00})=1,\\[0.1cm]
M_{10} = & K_{10}=u_x,  \\[0.1cm]
M_{01} = & K_{01}=u_y,\\[0.1cm]
M_{11} = & K_{11} + u_x u_y, \\[0.1cm] 
M_{20} = & K_{20} + u_x^2, \\[0.1cm] 
M_{02} = & K_{02} + u_y^2,\\[0.1cm]
M_{21} = & K_{21} + u_yK_{20} + 2u_xK_{11} + u_x^2u_y, \\[0.1cm]
M_{12} = & K_{12} + u_xK_{02} + 2u_yK_{11} + u_x u_y^2,\\[0.1cm]
M_{22} = & K_{22} + 2K_{21}u_y + 2K_{12}u_x + K_{20}K_{02}  \\[0.1cm]
              &+ K_{20}u_y^2 +K_{02} u_x^2 + 2K_{11}^2 \\[0.1cm]
              &+4K_{11}u_x u_y +  u_x^2 u_y^2,
\end{array}
\end{equation}
and
\begin{equation}
\begin{array}{l @{\:} l}
K_{00} = & \ln(M_{00})=0, \\[0.1cm]
K_{10} = & M_{10}=u_x,\\[0.1cm]
K_{01} = & M_{01}=u_y,\\[0.1cm]
K_{11} = & M_{11} - u_xu_y, \\[0.1cm]
K_{20} = & M_{20} - u_x^2, \\[0.1cm]
K_{02} = & M_{02} - u_y^2,\\[0.1cm]
K_{21} = & M_{21} - u_y M_{20} - 2u_x M_{11} + 2u_x^2 u_y, \\[0.1cm]
K_{12} = & M_{12} - u_x M_{02} - 2u_y M_{11} + 2u_x u_y^2,\\[0.1cm]
K_{22} = & M_{22} - 2u_y M_{21} -2 u_x M_{12} - M_{20}M_{02}  \\[0.1cm]
             &+ 2u_x^2M_{02} + 2M_{20}u_y^2  - 2M_{11}^2\\[0.1cm]
             &+ 8 M_{11}u_x u_y- 6u_x^2 u_y^2.
\end{array}
\end{equation}
For central moments, one can again discard terms proportional to $u_x$ and $u_y$. Doing so, one obtains 
\begin{equation}
\begin{array}{r @{\:} l}
\widetilde{M}_{00} &= \exp(K_{00})=1,\\
\widetilde{M}_{10} &= K_{10} - M_{10}=0, \\
\widetilde{M}_{01} &= K_{01} + M_{01}=0,\\
\widetilde{M}_{11} &= K_{11},\\
\widetilde{M}_{20} &= K_{20},\\
\widetilde{M}_{02} &= K_{02},\\
\widetilde{M}_{21} &= K_{21},\\
\widetilde{M}_{12} &= K_{12},\\
\widetilde{M}_{22} &= K_{22} + (K_{20}K_{02} + 2K_{11}^2),
\end{array}
\label{eq:K2CM_Q9}
\end{equation}
and
\begin{equation}
\begin{array}{r @{\:} l}
K_{00} &=\ln(\widetilde{M}_{00})=0\\
K_{10} &= \widetilde{M}_{10} + M_{10}=u_x,\\
K_{01} &= \widetilde{M}_{01} + M_{01}=u_y,\\
K_{11} &= \widetilde{M}_{11},\\
K_{20} &= \widetilde{M}_{20},\\
K_{02} &= \widetilde{M}_{02},\\
K_{21} &= \widetilde{M}_{21},\\
K_{12} &= \widetilde{M}_{12},\\
K_{22} &= \widetilde{M}_{22} - (\widetilde{M}_{20}\widetilde{M}_{02} + 2\widetilde{M}_{11}^2),
\end{array}
\end{equation}
where it is clear that deviations from central moments start appearing in fourth order cumulants.

From the implementation point of view, the conversion from central moments to cumulants is clearly simpler than the one using raw moments. Apart from that, there is no fundamental reason for the use of central moments instead of raw ones. Furthermore, formulas for post collision populations are more complex in the CM framework~(\ref{eq:Q9postCollVDF_CM}) than in the RM one~(\ref{eq:Q9postCollVDF_RM}). In the 2D case, it is then not clear if the use of central moments should be preferred or not.

\section{Trivariate formulas\label{sec:RelationshipsTrivariate}}
Hereafter, the methodology introduced in App.~\ref{sec:RelationshipsBivariate} is further extended to the trivariate case. Starting with raw and Hermite moments, orthogonality properties of Hermite polynomials are, once again, used to relate these two families of moments. For the D3Q27 lattice, corresponding formulas are

\onecolumngrid

\begin{equation}
\left.
\begin{array}{l @{\:} l}
M_{000} = & A_{000}=1, \\\\[0.1cm]
M_{100} = & A_{100}=u_x,  \\[0.1cm]
M_{010} = & A_{010}=u_y,  \\[0.1cm]
M_{001} = & A_{001}=u_z,  \\\\[0.1cm]
M_{110} = & A_{110},  \\[0.1cm]
M_{101} = & A_{101},  \\[0.1cm]
M_{011} = & A_{011},  \\[0.1cm]
M_{200} = & A_{200} + c_s^2,  \\[0.1cm]
M_{020} = & A_{020} + c_s^2,  \\[0.1cm]
M_{002} = & A_{002} + c_s^2,  \\\\[0.1cm]
M_{210} = & A_{210} + c_s^2 u_y,  \\[0.1cm]
M_{201} = & A_{201} + c_s^2 u_z,  \\[0.1cm]
M_{021} = & A_{021} + c_s^2 u_z,  \\[0.1cm]
M_{120} = & A_{120} + c_s^2 u_x,  \\[0.1cm]
M_{102} = & A_{102} + c_s^2 u_x,  \\[0.1cm]
M_{012} = & A_{012} + c_s^2 u_y,  \\[0.1cm]
M_{111} = & A_{111},  \\\\[0.1cm]
M_{220} = & A_{220} + c_s^2 (A_{200} + A_{020}) + c_s^4, \\[0.1cm]
M_{202} = & A_{202} + c_s^2 (A_{200} + A_{002}) + c_s^4, \\[0.1cm]
M_{022} = & A_{022} + c_s^2 (A_{020} + A_{002}) + c_s^4, \\[0.1cm]
M_{211} = & A_{211} + c_s^2 A_{011}, \\[0.1cm]
M_{121} = & A_{121} + c_s^2 A_{101}, \\[0.1cm]
M_{112} = & A_{112} + c_s^2 A_{110}, \\\\[0.1cm]
M_{221} = & A_{221} + c_s^2(A_{201}+A_{021}) + c_s^4 u_z,  \\[0.1cm]
M_{212} = & A_{212} + c_s^2(A_{210}+A_{012}) + c_s^4 u_y,  \\[0.1cm]
M_{122} = & A_{122} + c_s^2(A_{120}+A_{102}) + c_s^4 u_x,  \\\\[0.1cm]
M_{222} = & A_{222} + c_s^2(A_{220} + A_{202} + A_{022}) \\[0.1cm]
                &+ c_s^4(A_{200}+A_{020}+A_{002}) + c_s^6, 
\end{array}
\right.
\hspace*{1.5cm}
\left.
\begin{array}{l @{\:} l}
A_{000} = & M_{000}=1,\\\\[0.1cm]
A_{100} = & M_{100}=u_x,\\[0.1cm]
A_{010} = & M_{010}=u_y,\\[0.1cm]
A_{001} = & M_{001}=u_z,\\\\[0.1cm]
A_{110} = & M_{110},\\[0.1cm]
A_{101} = & M_{101},\\[0.1cm]
A_{011} = & M_{011},\\[0.1cm]
A_{200} = & M_{200} - c_s^2,\\[0.1cm]
A_{020} = & M_{020} - c_s^2,\\[0.1cm]
A_{002} = & M_{002} - c_s^2,\\\\[0.1cm]
A_{210} = & M_{210} - c_s^2 u_y,\\[0.1cm]
A_{201} = & M_{201} - c_s^2 u_z,\\[0.1cm]
A_{021} = & M_{021} - c_s^2 u_z,\\[0.1cm]
A_{120} = & M_{120} - c_s^2 u_x,\\[0.1cm]
A_{102} = & M_{102} - c_s^2 u_x,\\[0.1cm]
A_{012} = & M_{012} - c_s^2 u_y,\\[0.1cm]
A_{111} = & M_{111},\\\\[0.1cm]
A_{220} = & M_{220} - c_s^2 (M_{200} + M_{020}) + c_s^4,\\[0.1cm]
A_{202} = & M_{202} - c_s^2 (M_{200} + M_{002}) + c_s^4,\\[0.1cm]
A_{022} = & M_{022} - c_s^2 (M_{020} + M_{002}) + c_s^4,\\[0.1cm]
A_{211} = & M_{211} - c_s^2 M_{011},\\[0.1cm]
A_{121} = & M_{121} - c_s^2 M_{101},\\[0.1cm]
A_{112} = & M_{112} - c_s^2 M_{110},\\\\[0.1cm]
A_{221} = & M_{221} - c_s^2(M_{201}+M_{021}) + c_s^4 u_z,\\[0.1cm]
A_{212} = & M_{212} - c_s^2(M_{210}+M_{012}) + c_s^4 u_y,\\[0.1cm]
A_{122} = & M_{122} - c_s^2(M_{120}+M_{102}) + c_s^4 u_x,\\\\[0.1cm]
A_{222} = & M_{222} - c_s^2 (M_{220} + M_{202} + M_{022})\\[0.1cm]
                &+ c_s^4 (M_{200}+M_{020}+M_{002}) - c_s^6.
\end{array}
\right.
\label{eq:RawHermite3D}
\end{equation}

For raw and central moments, 
\begin{equation*}
\widetilde{M}_{pqr} = \sum_{k_x=0}^{p}\sum_{k_y=0}^{q}\sum_{k_z=0}^{r} \binom{p}{k_x}\binom{q}{k_y}\binom{r}{k_z} (-u_x)^{p-k_x}(-u_y)^{q-k_y}(-u_z)^{r-k_z} M_{pqr}.
 \end{equation*}
Inversion formulas are once again obtained discarding the minus sign, 
\begin{equation*}
{M}_{pqr} = \sum_{k_x=0}^{p}\sum_{k_y=0}^{q}\sum_{k_z=0}^{r} \binom{p}{k_x}\binom{q}{k_y}\binom{r}{k_z} u_x^{p-k_x}u_y^{q-k_y}u_z^{r-k_z} \widetilde{M}_{pqr}.
 \end{equation*}
Thus, relationships obtained for the D3Q27 lattice are
\begin{equation}
\left.
\begin{array}{l @{\:} l}
M_{000}= & \widetilde{M}_{000}=1,\\\\[0.1cm]
M_{100}= & \widetilde{M}_{100}+u_x=u_x,\\[0.1cm]
M_{010}= & \widetilde{M}_{010}+u_y=u_y,\\[0.1cm]
M_{001}= & \widetilde{M}_{001}+u_z=u_z,\\\\[0.1cm]
M_{200}= & \widetilde{M}_{200}+u_x^2,\\[0.1cm]
M_{020}= & \widetilde{M}_{020}+u_y^2,\\[0.1cm]
M_{002}= & \widetilde{M}_{002}+u_z^2,\\[0.1cm]
M_{110}= & \widetilde{M}_{110}+u_xu_y ,\\[0.1cm]
M_{101}= & \widetilde{M}_{101}+u_xu_z ,\\[0.1cm]
M_{011}= & \widetilde{M}_{011}+u_yu_z ,\\\\[0.1cm]
M_{210}= & \widetilde{M}_{210}+u_y \widetilde{M}_{200}+2 u_x \widetilde{M}_{110}+u_x^2 u_y,\\[0.1cm]
M_{201}= & \widetilde{M}_{201}+u_z \widetilde{M}_{200}+2 u_x \widetilde{M}_{101}+u_x^2u_z ,\\[0.1cm]
M_{120}= & \widetilde{M}_{120}+u_x \widetilde{M}_{020}+2 u_y \widetilde{M}_{110}+u_x u_y^2,\\[0.1cm]
M_{021}= & \widetilde{M}_{021}+u_z \widetilde{M}_{020}+2 u_y \widetilde{M}_{011}+u_y^2 u_z ,\\[0.1cm]
M_{102}= & \widetilde{M}_{102}+u_x \widetilde{M}_{002}+2 u_z \widetilde{M}_{101}+u_x u_z^2,\\[0.1cm]
M_{012}= & \widetilde{M}_{012}+u_y \widetilde{M}_{002}+2 u_z \widetilde{M}_{011}+u_y u_z^2,\\[0.1cm]
M_{111}= & \widetilde{M}_{111}+u_z \widetilde{M}_{110}+u_y \widetilde{M}_{101}+u_x \widetilde{M}_{011}+u_x u_y u_z,\\\\[0.1cm]
M_{220}= & \widetilde{M}_{220}+2 u_y \widetilde{M}_{210}+2 u_x \widetilde{M}_{120}+u_y^2 \widetilde{M}_{200} +u_x^2 \widetilde{M}_{020}+4 u_x u_y \widetilde{M}_{110}+u_x^2 u_y^2,\\[0.1cm]
M_{202}= & \widetilde{M}_{202}+2 u_z \widetilde{M}_{201}+2 u_x \widetilde{M}_{102}+u_z^2 \widetilde{M}_{200}+u_x^2 \widetilde{M}_{002}+4 u_x u_z \widetilde{M}_{101}+u_x^2 u_z^2,\\[0.1cm]
M_{022}= & \widetilde{M}_{022}+2 u_z \widetilde{M}_{021}+2 u_y \widetilde{M}_{012}+u_z^2 \widetilde{M}_{020}+u_y^2 \widetilde{M}_{002}+4 u_y u_z \widetilde{M}_{011}+u_y^2 u_z^2,\\[0.1cm]
M_{211}= & \widetilde{M}_{211}+u_z \widetilde{M}_{210} + u_y \widetilde{M}_{201}+ 2 u_x \widetilde{M}_{111} +u_y u_z\widetilde{M}_{200} +2 u_x u_z\widetilde{M}_{110}+2 u_x u_y\widetilde{M}_{101}+u_x^2 \widetilde{M}_{011} +u_x^2u_yu_z,\\[0.1cm]
M_{121}= & \widetilde{M}_{121}+u_z \widetilde{M}_{120} + u_x \widetilde{M}_{021}+ 2 u_y \widetilde{M}_{111} +u_x u_z\widetilde{M}_{020} +2 u_y u_z\widetilde{M}_{110}+2 u_x u_y\widetilde{M}_{011}+u_y^2 \widetilde{M}_{101}+u_x u_y^2 u_z ,\\[0.1cm]
M_{112}= & \widetilde{M}_{112}+u_y \widetilde{M}_{102} + u_x \widetilde{M}_{012}+ 2 u_z \widetilde{M}_{111} +u_x u_y\widetilde{M}_{002} +2 u_y u_z\widetilde{M}_{101}+2 u_x u_z\widetilde{M}_{011}+u_z^2 \widetilde{M}_{110}+u_xu_y u_z^2,\\\\[0.1cm]
M_{221}= & \widetilde{M}_{221}+u_z\widetilde{M}_{220}+2 u_y\widetilde{M}_{211}+2 u_x\widetilde{M}_{121}+2 u_y u_z\widetilde{M}_{210} +u_y^2\widetilde{M}_{201} +u_x^2\widetilde{M}_{021}+2 u_xu_z\widetilde{M}_{120}+4 u_x u_y\widetilde{M}_{111}\\[0.1cm]
         &+u_y^2 u_z\widetilde{M}_{200} +u_x^2 u_z\widetilde{M}_{020} +4 u_x u_y u_z\widetilde{M}_{110}+2 u_x u_y^2\widetilde{M}_{101} +2 u_x^2 u_y\widetilde{M}_{011} + u_x^2 u_y^2 u_z ,\\[0.1cm]
M_{212}= & \widetilde{M}_{212}+u_y\widetilde{M}_{202}+2 u_z\widetilde{M}_{211}+2 u_x\widetilde{M}_{112} +u_z^2\widetilde{M}_{210}+2 u_y u_z\widetilde{M}_{201}+2 u_x u_y\widetilde{M}_{102}+u_x^2\widetilde{M}_{012}+4 u_x u_z\widetilde{M}_{111}\\[0.1cm]
         &+u_y u_z^2\widetilde{M}_{200}+u_x^2 u_y\widetilde{M}_{002}+2 u_x u_z^2\widetilde{M}_{110}+2 u_x^2 u_z\widetilde{M}_{011}+4 u_x u_y u_z \widetilde{M}_{101}+ u_x^2 u_y u_z^2,\\[0.1cm]
M_{122}= & \widetilde{M}_{122}+u_x \widetilde{M}_{022}+2 u_z \widetilde{M}_{121}+2 u_y\widetilde{M}_{112} +u_z^2\widetilde{M}_{120} +u_y^2 \widetilde{M}_{102}+2 u_x u_z\widetilde{M}_{021}+2 u_x u_y\widetilde{M}_{012} +4 u_y u_z\widetilde{M}_{111}\\[0.1cm]
         &+u_x u_z^2\widetilde{M}_{020}+u_x u_y^2\widetilde{M}_{002}+2 u_y u_z^2 \widetilde{M}_{110} +2 u_y^2 u_z \widetilde{M}_{101} +4 u_xu_y u_z\widetilde{M}_{011}  +u_xu_y^2 u_z^2,\\\\[0.1cm]
M_{222}= & \widetilde{M}_{222}+2 u_z\widetilde{M}_{221} +2 u_y\widetilde{M}_{212} +2 u_x\widetilde{M}_{122} + u_z^2\widetilde{M}_{220}+u_y^2\widetilde{M}_{202}+u_x^2\widetilde{M}_{022}+4 u_y u_z\widetilde{M}_{211}+4 u_x u_z\widetilde{M}_{121}\\[0.1cm]
         &+4 u_x u_y\widetilde{M}_{112} +2 u_y u_z^2\widetilde{M}_{210}+2 u_y^2 u_z\widetilde{M}_{201}+2 u_x u_z^2\widetilde{M}_{120}+2 u_y^2 u_x\widetilde{M}_{102}+2 u_x^2 u_z\widetilde{M}_{021}+2 u_x^2 u_y\widetilde{M}_{012}\\[0.1cm]
         &+8 u_x u_y  u_z\widetilde{M}_{111}+u_y^2 u_z^2\widetilde{M}_{200}+u_x^2 u_z^2\widetilde{M}_{020}+u_x^2 u_y^2\widetilde{M}_{002} +4 u_x u_y u_z^2\widetilde{M}_{110}+4 u_x u_y^2 u_z\widetilde{M}_{101}\\[0.1cm]
         &+4 u_x^2 u_y u_z\widetilde{M}_{011} +u_x^2 u_y^2  u_z^2.
\end{array}
\right.
\label{eq:RawCentral3D}
\end{equation}
and
\begin{equation}
\left.
\begin{array}{l @{\:} l}
\widetilde{M}_{000}= & M_{000}=1,\\\\[0.1cm]
\widetilde{M}_{100}= & M_{100}-u_x=0,\\[0.1cm]
\widetilde{M}_{010}= & M_{010}-u_y=0,\\[0.1cm]
\widetilde{M}_{001}= & M_{001}-u_z=0,\\\\[0.1cm]
\widetilde{M}_{200}= & M_{200}-u_x^2,\\[0.1cm]
\widetilde{M}_{020}= & M_{020}-u_y^2,\\[0.1cm]
\widetilde{M}_{002}= & M_{002}-u_z^2,\\[0.1cm]
\widetilde{M}_{110}= & M_{110}-u_xu_y ,\\[0.1cm]
\widetilde{M}_{101}= & M_{101}-u_xu_z ,\\[0.1cm]
\widetilde{M}_{011}= & M_{011}-u_yu_z ,\\\\[0.1cm]
\widetilde{M}_{210}= & M_{210}-u_y M_{200}-2 u_x M_{110}+2 u_x^2 u_y,\\[0.1cm]
\widetilde{M}_{201}= & M_{201}-u_z M_{200}-2 u_x M_{101}+2 u_x^2u_z ,\\[0.1cm]
\widetilde{M}_{120}= & M_{120}-u_x M_{020}-2 u_y M_{110}+2 u_x u_y^2,\\[0.1cm]
\widetilde{M}_{021}= & M_{021}-u_z M_{020}-2 u_y M_{011}+2 u_y^2 u_z ,\\[0.1cm]
\widetilde{M}_{102}= & M_{102}-u_x M_{002}-2 u_z M_{101}+2 u_x u_z^2,\\[0.1cm]
\widetilde{M}_{012}= & M_{012}-u_y M_{002}-2 u_z M_{011}+2 u_y u_z^2,\\[0.1cm]
\widetilde{M}_{111}= & M_{111}-u_z M_{110}-u_y M_{101}-u_x M_{011}+2 u_x u_y u_z,\\\\[0.1cm]
\widetilde{M}_{220}= & M_{220}-2 u_y M_{210}-2 u_x M_{120}+u_y^2 M_{200}+u_x^2 M_{020}+4 u_x u_y M_{110}-3 u_x^2 u_y^2,\\[0.1cm]
\widetilde{M}_{202}= & M_{202}-2 u_z M_{201}-2 u_x M_{102}+u_z^2 M_{200}+u_x^2 M_{002}+4 u_x u_z M_{101}-3 u_x^2 u_z^2,\\[0.1cm]
\widetilde{M}_{022}= & M_{022}-2 u_z M_{021}-2 u_y M_{012}+u_z^2 M_{020}+u_y^2 M_{002}+4 u_y u_z M_{011}-3 u_y^2 u_z^2,\\[0.1cm]
\widetilde{M}_{211}= & M_{211}-u_z M_{210} - u_y M_{201}- 2 u_x M_{111} +u_y u_zM_{200} +2 u_x u_zM_{110}+2 u_x u_yM_{101}+u_x^2 M_{011} -3 u_x^2u_yu_z,\\[0.1cm]
\widetilde{M}_{121}= & M_{121}-u_z M_{120} - u_x M_{021}- 2 u_y M_{111} +u_x u_zM_{020} +2 u_y u_zM_{110}+2 u_x u_yM_{011}+u_y^2 M_{101} -3 u_x u_y^2 u_z ,\\[0.1cm]
\widetilde{M}_{112}= & M_{112}-u_y M_{102} - u_x M_{012}- 2 u_z M_{111} +u_x u_yM_{002} +2 u_y u_zM_{101}+2 u_x u_zM_{011}+u_z^2 M_{110} -3 u_xu_y u_z^2,\\\\[0.1cm]
\widetilde{M}_{221}= & M_{221}-u_z M_{220}-2 u_y M_{211}-2 u_xM_{121}+2 u_y u_zM_{210} +u_y^2M_{201} +u_x^2M_{021}+2 u_xu_zM_{120}+4 u_x u_yM_{111}\\[0.1cm]
         &-u_y^2 u_zM_{200} -u_x^2 u_zM_{020} -4 u_x u_y u_zM_{110}-2 u_x u_y^2M_{101} -2 u_x^2 u_yM_{011} + 4 u_x^2 u_y^2 u_z ,\\[0.1cm]
\widetilde{M}_{212}= & M_{212}-u_y M_{202}-2 u_z M_{211}-2 u_xM_{112} +u_z^2M_{210}+2 u_y u_zM_{201}+2 u_x u_yM_{102}+u_x^2M_{012}+4 u_x u_zM_{111}\\[0.1cm]
         &-u_y u_z^2M_{200}-u_x^2 u_yM_{002}-2 u_x u_z^2M_{110}-2 u_x^2 u_zM_{011}-4 u_x u_y u_z M_{101}+ 4 u_x^2 u_y u_z^2,\\[0.1cm]
\widetilde{M}_{122}= & M_{122}-u_x M_{022}-2 u_z M_{121}-2 u_yM_{112} +u_z^2M_{120} +u_y^2 M_{102}+2 u_x u_zM_{021}+2 u_x u_yM_{012} +4 u_y u_zM_{111}\\[0.1cm]
         &-u_x u_z^2M_{020}-u_x u_y^2M_{002}-2 u_y u_z^2 M_{110} -2 u_y^2 u_z M_{101} -4 u_xu_y u_zM_{011}  +4 u_xu_y^2 u_z^2,\\\\[0.1cm]
\widetilde{M}_{222}= & M_{222}-2 u_zM_{221} -2 u_yM_{212} -2 u_xM_{122} + u_z^2M_{220}+u_y^2M_{202}+u_x^2M_{022}+4 u_y u_zM_{211}+4 u_x u_zM_{121}\\[0.1cm]
         &+4 u_x u_yM_{112} -2 u_y u_z^2M_{210}-2 u_y^2 u_zM_{201}-2 u_x u_z^2M_{120}-2 u_y^2 u_xM_{102}-2 u_x^2 u_zM_{021}-2 u_x^2 u_yM_{012}\\[0.1cm]
         &-8 u_x u_y  u_zM_{111}+u_y^2 u_z^2M_{200}+u_x^2 u_z^2M_{020}+u_x^2 u_y^2M_{002} +4 u_x u_y u_z^2M_{110}+4 u_x u_y^2 u_zM_{101}\\[0.1cm]
         &+4 u_x^2 u_y u_zM_{011} -5 u_x^2 u_y^2  u_z^2.
\end{array}
\right.
\label{eq:CentralRaw3D}
\end{equation}

When it comes to raw moments and cumulants, Kendall's differential operators are used to derive trivariate formulas from the univariate case~\cite{KENDALL_AE_10_1940,COOK_Bio_38_1951}. Up to $K_{222}$ and $M_{222}$, this leads to

\begin{equation}
\hspace*{-1.25cm}
\left.
\begin{array}{l @{\:} l}
M_{000}= &\exp(K_{000})=1, \\\\[0.1cm] 
M_{100}= &K_{100}=u_x ,\: M_{010}= K_{010}=u_y,\:M_{001}= K_{001}=u_z, \\\\[0.1cm]
M_{200}= &K_{200} +u_x^2,\: M_{020}= K_{020} +u_y^2,\: M_{002}= K_{002} +u_z^2,\\[0.1cm]
M_{110}= &K_{110} +u_x u_y,\: M_{101}= K_{101} +u_x u_z,\: M_{011}= K_{011} +u_y u_z,\\\\[0.1cm]
M_{210}= &K_{210}+u_y K_{200}+2 u_x K_{110}+ u_x^2 u_y,\\[0.1cm]
M_{201}= &K_{201}+u_z K_{200}+2 u_x K_{101}+ u_x^2 u_z,\\[0.1cm]
M_{021}= &K_{021}+u_z K_{020}+2 u_y K_{011}+ u_y^2 u_z,\\[0.1cm]
M_{120}= &K_{120}+u_x K_{020}+2 u_y K_{110}+ u_x u_y^2,\\[0.1cm]
M_{102}= &K_{102}+u_x K_{002}+2 u_z K_{101}+ u_x u_z^2,\\[0.1cm]
M_{012}= &K_{012}+u_y K_{002}+2 u_z K_{011}+ u_y u_z^2,\\[0.1cm]
M_{111}= &K_{111}+u_z K_{110}+u_y K_{101}+u_x K_{011}+ u_x u_y u_z,\\\\[0.1cm]
M_{220}= & K_{220} +2(u_y K_{210}+  u_x K_{120})+u_x^2 K_{020}+K_{020} K_{200} +u_y^2 K_{200}+2 K_{110}^2+4 u_x u_y K_{110}+ u_x^2 u_y^2,\\[0.1cm]
M_{202}= & K_{202} +2(u_z K_{201}+ u_x K_{102})+u_x^2 K_{002}+K_{002} K_{200} +u_z^2 K_{200}+2 K_{101}^2+4 u_x u_z K_{101}+ u_x^2 u_z^2,\\[0.1cm]
M_{022}= & K_{022} +2(u_z K_{021}+ u_y K_{012})+u_y^2 K_{002}+K_{002} K_{020} +u_z^2 K_{020}+2 K_{011}^2+4 u_y u_z K_{011}+ u_y^2 u_z^2,\\[0.1cm]
M_{211}= &K_{211}+ u_z K_{210}+u_y K_{201} +2 u_x K_{111}+K_{011} K_{200}+2 K_{101} K_{110}+ u_y u_z K_{200}
+2 (u_x u_z K_{110}\\[0.1cm]
    &+u_x u_y K_{101})+ u_x^2 K_{011}+ u_x^2 u_y u_z,\\[0.1cm]
M_{121}= &K_{121}+ u_z K_{120}+u_x K_{021} +2 u_y K_{111}+K_{101} K_{020}+2 K_{011} K_{110}+ u_x u_z K_{020}
+2 (u_y u_z K_{110}\\[0.1cm]
    &+u_x u_y K_{011})+ u_y^2 K_{101}+ u_x u_y^2 u_z,\\[0.1cm]
M_{112}= &K_{112}+ u_y K_{102}+u_x K_{012} +2 u_z K_{111}+K_{110} K_{002}+2 K_{011} K_{101}+ u_x u_y K_{002}
+2 (u_y u_z K_{101}\\[0.1cm]
    &+u_x u_z K_{011})+ u_z^2 K_{110}+ u_x u_y u_z^2,\\\\[0.1cm]
M_{221}= &K_{221}+u_z K_{220} +2 (u_x K_{121}+u_y K_{211})+ K_{020} K_{201}+ K_{200}K_{021} +2 (K_{101} K_{120}+K_{011} K_{210}) \\[0.1cm] 
    &+4 K_{110} K_{111}+2 (u_y u_z K_{210}+u_x u_z K_{120})+ u_y^2 K_{201}+u_x^2 K_{021} +4 u_x u_y K_{111} + u_z K_{020} K_{200}\\[0.1cm] 
    &+2 (u_y K_{011} K_{200}+ u_x K_{101} K_{020}+ u_z K_{110}^2) +4(u_y K_{101} K_{110}+u_x K_{011} K_{110})+u_y^2 u_z K_{200}\\[0.1cm]
    &+u_x^2 u_z K_{020} +2(u_x u_y^2 K_{101}+u_x^2 u_y K_{011})+4 u_x u_y u_z K_{110}+ u_x^2 u_y^2 u_z,\\[0.1cm] 
M_{212}= &K_{212}+u_y K_{202} +2 (u_z K_{211}+u_x K_{112})+ K_{002} K_{210}+ K_{200}K_{012} +2 (K_{011} K_{201}+K_{110} K_{102}) \\[0.1cm] 
    &+4 K_{101} K_{111}+2 (u_y u_z K_{201}+u_x u_y K_{102})+ u_z^2 K_{210}+u_x^2 K_{012} +4 u_x u_z K_{111} + u_y K_{002} K_{200}\\[0.1cm] 
    &+2 (u_z K_{011} K_{200}+ u_x K_{110} K_{002}+ u_y K_{101}^2) +4(u_z K_{110} K_{101}+u_x K_{011} K_{101})+u_y u_z^2 K_{200}\\[0.1cm]
    &+u_x^2 u_y K_{002} +2(u_x u_z^2 K_{110}+u_x^2 u_z K_{011})+4 u_x u_y u_z K_{101}+ u_x^2 u_y u_z^2,\\[0.1cm]
M_{122}= &K_{122}+u_x K_{022} +2 (u_z K_{121}+u_y K_{112})+ K_{002} K_{120}+ K_{020}K_{102} +2 (K_{101} K_{021}+K_{110} K_{012}) \\[0.1cm] 
    &+4 K_{011} K_{111}+2 (u_x u_z K_{021}+u_x u_y K_{012})+ u_z^2 K_{120}+u_y^2 K_{102} +4 u_y u_z K_{111} + u_x K_{002} K_{020}\\[0.1cm] 
    &+2 (u_z K_{101} K_{020}+ u_y K_{110} K_{002}+ u_x K_{011}^2) +4(u_z K_{110} K_{011}+u_y K_{011} K_{101})+u_x u_z^2 K_{020}\\[0.1cm]
    &+u_x u_y^2 K_{002} +2(u_y u_z^2 K_{110}+u_y^2 u_z K_{101})+4 u_x u_y u_z K_{011}+ u_x u_y^2 u_z^2,\\\\[0.1cm]
M_{222}= &K_{222} +2 (u_z K_{221} + u_y K_{212} + u_x K_{122}) + K_{002} K_{220} + K_{020} K_{202} + K_{200}K_{022} +4(K_{011} K_{211} \\[0.1cm] 
    &+ K_{101} K_{121} + K_{110} K_{112})+ u_z^2 K_{220} + u_y^2 K_{202} + u_x^2 K_{022} +4 (u_y u_z K_{211} + u_x u_z K_{121} + u_x u_y K_{112})\\[0.1cm] 
    &+2 (K_{012} K_{210} + K_{021} K_{201} + K_{102} K_{120})+4 K_{111}^2 +2 (u_y K_{002} K_{210}+u_z K_{020} K_{201}+u_x K_{002} K_{120}\\[0.1cm] 
    &+u_y K_{200} K_{012}+u_z K_{200} K_{021}+u_x K_{020} K_{102})+4 (u_z K_{011} K_{210}+u_y K_{011} K_{201}+u_x K_{101} K_{021} \\[0.1cm] 
    &+u_z K_{101} K_{120}+u_y K_{110} K_{102}+u_x K_{110} K_{012})+8(u_z K_{110} K_{111}+u_y K_{101} K_{111}+u_x K_{011} K_{111})\\[0.1cm] 
    &+2 (u_y u_z^2 K_{210}+u_y^2 u_z K_{201}+u_x^2 u_z K_{021}+u_x u_z^2 K_{120}+u_x u_y^2 K_{102}+u_x^2 u_y K_{012})+8 u_x u_y u_z K_{111}\\[0.1cm] 
    &+ K_{200} K_{020} K_{002}+2(K_{110}^2 K_{002}+K_{011}^2 K_{200}+K_{101}^2 K_{020})+8 K_{011} K_{101} K_{110}+ u_z^2 K_{200} K_{020}\\[0.1cm] 
    &+u_y^2 K_{200} K_{002}+u_x^2 K_{020} K_{002}+4(u_y u_z K_{011} K_{200}+u_x u_z K_{101} K_{020}+u_x u_y K_{110} K_{002})\\[0.1cm] 
    &+2(u_z^2 K_{110}^2+u_y^2 K_{101}^2+u_x^2 K_{011}^2)+8(u_x u_z K_{011} K_{110}+u_y u_z K_{101} K_{110}+u_x u_y K_{011} K_{101})\\[0.1cm]
    &+ u_y^2 u_z^2 K_{200}+u_x^2 u_z^2 K_{020}+u_x^2 u_y^2 K_{002}+4(u_x u_y u_z^2 K_{110}+u_x u_y^2 u_z K_{101}+u_x^2 u_y u_z K_{011})+ u_y^2 u_z^2 u_x^2,
\end{array}
\right.
\label{eq:RawCumul3D}
\end{equation}
and
\begin{equation}
\hspace*{-.25cm}
\left.
\begin{array}{l @{\:} l}
K_{000}= &\ln(M_{000})=0, \\\\[0.1cm] 
K_{100}= &M_{100}=u_x ,\: K_{010}= M_{010}=u_y,\:K_{001}= M_{001}=u_z, \\\\[0.1cm]
K_{200}= &M_{200} -u_x^2,\: K_{020}= M_{020} -u_y^2,\: K_{002}= M_{002} -u_z^2,\\[0.1cm]
K_{110}= &M_{110} -u_x u_y,\: K_{101}= M_{101} -u_x u_z,\: K_{011}= M_{011} -u_y u_z,\\\\[0.1cm]
K_{210}= &M_{210}-u_y M_{200}-2 u_x M_{110}+2 u_x^2 u_y,\\[0.1cm]
K_{201}= &M_{201}-u_z M_{200}-2 u_x M_{101}+2 u_x^2 u_z,\\[0.1cm]
K_{021}= &M_{021}-u_z M_{020}-2 u_y M_{011}+2 u_y^2 u_z,\\[0.1cm]
K_{120}= &M_{120}-u_x M_{020}-2 u_y M_{110}+2 u_x u_y^2,\\[0.1cm]
K_{102}= &M_{102}-u_x M_{002}-2 u_z M_{101}+2 u_x u_z^2,\\[0.1cm]
K_{012}= &M_{012}-u_y M_{002}-2 u_z M_{011}+2 u_y u_z^2,\\[0.1cm]
K_{111}= &M_{111} -u_z M_{110} -u_y M_{101} -u_x M_{011}+2 u_x u_y u_z,\\\\[0.1cm]
K_{220}= & M_{220} -2 (u_y M_{210} + u_x M_{120})-M_{200}M_{020}+2 u_x^2 M_{020} +2 u_y^2 M_{200}-2 M_{110}^2+8 u_x u_y M_{110}-6 u_x^2 u_y^2,\\[0.1cm]
K_{202}= & M_{202} -2 (u_z M_{201} + u_x M_{102})-M_{200}M_{002}+2 u_x^2 M_{002} +2 u_z^2 M_{200}-2 M_{101}^2+8 u_x u_z M_{101}-6 u_x^2 u_z^2,\\[0.1cm]
K_{022}= & M_{022} -2 (u_z M_{021} + u_y M_{012})-M_{020}M_{002}+2 u_y^2 M_{002} +2 u_z^2 M_{020}-2 M_{011}^2+8 u_y u_z M_{011}-6 u_y^2 u_z^2,\\[0.1cm]
K_{211}= &M_{211}-(u_z M_{210}+u_y M_{201})-2  u_x M_{111}-M_{011} M_{200}-2 M_{101} M_{110}+2 u_y u_z M_{200}
\\[0.1cm]
    &+4 (u_x u_z M_{110}+u_x u_y M_{101})+2 u_x^2 M_{011}-6 u_x^2 u_y u_z,\\[0.1cm]
K_{121}= &M_{121}-(u_z M_{120}+u_x M_{021})-2 u_y M_{111}-M_{101} M_{020}-2 M_{011} M_{110}+2 u_x u_z M_{020}
\\[0.1cm]
    &+4 (u_y u_z M_{110}+u_x u_y M_{011})+2 u_y^2 M_{101} -6 u_x u_y^2 u_z,\\[0.1cm]
K_{112}= &M_{112}-(u_y M_{102}+u_x M_{012})-2 u_z M_{111}-M_{110} M_{002}-2 M_{011} M_{101}+2 u_x u_y M_{002}
\\[0.1cm]
    &+4 (u_y u_z M_{101}+u_x u_z M_{011})+2 u_z^2 M_{110} -6 u_x u_y u_z^2,\\\\[0.1cm]
K_{221}= &M_{221}-u_z M_{220} -2 (u_x M_{121}+u_y M_{211})-(M_{020} M_{201}+ M_{200}M_{021})-2 (M_{101} M_{120}+M_{011} M_{210})\\[0.1cm] 
    &-4 M_{110} M_{111}+4 (u_y u_z M_{210}+u_x u_z M_{120})+2 (u_y^2 M_{201}+u_x^2 M_{021}) +8 u_x u_y M_{111} \\[0.1cm] 
    &+2 u_z M_{020} M_{200}+4 (u_y M_{011} M_{200}+ u_x M_{101} M_{020}+ u_z M_{110}^2)+8(u_y M_{101} M_{110}+u_x M_{011} M_{110}) \\[0.1cm]
    &-6 (u_y^2 u_z M_{200}+u_x^2 u_z M_{020})-12(u_x u_y^2 M_{101}+u_x^2 u_y M_{011})-24 u_x u_y u_z M_{110}+24 u_x^2 u_y^2 u_z,\\[0.1cm]
K_{212}= &M_{212}- u_y M_{202} -2 (u_z M_{211}+u_x M_{112})-(M_{002} M_{210}+ M_{200}M_{012})-2 (M_{011} M_{201}+M_{110} M_{102}) \\[0.1cm] 
    &-4 M_{101} M_{111}+4 (u_y u_z M_{201}+u_x u_y M_{102})+2 (u_z^2 M_{210}+u_x^2 M_{012}) +8 u_x u_z M_{111} \\[0.1cm] 
    &+2 u_y M_{002} M_{200}+4 (u_z M_{011} M_{200}+u_x M_{110} M_{002}+u_y M_{101}^2) +8(u_z M_{110} M_{101}+u_x M_{011} M_{101})\\[0.1cm]
    &-6 (u_y u_z^2 M_{200}+u_x^2 u_y M_{002})-12(u_x u_z^2 M_{110}+u_x^2 u_z M_{011})-24 u_x u_y u_z M_{101}+24 u_x^2 u_y u_z^2,\\[0.1cm]
K_{122}= &M_{122}- u_x M_{022} -2 (u_z M_{121}+u_y M_{112})-(M_{002} M_{120}+ M_{020}M_{102})-2 (M_{101} M_{021}+M_{110} M_{012}) \\[0.1cm] 
    &-4 M_{011} M_{111}+4 (u_x u_z M_{021}+u_x u_y M_{012})+2 (u_z^2 M_{120}+u_y^2 M_{102}) +8 u_y u_z M_{111} \\[0.1cm] 
    &+2 u_x M_{002} M_{020}+4 (u_z M_{101} M_{020}+ u_y M_{110} M_{002}+ u_x M_{011}^2) +8(u_z M_{110} M_{011}+u_y M_{011} M_{101})\\[0.1cm]
    &-6 (u_x u_z^2 M_{020}+u_x u_y^2 M_{002})-12(u_y u_z^2 M_{110}+u_y^2 u_z M_{101})-24 u_x u_y u_z M_{011}+24 u_x u_y^2 u_z^2,\\\\[0.1cm] 
K_{222}= &M_{222} -2 (u_z M_{221} + u_y M_{212} + u_x M_{122}) -(M_{002} M_{220} + M_{020} M_{202} + M_{200}M_{022})-4(M_{011} M_{211} \\[0.1cm] 
    &+ M_{101} M_{121} + M_{110} M_{112})+2 (u_z^2 M_{220} + u_y^2 M_{202} + u_x^2 M_{022})+8 (u_y u_z M_{211} + u_x u_z M_{121} + u_x u_y M_{112})\\[0.1cm] 
    &-2 (M_{012} M_{210} + M_{021} M_{201} + M_{102} M_{120})-4 M_{111}^2+4 (u_y M_{002} M_{210}+u_z M_{020} M_{201}+u_x M_{002} M_{120}\\[0.1cm] 
    &+u_y M_{200} M_{012}+u_z M_{200} M_{021}+u_x M_{020} M_{102})+8 (u_z M_{011} M_{210}+u_y M_{011} M_{201}+u_x M_{101} M_{021} \\[0.1cm] 
    &+u_z M_{101} M_{120}+u_y M_{110} M_{102}+u_x M_{110} M_{012})+16(u_z M_{110} M_{111}+u_y M_{101} M_{111}+u_x M_{011} M_{111})\\[0.1cm] 
    &-12 (u_y u_z^2 M_{210}+u_y^2 u_z M_{201}+u_x^2 u_z M_{021}+u_x u_z^2 M_{120}+u_x u_y^2 M_{102}+u_x^2 u_y M_{012})-48 u_x u_y u_z M_{111}\\[0.1cm] 
    &+2 M_{200} M_{020} M_{002}+4(M_{110}^2 M_{002}+M_{011}^2 M_{200}+M_{101}^2 M_{020})+16 M_{011} M_{101} M_{110}-6(u_z^2 M_{200} M_{020}\\[0.1cm] 
    &+u_y^2 M_{200} M_{002}+u_x^2 M_{020} M_{002})-24(u_y u_z M_{011} M_{200}+u_x u_z M_{101} M_{020}+u_x u_y M_{110} M_{002})-12(u_z^2 M_{110}^2\\[0.1cm] 
    &+u_y^2 M_{101}^2+u_x^2 M_{011}^2)-48(u_x u_z M_{011} M_{110}+u_y u_z M_{101} M_{110}+u_x u_y M_{011} M_{101})+24(u_y^2 u_z^2 M_{200}\\[0.1cm]
    &+u_x^2 u_z^2 M_{020}+u_x^2 u_y^2 M_{002})+96(u_x u_y u_z^2 M_{110}+u_x u_y^2 u_z M_{101}+u_x^2 u_y u_z M_{011})-120 u_y^2 u_z^2 u_x^2. 
\end{array}
\right.
\label{eq:CumulRaw3D}
\end{equation}


Relationships between central moments and cumulants are finally obtained discarding velocity dependent terms in the above formulas, 

\begin{equation}
\left.
\begin{array}{l @{\:} l @{\quad} r @{\:} l}
K_{000}= & \ln(\widetilde{M}_{000})=0, & \widetilde{M}_{000}= & \exp(\widetilde{K}_{000})=1,\\\\[0.1cm]
K_{100}= & \widetilde{M}_{100}+u_x=u_x, & \widetilde{M}_{100}= & K_{100} - u_x=0,\\[0.1cm]
K_{010}= & \widetilde{M}_{010}+u_y=u_y, & \widetilde{M}_{010}= & K_{010} - u_y=0,\\[0.1cm]
K_{001}= & \widetilde{M}_{001}+u_z=u_z, & \widetilde{M}_{001}= & K_{001} - u_z=0,\\\\[0.1cm]
K_{200}= & \widetilde{M}_{200}, & \widetilde{M}_{200}= & K_{200},\\[0.1cm]
K_{020}= & \widetilde{M}_{020}, & \widetilde{M}_{020}= & K_{020},\\[0.1cm]
K_{002}= & \widetilde{M}_{002}, & \widetilde{M}_{002}= & K_{002},\\[0.1cm]
K_{110}= & \widetilde{M}_{110}, & \widetilde{M}_{110}= & K_{110},\\[0.1cm]
K_{101}= & \widetilde{M}_{101}, & \widetilde{M}_{101}= & K_{101},\\[0.1cm]
K_{011}= & \widetilde{M}_{011}, & \widetilde{M}_{011}= & K_{011},\\\\[0.1cm]
K_{210}= & \widetilde{M}_{210}, & \widetilde{M}_{210}= & K_{210},\\[0.1cm]
K_{201}= & \widetilde{M}_{201}, & \widetilde{M}_{201}= & K_{201},\\[0.1cm]
K_{120}= & \widetilde{M}_{120}, & \widetilde{M}_{120}= & K_{120},\\[0.1cm]
K_{021}= & \widetilde{M}_{021}, & \widetilde{M}_{021}= & K_{021},\\[0.1cm]
K_{102}= & \widetilde{M}_{102}, & \widetilde{M}_{102}= & K_{102},\\[0.1cm]
K_{012}= & \widetilde{M}_{012}, & \widetilde{M}_{012}= & K_{012},\\[0.1cm]
K_{111}= & \widetilde{M}_{111}, & \widetilde{M}_{111}= & K_{111},\\\\[0.1cm]
K_{220}= & \widetilde{M}_{220}-\widetilde{M}_{200}\widetilde{M}_{020}-2 \widetilde{M}_{110}^2, & \widetilde{M}_{220}= & K_{220}+K_{200}K_{020}+2 K_{110}^2,\\[0.1cm]
K_{202}= & \widetilde{M}_{202}-\widetilde{M}_{200}\widetilde{M}_{002}-2 \widetilde{M}_{101}^2, & \widetilde{M}_{202}= & K_{202}+K_{200}K_{002}+2 K_{101}^2,\\[0.1cm]
K_{022}= & \widetilde{M}_{022}-\widetilde{M}_{020}\widetilde{M}_{002}-2 \widetilde{M}_{011}^2, & \widetilde{M}_{022}= & K_{022}+K_{020}K_{002}+2 K_{011}^2,\\[0.1cm]
K_{211}= & \widetilde{M}_{211}-\widetilde{M}_{200}\widetilde{M}_{011}-2 \widetilde{M}_{110}\widetilde{M}_{101}, & \widetilde{M}_{211}= & K_{211}+K_{200}K_{011}+2 K_{110}K_{101},\\[0.1cm]
K_{121}= & \widetilde{M}_{121}-\widetilde{M}_{020}\widetilde{M}_{101}-2 \widetilde{M}_{110}\widetilde{M}_{011}, & \widetilde{M}_{121}= & K_{121}+K_{020}K_{101}+2 K_{110}K_{011},\\[0.1cm]
K_{112}= & \widetilde{M}_{112}-\widetilde{M}_{002}\widetilde{M}_{110}-2 \widetilde{M}_{101}\widetilde{M}_{011}, & \widetilde{M}_{112}= & K_{112}+K_{002}K_{110}+2 K_{101}K_{011},\\\\[0.1cm]
K_{221}= & \widetilde{M}_{221}-\widetilde{M}_{201}\widetilde{M}_{020}-\widetilde{M}_{021}\widetilde{M}_{200}-2 \widetilde{M}_{210}\widetilde{M}_{011} & \widetilde{M}_{221}= & K_{221}+K_{201}K_{020}+K_{021}K_{200}+2 K_{210}K_{011}\\[0.1cm]
         &-2 \widetilde{M}_{120} \widetilde{M}_{101}-4 \widetilde{M}_{111}\widetilde{M}_{110}, & &+2 K_{120} K_{101}+4 K_{111}K_{110},\\[0.1cm]
K_{212}= & \widetilde{M}_{212}-\widetilde{M}_{210}\widetilde{M}_{002}-\widetilde{M}_{012}\widetilde{M}_{200}-2 \widetilde{M}_{201}\widetilde{M}_{011} & \widetilde{M}_{212}= & K_{212}+K_{210}K_{002}+K_{012}K_{200}+2 K_{201}K_{011}\\[0.1cm]
         &-2 \widetilde{M}_{102} \widetilde{M}_{110}-4 \widetilde{M}_{111}\widetilde{M}_{101}, & &+2 K_{102} K_{110}+4 K_{111}K_{101},\\[0.1cm]
K_{122}= & \widetilde{M}_{122}-\widetilde{M}_{120}\widetilde{M}_{002}-\widetilde{M}_{102}\widetilde{M}_{020}-2 \widetilde{M}_{021}\widetilde{M}_{101} & \widetilde{M}_{122}= & K_{122}+K_{120}K_{002}+K_{102}K_{020}+2 K_{021}K_{101}\\[0.1cm]
         &-2 \widetilde{M}_{012} \widetilde{M}_{110}-4 \widetilde{M}_{111}\widetilde{M}_{011}, & &+2 K_{012} K_{110}+4 K_{111}K_{011},\\\\[0.1cm]
K_{222}= & \widetilde{M}_{222}-(\widetilde{M}_{220}\widetilde{M}_{002}+\widetilde{M}_{202}\widetilde{M}_{020}+\widetilde{M}_{022}\widetilde{M}_{200}) & \widetilde{M}_{222}= & K_{222}+K_{220}K_{002}+K_{202}K_{020}+K_{022}K_{200}\\[0.1cm]
         &-4(\widetilde{M}_{211} \widetilde{M}_{011}+ \widetilde{M}_{121}\widetilde{M}_{101}+ \widetilde{M}_{112}\widetilde{M}_{110}) & &+4 (K_{211} K_{011}+ K_{121}K_{101}+ K_{112}K_{110})\\[0.1cm]
         &-2(\widetilde{M}_{210}\widetilde{M}_{012}+ \widetilde{M}_{201}\widetilde{M}_{021}+ \widetilde{M}_{120}\widetilde{M}_{102})-4 \widetilde{M}_{111}^2 & &+2 (K_{210}K_{012}+ K_{201}K_{021}+ K_{120}K_{102})+4 K_{111}^2\\[0.1cm]
         &+4(\widetilde{M}_{200} \widetilde{M}_{011}^2+ \widetilde{M}_{020} \widetilde{M}_{101}^2+ \widetilde{M}_{002} \widetilde{M}_{110}^2) & &+2 (K_{200} K_{011}^2+ K_{020} K_{101}^2+ K_{002} K_{110}^2)\\[0.1cm]
         &+16 \widetilde{M}_{110}\widetilde{M}_{101}\widetilde{M}_{011}+2 \widetilde{M}_{002} \widetilde{M}_{020} \widetilde{M}_{200}, & &+8 K_{110}K_{101}K_{011}+ K_{002} K_{020} K_{200}.
\end{array}
\right.
\label{eq:CumulCentral3D}
\end{equation}

\section{Linear transformation matrices\label{sec:LinearTransMatrix}}

This appendix compiles all linear transformations allowing us to write the collision step in a matrix form in the context of the D2Q9 velocity discretization, i.e,
$$\textcolor{black}{\bm{f}^* = \bm{f}^{eq} + (\bm{I} - \bm{M^{-1}_{\mathrm{HM}}S_{\mathrm{HM}}M_{\mathrm{HM}}}) \bm{f}^{neq}},$$
where $\textcolor{black}{\bm{f}^{*}}$, $\textcolor{black}{\bm{f}^{eq}}$ and $\textcolor{black}{\bm{f}^{neq}}$ are vectors composed of all nine post collision, \textcolor{black}{equilibrium} and nonequilibrium populations respectively.
Starting with the set of orthogonal polynomials~(\ref{eq:MRTBasis}), it is explained how to compute $\bm{M}$ and $\bm{M^{-1}}$ from a general point of view~\cite{LALLEMAND_PRE_61_2000}. This allows a straightforward extension to any kind of lattice, in either 2D or 3D. These linear transformation matrices (LTMs) are then given for RM, HM, CM and CHM frameworks. For the last two, the concept of shifting matrices is used to move from the reference frame at rest to the comoving one~\cite{FEI_PRE_96_2017}. Eventually, these matrices are provided in the context of the Gauss-Hermite quadrature on which regularized collision models are based. Hereafter, the single index $i$ is used to describe each discrete velocity $\bm{\xi}_i$ and its related population $f_i$.

\subsection{Construction of LTMs}

Let us start with the orthogonal basis proposed by Lallemand and Luo~\cite{LALLEMAND_PRE_61_2000},
$$(M_{\rho},M_{j_x},M_{j_y},M_{e},M_{p_{xx}},M_{p_{xy}},M_{q_x},M_{q_y},M_{\varepsilon}),
$$
where each of its component are defined as,
\begin{align*}
M_{\rho} &= 1, \: M_{j_x} = \xi_{i,x}, \:M_{j_y} = \xi_{i,y}, \:M_{e} = -4 + 3(\xi_{i,x}^2+\xi_{i,y}^2), \:M_{p_{xy}} = \xi_{i,x}^2-\xi_{i,y}^2, M_{p_{xx}} = \xi_{i,x}\xi_{i,y}, \\
M_{q_{x}} &= [-5 + 3(\xi_{i,x}^2+\xi_{i,y}^2)]\xi_{i,x}, \:M_{q_{y}} = [-5 + 3(\xi_{i,x}^2+\xi_{i,y}^2)]\xi_{i,y}, \:M_{\varepsilon} = 4 - \frac{21}{2}(\xi_{i,x}^2+\xi_{i,y}^2) + \frac{9}{2}(\xi_{i,x}^2+\xi_{i,y}^2)^2. \\
\end{align*}
Due to the aliasing defect of the D2Q9 lattice ($\xi_{i,x}^3=\xi_{i,x}$ and $\xi_{i,y}^3=\xi_{i,y}$), $M_{q_{x}}$, $M_{q_{y}}$ and $M_{q_{\varepsilon}}$ can further be simplified into
 \begin{align*}
M_{q_{x}} &= [-5 + 3(1+\xi_{i,y}^2)]\xi_{i,x}, \: M_{q_{y}} = [-5 + 3(\xi_{i,x}^2+1)]\xi_{i,y},\: M_{\varepsilon} = 4 - 6(\xi_{i,x}^2+\xi_{i,y}^2) + 9\xi_{i,x}^2\xi_{i,y}^2.
\end{align*}
On can see that the ordering of polynomials proposed here is different from the one introduced in the original paper. The present ordering is based on a progressive increase of the degree of each polynomial. Doing so, the comparison with other frameworks, such as cascaded or regularized collision models, is simplified.

In addition, the standard ordering of discrete velocities is considered. In other words, velocities are gathered within several groups depending on their norm, and they are then read `counterclockwise'. This leads to 
\begin{equation}
\begin{array}{r @{\: = \:} r r r r r r r r r}
M_{j_x} & (0 & \phantom{-} 1 & \phantom{-} 0 &-1 & \phantom{-} 0 & \phantom{-} 1 &-1 &-1 & \phantom{-} 1),\\[0.1cm]
M_{j_y} & (0 & \phantom{-} 0 & \phantom{-} 1 & \phantom{-} 0 &-1 & \phantom{-} 1 & \phantom{-} 1 &-1 &-1).
\end{array}
\label{eq:OrderingVelocities}
\end{equation}
The ordering of both sets of discrete velocities and moments is the first source of mistakes when deriving LTMs. It is then of uttermost importance to check them before moving to the construction of $\bm{M}$ and $\bm{M^{-1}}$. In the present work, chosen orderings are described in Eqs.~(\ref{eq:OrderingVelocities}) and~(\ref{eq:MRTBasis}). While the LTM to change from populations to moments is derived replacing each $\bm{\xi}_i$ by its value in the chosen set of polynomials, its inverse is simply derived using standard linear algebra libraries. This eventually leads to
\begin{equation}
\bm{M_{\mathrm{LL}}}=\left[
\begin{array}{c}
M_{\rho}\\[0.1cm]
M_{j_x}\\[0.1cm]
M_{j_y}\\[0.1cm]
M_{e}\\[0.1cm]
M_{p_{xx}}\\[0.1cm]
M_{p_{xy}}\\[0.1cm]
M_{q_x}\\[0.1cm]
M_{q_y}\\[0.1cm]
M_{\varepsilon}
\end{array}
\right]=
\left[
\begin{array}{r r r r r r r r r}
\phantom{-}1 & \phantom{-}1 & \phantom{-}1 & \phantom{-}1 & \phantom{-}1 & \phantom{-}1 & \phantom{-}1 & \phantom{-}1 & \phantom{-}1 \phantom{-}\\
0 & 1 & 0 &-1 & 0 & 1 &-1 &-1 & 1 \phantom{-}\\[0.1cm]
0 & 0 & 1 & 0 &-1 & 1 & 1 &-1 &-1 \phantom{-}\\[0.1cm]
-4 &-1 &-1 &-1 &-1 & 2 & 2 & 2 & 2 \phantom{-}\\[0.1cm]
0 & 1 &-1 & 1 &-1 & 0 & 0 & 0 & 0 \phantom{-}\\[0.1cm]
0 & 0 & 0 & 0 & 0 & 1 &-1 & 1 &-1 \phantom{-}\\[0.1cm]
0 & -2 & 0 & 2 & 0 & 1 &-1 &-1 &1 \phantom{-}\\[0.1cm]
0 & 0 & -2 & 0 & 2 & 1 & 1 &-1 &-1 \phantom{-}\\[0.1cm]
4 &-2 &-2 &-2 &-2 & 1 & 1 & 1 & 1 \phantom{-}
\end{array}
\right],
\end{equation}
and
\begin{equation}
\bm{M^{-1}_{\mathrm{LL}}}=\dfrac{1}{36}\left[
\begin{array}{r r r r r r r r r}
\phantom{-}4 & \phantom{-}0 & \phantom{-}0 & -4 & \phantom{-}0 & \phantom{-}0 & \phantom{-}0 & \phantom{-}0 & \phantom{-}4\phantom{-} \\[0.1cm]
4 &  6 &  0 & -1 &  9 &  0 & -6 &  0 & -2\phantom{-} \\[0.1cm]
4 &  0 &  6 & -1 & -9 &  0 &  0 & -6 & -2\phantom{-} \\[0.1cm]
4 & -6 &  0 & -1 &  9 &  0 &  6 &  0 & -2\phantom{-} \\[0.1cm]
4 &  0 & -6 & -1 & -9 &  0 &  0 &  6 & -2\phantom{-} \\[0.1cm]
4 &  6 &  6 &  2 &  0 &  9 &  3 &  3 &  1\phantom{-} \\[0.1cm]
4 & -6 &  6 &  2 &  0 & -9 & -3 &  3 &  1\phantom{-} \\[0.1cm]
4 & -6 & -6 &  2 &  0 &  9 & -3 & -3 &  1\phantom{-} \\[0.1cm]
4 &  6 & -6 &  2 &  0 & -9 &  3 & -3 &  1\phantom{-}
\end{array}
\right],
\end{equation}
where the subscript LL stands for the orthogonal basis proposed by Lallemand and Luo. The corresponding collision matrix is diagonal, and it reads as $\bm{S_{\mathrm{LL}}} =\mathrm{diag}(0,0,0,\omega_e,\omega_{\nu},\omega_{\nu},\omega_q,\omega_q,\omega_{\varepsilon})$. The collision frequency $\omega_{\nu}$ controls the relaxation of shear related phenomena, whereas $\omega_e$ is related to the attenuation of acoustic waves. $\omega_q$ and $\omega_{\varepsilon}$ further control the dissipation of third and fourth order moments respectively. The latter are usually considered as free parameters that can be tuned to increase the stability of the LBM without impacting the resulting physics.

\subsection{Application to RM and HM frameworks}
In the RM framework, monomials compose the polynomials basis
$$(1,\xi_{i,x},\xi_{i,y},\xi_{i,x}^2,\xi_{i,y}^2,\xi_{i,x}\xi_{i,y},\xi_{i,x}^2\xi_{i,y},\xi_{i,x}\xi_{i,y}^2,\xi_{i,x}^2\xi_{i,y}^2),$$ 
which is now nonorthogonal. Following the same steps as before, new LTMs are obtained. They read as
\begin{equation}
\bm{M_{\mathrm{RM}}}=
\left[
\begin{array}{r r r r r r r r r}
1 & \phantom{-}1 & \phantom{-}1 & \phantom{-}1 & \phantom{-}1 & \phantom{-}1 & \phantom{-}1 & \phantom{-}1 & \phantom{-}1\\[0.1cm]
 0 & 1 & 0 &-1 & 0 & 1 &-1 &-1 & 1\\[0.1cm]
 0 & 0 & 1 & 0 &-1 & 1 & 1 &-1 &-1\\[0.1cm]
 0 & 1 & 0 & 1 & 0 & 1 & 1 & 1 & 1\\[0.1cm]
 0 & 0 & 1 & 0 & 1 & 1 & 1 & 1 & 1\\[0.1cm]
 0 & 0 & 0 & 0 & 0 & 1 &-1 & 1 &-1\\[0.1cm]
 0 & 0 & 0 & 0 & 0 & 1 & 1 &-1 &-1\\[0.1cm]
 0 & 0 & 0 & 0 & 0 & 1 &-1 &-1 & 1\\[0.1cm]
 0 & 0 & 0 & 0 & 0 & 1 & 1 & 1 & 1
\end{array}
\right]\:\:\text{and}\:\:
\bm{M^{-1}_{\mathrm{RM}}}=\dfrac{1}{4}\left[
\begin{array}{r r r r r r r r r}
 4 & \phantom{-}0 & \phantom{-}0 &-4 &-4 & \phantom{-}0 & \phantom{-}0 & \phantom{-}0 & \phantom{-}4\\[0.1cm]
 0 & 2 & 0 & 2 & 0 & 0 & 0 &-2 &-2\\[0.1cm]
 0 & 0 & 2 & 0 & 2 & 0 &-2 & 0 &-2\\[0.1cm]
 0 &-2 & 0 & 2 & 0 & 0 & 0 & 2 &-2\\[0.1cm]
 0 & 0 &-2 & 0 & 2 & 0 & 2 & 0 &-2\\[0.1cm]
 0 & 0 & 0 & 0 & 0 & 1 & 1 & 1 & 1\\[0.1cm]
 0 & 0 & 0 & 0 & 0 &-1 & 1 &-1 & 1\\[0.1cm]
 0 & 0 & 0 & 0 & 0 & 1 &-1 &-1 & 1\\[0.1cm]
 0 & 0 & 0 & 0 & 0 &-1 &-1 & 1 & 1
\end{array}
\right].
\label{eq:LTM_RM}
\end{equation}
In addition, the collision matrix can have two forms depending on the prerequisites of the reader. If one is not interested in acoustically related phenomena, one can impose $\omega_{\nu_b}=\omega_{\nu}$ which leads to $\bm{S_{RM}}=\rm{diag}(0,0,0,\omega_{\nu},\omega_{\nu},\omega_{\nu},\omega_3,\omega_3,\omega_4)$. Otherwise, the collision matrix is only block-diagonal~\cite{KRUGER_Book_2017,FEI_PRE_97_2018}
\begin{equation}
\bm{S_{RM}}=\mathrm{diag}(0,0,0,\bm{C},\omega_{\nu},\omega_3,\omega_3,\omega_4),
\end{equation}
with
\begin{equation*}
\bm{C}=\begin{bmatrix}
\frac{\omega_{\nu_b}+\omega_{\nu}}{2} & \frac{\omega_{\nu_b}-\omega_{\nu}}{2} \\[0.1cm]
\frac{\omega_{\nu_b}-\omega_{\nu}}{2} & \frac{\omega_{\nu_b}+\omega_{\nu}}{2}
\end{bmatrix}.
\end{equation*} 

For the HM framework, one can either use the above method to derive LTMs, or rely on relationships obtained for raw and Hermite moments (Eqs.~(\ref{eq:HM2RM_Q9}) and~(\ref{eq:RM2HM_Q9})). The second method was originally used to build matrices allowing the shift from raw to central moments, and vice versa~\cite{FEI_PRE_96_2017}. Hereafter, it will be employed to derive matrices allowing to link raw and Hermite moments. To do so, one simply needs to rewrite relationships between both kinds of moment in the form of a row-column product. As an example, 
$$A_{22} = M_{22}-c_s^2(M_{20}+M_{02}) +c_s^4 M_{00} = \big[ c_s^4 \:\: 0 \:\: 0 \:\: -c_s^2 \:\: -c_s^2 \:\: 0 \:\: 0 \:\: 0 \:\: 1 \big]\cdot \big[ M_{00} \:\: M_{10} \:\: M_{01} \:\: M_{11} \:\: M_{20} \:\: M_{02} \:\: M_{21} \:\: M_{12} \:\: M_{22} \big]^T,$$
and 
$$M_{22} = A_{22}+c_s^2(A_{20}+A_{02}) +c_s^4 A_{00} = \big[ c_s^4 \:\: 0 \:\: 0 \:\: c_s^2 \:\: c_s^2 \:\: 0 \:\: 0 \:\: 0 \:\: 1 \big]\cdot \big[ A_{00} \:\: A_{10} \:\: A_{01} \:\: A_{11} \:\: A_{20} \:\: A_{02} \:\: A_{21} \:\: A_{12} \:\: A_{22} \big]^T.$$
Eventually,
\begin{equation}
\bm{B_{\mathrm{RM}}^{\mathrm{HM}}}=
\left[
\begin{array}{r r r r r r r r r}
 1 & 0 & 0 & 0 & 0 & 0 & 0 & 0 & 0\\[0.1cm]
 0 & 1 & 0 & 0 & 0 & 0 & 0 & 0 & 0\\[0.1cm]
 0 & 0 & 1 & 0 & 0 & 0 & 0 & 0 & 0\\[0.1cm]
 -c_s^2 & 0 & 0 & 1 & 0 & 0 & 0 & 0 & 0\\[0.1cm]
 -c_s^2 & 0 & 0 & 0 & 1 & 0 & 0 & 0 & 0\\[0.1cm]
 0 & 0 & 0 & 0 & 0 & 1 & 0 & 0 & 0\\[0.1cm]
 0 & 0 & -c_s^2 & 0 & 0 & 0 & 1 & 0 & 0\\[0.1cm]
 0 & -c_s^2 & 0 & 0 & 0 & 0 & 0 & 1 & 0\\[0.1cm]
 c_s^4 & 0 & 0 & -c_s^2 & -c_s^2 & 0 & 0 & 0 & 1
\end{array}
\right]
\:\text{and}\:\:
\bm{B_{\mathrm{HM}}^{\mathrm{RM}}}=\left[
\begin{array}{r r r r r r r r r}
 1 & 0 & 0 & 0 & 0 & 0 & 0 & 0 & 0\\[0.1cm]
 0 & 1 & 0 & 0 & 0 & 0 & 0 & 0 & 0\\[0.1cm]
 0 & 0 & 1 & 0 & 0 & 0 & 0 & 0 & 0\\[0.1cm]
 c_s^2 & 0 & 0 & 1 & 0 & 0 & 0 & 0 & 0\\[0.1cm]
 c_s^2 & 0 & 0 & 0 & 1 & 0 & 0 & 0 & 0\\[0.1cm]
 0 & 0 & 0 & 0 & 0 & 1 & 0 & 0 & 0\\[0.1cm]
 0 & 0 & c_s^2 & 0 & 0 & 0 & 1 & 0 & 0\\[0.1cm]
 0 & c_s^2 & 0 & 0 & 0 & 0 & 0 & 1 & 0\\[0.1cm]
 c_s^4 & 0 & 0 & c_s^2 & c_s^2 & 0 & 0 & 0 & 1
\end{array}
\right],
\end{equation}
where $\bm{B_{\mathrm{RM}}^{\mathrm{HM}}}$ and $\bm{B_{\mathrm{HM}}^{\mathrm{RM}}}=\bm{(B_{\mathrm{RM}}^{\mathrm{HM}})^{-1}}$ allow us to move from raw to Hermite moments, and vice versa. The lower triangular form of these matrices translates the dependency of high order Hermite moments with respects to lower order raw moments and conversely.

Post collision populations are finally obtained through
$$\textcolor{black}{\bm{f}^* = \bm{f}^{eq} + (\bm{I} - \bm{M^{-1}_{\mathrm{HM}}S_{\mathrm{HM}}M_{\mathrm{HM}}}) \bm{f}^{neq}}$$
with $\bm{M_{\mathrm{HM}}}=\bm{B_{\mathrm{RM}}^{\mathrm{HM}}M_{\mathrm{RM}}}$ and $\bm{M_{\mathrm{HM}}^{-1}}=\bm{(B_{\mathrm{RM}}^{\mathrm{HM}}M_{\mathrm{RM}})^{-1}}=\bm{M_{\mathrm{RM}}^{-1}B_{\mathrm{HM}}^{\mathrm{RM}}}$. Once again, $\bm{S_{\mathrm{HM}}}$ is diagonal if one assumes $\omega_{\nu_b}=\omega_{\nu}$, and block-diagonal otherwise.

\subsection{Change of reference frame}
To derive LTMs corresponding to the CM framework, one starts from matrices relating populations to raw moments~(\ref{eq:LTM_RM}). Then, one further uses shifting matrices to move from raw to central moments. These shifting matrices are obtained thanks to relationships compiled in Eqs.~(\ref{eq:CM2RM_Q9}) and~(\ref{eq:RM2CM_Q9}). One must be careful and use binomial formulas where zeroth and first order moments are not replaced by their values, i.e, $M_{00}=\widetilde{M}_{00}=1$, $M_{10}=u_x$, $\widetilde{M}_{10}=0$, etc. As an example,
\begin{align*}
\widetilde{M}_{22} &= M_{22}-2 u_y M_{21}-2 u_x M_{12}+u_y^2 M_{20}+u_x^2 M_{02}+4 u_x u_y M_{11}-2 u_x^2 u_yM_{01}-2 u_x u_y^2 M_{10} + u_x^2 u_y^2 M_{00}\\
 &= \big[ u_x^2 u_y^2 \:\: -2 u_x u_y^2 \:\: -2 u_x^2 u_y \:\: u_y^2 \:\: u_x^2 \:\: 4 u_x u_y \:\: -2 u_y \:\: -2 u_x \:\: 1\big]\cdot \big[ M_{00} \:\: M_{10} \:\: M_{01} \:\: M_{11} \:\: M_{20} \:\: M_{02} \:\: M_{21} \:\: M_{12} \:\: M_{22}\big]^T,
\end{align*}
and
\begin{align*}
M_{22} &= \widetilde{M}_{22}+2 u_y \widetilde{M}_{21}+2 u_x \widetilde{M}_{12}+u_y^2 \widetilde{M}_{20} +u_x^2 \widetilde{M}_{02}+4 u_x u_y \widetilde{M}_{11}+2 u_x^2 u_y\widetilde{M}_{01}+2 u_x u_y^2 \widetilde{M}_{10} + u_x^2 u_y^2 \widetilde{M}_{00} \\
&= \big[ u_x^2 u_y^2 \:\: 2 u_x u_y^2 \:\: 2 u_x^2 u_y \:\: u_y^2 \:\: u_x^2 \:\: 4 u_x u_y \:\: 2 u_y \:\: 2 u_x \:\: 1\big]\cdot \big[ \widetilde{M}_{00} \:\: \widetilde{M}_{10} \:\: \widetilde{M}_{01} \:\: \widetilde{M}_{11} \:\: \widetilde{M}_{20} \:\: \widetilde{M}_{02} \:\: \widetilde{M}_{21} \:\: \widetilde{M}_{12} \:\: \widetilde{M}_{22}\big]^T.
\end{align*}
By rewriting zeroth through third order moment in the same spirit, one ends up with the following shifting matrices
\begin{equation}
\bm{B_{\mathrm{RM}}^{\mathrm{CM}}}=\left[
\begin{array}{c c c c c c c c c}
1 & 0 & 0 & 0 & 0 & 0 & 0 & 0 & 0 \\[0.1cm]
-u_x & 1 & 0 & 0 & 0 & 0 & 0 & 0 & 0 \\[0.1cm]
-u_y & 0 & 1 & 0 & 0 & 0 & 0 & 0 & 0 \\[0.1cm]
u_x^2 & -2u_x & 0 & 1 & 0 & 0 & 0 & 0 & 0 \\[0.1cm]
u_y^2 & 0 & -2u_y & 0 & 1 & 0 & 0 & 0 & 0 \\[0.1cm]
u_x u_y & -u_y & -u_x & 0 & 0 & 1 & 0 & 0 & 0 \\[0.1cm]
-u_x^2 u_y & 2u_x u_y & u_x^2 & -u_y & 0 & -2 u_x & 1 & 0 & 0 \\[0.1cm]
-u_x u_y^2 & u_y^2 & 2u_x u_y & 0 & -u_x & -2 u_y & 0 & 1 & 0 \\[0.1cm]
u_x^2 u_y^2 & -2 u_x u_y^2 & -2 u_x^2 u_y & u_y^2 & u_x^2 & 4 u_x u_y & -2 u_y & -2 u_x & 1 
\end{array}
\right],
\end{equation}
\begin{equation}
\bm{B_{\mathrm{CM}}^{\mathrm{RM}}}=\bm{\big(B_{\mathrm{RM}}^{\mathrm{CM}}\big)^{-1}}=\left[
\begin{array}{c c c c c c c c c}
1 & 0 & 0 & 0 & 0 & 0 & 0 & 0 & 0 \\[0.1cm]
u_x & 1 & 0 & 0 & 0 & 0 & 0 & 0 & 0 \\[0.1cm]
u_y & 0 & 1 & 0 & 0 & 0 & 0 & 0 & 0 \\[0.1cm]
u_x^2 & 2u_x & 0 & 1 & 0 & 0 & 0 & 0 & 0 \\[0.1cm]
u_y^2 & 0 & 2u_y & 0 & 1 & 0 & 0 & 0 & 0 \\[0.1cm]
u_x u_y & u_y & u_x & 0 & 0 & 1 & 0 & 0 & 0 \\[0.1cm]
u_x^2 u_y & 2u_x u_y & u_x^2 & u_y & 0 & 2 u_x & 1 & 0 & 0 \\[0.1cm]
u_x u_y^2 & u_y^2 & 2u_x u_y & 0 & u_x & 2 u_y & 0 & 1 & 0 \\[0.1cm]
u_x^2 u_y^2 & 2 u_x u_y^2 & 2 u_x^2 u_y & u_y^2 & u_x^2 & 4 u_x u_y & 2 u_y & 2 u_x & 1 
\end{array}
\right].
\end{equation}
LTMs for the CM framework then read as
$$\bm{M_{\mathrm{CM}}}=\bm{B_{\mathrm{RM}}^{\mathrm{CM}}}\bm{M_{\mathrm{RM}}},$$
and
$$\bm{M_{\mathrm{CM}}^{-1}}=\bm{M_{\mathrm{RM}}^{-1}}\bm{B_{\mathrm{CM}}^{\mathrm{RM}}}.$$

To derive the LTMs for the CHM framework, one simply needs to start from $\bm{M_{\mathrm{CM}}}$ and  $\bm{M_{\mathrm{CM}}^{-1}}$ and then to switch from CMs to CHMs. The latter transformation can be done using $$\bm{B_{\mathrm{CM}}^{\mathrm{CHM}}}=\bm{B_{\mathrm{RM}}^{\mathrm{HM}}},\quad \bm{B_{\mathrm{CHM}}^{\mathrm{CM}}}=\bm{B_{\mathrm{HM}}^{\mathrm{RM}}}$$ since relationships between RMs and HMs are the same as those between their counterparts in the comoving reference frame. Hence,
$$\bm{M_{\mathrm{CHM}}}=\bm{B_{\mathrm{CM}}^{\mathrm{CHM}}}\bm{M_{\mathrm{CM}}},$$
and
$$\bm{M_{\mathrm{CHM}}^{-1}}=\bm{M_{\mathrm{CM}}^{-1}}\bm{B_{\mathrm{CHM}}^{\mathrm{CM}}}.$$

\section{D3Q27 and D3Q19 formulations of populations \label{sec:3Dextension}}

This appendix is dedicated to the derivation of 3D populations for raw, Hermite, central and central Hermite moments. They are obtained enforcing the isotropy of $f_i$ up to the second order (in each direction),
$$M_{p}^{\mathrm{1D}}M_{q}^{\mathrm{1D}}M_{r}^{\mathrm{1D}}=M_{p00}^{\mathrm{3D}}M_{0q0}^{\mathrm{3D}}M_{00r}^{\mathrm{3D}}=M_{pqr},$$ 
with $(p,q,r)\in\{0,1,2\}^3$. This isotropy condition is valid for all kinds of moments considered hereafter, with the exception of post collision cumulants due to their nonlinear relationship with central moments. 

Assuming that $(\sigma,\lambda,\delta)\in \{\pm 1\}^3$, populations evolving in the raw moment space read as
\begin{subequations}
\label{eq:RMQ27}
\begin{align}
f^{\mathrm{RM}}_{(0, 0, 0)}&=\rho [1-M_{200}-M_{020}-M_{002}+M_{220}+M_{202}+M_{022}-M_{222}],\\
f^{\mathrm{RM}}_{(\sigma, 0, 0)}&= \dfrac{\rho}{2} [ \sigma  u_x +M_{200} -\sigma  M_{120} -\sigma  M_{102} -M_{220} -M_{202} +\sigma  M_{122} +M_{222}],\\
f^{\mathrm{RM}}_{(0, \lambda, 0)}&= \dfrac{\rho}{2} [\lambda  u_y +M_{020} -\lambda  M_{210} -\lambda  M_{012} -M_{220} -M_{022} +\lambda  M_{212} +M_{222}],\\
f^{\mathrm{RM}}_{(0, 0, \delta)}&= \dfrac{\rho}{2} [  \delta  u_z +M_{002} -\delta  M_{201} -\delta  M_{021} -M_{202} -M_{022} +\delta  M_{221} +M_{222} ],\\
f^{\mathrm{RM}}_{(\sigma, \lambda, 0)}&= \dfrac{\rho}{4} [ \sigma\lambda  M_{110}+\lambda  M_{210}+\sigma  M_{120}+M_{220}-\sigma\lambda    M_{112}-\lambda  M_{212}-\sigma  M_{122}-M_{222}  ],\\
f^{\mathrm{RM}}_{(\sigma, 0, \delta)}&= \dfrac{\rho}{4} [\sigma\delta  M_{101} +\delta  M_{201} +\sigma  M_{102} +M_{202}-\sigma \delta M_{121} -\delta  M_{221} -\sigma  M_{122} -M_{222}   ],\\
f^{\mathrm{RM}}_{(0, \lambda, \delta)}&= \dfrac{\rho}{4} [\lambda\delta  M_{011} +\delta  M_{021} +\lambda  M_{012} +M_{022} -\lambda\delta  M_{211}  -\delta  M_{221} -\lambda  M_{212} -M_{222}   ],\\
f^{\mathrm{RM}}_{(\sigma, \lambda, \delta)}&= \dfrac{\rho}{8} [\sigma\lambda\delta M_{111}+ \lambda\delta  M_{211}+\sigma\delta  M_{121}+\sigma\lambda  M_{112}+\delta  M_{221}+\lambda  M_{212}+\sigma  M_{122}+M_{222}],
\end{align}
\end{subequations}
whereas in the case of Hermite moments they read as
\begin{subequations}
\begin{align}
f^{\mathrm{HM}}_{(0, 0, 0)}&=\rho  \big[C^3-C^2  A_{200}-C^2  A_{020}-C^2  A_{002}+C  A_{220}+C  A_{202}
+C  A_{022}- A_{222}\big],\\
f^{\mathrm{HM}}_{(\sigma, 0, 0)}&= \dfrac{\rho}{2}  \big[ C^2  (c_s^2+\sigma  u_x)+C^2  A_{200}-c_s^2 C  A_{020}-c_s^2 C  A_{002}-\sigma  C  A_{120}-\sigma  C  A_{102}-C  A_{220}-C  A_{202}+c_s^2  A_{022}\notag\\
&\quad+\sigma   A_{122}+ A_{222}\big],\\
f^{\mathrm{HM}}_{(0, \lambda, 0)}&= \dfrac{\rho}{2} \big[C^2 (c_s^2+\lambda  u_y)- c_s^2 C  A_{200}+ C^2  A_{020}- c_s^2  C  A_{002}- \lambda  C  A_{210}- \lambda  C  A_{012}- C  A_{220}+ c_s^2  A_{202}- C  A_{022} \notag\\ 
&\quad+ \lambda   A_{212}+  A_{222}\big],\\
f^{\mathrm{HM}}_{(0, 0, \delta)}&= \dfrac{\rho}{2} \big[ C^2 (c_s^2+\delta  u_z)- c_s^2 C  A_{200}- c_s^2 C  A_{020}+ C^2  A_{002}- \delta  C  A_{201}- \delta  C  A_{021}+ c_s^2  A_{220}- C  A_{202}- C  A_{022}\notag\\
&\quad+ \delta   A_{221}+  A_{222}\big],\\
f^{\mathrm{HM}}_{(\sigma, \lambda, 0)}&= \dfrac{\rho}{4} \big[ c_s^2 C (c_s^2+\sigma  u_x+\lambda  u_y)+\sigma \lambda   C  A_{110}+c_s^2C  A_{200}+c_s^2C  A_{020}-c_s^4  A_{002}+\lambda C  A_{210}+\sigma C  A_{120}-\sigma  c_s^2  A_{102}\notag\\
&\quad-\lambda  c_s^2  A_{012}-\textcolor{black}{\sigma\lambda}  A_{112}+C  A_{220}-c_s^2  A_{202}-c_s^2  A_{022}-\lambda   A_{212}-\sigma   A_{122}- A_{222}\big],\\
f^{\mathrm{HM}}_{(\sigma, 0, \delta)}&= \dfrac{\rho}{4} \big[ c_s^2 C (c_s^2+\sigma  u_x+\delta  u_z)+ \sigma  \delta  C  A_{101}+c_s^2C  A_{200}-c_s^4  A_{020}+c_s^2C  A_{002}+\delta C  A_{201}-\delta  c_s^2  A_{021}-\sigma  c_s^2  A_{120}\notag\\
&\quad+\sigma C  A_{102}-\textcolor{black}{\sigma\delta}  A_{121}-c_s^2  A_{220}+C  A_{202}-c_s^2  A_{022}-\delta   A_{221}-\sigma   A_{122}- A_{222}\big],\\
f^{\mathrm{HM}}_{(0, \lambda, \delta)}&= \dfrac{\rho}{4} \big[ c_s^2 C (c_s^2+\lambda  u_y+\delta  u_z)+\lambda \delta   C  A_{011}-c_s^4  A_{200}+c_s^2C  A_{020}+c_s^2C  A_{002}-\lambda  c_s^2  A_{210}-\delta  c_s^2  A_{201}+\delta C  A_{021}\notag\\
&\quad+\lambda C  A_{012}-\lambda   \delta  A_{211}-c_s^2  A_{220}-c_s^2  A_{202}+C  A_{022}-\delta   A_{221}-\lambda   A_{212}- A_{222}\big],\\
f^{\mathrm{HM}}_{(\sigma, \lambda, \delta)}&= \dfrac{\rho}{8} \big[c_s^4 (c_s^2+\sigma  u_x+\lambda  u_y+\delta  u_z)+c_s^2 (\sigma \lambda    A_{110}+ \sigma  \delta  A_{101}+ \lambda  \delta  A_{011})+c_s^4 ( A_{200}+ A_{020}+ A_{002})+\sigma   \lambda  \delta  A_{111}\notag\\
&\quad+c_s^2(\lambda  A_{210}+\delta  A_{201}+\delta  A_{021}+\sigma  A_{120}+\sigma A_{102}+ \lambda  A_{012})+\lambda  \delta  A_{211}+\sigma   \delta  A_{121}+\sigma  \lambda  A_{112}+c_s^2  (A_{220}\notag\\
&\quad+A_{202}+A_{022})+\delta   A_{221}+\lambda   A_{212}+\sigma   A_{122}+ A_{222}\big],
\end{align}
\end{subequations}
with $C=1-c_s^2$.
It is interesting to note that $f_i^{\mathrm{HM}}$ reduce to $f_i^{\mathrm{RM}}$ imposing $c_s=0$ in the above formulas. This is due to the fact that $H_{i,pqr}=\xi_{i,x}^{p}\xi_{i,y}^{q}\xi_{i,z}^{r}$ when terms proportional to $c_s$ are discarded in the definition of Hermite polynomials.
\begin{subequations}
\begin{align}
f^{\mathrm{CM}}_{(0,0,0)}&=\rho \big[U_x U_y U_z+4 u_x u_y U_z \widetilde{M}_{110}+4 u_x U_y u_z \widetilde{M}_{101}+4 U_x u_y u_z \widetilde{M}_{011}-U_y U_z \widetilde{M}_{200}-U_x U_z \widetilde{M}_{020}\notag\\
&-U_x U_y \widetilde{M}_{002}-8 u_x u_y u_z \widetilde{M}_{111}+2 u_y U_z \widetilde{M}_{210}+2 U_y u_z \widetilde{M}_{201}+2 U_x u_z \widetilde{M}_{021}+2 u_x U_z \widetilde{M}_{120}+2 u_x U_y \widetilde{M}_{102}\notag\\
&+2 U_x u_y \widetilde{M}_{012}-4 u_y u_z \widetilde{M}_{211}-4 u_x u_z \widetilde{M}_{121}-4 u_x u_y \widetilde{M}_{112}+U_z \widetilde{M}_{220}+U_y \widetilde{M}_{202}+U_x \widetilde{M}_{022}-2 u_z \widetilde{M}_{221}\notag\\
&-2 u_y \widetilde{M}_{212}-2 u_x \widetilde{M}_{122}-\widetilde{M}_{222}\big],\\
f^{\mathrm{CM}}_{(\sigma,0,0)}&=\dfrac{\rho}{2} \big[\textcolor{black}{u_x \sigma_x}U_y U_z-2 \sigma_{2x} u_y U_z \widetilde{M}_{110}-2 \sigma_{2x} U_y u_z \widetilde{M}_{101}+4 u_x \sigma_x u_y u_z \widetilde{M}_{011}+U_y U_z \widetilde{M}_{200}-u_x \sigma_x U_z \widetilde{M}_{020}\notag\\
&-u_x \sigma_x U_y \widetilde{M}_{002}+4 \sigma_{2x} u_y u_z \widetilde{M}_{111}-2 u_y U_z \widetilde{M}_{210}-2 U_y u_z \widetilde{M}_{201}+2 u_x \sigma_x u_z \widetilde{M}_{021}-\sigma_{2x} U_z \widetilde{M}_{120}\notag\\
&-\sigma_{2x} U_y \widetilde{M}_{102}+2 u_x \sigma_x u_y \widetilde{M}_{012}+4 u_y u_z \widetilde{M}_{211}+2 \sigma_{2x} u_z \widetilde{M}_{121}+2 \sigma_{2x} u_y \widetilde{M}_{112}-U_z \widetilde{M}_{220}-U_y \widetilde{M}_{202}\notag\\
&+u_x \sigma_x \widetilde{M}_{022}+2 u_z \widetilde{M}_{221}+2 u_y \widetilde{M}_{212}+\sigma_{2x} \widetilde{M}_{122}+\widetilde{M}_{222}\big],\\
f^{\mathrm{CM}}_{(0,\lambda,0)}&=\dfrac{\rho}{2} \big[U_x \textcolor{black}{u_y\lambda_y} U_z-2 u_x \lambda_{2y} U_z \widetilde{M}_{110}+4 u_x u_y \lambda_y u_z \widetilde{M}_{101}-2 U_x \lambda_{2y} u_z \widetilde{M}_{011}-u_y \lambda_y U_z \widetilde{M}_{200}+U_x U_z \widetilde{M}_{020}\notag\\
&-U_x u_y \lambda_y \widetilde{M}_{002}+4 u_x \lambda_{2y} u_z \widetilde{M}_{111}-\lambda_{2y} U_z \widetilde{M}_{210}+2 u_y \lambda_y u_z \widetilde{M}_{201}-2 U_x u_z \widetilde{M}_{021}-2 u_x U_z \widetilde{M}_{120}\notag\\
&+2 u_x u_y \lambda_y \widetilde{M}_{102}-U_x \lambda_{2y} \widetilde{M}_{012}+2 \lambda_{2y} u_z \widetilde{M}_{211}+4 u_x u_z \widetilde{M}_{121}+2 u_x \lambda_{2y} \widetilde{M}_{112}-U_z \widetilde{M}_{220}+u_y \lambda_y \widetilde{M}_{202}\notag\\
&-U_x \widetilde{M}_{022}+2 u_z \widetilde{M}_{221}+\lambda_{2y} \widetilde{M}_{212}+2 u_x \widetilde{M}_{122}+\widetilde{M}_{222}
\big],\\
f^{\mathrm{CM}}_{(0,0,\delta)}&=\dfrac{\rho}{2} \big[U_x U_y u_z \delta_z+4 u_x u_y u_z \delta_z \widetilde{M}_{110}-2 u_x U_y \delta_{2z} \widetilde{M}_{101}-2 U_x u_y \delta_{2z} \widetilde{M}_{011}-U_y u_z \delta_z \widetilde{M}_{200}-U_x u_z \delta_z \widetilde{M}_{020}\notag\\
&+U_x U_y \widetilde{M}_{002}+4 u_x u_y \delta_{2z} \widetilde{M}_{111}+2 u_y u_z \delta_z \widetilde{M}_{210}-U_y \delta_{2z} \widetilde{M}_{201}-U_x \delta_{2z} \widetilde{M}_{021}+2 u_x u_z \delta_z \widetilde{M}_{120}\notag\\
&-2 u_x U_y \widetilde{M}_{102}-2 U_x u_y \widetilde{M}_{012}+2 u_y \delta_{2z} \widetilde{M}_{211}+2 u_x \delta_{2z} \widetilde{M}_{121}+4 u_x u_y \widetilde{M}_{112}+u_z \delta_z \widetilde{M}_{220}-U_y \widetilde{M}_{202}\notag\\
&-U_x \widetilde{M}_{022}+\delta_{2z} \widetilde{M}_{221}+2 u_y \widetilde{M}_{212}+2 u_x \widetilde{M}_{122}+\widetilde{M}_{222}
\big],\\
f^{\mathrm{CM}}_{(\sigma,\lambda,0)}&=\dfrac{\rho}{4} \big[\textcolor{black}{u_x \sigma_x u_y \lambda_y}U_z +\sigma_{2x} \lambda_{2y} U_z \widetilde{M}_{110}-2 \sigma_{2x} u_y \lambda_y u_z \widetilde{M}_{101}-2 u_x \sigma_x \lambda_{2y} u_z \widetilde{M}_{011}+u_y \lambda_y U_z \widetilde{M}_{200}\notag\\
&+u_x \sigma_x U_z \widetilde{M}_{020}-u_x \sigma_x u_y \lambda_y \widetilde{M}_{002}-2 \sigma_{2x} \lambda_{2y} u_z \widetilde{M}_{111}+\lambda_{2y} U_z \widetilde{M}_{210}-2 u_y \lambda_y u_z \widetilde{M}_{201}-2 u_x \sigma_x u_z \widetilde{M}_{021}\notag\\
&+\sigma_{2x} U_z \widetilde{M}_{120}-\sigma_{2x} u_y \lambda_y \widetilde{M}_{102}-u_x \sigma_x \lambda_{2y} \widetilde{M}_{012}-2 \lambda_{2y} u_z \widetilde{M}_{211}-2 \sigma_{2x} u_z \widetilde{M}_{121}-\sigma_{2x} \lambda_{2y} \widetilde{M}_{112}\notag\\
&+U_z \widetilde{M}_{220}-u_y \lambda_y \widetilde{M}_{202}-u_x \sigma_x \widetilde{M}_{022}-2 u_z \widetilde{M}_{221}-\lambda_{2y} \widetilde{M}_{212}-\sigma_{2x} \widetilde{M}_{122}-\widetilde{M}_{222}
\big],\\
f^{\mathrm{CM}}_{(\sigma,0,\delta)}&=\dfrac{\rho}{4} \big[\textcolor{black}{u_x \sigma_x} U_y u_z \delta_z-2 \sigma_{2x} u_y u_z \delta_z \widetilde{M}_{110}+\sigma_{2x} U_y \delta_{2z} \widetilde{M}_{101}-2 u_x \sigma_x u_y \delta_{2z} \widetilde{M}_{011}+U_y u_z \delta_z \widetilde{M}_{200}\notag\\
&-u_x \sigma_x u_z \delta_z \widetilde{M}_{020}+u_x \sigma_x U_y \widetilde{M}_{002}-2 \sigma_{2x} u_y \delta_{2z} \widetilde{M}_{111}-2 u_y u_z \delta_z \widetilde{M}_{210}+U_y \delta_{2z} \widetilde{M}_{201}-u_x \sigma_x \delta_{2z} \widetilde{M}_{021}\notag\\
&-\sigma_{2x} u_z \delta_z \widetilde{M}_{120}+\sigma_{2x} U_y \widetilde{M}_{102}-2 u_x \sigma_x u_y \widetilde{M}_{012}-2 u_y \delta_{2z} \widetilde{M}_{211}-\sigma_{2x} \delta_{2z} \widetilde{M}_{121}-2 \sigma_{2x} u_y \widetilde{M}_{112}\notag\\
&-u_z \delta_z \widetilde{M}_{220}+U_y \widetilde{M}_{202}-u_x \sigma_x \widetilde{M}_{022}-\delta_{2z} \widetilde{M}_{221}-2 u_y \widetilde{M}_{212}-\sigma_{2x} \widetilde{M}_{122}-\widetilde{M}_{222}
\big],\\
f^{\mathrm{CM}}_{(0,\lambda,\delta)}&=\dfrac{\rho}{4} \big[U_x \textcolor{black}{u_y \lambda_y} u_z \delta_z-2 u_x \lambda_{2y} u_z \delta_z \widetilde{M}_{110}-2 u_x u_y \lambda_y \delta_{2z} \widetilde{M}_{101}+U_x \lambda_{2y} \delta_{2z} \widetilde{M}_{011}-u_y \lambda_y u_z \delta_z \widetilde{M}_{200}\notag\\
&+U_x u_z \delta_z \widetilde{M}_{020}+U_x u_y \lambda_y \widetilde{M}_{002}-2 u_x \lambda_{2y} \delta_{2z} \widetilde{M}_{111}-\lambda_{2y} u_z \delta_z \widetilde{M}_{210}-u_y \lambda_y \delta_{2z} \widetilde{M}_{201}+U_x \delta_{2z} \widetilde{M}_{021}\notag\\
&-2 u_x u_z \delta_z \widetilde{M}_{120}-2 u_x u_y \lambda_y \widetilde{M}_{102}+U_x \lambda_{2y} \widetilde{M}_{012}-\lambda_{2y} \delta_{2z} \widetilde{M}_{211}-2 u_x \delta_{2z} \widetilde{M}_{121}-2 u_x \lambda_{2y} \widetilde{M}_{112}\notag\\
&-u_z \delta_z \widetilde{M}_{220}-u_y \lambda_y \widetilde{M}_{202}+U_x \widetilde{M}_{022}-\delta_{2z} \widetilde{M}_{221}-\lambda_{2y} \widetilde{M}_{212}-2 u_x \widetilde{M}_{122}-\widetilde{M}_{222}
\big],\\
f^{\mathrm{CM}}_{(\sigma,\lambda,\delta)}&=\dfrac{\rho}{8} \big[\textcolor{black}{u_x \sigma_x u_y \lambda_y u_z \delta_z}+\sigma_{2x} \lambda_{2y} u_z \delta_z \widetilde{M}_{110}+\sigma_{2x} u_y \lambda_y \delta_{2z} \widetilde{M}_{101}+u_x \sigma_x \lambda_{2y} \delta_{2z} \widetilde{M}_{011}+u_y \lambda_y u_z \delta_z \widetilde{M}_{200}\notag\\
&+u_x \sigma_x u_z \delta_z \widetilde{M}_{020}+u_x \sigma_x u_y \lambda_y \widetilde{M}_{002}+\sigma_{2x} \lambda_{2y} \delta_{2z} \widetilde{M}_{111}+\lambda_{2y} u_z \delta_z \widetilde{M}_{210}+u_y \lambda_y \delta_{2z} \widetilde{M}_{201}\notag\\
&+u_x \sigma_x \delta_{2z} \widetilde{M}_{021}+\sigma_{2x} u_z \delta_z \widetilde{M}_{120}+\sigma_{2x} u_y \lambda_y \widetilde{M}_{102}+u_x \sigma_x \lambda_{2y} \widetilde{M}_{012}+\lambda_{2y} \delta_{2z} \widetilde{M}_{211}+\sigma_{2x} \delta_{2z} \widetilde{M}_{121}\notag\\
&+\sigma_{2x} \lambda_{2y} \widetilde{M}_{112}+u_z \delta_z \widetilde{M}_{220}+u_y \lambda_y \widetilde{M}_{202}+u_x \sigma_x \widetilde{M}_{022}+\delta_{2z} \widetilde{M}_{221}+\lambda_{2y} \widetilde{M}_{212}+\sigma_{2x} \widetilde{M}_{122}+\widetilde{M}_{222}
\big],
\end{align}
\end{subequations}
with $U_x = 1-u_x^2$, $U_y = 1-u_y^2$, $U_z = 1-u_z^2$, $\sigma_x=\sigma + u_x$, $\sigma_{2x}=\sigma + 2u_x$, $\lambda_y=\lambda + u_y$, $\lambda_{2y}=\lambda + 2u_y$, $\delta_z=\delta + u_z$, $\delta_{2z}=\delta + 2u_z$.
Discarding velocity dependent terms in $f_i^{\mathrm{CM}}$ allows us to recover the definitions of $f_i^{\mathrm{RM}}$. This is another simple way that permits to easily check the validity of the above formulas.
\begin{subequations}
\begin{align}
f^{\mathrm{CHM}}_{(0,0,0)}&=\rho\big[C_xC_yC_z+4(u_xu_yC_z\widetilde{A}_{110}+u_xC_yu_z\widetilde{A}_{101}+C_xu_yu_z\widetilde{A}_{011})-C_yC_z\widetilde{A}_{200}-C_xC_z\widetilde{A}_{020}-C_xC_y\widetilde{A}_{002}\notag\\
&-8u_xu_yu_z\widetilde{A}_{111}+2(u_yC_z\widetilde{A}_{210}+C_yu_z\widetilde{A}_{201}+C_xu_z\widetilde{A}_{021}+u_xC_z\widetilde{A}_{120}+u_xC_y\widetilde{A}_{102}+C_xu_y\widetilde{A}_{012})-4(u_yu_z\widetilde{A}_{211}\notag\\
&+u_xu_z\widetilde{A}_{121}+u_xu_y\widetilde{A}_{112})+C_z\widetilde{A}_{220}+C_y\widetilde{A}_{202}+C_x\widetilde{A}_{022}-2u_z\widetilde{A}_{221}-2u_y\widetilde{A}_{212}-2u_x\widetilde{A}_{122}-\widetilde{A}_{222}\big],\\
f^{\mathrm{CHM}}_{(\sigma,0,0)}&=\frac{\rho}{2}\big[C_yC_zC_{\sigma}-2\sigma_{2x}u_yC_z\widetilde{A}_{110}-2\sigma_{2x}C_yu_z\widetilde{A}_{101}+4C_{\sigma}u_yu_z\widetilde{A}_{011}+C_yC_z\widetilde{A}_{200}-C_{\sigma}C_z\widetilde{A}_{020}-C_{\sigma}C_y\widetilde{A}_{002}\notag\\
&+4\sigma_{2x}u_yu_z\widetilde{A}_{111}-2u_yC_z\widetilde{A}_{210}-2C_yu_z\widetilde{A}_{201}+2C_{\sigma}u_z\widetilde{A}_{021}-\sigma_{2x}C_z\widetilde{A}_{120}-\sigma_{2x}C_y\widetilde{A}_{102}+2C_{\sigma}u_y\widetilde{A}_{012}+4u_yu_z\widetilde{A}_{211}\notag\\
&+2\sigma_{2x}u_z\widetilde{A}_{121}+2\sigma_{2x}u_y\widetilde{A}_{112}-C_z\widetilde{A}_{220}-C_y\widetilde{A}_{202}+C_{\sigma}\widetilde{A}_{022}+2u_z\widetilde{A}_{221}+2u_y\widetilde{A}_{212}+\sigma_{2x}\widetilde{A}_{122}+\widetilde{A}_{222}\big],\\
f^{\mathrm{CHM}}_{(0,\lambda,0)}&=\frac{\rho}{2}\big[C_xC_zC_{\lambda}-2u_x\lambda_{2y}C_z\widetilde{A}_{110}+4u_xC_{\lambda}u_z\widetilde{A}_{101}-2C_x\lambda_{2y}u_z\widetilde{A}_{011}-C_{\lambda}C_z\widetilde{A}_{200}+C_xC_z\widetilde{A}_{020}-C_xC_{\lambda}\widetilde{A}_{002}\notag\\
&+4u_x\lambda_{2y}u_z\widetilde{A}_{111}-\lambda_{2y}C_z\widetilde{A}_{210}+2C_{\lambda}u_z\widetilde{A}_{201}-2C_xu_z\widetilde{A}_{021}-2u_xC_z\widetilde{A}_{120}+2u_xC_{\lambda}\widetilde{A}_{102}-C_x\lambda_{2y}\widetilde{A}_{012}+2\lambda_{2y}u_z\widetilde{A}_{211}\notag\\
&+4u_xu_z\widetilde{A}_{121}+2u_x\lambda_{2y}\widetilde{A}_{112}-C_z\widetilde{A}_{220}+C_{\lambda}\widetilde{A}_{202}-C_x\widetilde{A}_{022}+2u_z\widetilde{A}_{221}+\lambda_{2y}\widetilde{A}_{212}+2u_x\widetilde{A}_{122}+\widetilde{A}_{222}\big],\\
f^{\mathrm{CHM}}_{(0,0,\delta)}&=\frac{\rho}{2}\big[C_xC_yC_{\delta}+4u_xu_yC_{\delta}\widetilde{A}_{110}-2u_xC_y\delta_{2z}\widetilde{A}_{101}-2C_xu_y\delta_{2z}\widetilde{A}_{011}-C_yC_{\delta}\widetilde{A}_{200}-C_xC_{\delta}\widetilde{A}_{020}+C_xC_y\widetilde{A}_{002}\notag\\
&+4u_xu_y\delta_{2z}\widetilde{A}_{111}+2u_yC_{\delta}\widetilde{A}_{210}-C_y\delta_{2z}\widetilde{A}_{201}-C_x\delta_{2z}\widetilde{A}_{021}+2u_xC_{\delta}\widetilde{A}_{120}-2u_xC_y\widetilde{A}_{102}-2C_xu_y\widetilde{A}_{012}+2u_y\delta_{2z}\widetilde{A}_{211}\notag\\
&+2u_x\delta_{2z}\widetilde{A}_{121}+4u_xu_y\widetilde{A}_{112}+C_{\delta}\widetilde{A}_{220}-C_y\widetilde{A}_{202}-C_x\widetilde{A}_{022}+\delta_{2z}\widetilde{A}_{221}+2u_y\widetilde{A}_{212}+2u_x\widetilde{A}_{122}+\widetilde{A}_{222}\big],\\
f^{\mathrm{CHM}}_{(\sigma,\lambda,0)}&=\frac{\rho}{4}\big[C_zC_{\sigma}C_{\lambda}+\sigma_{2x}\lambda_{2y}C_z\widetilde{A}_{110}-2\sigma_{2x}C_{\lambda}u_z\widetilde{A}_{101}-2C_{\sigma}\lambda_{2y}u_z\widetilde{A}_{011}+C_{\lambda}C_z\widetilde{A}_{200}+C_{\sigma}C_z\widetilde{A}_{020}-C_{\sigma}C_{\lambda}\widetilde{A}_{002}\notag\\
&-2\sigma_{2x}\lambda_{2y}u_z\widetilde{A}_{111}+\lambda_{2y}C_z\widetilde{A}_{210}-2C_{\lambda}u_z\widetilde{A}_{201}-2C_{\sigma}u_z\widetilde{A}_{021}+\sigma_{2x}C_z\widetilde{A}_{120}-\sigma_{2x}C_{\lambda}\widetilde{A}_{102}-C_{\sigma}\lambda_{2y}\widetilde{A}_{012}-2\lambda_{2y}u_z\widetilde{A}_{211}\notag\\
&-2\sigma_{2x}u_z\widetilde{A}_{121}-\sigma_{2x}\lambda_{2y}\widetilde{A}_{112}+C_z\widetilde{A}_{220}-C_{\lambda}\widetilde{A}_{202}-C_{\sigma}\widetilde{A}_{022}-2u_z\widetilde{A}_{221}-\lambda_{2y}\widetilde{A}_{212}-\sigma_{2x}\widetilde{A}_{122}-\widetilde{A}_{222}\big],\\
f^{\mathrm{CHM}}_{(\sigma,0,\delta)}&=\frac{\rho}{4}\big[C_yC_{\sigma}C_{\delta}-2\sigma_{2x}u_yC_{\delta}\widetilde{A}_{110}+\sigma_{2x}C_y\delta_{2z}\widetilde{A}_{101}-2C_{\sigma}u_y\delta_{2z}\widetilde{A}_{011}+C_yC_{\delta}\widetilde{A}_{200}-C_{\sigma}C_{\delta}\widetilde{A}_{020}+C_{\sigma}C_y\widetilde{A}_{002}\notag\\
&-2\sigma_{2x}u_y\delta_{2z}\widetilde{A}_{111}-2u_yC_{\delta}\widetilde{A}_{210}+C_y\delta_{2z}\widetilde{A}_{201}-C_{\sigma}\delta_{2z}\widetilde{A}_{021}-\sigma_{2x}C_{\delta}\widetilde{A}_{120}+\sigma_{2x}C_y\widetilde{A}_{102}-2C_{\sigma}u_y\widetilde{A}_{012}-2u_y\delta_{2z}\widetilde{A}_{211}\notag\\
&-\sigma_{2x}\delta_{2z}\widetilde{A}_{121}-2\sigma_{2x}u_y\widetilde{A}_{112}-C_{\delta}\widetilde{A}_{220}+C_y\widetilde{A}_{202}-C_{\sigma}\widetilde{A}_{022}-\delta_{2z}\widetilde{A}_{221}-2u_y\widetilde{A}_{212}-\sigma_{2x}\widetilde{A}_{122}-\widetilde{A}_{222}\big],\\
f^{\mathrm{CHM}}_{(0,\lambda,\delta)}&=\frac{\rho}{4}\big[C_xC_{\lambda}C_{\delta}-2u_x\lambda_{2y}C_{\delta}\widetilde{A}_{110}-2u_xC_{\lambda}\delta_{2z}\widetilde{A}_{101}+C_x\lambda_{2y}\delta_{2z}\widetilde{A}_{011}-C_{\lambda}C_{\delta}\widetilde{A}_{200}+C_xC_{\delta}\widetilde{A}_{020}+C_xC_{\lambda}\widetilde{A}_{002}\notag\\
&-2u_x\lambda_{2y}\delta_{2z}\widetilde{A}_{111}-\lambda_{2y}C_{\delta}\widetilde{A}_{210}-C_{\lambda}\delta_{2z}\widetilde{A}_{201}+C_x\delta_{2z}\widetilde{A}_{021}-2u_xC_{\delta}\widetilde{A}_{120}-2u_xC_{\lambda}\widetilde{A}_{102}+C_x\lambda_{2y}\widetilde{A}_{012}-\lambda_{2y}\delta_{2z}\widetilde{A}_{211}\notag\\
&-2u_x\delta_{2z}\widetilde{A}_{121}-2u_x\lambda_{2y}\widetilde{A}_{112}-C_{\delta}\widetilde{A}_{220}-C_{\lambda}\widetilde{A}_{202}+C_x\widetilde{A}_{022}-\delta_{2z}\widetilde{A}_{221}-\lambda_{2y}\widetilde{A}_{212}-2u_x\widetilde{A}_{122}-\widetilde{A}_{222}\big],\\
f^{\mathrm{CHM}}_{(\sigma,\lambda,\delta)}&=\frac{\rho}{8}\big[C_{\sigma}C_{\lambda}C_{\delta}+\sigma_{2x}\lambda_{2y}C_{\delta}\widetilde{A}_{110}+\sigma_{2x}C_{\lambda}\delta_{2z}\widetilde{A}_{101}+C_{\sigma}\lambda_{2y}\delta_{2z}\widetilde{A}_{011}+C_{\lambda}C_{\delta}\widetilde{A}_{200}+C_{\sigma}C_{\delta}\widetilde{A}_{020}+C_{\sigma}C_{\lambda}\widetilde{A}_{002}\notag\\
&+\sigma_{2x}\lambda_{2y}\delta_{2z}\widetilde{A}_{111}+\lambda_{2y}C_{\delta}\widetilde{A}_{210}+C_{\lambda}\delta_{2z}\widetilde{A}_{201}+C_{\sigma}\delta_{2z}\widetilde{A}_{021}+\sigma_{2x}C_{\delta}\widetilde{A}_{120}+\sigma_{2x}C_{\lambda}\widetilde{A}_{102}+C_{\sigma}\lambda_{2y}\widetilde{A}_{012}+\lambda_{2y}\delta_{2z}\widetilde{A}_{211}\notag\\
&+\sigma_{2x}\delta_{2z}\widetilde{A}_{121}+\sigma_{2x}\lambda_{2y}\widetilde{A}_{112}+C_{\delta}\widetilde{A}_{220}+C_{\lambda}\widetilde{A}_{202}+C_{\sigma}\widetilde{A}_{022}+\delta_{2z}\widetilde{A}_{221}+\lambda_{2y}\widetilde{A}_{212}+\sigma_{2x}\widetilde{A}_{122}+\widetilde{A}_{222}\big],
\end{align}
\end{subequations}
with 
\begin{equation}
\begin{array}{c @{\: = \:} c @{\: + \:} c}
\sigma_x & \sigma & u_x,\\[0.1cm] 
\lambda_y & \lambda & u_y,\\[0.1cm]
\delta_z & \delta & u_z, 
\end{array}
\quad
\begin{array}{c @{\: = \:} c @{\: + \:} c}
\sigma_{2x} & \sigma & 2u_x,\\[0.1cm]
\lambda_{2y} & \lambda & 2u_y,\\[0.1cm] 
\delta_{2z} & \delta & 2u_z,
\end{array}
\quad
\begin{array}{c @{\: = \:} c @{\: - \:} c @{\: - \:} c}
C_x & 1 & c_s^2 & u_x^2,\\[0.1cm] 
C_y & 1 & c_s^2 & u_y^2,\\[0.1cm] 
C_z & 1 & c_s^2 & u_z^2,
\end{array}
\quad
\begin{array}{c @{\: = \:} c @{\: + \:} l @{\: + \:} l}
C_{\sigma} & c_s^2 & u_x (\sigma & u_x),\\[0.1cm] 
C_{\lambda} & c_s^2 & u_y (\lambda & u_y),\\[0.1cm]
C_{\delta} & c_s^2 & u_z (\delta & u_z).
\end{array}
\end{equation}
Here, if one neglects $c_s$-dependent terms, the definition of $f_i^{\mathrm{CM}}$ is recovered. On the contrary, discarding velocity dependent terms leads to the definition of $f_i^{\mathrm{HM}}$. Eventually, neglecting both kinds of terms allows us to recover the definition of $f_i^{\mathrm{RM}}$. Regarding regularized LBMs, their populations are exactly the same as those expressed within the HM framework. For the K-LBM, one simply needs to replace central moments of order $n\geq 4$ in $f_i^{\mathrm{CM}}$ using corresponding formulas detailed in Eq.~(\ref{eq:CumulCentral3D}).

In the particular case of the D3Q19 velocity discretization, the easiest way to derive populations is to start from their D3Q27 counterparts and then to neglect moments that are not compliant with this lattice. One further needs to choose to either rely on the Gauss-Hermite formalism or not since this choice directly impacts the form of the equilibrium state (see App.~\ref{sec:EqFunction} and more specifically Eqs.~(\ref{eq:EqQ19RM1})-(\ref{eq:EqQ19RM2}) and~(\ref{eq:EqQ19GH1})-(\ref{eq:EqQ19GH2})). Once this is done, one simply needs to use relationships derived in App.~\ref{sec:RelationshipsTrivariate} to switch from one framework to another one. Taking the example of $f_i^{\mathrm{CM}}$~(\ref{eq:RMQ27}), one needs to discard $M_{111}$, $M_{211}$, $M_{121}$, $M_{112}$, $M_{221}$, $M_{212}$, $M_{112}$, $M_{222}$ as explained in Sec.~\ref{sec:D3Q19}. This leads to
\begin{subequations}
\label{eq:Q19RM}
\begin{align}
f^{\mathrm{RM,Q19}}_{(0, 0, 0)}&=\rho [1-M_{200}-M_{020}-M_{002}+M_{220}+M_{202}+M_{022}],\\
f^{\mathrm{RM,Q19}}_{(\sigma, 0, 0)}&= \dfrac{\rho}{2} [ \sigma  u_x +M_{200} -\sigma  M_{120} -\sigma  M_{102} -M_{220} -M_{202}],\\
f^{\mathrm{RM,Q19}}_{(0, \lambda, 0)}&= \dfrac{\rho}{2} [\lambda  u_y +M_{020} -\lambda  M_{210} -\lambda  M_{012} -M_{220} -M_{022}],\\
f^{\mathrm{RM,Q19}}_{(0, 0, \delta)}&= \dfrac{\rho}{2} [  \delta  u_z +M_{002} -\delta  M_{201} -\delta  M_{021} -M_{202} -M_{022}],\\
f^{\mathrm{RM,Q19}}_{(\sigma, \lambda, 0)}&= \dfrac{\rho}{4} [ \sigma\lambda  M_{110}+\lambda  M_{210}+\sigma  M_{120}+M_{220}],\\
f^{\mathrm{RM,Q19}}_{(\sigma, 0, \delta)}&= \dfrac{\rho}{4} [\sigma\delta  M_{101} +\delta  M_{201} +\sigma  M_{102} +M_{202}],\\
f^{\mathrm{RM,Q19}}_{(0, \lambda, \delta)}&= \dfrac{\rho}{4} [\lambda\delta  M_{011} +\delta  M_{021} +\lambda  M_{012} +M_{022}],\\
f^{\mathrm{RM,Q19}}_{(\sigma, \lambda, \delta)}&= 0.
\end{align}
\end{subequations}
One can notice that populations corresponding to discrete velocities $(\pm 1,\pm 1,\pm 1)$ are null. This confirms the validity of the proposed choice of raw moments. Replacing raw moments by their Hermite counterpart further results in
\begin{subequations}
\label{eq:Q19HM}
\begin{align}
f^{\mathrm{HM,Q19}}_{(0, 0, 0)}&=\rho \big[(1-c_s^2)^3 + c_s^6 -C_2(\textcolor{black}{A_{002}}+\textcolor{black}{A_{020}}+\textcolor{black}{A_{200}})+\textcolor{black}{A_{022}}+\textcolor{black}{A_{202}}+\textcolor{black}{A_{220}}\big],\\
f^{\mathrm{HM,Q19}}_{(\sigma, 0, 0)}&=\frac{\rho}{2}\big[\textcolor{black}{C_2(c_s^2 +\sigma u_x)}+C_2\textcolor{black}{A_{200}}-c_s^2 \textcolor{black}{A_{020}}-c_s^2 \textcolor{black}{A_{002}}-\sigma  (\textcolor{black}{A_{120}}+\textcolor{black}{A_{102}})-\textcolor{black}{A_{202}}-\textcolor{black}{A_{220}}\big],\\
f^{\mathrm{HM,Q19}}_{(0, \lambda, 0)}&=\frac{\rho}{2}\big[\textcolor{black}{C_2(c_s^2 +\lambda u_y)}-c_s^2 \textcolor{black}{A_{200}}+C_2\textcolor{black}{A_{020}}-c_s^2 \textcolor{black}{A_{002}}-\lambda  (\textcolor{black}{A_{210}}+\textcolor{black}{A_{012}})-\textcolor{black}{A_{022}}-\textcolor{black}{A_{220}}\big],\\
f^{\mathrm{HM,Q19}}_{(0, 0, \delta)}&=\frac{\rho}{2}\big[\textcolor{black}{C_2(c_s^2 +\delta u_z)}-c_s^2 \textcolor{black}{A_{200}}-c_s^2 \textcolor{black}{A_{020}}+C_2 \textcolor{black}{A_{002}}-\delta  (\textcolor{black}{A_{201}}+\textcolor{black}{A_{021}})-\textcolor{black}{A_{022}}-\textcolor{black}{A_{202}}\big],\\
f^{\mathrm{HM,Q19}}_{(\sigma, \lambda, 0)}&=\frac{\rho}{4}\big[c_s^2(c_s^2 + \sigma   u_x+\lambda   u_y) + c_s^2(\textcolor{black}{A_{200}}+\textcolor{black}{A_{020}})+\lambda  \sigma  \textcolor{black}{A_{110}}+\lambda \textcolor{black}{A_{210}}+\sigma  \textcolor{black}{A_{120}}+\textcolor{black}{A_{220}}\big],\\
f^{\mathrm{HM,Q19}}_{(\sigma, 0, \delta)}&=\frac{\rho}{4}\big[c_s^2(c_s^2 + \sigma   u_x+\delta   u_z) + c_s^2(\textcolor{black}{A_{200}}+\textcolor{black}{A_{002}})+\delta  \sigma  \textcolor{black}{A_{101}}+\delta \textcolor{black}{A_{201}}+\sigma  \textcolor{black}{A_{102}}+\textcolor{black}{A_{202}}\big],\\
f^{\mathrm{HM,Q19}}_{(0, \lambda, \delta)}&=\frac{\rho}{4}\big[c_s^2(c_s^2 + \lambda   u_y+\delta   u_z) + c_s^2(\textcolor{black}{A_{020}}+\textcolor{black}{A_{002}})+\delta  \lambda  \textcolor{black}{A_{011}}+\delta \textcolor{black}{A_{021}}+\lambda  \textcolor{black}{A_{012}}+\textcolor{black}{A_{022}}\big],
\end{align}
\end{subequations}
with $C_2 = 1-2 c_s^2$. For the CM framework, populations read as
\begin{subequations}
\label{eq:Q19CM}
\begin{align}
f^{\mathrm{CM,Q19}}_{(0, 0, 0)}&=\rho  \big[U_{x}U_{y} U_{z}+u_x^2 u_y^2 u_z^2 - U_{yz} \widetilde{M}_{200} - U_{xz} \widetilde{M}_{020} - U_{xy} \widetilde{M}_{002} + 4 (u_x u_y \widetilde{M}_{110} + u_x u_z \widetilde{M}_{101} + u_y u_z \widetilde{M}_{011} ) \notag\\
    &\quad\quad + 2 (u_y \widetilde{M}_{210} + u_z \widetilde{M}_{201} + u_z \widetilde{M}_{021} + u_x \widetilde{M}_{120} + u_x \widetilde{M}_{102} + u_y \widetilde{M}_{012}) + \widetilde{M}_{220} + \widetilde{M}_{202} + \widetilde{M}_{022}\big],\\
f^{\mathrm{CM,Q19}}_{(\sigma, 0, 0)}&=\frac{\rho}{2}  \big[u_x \sigma_x U_{yz} + U_{yz} \widetilde{M}_{200} -u_x \sigma_x \widetilde{M}_{020} -u_x \sigma_x \widetilde{M}_{002} -2 \sigma_{2x} (u_y \widetilde{M}_{110} + u_z \widetilde{M}_{101})-2 u_y \widetilde{M}_{210}\notag\\
    &\quad\quad -2 u_z \widetilde{M}_{201} - \sigma_{2x} \widetilde{M}_{120} - \sigma_{2x} \widetilde{M}_{102} -\widetilde{M}_{220} -\widetilde{M}_{202}\big],\\
f^{\mathrm{CM,Q19}}_{(0, \lambda, 0)}&=\frac{\rho}{2}  \big[u_y \lambda_y U_{xz} -u_y \lambda_y \widetilde{M}_{200} + U_{xz} \widetilde{M}_{020} -u_y \lambda_y \widetilde{M}_{002} -2 \lambda_{2y}(u_x  \widetilde{M}_{110} + u_z \widetilde{M}_{011})- \lambda_{2y} \widetilde{M}_{210} \notag\\ 
    &\quad\quad-2 u_z \widetilde{M}_{021} -2 u_x \widetilde{M}_{120} - \lambda_{2y} \widetilde{M}_{012} -\widetilde{M}_{220} -\widetilde{M}_{022}\big],\\
f^{\mathrm{CM,Q19}}_{(0, 0, \delta)}&=\frac{\rho}{2}  \big[u_z \delta_z U_{xy} -u_z \delta_z \widetilde{M}_{200} -u_z \delta_z \widetilde{M}_{020} + U_{xy} \widetilde{M}_{002} -2 \delta_{2z}(u_x  \widetilde{M}_{101} + u_y \widetilde{M}_{011})- \delta_{2z} \widetilde{M}_{201}\notag\\ 
    &\quad\quad - \delta_{2z} \widetilde{M}_{021} -2 u_x \widetilde{M}_{102} -2 u_y \widetilde{M}_{012} -\widetilde{M}_{202} -\widetilde{M}_{022}\big],\\
f^{\mathrm{CM,Q19}}_{(\sigma, \lambda, 0)}&=\frac{\rho}{4}  \big[u_x u_y \sigma_x \lambda_y + \sigma_{2x} \lambda_{2y} \widetilde{M}_{110} + u_y \lambda_y \widetilde{M}_{200} + u_x \sigma_x \widetilde{M}_{020} + \lambda_{2y} \widetilde{M}_{210} + \sigma_{2x} \widetilde{M}_{120} + \widetilde{M}_{220}\big],\\
f^{\mathrm{CM,Q19}}_{(\sigma, 0, \delta)}&=\frac{\rho}{4}  \big[u_x u_z \sigma_x \delta_z + \sigma_{2x} \delta_{2z} \widetilde{M}_{101} + u_z \delta_z \widetilde{M}_{200} + u_x \sigma_x \widetilde{M}_{002} + \delta_{2z} \widetilde{M}_{201} + \sigma_{2x} \widetilde{M}_{102} + \widetilde{M}_{202}\big],\\
f^{\mathrm{CM,Q19}}_{(0, \lambda, \delta)}&=\frac{\rho}{4}  \big[u_y u_z \lambda_y \delta_z + \lambda_{2y} \delta_{2z} \widetilde{M}_{011} + u_z \delta_z \widetilde{M}_{020} + u_y \lambda_y \widetilde{M}_{002} + \delta_{2z} \widetilde{M}_{021} + \lambda_{2y} \widetilde{M}_{012} + \widetilde{M}_{022}\big],
\end{align}
\end{subequations}
where $U_{xy} = 1-u_x^2-u_y^2$, $U_{xz} = 1-u_x^2-u_z^2$ and $U_{yz} = 1-u_y^2-u_z^2$. To derive populations corresponding to the K-LBM, one simply needs to replace $\widetilde{M}_{220}$, $\widetilde{M}_{202}$ and $\widetilde{M}_{022}$ by their cumulant counterparts using Eq.~(\ref{eq:CumulCentral3D}). Eventually, populations defined in the CHM framework are
\begin{subequations}
\label{eq:Q19CHM}
\begin{align}
f^{\mathrm{CHM,Q19}}_{(0, 0, 0)}&=\rho  \big[C_xC_yC_z+c_{x}c_{y}c_{z} - C_{yz} \textcolor{black}{\tilde{A}_{200}} - C_{xz} \textcolor{black}{\tilde{A}_{020}} - C_{xy} \textcolor{black}{\tilde{A}_{002}}+ 4 (u_x u_y \textcolor{black}{\tilde{A}_{110}} + u_x u_z \textcolor{black}{\tilde{A}_{101}} + u_y u_z \textcolor{black}{\tilde{A}_{011}})\notag \\
    &\quad\:\:\:+ 2 (u_y \textcolor{black}{\tilde{A}_{210}} + u_z \textcolor{black}{\tilde{A}_{201}} + u_z \textcolor{black}{\tilde{A}_{021}} + u_x \textcolor{black}{\tilde{A}_{120}} + u_x \textcolor{black}{\tilde{A}_{102}} + u_y \textcolor{black}{\tilde{A}_{012}}) + \textcolor{black}{\tilde{A}_{220}} + \textcolor{black}{\tilde{A}_{202}} + \textcolor{black}{\tilde{A}_{022}}\big], \\
f^{\mathrm{CHM,Q19}}_{(\sigma, 0, 0)}&=\frac{\rho}{2}  \big[C_{\sigma} C_{yz} + C_{yz} \textcolor{black}{\tilde{A}_{200}} - C_{\sigma} \textcolor{black}{\tilde{A}_{020}} - C_{\sigma} \textcolor{black}{\tilde{A}_{002}} -2 \sigma_{2x} (u_y \textcolor{black}{\tilde{A}_{110}} + u_z \textcolor{black}{\tilde{A}_{101}}) -2 u_y \textcolor{black}{\tilde{A}_{210}} -2 u_z \textcolor{black}{\tilde{A}_{201}} \notag\\
    &\quad\:\:\:- \sigma_{2x} \textcolor{black}{\tilde{A}_{120}} - \sigma_{2x} \textcolor{black}{\tilde{A}_{102}} -\textcolor{black}{\tilde{A}_{220}} -\textcolor{black}{\tilde{A}_{202}}\big],\\
f^{\mathrm{CHM,Q19}}_{(0, \lambda, 0)}&=\frac{\rho}{2}  \big[C_{\lambda}C_{xz}  -C_{\lambda} \textcolor{black}{\tilde{A}_{200}} + C_{xz} \textcolor{black}{\tilde{A}_{020}} -C_{\lambda} \textcolor{black}{\tilde{A}_{002}} -2 \lambda_{2y} (u_x \textcolor{black}{\tilde{A}_{110}} + u_z \textcolor{black}{\tilde{A}_{011}}) -\lambda_{2y} \textcolor{black}{\tilde{A}_{210}} -2 u_z \textcolor{black}{\tilde{A}_{021}}\notag\\
    &\quad\:\:\: -2 u_x \textcolor{black}{\tilde{A}_{120}} -\lambda_{2y} \textcolor{black}{\tilde{A}_{012}} -\textcolor{black}{\tilde{A}_{220}} -\textcolor{black}{\tilde{A}_{022}}\big], \\ 
f^{\mathrm{CHM,Q19}}_{(0, 0, \delta)}&=\frac{\rho}{2}  \big[C_{\delta}C_{xy} -C_{\delta} \textcolor{black}{\tilde{A}_{200}} -C_{\delta} \textcolor{black}{\tilde{A}_{020}} + C_{xy} \textcolor{black}{\tilde{A}_{002}} -2 \delta_{2z} (u_x \textcolor{black}{\tilde{A}_{101}} + u_y \textcolor{black}{\tilde{A}_{011}}) -\delta_{2z} \textcolor{black}{\tilde{A}_{201}} -\delta_{2z} \textcolor{black}{\tilde{A}_{021}}\notag\\
    &\quad\:\:\: -2 u_x \textcolor{black}{\tilde{A}_{102}} -2 u_y \textcolor{black}{\tilde{A}_{012}} -\textcolor{black}{\tilde{A}_{202}} -\textcolor{black}{\tilde{A}_{022}}\big], \\ 
f^{\mathrm{CHM,Q19}}_{(\sigma, \lambda, 0)}&=\frac{\rho}{4}  \big[C_{\sigma} C_{\lambda} + C_{\lambda} \textcolor{black}{\tilde{A}_{200}} + C_{\sigma} \textcolor{black}{\tilde{A}_{020}} + \sigma_{2x} \lambda_{2y} \textcolor{black}{\tilde{A}_{110}} + \lambda_{2y} \textcolor{black}{\tilde{A}_{210}} + \sigma_{2x} \textcolor{black}{\tilde{A}_{120}} + \textcolor{black}{\tilde{A}_{220}}\big],\\
f^{\mathrm{CHM,Q19}}_{(\sigma, 0, \delta)}&=\frac{\rho}{4}  \big[C_{\sigma} C_{\delta} + C_{\delta} \textcolor{black}{\tilde{A}_{200}} + C_{\sigma} \textcolor{black}{\tilde{A}_{002}} + \sigma_{2x} \delta_{2z} \textcolor{black}{\tilde{A}_{101}} + \delta_{2z} \textcolor{black}{\tilde{A}_{201}} + \sigma_{2x} \textcolor{black}{\tilde{A}_{102}} + \textcolor{black}{\tilde{A}_{202}} \big],\\
f^{\mathrm{CHM,Q19}}_{(0, \lambda, \delta)}&=\frac{\rho}{4}  \big[C_{\lambda} C_{\delta} + C_{\delta} \textcolor{black}{\tilde{A}_{020}} + C_{\lambda} \textcolor{black}{\tilde{A}_{002}} + \lambda_{2y} \delta_{2z} \textcolor{black}{\tilde{A}_{011}} + \delta_{2z} \textcolor{black}{\tilde{A}_{021}} + \lambda_{2y} \textcolor{black}{\tilde{A}_{012}} + \textcolor{black}{\tilde{A}_{022}}\big],
\end{align}
\end{subequations}
with 
\begin{equation}
\begin{array}{c @{\: = \:} c @{\: + \:} c}
c_{x} & c_s^2 & u_x^2,\\[0.1cm]
c_{y} & c_s^2 & u_y^2,\\[0.1cm] 
c_{z} & c_s^2 & u_z^2,
\end{array}
\:\:\:
\begin{array}{c @{\: = \:} c @{\: - \:} c @{\: - \:} c}
C_{x} & 1 & c_s^2 & u_x^2,\\[0.1cm] 
C_{y} & 1 & c_s^2 & u_y^2,\\[0.1cm] 
C_{z} & 1 & c_s^2 & u_z^2,
\end{array}
\:\:\:
\begin{array}{c @{\: = \:} c @{\: - \:} c @{\: - \:} c @{\: - \:} c}
C_{xy} & 1 & 2 c_s^2 & u_x^2 & u_y^2,\\[0.1cm] 
C_{xz} & 1 & 2 c_s^2 & u_x^2 & u_z^2,\\[0.1cm] 
C_{yz} & 1 & 2 c_s^2 & u_y^2 & u_z^2,
\end{array}
\:\:\:
\begin{array}{c @{\: = \:} c @{\: + \:} l @{\: + \:} l}
C_{\sigma} & c_s^2 & u_x (\sigma & u_x),\\[0.1cm] 
C_{\lambda} & c_s^2 & u_y (\lambda & u_y),\\[0.1cm]
C_{\delta} & c_s^2 & u_z (\delta & u_z),
\end{array}
\:\:\:
\begin{array}{c @{\: = \:} c @{\: + \:} c}
\sigma_{2x} & \sigma & 2 u_x,\\[0.1cm] 
\lambda_{2y} & \lambda & 2 u_y,\\[0.1cm]
\delta_{2z} & \delta & 2 u_z. 
\end{array}
\quad
\end{equation}

If one chooses to start from the Gauss-Hermite formalism, \textcolor{black}{then populations read as}
\begin{equation}
f^{\mathrm{GH,Q19}}_{(i,j,k)} = w_{(i,j,k)}\sum_{n=0}^N \dfrac{\rho}{n! c_s^{2n}}\bm{A}_{pqr} : \bm{H}_{i,pqr},
\end{equation}
with $n=p+q+r$. It is preferred here to rely on $\bm{A}_{pqr}$ instead of $\bm{a}_{pqr}=\rho \bm{A}_{pqr}$ to ease the change of framework through formulas compiled in App.~\ref{sec:RelationshipsTrivariate}. From this, it is clear that the D3Q19 formulation of both regularization steps is based on the following populations
\begin{subequations}
\label{eq:Q19GH}
\begin{align}
f^{\mathrm{GH,Q19}}_{(0, 0, 0)}&=\frac{\rho}{3} \big[1-\tfrac{3 }{2}(A_{200} + A_{020} + A_{002})+\tfrac{9}{4} (A_{220} + A_{202} + A_{022})\big],\\[0.1cm]
f^{\mathrm{GH,Q19}}_{(\sigma, 0, 0)}&= \frac{\rho}{18} \big[1+3 \sigma  u_x+\tfrac{3}{2} (2 A_{200}-A_{020}-A_{002}) -\tfrac{9}{2} \sigma  (A_{120} + A_{102})-\tfrac{9}{4} (2 A_{220} + 2 A_{202} - A_{022})\big],\\[0.1cm]
f^{\mathrm{GH,Q19}}_{(0, \lambda, 0)}&= \frac{\rho}{18} \big[1+3 \lambda  u_y+\tfrac{3}{2} (2 A_{020}-A_{200}-A_{002}) -\tfrac{9}{2} \lambda  (A_{210} + A_{012})-\tfrac{9}{4} (2 A_{220} + 2 A_{022} - A_{202})\big],\\[0.1cm]
f^{\mathrm{GH,Q19}}_{(0, 0, \delta)}&= \frac{\rho}{18} \big[1+3 \delta  u_z+\tfrac{3}{2} (2 A_{002}-A_{200}-A_{020}) -\tfrac{9}{2} \delta  (A_{201} + A_{021})-\tfrac{9}{4} (2 A_{202} + 2 A_{022} - A_{220})\big],\\[0.1cm]
f^{\mathrm{GH,Q19}}_{(\sigma, \lambda, 0)} &= \frac{\rho}{36} \big[1+3 (\sigma  u_x + \lambda  u_y) +\tfrac{3}{2} (2 A_{200}+ 2 A_{020} - A_{002}) +9 \sigma\lambda A_{110} +\tfrac{9}{2}(2 \lambda  A_{210}+ 2  \sigma  A_{120} - \sigma  A_{102} - \lambda  A_{012}) \notag\\[0.1cm] 
    &\quad\quad\:\:\: + \tfrac{9}{2}( 2 A_{220}-A_{202}-A_{022})\big],\\[0.1cm]
f^{\mathrm{GH,Q19}}_{(\sigma, 0, \delta)} &= \frac{\rho}{36} \big[1+3 (\sigma  u_x + \delta  u_z) +\tfrac{3}{2} (2 A_{200}+ 2 A_{002} - A_{020}) +9 \sigma\delta A_{101} +\tfrac{9}{2}(2 \delta  A_{201}+ 2  \sigma  A_{102} - \sigma  A_{120} - \delta  A_{021}) \notag\\[0.1cm] 
    &\quad\quad\:\:\: + \tfrac{9}{2}( 2 A_{202}-A_{220}-A_{022})\big],\\[0.1cm]
f^{\mathrm{GH,Q19}}_{(0, \lambda, \delta)}&= \frac{\rho}{36} \big[1+3 (\lambda u_y + \delta u_z) +\tfrac{3}{2} (2 A_{020}+ 2 A_{002} - A_{200}) +9 \lambda\delta A_{011} +\tfrac{9}{2}(2 \delta  A_{021}+ 2  \lambda  A_{012} - \lambda  A_{210} - \delta  A_{201}) \notag\\[0.1cm] 
    &\quad\quad\:\:\: + \tfrac{9}{2}( 2 A_{022}-A_{220}-A_{202})\big].
\end{align}
\end{subequations}  
where both the lattice constant and the weights have been replaced by their values (Tab.~\ref{tab:Lattices}). Depending on the way $A_{pqr}^{neq}$ will be computed, Eq.~(\ref{eq:Q19GH}) will either lead to the PR ($A_{pqr}^{neq}= A_{pqr}-A_{pqr}^{eq}$) or to the RR ($A_{pqr}^{neq}=A_{pqr}^{(1)}$ computed via Eq.~(\ref{eq:RRformulas_D3Q27})) collision models. 

One can see that $f^{\mathrm{HM,Q19}}_i$ and $f^{\mathrm{GH,Q19}}_i$ lead to two families of LBMs that are based on two different kinds of equilibrium states (see App.~\ref{sec:EqFunction}). \textcolor{black}{The origin of this mismatch comes from (1) the weights which differ between the D3Q19 and the D3Q27 lattices, and (2) the difference between high-order contributions that are discarded. The former point is crystal clear when one looks at the values of the weights (see Tab.~\ref{tab:Lattices}). Regarding the latter point, it flows from the fact that when neglecting $M_{222}$ in the unweighted formalism, then it affects, for example, $A_{222}$ and all other even Hermite moments ($A_{220}$, $A_{200}$, $A_{000}$, and their cyclic permutations). On the contrary, only the contribution of $A_{222}$ is discarded in the Gauss-Hermite formalism. The same reasoning applies to all moments that are neglected when moving from the D3Q27 formulation to the D3Q19 one.} 

What is interesting about these two families of LBMs is that one can derive new collision models by simply switching from $A_{pqr}$ to the statistical quantity of interest. As an example, one could derive a new D3Q19-K-LBM based on either $f^{\mathrm{HM,Q19}}_i$ or $f^{\mathrm{GH,Q19}}_i$ by first moving from $A_{pqr}$ to $M_{pqr}$, then to $\widetilde{M}_{pqr}$, and finally to $K_{pqr}$ using successively Eqs.~(\ref{eq:RawHermite3D}),~(\ref{eq:RawCentral3D}) and~(\ref{eq:CumulCentral3D}).

\textcolor{black}{In the end, one may wonder what formalism should be preferred when it comes to the D3Q19 lattice. In fact, populations obtained in the unweighted formalism~(\ref{eq:Q19RM})-(\ref{eq:Q19CHM}) can be shown to be solutions of the problem~(\ref{eq:MomentConservation}). This leads to the recovery of the proper set of $19$ moments, and as a consequence, these models have the correct macroscopic behavior. On the contrary, the \emph{present} Gauss-Hermite formulation~(\ref{eq:Q19GH}) leads to spurious coupling between hydrodynamic and high-order moments, and a fortiori, to an improper macroscopic behavior. Indeed, 
\begin{subequations}
\begin{align}
\sum_i f^{\mathrm{GH,Q19}}_i H_{i,200} = \rho A_{200} -\dfrac{3}{4}\rho A_{022} \neq \displaystyle\int{f H_{200} \mathrm{d}\xi}, \\[0.1cm]
\sum_i f^{\mathrm{GH,Q19}}_i H_{i,020} = \rho A_{020} -\dfrac{3}{4}\rho A_{202} \neq \displaystyle\int{f H_{020} \mathrm{d}\xi}, \\[0.1cm]
\sum_i f^{\mathrm{GH,Q19}}_i H_{i,002} = \rho A_{002} -\dfrac{3}{4}\rho A_{220} \neq \displaystyle\int{f H_{002} \mathrm{d}\xi},
\end{align}
\end{subequations}
meaning the viscous stress tensor will depend on fourth-order nonequilibrium moments. Further couplings can be observed among third-order moments,
\begin{subequations}
\begin{align}
\sum_i f^{\mathrm{GH,Q19}}_i H_{i,210} = \rho A_{210} -\dfrac{1}{2}\rho A_{012} \neq \displaystyle\int{f H_{210} \mathrm{d}\xi}, \\[0.1cm]
\sum_i f^{\mathrm{GH,Q19}}_i H_{i,201} = \rho A_{201} -\dfrac{1}{2}\rho A_{021} \neq \displaystyle\int{f H_{201} \mathrm{d}\xi}, \\[0.1cm]
\sum_i f^{\mathrm{GH,Q19}}_i H_{i,021} = \rho A_{021} -\dfrac{1}{2}\rho A_{201} \neq \displaystyle\int{f H_{021} \mathrm{d}\xi}, \\[0.1cm]
\sum_i f^{\mathrm{GH,Q19}}_i H_{i,120} = \rho A_{120} -\dfrac{1}{2}\rho A_{102} \neq \displaystyle\int{f H_{120} \mathrm{d}\xi}, \\[0.1cm]
\sum_i f^{\mathrm{GH,Q19}}_i H_{i,102} = \rho A_{102} -\dfrac{1}{2}\rho A_{120} \neq \displaystyle\int{f H_{102} \mathrm{d}\xi}, \\[0.1cm]
\sum_i f^{\mathrm{GH,Q19}}_i H_{i,012} = \rho A_{012} -\dfrac{1}{2}\rho A_{210} \neq \displaystyle\int{f H_{012} \mathrm{d}\xi},
\end{align}
\end{subequations}
and also fourth-order moments,
\begin{subequations}
\begin{align}
\sum_i f^{\mathrm{GH,Q19}}_i H_{i,220} = \rho A_{220} -\dfrac{\rho}{4}(\rho A_{202}+\rho A_{022}) -\dfrac{\rho}{6}A_{002} \neq \displaystyle\int{f H_{220} \mathrm{d}\xi}, \\[0.1cm]
\sum_i f^{\mathrm{GH,Q19}}_i H_{i,202} = \rho A_{202} -\dfrac{\rho}{4}(\rho A_{220}+\rho A_{022}) -\dfrac{\rho}{6}A_{020} \neq \displaystyle\int{f H_{202} \mathrm{d}\xi}, \\[0.1cm]
\sum_i f^{\mathrm{GH,Q19}}_i H_{i,022} = \rho A_{022} -\dfrac{\rho}{4}(\rho A_{220}+\rho A_{202}) -\dfrac{\rho}{6}A_{200} \neq \displaystyle\int{f H_{022} \mathrm{d}\xi}.
\end{align}
\end{subequations}
These coupling defects are usually not encountered since, in most cases, the Hermite polynomial expansion is done up to the second-order for the D3Q19 lattice~\cite{SHAN_JFM_550_2006,KRUGER_Book_2017}. Nevertheless, a few authors have also proposed to include third-order terms in its equilibrium state to improve the macroscopic behavior of the resulting LBM~\cite{HAZI_JPA_39_2006,DELLAR_JCP_259_2014,FENG_CF_131_2016}.} 

\textcolor{black}{There are several possibilities to deal with these spurious couplings. First, one can impose third- and fourth-order terms to their equilibrium value, which leads to a PR collision model based on an extended equilibrium state. One then ends up with the correct viscous stress tensor. The second option consists in discarding fourth-order moments, and using orthogonal third-order Hermite polynomials~\cite{JACOB_JT_0_2018},
\begin{equation}
(H_{i,210}+H_{i,012}, H_{i,201}+H_{i,021}, H_{i,120}+H_{i,102},H_{i,210}-H_{i,012}, H_{i,201}-H_{i,021}, H_{i,120}-H_{i,102}).
\label{eq:BasisOrtho3} 
\end{equation}
Equilibriating fourth-order moments also works here. The last way to obtain a stable numerical scheme is to rely on a fully orthogonal basis. By orthogonalizing fourth-order Hermite polynomials with second-order ones using the weighted scalar product, the following formulas are derived
\begin{equation}
H_{i,220}^{4o2}=H_{i,220}+(1/6)H_{i,002}, \quad H_{i,202}^{4o2}=H_{i,202}+(1/6)H_{i,020}, \quad H_{i,022}^{4o2}=H_{i,022}+(1/6)H_{i,200},
\label{eq:BasisOrtho42} 
\end{equation}
with the superscript $4o2$ standing for `order $4$ orthogonal with respect to order $2$'. By further orthogonalizing them between each other, one obtains
\begin{equation}
H_{i,220}^{\mathrm{FO}}=H_{i,220}^{4o2}, \quad H_{i,202}^{\mathrm{FO}}=H_{i,202}^{4o2}+(2/7)H_{i,220}^{\mathrm{FO}}, \quad H_{i,022}^{\mathrm{FO}}=H_{i,022}^{4o2}+(2/7)H_{i,220}^{\mathrm{FO}}+(2/5)H_{i,202}^{\mathrm{FO}},
\label{eq:BasisFullOrtho} 
\end{equation}
where the superscript $\mathrm{FO}$ means `Fully Orthogonal'. Eventually, by combining Eqs.~(\ref{eq:BasisOrtho3}) and~(\ref{eq:BasisFullOrtho}) with the second order polynomial basis, the new D3Q19-GH formulation of the populations reads
\begin{align}
f^{\mathrm{GH,Q19}}_{i} &= w_{i}\rho \bigg\{\sum_{n=0}^2 \dfrac{1}{n! c_s^{2n}}{H}_{i,pqr} : {A}_{pqr} +\frac{1}{6 c_s^6}\Big[3(H_{i,210}+H_{i,012})(A_{210}+A_{012}) + 3(H_{i,201}+H_{i,021})(A_{201}+A_{021})\nonumber\\
&\quad  + 3(H_{i,120}+H_{i,102})(A_{120}+A_{102}) + (H_{i,210}-H_{i,012})(A_{210}-A_{012}) + (H_{i,201}-H_{i,021})(A_{201}-A_{021})  \nonumber\\
&\quad + (H_{i,120}-H_{i,102})(A_{120}-A_{102}) \Big] +\frac{1}{4 c_s^8} \Big[ (8/7)H_{i,220}^{\mathrm{FO}} A_{220}^{\mathrm{FO}} + (56/45) H_{i,202}^{\mathrm{FO}} A_{202}^{\mathrm{FO}} + (40/27) H_{i,022}^{\mathrm{FO}} A_{022}^{\mathrm{FO}}\Big] \bigg\},
\end{align}
where $A_{pqr}^{\mathrm{FO}}$ are Hermite polynomial coefficients corresponding to the fully orthogonal set of Hermite polynomials $H_{i,pqr}^{\mathrm{FO}}$. Among the three above propositions, this last model was identified as the most (linearly) stable one when coupled with the BGK collision operator (all relaxation frequencies equal to $\omega_{\nu}$).  
}

\section{Equilibrium distribution functions \label{sec:EqFunction}}
This appendix is dedicated to the equilibrium state recovered by all LBMs considered in the present work. Starting from the one-dimensional case, it was demonstrated that all LBMs recovered the following formulas (Sec.~\ref{sec:CompStudy1})

\begin{align}
f_{0}^{eq} &= \rho (C-u_x^2),\label{eq:feq1Da}\\[0.1cm]
f^{eq}_{\sigma} &= \dfrac{\rho}{2}(c_s^2 + \sigma u_x + u_x^2),\label{eq:feq1Db}
\intertext{with $C=(1-c_s^2)$. Since the continuous Maxwell-Boltzmann equilibrium distribution is by definition isotropic, it is further proposed to enforce the isotropy of the discrete equilibrium state through the use of tensor product rules~\cite{KARLIN_PA_389_2010}. This $D$-dimensional extension naturally leads to}
f^{eq}_{(0, 0)} &= \rho (C-u_x^2)(C-u_y^2),\\[0.1cm] 
f^{eq}_{(\sigma, 0)}&= \dfrac{\rho}{2}(c_s^2 + \sigma u_x + u_x^2)(C-u_y^2),\\[0.1cm]
f^{eq}_{(0, \lambda)}&= \dfrac{\rho}{2}(C-u_x^2)(c_s^2 + \lambda u_y + u_y^2),\\[0.1cm]
f^{eq}_{(\sigma, \lambda)}&=\dfrac{\rho}{4}(c_s^2 + \sigma u_x + u_x^2)(c_s^2 + \lambda u_y + u_y^2),
\intertext{for the D2Q9 lattice, and}
f^{eq}_{(0, 0, 0)}&=\rho (C-u_x^2)(C-u_y^2)(C-u_z^2),\\[0.1cm]
f^{eq}_{(\sigma, 0, 0)}&= \dfrac{\rho}{2}(c_s^2 + \sigma u_x + u_x^2)(C-u_y^2)(C-u_z^2),\\[0.1cm]
f^{eq}_{(0, \lambda, 0)}&= \dfrac{\rho}{2}(C-u_x^2)(c_s^2 + \lambda u_y + u_y^2)(C-u_z^2),\\[0.1cm]
f^{eq}_{(0, 0, \delta)}&= \dfrac{\rho}{2}(C-u_x^2)(C-u_y^2)(c_s^2 + \delta u_z + u_z^2),\\[0.1cm]
f^{eq}_{(\sigma, \lambda, 0)} &= \dfrac{\rho}{4}(c_s^2 + \sigma u_x + u_x^2)(c_s^2 + \lambda u_y + u_y^2)(C-u_z^2),\\[0.1cm]
f^{eq}_{(\sigma, 0, \delta)} &= \dfrac{\rho}{4}(c_s^2 + \sigma u_x + u_x^2)(C-u_y^2)(c_s^2 + \delta u_z + u_z^2),\\[0.1cm]
f^{eq}_{(0, \lambda, \delta)}&= \dfrac{\rho}{4}(C-u_x^2)(c_s^2 + \lambda u_y + u_y^2)(c_s^2 + \delta u_z + u_z^2),\\[0.1cm]
f^{eq}_{(\sigma, \lambda, \delta)}&= \dfrac{\rho}{8}(c_s^2 + \sigma u_x + u_x^2)(c_s^2 + \lambda u_y + u_y^2)(c_s^2 + \delta u_z + u_z^2).
\end{align}
for the D3Q27 velocity discretization, where $C=1-c_s^2$. For all above formulas, it is considered that $(\sigma,\lambda,\delta)=\{\pm 1\}^3$. 

Knowing that all models recover the same equilibrium state in the one-dimensional case, the above construction then ensures that all LBMs also share the exact same equilibrium state for both two- and three-dimensional cases. In particular, this is true for models based on orthogonal bases (MRT and TRT), the regularized LBMs and the K-LBM. For regularized LBMs, it is explained by the fact that in the one-dimensional case, either solving Eq.~(\ref{eq:}) or using the Gauss-Hermite quadrature formulation~(\ref{eq:VDF_GH_Quad}) leads to the very same formalism. Tensor product rules compiled in Eqs.~(\ref{eq:TPvel})-(\ref{eq:TPweight}) allow us to extend the previous result to both two- and three-dimensional cases. For cumulants, deviations with respect to the CM formalism can only come from equilibrium cumulants of order $n\geq 4$, but they are null by definition if $n\geq 3$.

One must be careful regarding the equilibrium state derived in the case of the D3Q19 lattice. Indeed, if one starts from the RM framework and discards equilibrium raw moments via instructions given in Sec.~\ref{sec:D3Q19}, then one obtains the following type of equilibrium state
\begin{align}
f^{eq,\mathrm{RM}}_{(0, 0, 0)}&=\frac{\rho}{3} \big[1 -(u_x^2+u_y^2+u_z^2)+ 3(u_x^2 u_y^2+u_x^2 u_z^2+u_y^2 u_z^2)\big],\label{eq:EqQ19RM1}\\[0.1cm]
f^{eq,\mathrm{RM}}_{(\sigma, 0, 0)}&= \frac{\rho}{18} \big[1 +3 \sigma u_x + 3 (u_x^2- u_y^2- u_z^2) - 9\sigma  (u_x u_y^2 + u_x u_z^2)-9 (u_x^2 u_y^2 + u_x^2 u_z^2)\big],\\[0.1cm]
f^{eq,\mathrm{RM}}_{(0, \lambda, 0)}&= \frac{\rho}{18} \big[1 +3 \lambda u_y +3 (- u_x^2+  u_y^2- u_z^2) -9\lambda  (u_x^2 u_y + u_y u_z^2)-9 (u_x^2 u_y^2+ u_y^2 u_z^2)\big],\\[0.1cm]
f^{eq,\mathrm{RM}}_{(0, 0, \delta)}&= \frac{\rho}{18} \big[1 +3 \delta u_z +3(- u_x^2- u_y^2+  u_z^2) -9\delta  (u_x^2 u_z + u_y^2 u_z)-9 (u_x^2 u_z^2+ u_y^2 u_z^2)\big],\\[0.1cm]
f^{eq,\mathrm{RM}}_{(\sigma, \lambda, 0)} &= \frac{\rho}{36}  \big[1+3(\sigma u_x + \lambda u_y) + 3 (u_x^2+ u_y^2)+9\sigma\lambda  u_x u_y+ 9(\lambda  u_x^2 u_y+\sigma u_x u_y^2)+9 u_x^2 u_y^2\big],\\[0.1cm]
f^{eq,\mathrm{RM}}_{(\sigma, 0, \delta)} &= \frac{\rho}{36}  \big[1+3(\sigma u_x + \delta u_z) + 3 (u_x^2+ u_z^2)+9\sigma\delta  u_x u_z+ 9(\delta  u_x^2 u_z+\sigma u_x u_z^2)+9 u_x^2 u_z^2\big],\\[0.1cm]
f^{eq,\mathrm{RM}}_{(0, \lambda, \delta)}&= \frac{\rho}{36}  \big[1+3(\lambda u_y + \delta u_z) + 3 (u_y^2+ u_z^2)+9\lambda\delta  u_y u_z+ 9(\delta  u_y^2 u_z+\lambda u_y u_z^2)+9 u_y^2 u_z^2\big]\label{eq:EqQ19RM2},
\end{align}
where the lattice constant has been replaced with its value $c_s^{\mathrm{Q19}}=1/\sqrt{3}$.
Unfortunately, these equilibrium VDFs do not match anymore those obtained through the Gauss-Hermite quadrature,
\begin{align}
f^{eq,\mathrm{GH}}_{(0, 0, 0)}&=\frac{\rho}{3} \big[1-\tfrac{3 }{2}(u_x^2+u_y^2+u_z^2)+\tfrac{9}{4} (u_x^2 u_y^2+ u_x^2 u_z^2+ u_y^2 u_z^2)\big],\label{eq:EqQ19GH1}\\[0.1cm]
f^{eq,\mathrm{GH}}_{(\sigma, 0, 0)}&= \frac{\rho}{18} \big[1+3 \sigma  u_x+\tfrac{3}{2} (2 u_x^2-u_y^2-u_z^2) -\tfrac{9}{2} \sigma  (u_x u_y^2 + u_x u_z^2)-\tfrac{9}{4} (2 u_x^2 u_y^2 + 2 u_x^2 u_z^2 - u_y^2 u_z^2)\big],\\[0.1cm]
f^{eq,\mathrm{GH}}_{(0, \lambda, 0)}&= \frac{\rho}{18} \big[1+3 \lambda  u_y+\tfrac{3}{2} (2 u_y^2-u_x^2-u_z^2) -\tfrac{9}{2} \lambda  (u_x^2 u_y + u_y u_z^2)-\tfrac{9}{4} (2 u_x^2 u_y^2 + 2 u_y^2 u_z^2 - u_x^2 u_z^2)\big],\\[0.1cm]
f^{eq,\mathrm{GH}}_{(0, 0, \delta)}&= \frac{\rho}{18} \big[1+3 \delta  u_z+\tfrac{3}{2} (2 u_z^2- u_x^2 - u_y^2) -\tfrac{9}{2} \delta  (u_x^2 u_z + u_y^2 u_z)-\tfrac{9}{4} (2 u_x^2 u_z^2 + 2 u_y^2 u_z^2 - u_x^2 u_y^2)\big],\\[0.1cm]
f^{eq,\mathrm{GH}}_{(\sigma, \lambda, 0)} &= \frac{\rho}{36} \big[1+3 (\sigma  u_x + \lambda  u_y) +3 (u_x^2+ u_y^2) - \tfrac{1}{2}u_z^2 +9 \sigma\lambda u_x u_y +\tfrac{9}{2}(2 \lambda  u_x^2 u_y+ 2  \sigma  u_x u_y^2 - \sigma  u_x u_z^2 - \lambda  u_y u_z^2) \notag\\[0.1cm] 
    &\quad\quad\:\:\: + \tfrac{9}{2}( 2 u_x^2 u_y^2-u_x^2 u_z^2-u_y^2 u_z^2)\big],\\[0.1cm]
f^{eq,\mathrm{GH}}_{(\sigma, 0, \delta)} &= \frac{\rho}{36} \big[1+3 (\sigma  u_x + \delta  u_z) +3 (u_x^2 + u_z^2) -\tfrac{1}{2} u_y^2 +9 \sigma\delta u_x u_z +\tfrac{9}{2}(2 \delta  u_x^2 u_z+ 2  \sigma  u_x u_z^2 - \sigma  u_x u_y^2 - \delta  u_y^2 u_z) \notag\\[0.1cm] 
    &\quad\quad\:\:\: + \tfrac{9}{2}( 2 u_x^2 u_z^2-u_x^2 u_y^2-u_y^2 u_z^2)\big],\\[0.1cm]
f^{eq,\mathrm{GH}}_{(0, \lambda, \delta)}&= \frac{\rho}{36} \big[1+3 (\lambda u_y + \delta u_z) +3 (u_y^2 + u_z^2) -\tfrac{1}{2} u_x^2 +9 \lambda\delta u_y u_z +\tfrac{9}{2}(2 \delta  u_y^2 u_z+ 2  \lambda  u_y u_z^2 - \lambda  u_x^2 u_y - \delta  u_x^2 u_z) \notag\\[0.1cm] 
    &\quad\quad\:\:\: + \tfrac{9}{2}( 2 u_y^2 u_z^2-u_x^2 u_y^2-u_x^2 u_z^2)\big].\label{eq:EqQ19GH2}
\end{align}
where both the lattice constant and weights have been replaced with their value (Tab.~\ref{tab:Lattices}) to ease the comparison with $f^{eq,\mathrm{RM}}_{i}$. 

From this, one obtains two different kinds of equilibrium states whose difference scales as $\mathcal{O}(u^2)$. This means that these equilibrium VDFs should lead to different results even for low Mach number flows. One can further expect that they will also result in different linear stability domains for the D3Q19 lattice, as it was already confirmed for the D2Q9 lattice~\cite{COREIXAS_PhD_2018}.

\twocolumngrid

\bibliography{COREIXAS_PRE_2018_Theo_Bibliography}

\begin{thebibliography}{134}%
\makeatletter
\providecommand \@ifxundefined [1]{%
 \@ifx{#1\undefined}
}%
\providecommand \@ifnum [1]{%
 \ifnum #1\expandafter \@firstoftwo
 \else \expandafter \@secondoftwo
 \fi
}%
\providecommand \@ifx [1]{%
 \ifx #1\expandafter \@firstoftwo
 \else \expandafter \@secondoftwo
 \fi
}%
\providecommand \natexlab [1]{#1}%
\providecommand \enquote  [1]{``#1''}%
\providecommand \bibnamefont  [1]{#1}%
\providecommand \bibfnamefont [1]{#1}%
\providecommand \citenamefont [1]{#1}%
\providecommand \href@noop [0]{\@secondoftwo}%
\providecommand \href [0]{\begingroup \@sanitize@url \@href}%
\providecommand \@href[1]{\@@startlink{#1}\@@href}%
\providecommand \@@href[1]{\endgroup#1\@@endlink}%
\providecommand \@sanitize@url [0]{\catcode `\\12\catcode `\$12\catcode
  `\&12\catcode `\#12\catcode `\^12\catcode `\_12\catcode `\%12\relax}%
\providecommand \@@startlink[1]{}%
\providecommand \@@endlink[0]{}%
\providecommand \url  [0]{\begingroup\@sanitize@url \@url }%
\providecommand \@url [1]{\endgroup\@href {#1}{\urlprefix }}%
\providecommand \urlprefix  [0]{URL }%
\providecommand \Eprint [0]{\href }%
\providecommand \doibase [0]{http://dx.doi.org/}%
\providecommand \selectlanguage [0]{\@gobble}%
\providecommand \bibinfo  [0]{\@secondoftwo}%
\providecommand \bibfield  [0]{\@secondoftwo}%
\providecommand \translation [1]{[#1]}%
\providecommand \BibitemOpen [0]{}%
\providecommand \bibitemStop [0]{}%
\providecommand \bibitemNoStop [0]{.\EOS\space}%
\providecommand \EOS [0]{\spacefactor3000\relax}%
\providecommand \BibitemShut  [1]{\csname bibitem#1\endcsname}%
\let\auto@bib@innerbib\@empty
\bibitem [{\citenamefont {Manoha}\ and\ \citenamefont
  {Caruelle}(2015)}]{MANOHA_AIAA_2846_2015}%
  \BibitemOpen
  \bibfield  {author} {\bibinfo {author} {\bibfnamefont {E.}~\bibnamefont
  {Manoha}}\ and\ \bibinfo {author} {\bibfnamefont {B.}~\bibnamefont
  {Caruelle}},\ }in\ \href {10.2514/6.2014-0952} {\emph {\bibinfo {booktitle}
  {21st AIAA/CEAS Aeroacoustics Conference}}}\ (\bibinfo {year} {2015})\ p.\
  \bibinfo {pages} {2846}\BibitemShut {NoStop}%
\bibitem [{\citenamefont {Rumsey}\ \emph {et~al.}(2018)\citenamefont {Rumsey},
  \citenamefont {Slotnick},\ and\ \citenamefont
  {Sclafani}}]{RUMSEY_AIAA_1258_2018}%
  \BibitemOpen
  \bibfield  {author} {\bibinfo {author} {\bibfnamefont {C.~L.}\ \bibnamefont
  {Rumsey}}, \bibinfo {author} {\bibfnamefont {J.~P.}\ \bibnamefont
  {Slotnick}}, \ and\ \bibinfo {author} {\bibfnamefont {A.~J.}\ \bibnamefont
  {Sclafani}},\ }in\ \href {\doibase 10.2514/6.2018-1258} {\emph {\bibinfo
  {booktitle} {2018 AIAA Aerospace Sciences Meeting}}}\ (\bibinfo {year}
  {2018})\ p.\ \bibinfo {pages} {1258}\BibitemShut {NoStop}%
\bibitem [{\citenamefont {Succi}(2018)}]{SUCCI_Book_2018}%
  \BibitemOpen
  \bibfield  {author} {\bibinfo {author} {\bibfnamefont {S.}~\bibnamefont
  {Succi}},\ }\href
  {https://global.oup.com/academic/product/the-lattice-boltzmann-equation-9780199592357?q=The%20Lattice%20Boltzmann%20Equation:%20For%20Complex%20States%20of%20Flowing%20Matter&lang=en&cc=ch}
  {\emph {\bibinfo {title} {The Lattice Boltzmann Equation: For Complex States
  of Flowing Matter}}}\ (\bibinfo  {publisher} {Oxford University Press},\
  \bibinfo {year} {2018})\BibitemShut {NoStop}%
\bibitem [{\citenamefont {Ricot}\ \emph {et~al.}(2009)\citenamefont {Ricot},
  \citenamefont {Marié}, \citenamefont {Sagaut},\ and\ \citenamefont
  {Bailly}}]{RICOT_JCP_228_2009}%
  \BibitemOpen
  \bibfield  {author} {\bibinfo {author} {\bibfnamefont {D.}~\bibnamefont
  {Ricot}}, \bibinfo {author} {\bibfnamefont {S.}~\bibnamefont {Marié}},
  \bibinfo {author} {\bibfnamefont {P.}~\bibnamefont {Sagaut}}, \ and\ \bibinfo
  {author} {\bibfnamefont {C.}~\bibnamefont {Bailly}},\ }\href {\doibase
  10.1016/j.jcp.2009.03.030} {\bibfield  {journal} {\bibinfo  {journal} {J.
  Comput. Phys.}\ }\textbf {\bibinfo {volume} {228}},\ \bibinfo {pages} {4478 }
  (\bibinfo {year} {2009})}\BibitemShut {NoStop}%
\bibitem [{\citenamefont {Mari\'{e}}\ and\ \citenamefont
  {Gloerfelt}(2017)}]{MARIE_JCP_333_2017}%
  \BibitemOpen
  \bibfield  {author} {\bibinfo {author} {\bibfnamefont {S.}~\bibnamefont
  {Mari\'{e}}}\ and\ \bibinfo {author} {\bibfnamefont {X.}~\bibnamefont
  {Gloerfelt}},\ }\href {\doibase 10.1016/j.jcp.2016.12.017} {\bibfield
  {journal} {\bibinfo  {journal} {J. Comput. Phys.}\ }\textbf {\bibinfo
  {volume} {333}},\ \bibinfo {pages} {212 } (\bibinfo {year}
  {2017})}\BibitemShut {NoStop}%
\bibitem [{\citenamefont {Sagaut}(2006)}]{SAGAUT_Book_2006}%
  \BibitemOpen
  \bibfield  {author} {\bibinfo {author} {\bibfnamefont {P.}~\bibnamefont
  {Sagaut}},\ }\href
  {https://www.springer.com/de/book/9783540263449?token=gbgen&wt_mc=GoogleBooks.GoogleBooks.3.EN#otherversion=9783540264033}
  {\emph {\bibinfo {title} {Large eddy simulation for incompressible flows: an
  introduction}}}\ (\bibinfo  {publisher} {Springer Science \& Business
  Media},\ \bibinfo {year} {2006})\BibitemShut {NoStop}%
\bibitem [{\citenamefont {Sagaut}(2010)}]{SAGAUT_CMA_59_2010}%
  \BibitemOpen
  \bibfield  {author} {\bibinfo {author} {\bibfnamefont {P.}~\bibnamefont
  {Sagaut}},\ }\href {\doibase 10.1016/j.camwa.2009.08.051} {\bibfield
  {journal} {\bibinfo  {journal} {Comput. Math. Appl.}\ }\textbf {\bibinfo
  {volume} {59}},\ \bibinfo {pages} {2194 } (\bibinfo {year}
  {2010})}\BibitemShut {NoStop}%
\bibitem [{\citenamefont {Karlin}\ \emph {et~al.}(1998)\citenamefont {Karlin},
  \citenamefont {Gorban}, \citenamefont {Succi},\ and\ \citenamefont
  {Boffi}}]{KARLIN_PRL_81_1998}%
  \BibitemOpen
  \bibfield  {author} {\bibinfo {author} {\bibfnamefont {I.~V.}\ \bibnamefont
  {Karlin}}, \bibinfo {author} {\bibfnamefont {A.~N.}\ \bibnamefont {Gorban}},
  \bibinfo {author} {\bibfnamefont {S.}~\bibnamefont {Succi}}, \ and\ \bibinfo
  {author} {\bibfnamefont {V.}~\bibnamefont {Boffi}},\ }\href {\doibase
  10.1103/PhysRevLett.81.6} {\bibfield  {journal} {\bibinfo  {journal} {Phys.
  Rev. Lett.}\ }\textbf {\bibinfo {volume} {81}},\ \bibinfo {pages} {6}
  (\bibinfo {year} {1998})}\BibitemShut {NoStop}%
\bibitem [{\citenamefont {Boghosian}\ \emph {et~al.}(2001)\citenamefont
  {Boghosian}, \citenamefont {Yepez}, \citenamefont {Coveney},\ and\
  \citenamefont {Wager}}]{BOGHOSIAN_PRSLA_457_2001a}%
  \BibitemOpen
  \bibfield  {author} {\bibinfo {author} {\bibfnamefont {B.~M.}\ \bibnamefont
  {Boghosian}}, \bibinfo {author} {\bibfnamefont {J.}~\bibnamefont {Yepez}},
  \bibinfo {author} {\bibfnamefont {P.~V.}\ \bibnamefont {Coveney}}, \ and\
  \bibinfo {author} {\bibfnamefont {A.}~\bibnamefont {Wager}},\ }\href
  {\doibase 10.1098/rspa.2000.0689} {\bibfield  {journal} {\bibinfo  {journal}
  {Proc. Royal Soc. A}\ }\textbf {\bibinfo {volume} {457}},\ \bibinfo {pages}
  {717} (\bibinfo {year} {2001})}\BibitemShut {NoStop}%
\bibitem [{\citenamefont {Malaspinas}\ \emph {et~al.}(2008)\citenamefont
  {Malaspinas}, \citenamefont {Deville},\ and\ \citenamefont
  {Chopard}}]{MALASPINAS_PRE_78_2008}%
  \BibitemOpen
  \bibfield  {author} {\bibinfo {author} {\bibfnamefont {O.}~\bibnamefont
  {Malaspinas}}, \bibinfo {author} {\bibfnamefont {M.}~\bibnamefont {Deville}},
  \ and\ \bibinfo {author} {\bibfnamefont {B.}~\bibnamefont {Chopard}},\ }\href
  {\doibase 10.1103/PhysRevE.78.066705} {\bibfield  {journal} {\bibinfo
  {journal} {Phys. Rev. E}\ }\textbf {\bibinfo {volume} {78}},\ \bibinfo
  {pages} {066705} (\bibinfo {year} {2008})}\BibitemShut {NoStop}%
\bibitem [{\citenamefont {Hamdi}\ \emph {et~al.}(2018)\citenamefont {Hamdi},
  \citenamefont {Elalimi},\ and\ \citenamefont
  {Nasrallah}}]{HAMDI_InBook_2018}%
  \BibitemOpen
  \bibfield  {author} {\bibinfo {author} {\bibfnamefont {M.}~\bibnamefont
  {Hamdi}}, \bibinfo {author} {\bibfnamefont {S.}~\bibnamefont {Elalimi}}, \
  and\ \bibinfo {author} {\bibfnamefont {S.~B.}\ \bibnamefont {Nasrallah}},\
  }\enquote {\bibinfo {title} {Large eddy simulation-based lattice {B}oltzmann
  method with different collision models},}\ in\ \href {\doibase
  10.1007/978-3-319-62572-0_43} {\emph {\bibinfo {booktitle} {Exergy for A
  Better Environment and Improved Sustainability 1: Fundamentals}}}\ (\bibinfo
  {publisher} {Springer International Publishing},\ \bibinfo {year} {2018})\
  pp.\ \bibinfo {pages} {661--683}\BibitemShut {NoStop}%
\bibitem [{\citenamefont {Karlin}\ \emph {et~al.}(2014)\citenamefont {Karlin},
  \citenamefont {B\"osch},\ and\ \citenamefont
  {Chikatamarla}}]{KARLIN_PRE_90_2014a}%
  \BibitemOpen
  \bibfield  {author} {\bibinfo {author} {\bibfnamefont {I.~V.}\ \bibnamefont
  {Karlin}}, \bibinfo {author} {\bibfnamefont {F.}~\bibnamefont {B\"osch}}, \
  and\ \bibinfo {author} {\bibfnamefont {S.~S.}\ \bibnamefont {Chikatamarla}},\
  }\href {\doibase 10.1103/PhysRevE.90.031302} {\bibfield  {journal} {\bibinfo
  {journal} {Phys. Rev. E}\ }\textbf {\bibinfo {volume} {90}},\ \bibinfo
  {pages} {031302} (\bibinfo {year} {2014})}\BibitemShut {NoStop}%
\bibitem [{\citenamefont {B{\"o}sch}\ \emph {et~al.}(2015)\citenamefont
  {B{\"o}sch}, \citenamefont {Chikatamarla},\ and\ \citenamefont
  {Karlin}}]{BOSCH_ESAIM_52_2015}%
  \BibitemOpen
  \bibfield  {author} {\bibinfo {author} {\bibfnamefont {F.}~\bibnamefont
  {B{\"o}sch}}, \bibinfo {author} {\bibfnamefont {S.~S.}\ \bibnamefont
  {Chikatamarla}}, \ and\ \bibinfo {author} {\bibfnamefont {I.}~\bibnamefont
  {Karlin}},\ }\href {\doibase 10.1051/proc/201552001} {\bibfield  {journal}
  {\bibinfo  {journal} {ESAIM Proc.}\ }\textbf {\bibinfo {volume} {52}},\
  \bibinfo {pages} {1} (\bibinfo {year} {2015})}\BibitemShut {NoStop}%
\bibitem [{\citenamefont {B\"osch}\ \emph {et~al.}(2015)\citenamefont
  {B\"osch}, \citenamefont {Chikatamarla},\ and\ \citenamefont
  {Karlin}}]{BOSCH_PRE_92_2015}%
  \BibitemOpen
  \bibfield  {author} {\bibinfo {author} {\bibfnamefont {F.}~\bibnamefont
  {B\"osch}}, \bibinfo {author} {\bibfnamefont {S.~S.}\ \bibnamefont
  {Chikatamarla}}, \ and\ \bibinfo {author} {\bibfnamefont {I.~V.}\
  \bibnamefont {Karlin}},\ }\href {\doibase 10.1103/PhysRevE.92.043309}
  {\bibfield  {journal} {\bibinfo  {journal} {Phys. Rev. E}\ }\textbf {\bibinfo
  {volume} {92}},\ \bibinfo {pages} {043309} (\bibinfo {year}
  {2015})}\BibitemShut {NoStop}%
\bibitem [{\citenamefont {Dorschner}\ \emph {et~al.}(2016)\citenamefont
  {Dorschner}, \citenamefont {Frapolli}, \citenamefont {Chikatamarla},\ and\
  \citenamefont {Karlin}}]{DORSCHNER_PRE_94_2016}%
  \BibitemOpen
  \bibfield  {author} {\bibinfo {author} {\bibfnamefont {B.}~\bibnamefont
  {Dorschner}}, \bibinfo {author} {\bibfnamefont {N.}~\bibnamefont {Frapolli}},
  \bibinfo {author} {\bibfnamefont {S.~S.}\ \bibnamefont {Chikatamarla}}, \
  and\ \bibinfo {author} {\bibfnamefont {I.~V.}\ \bibnamefont {Karlin}},\
  }\href {\doibase 10.1103/PhysRevE.94.053311} {\bibfield  {journal} {\bibinfo
  {journal} {Phys. Rev. E}\ }\textbf {\bibinfo {volume} {94}},\ \bibinfo
  {pages} {053311} (\bibinfo {year} {2016})}\BibitemShut {NoStop}%
\bibitem [{\citenamefont {Dorschner}\ \emph {et~al.}(2018)\citenamefont
  {Dorschner}, \citenamefont {Chikatamarla},\ and\ \citenamefont
  {Karlin}}]{DORSCHNER_PRE_97_2018}%
  \BibitemOpen
  \bibfield  {author} {\bibinfo {author} {\bibfnamefont {B.}~\bibnamefont
  {Dorschner}}, \bibinfo {author} {\bibfnamefont {S.~S.}\ \bibnamefont
  {Chikatamarla}}, \ and\ \bibinfo {author} {\bibfnamefont {I.~V.}\
  \bibnamefont {Karlin}},\ }\href {\doibase 10.1103/PhysRevE.97.023305}
  {\bibfield  {journal} {\bibinfo  {journal} {Phys. Rev. E}\ }\textbf {\bibinfo
  {volume} {97}},\ \bibinfo {pages} {023305} (\bibinfo {year}
  {2018})}\BibitemShut {NoStop}%
\bibitem [{\citenamefont {Flint}\ and\ \citenamefont
  {Vahala}(2018)}]{FLINT_PRE_97_2018}%
  \BibitemOpen
  \bibfield  {author} {\bibinfo {author} {\bibfnamefont {C.}~\bibnamefont
  {Flint}}\ and\ \bibinfo {author} {\bibfnamefont {G.}~\bibnamefont {Vahala}},\
  }\href {\doibase 10.1103/PhysRevE.97.013302} {\bibfield  {journal} {\bibinfo
  {journal} {Phys. Rev. E}\ }\textbf {\bibinfo {volume} {97}},\ \bibinfo
  {pages} {013302} (\bibinfo {year} {2018})}\BibitemShut {NoStop}%
\bibitem [{\citenamefont {B\"{o}sch}\ \emph {et~al.}(2018)\citenamefont
  {B\"{o}sch}, \citenamefont {Dorschner},\ and\ \citenamefont
  {Karlin}}]{BOSCH_EPL_122_2018}%
  \BibitemOpen
  \bibfield  {author} {\bibinfo {author} {\bibfnamefont {F.}~\bibnamefont
  {B\"{o}sch}}, \bibinfo {author} {\bibfnamefont {B.}~\bibnamefont
  {Dorschner}}, \ and\ \bibinfo {author} {\bibfnamefont {I.}~\bibnamefont
  {Karlin}},\ }\href {http://stacks.iop.org/0295-5075/122/i=1/a=14002}
  {\bibfield  {journal} {\bibinfo  {journal} {Europhys. Lett.}\ }\textbf
  {\bibinfo {volume} {122}},\ \bibinfo {pages} {14002} (\bibinfo {year}
  {2018})}\BibitemShut {NoStop}%
\bibitem [{\citenamefont {Mattila}\ \emph {et~al.}(2015)\citenamefont
  {Mattila}, \citenamefont {Hegele},\ and\ \citenamefont
  {Philippi}}]{MATTILA_PRE_91_2015}%
  \BibitemOpen
  \bibfield  {author} {\bibinfo {author} {\bibfnamefont {K.~K.}\ \bibnamefont
  {Mattila}}, \bibinfo {author} {\bibfnamefont {L.~A.}\ \bibnamefont {Hegele}},
  \ and\ \bibinfo {author} {\bibfnamefont {P.~C.}\ \bibnamefont {Philippi}},\
  }\href {\doibase 10.1103/PhysRevE.91.063010} {\bibfield  {journal} {\bibinfo
  {journal} {Phys. Rev. E}\ }\textbf {\bibinfo {volume} {91}},\ \bibinfo
  {pages} {063010} (\bibinfo {year} {2015})}\BibitemShut {NoStop}%
\bibitem [{\citenamefont {Brownlee}\ \emph {et~al.}(2006)\citenamefont
  {Brownlee}, \citenamefont {Gorban},\ and\ \citenamefont
  {Levesley}}]{BROWNLEE_PRE_74_2006}%
  \BibitemOpen
  \bibfield  {author} {\bibinfo {author} {\bibfnamefont {R.~A.}\ \bibnamefont
  {Brownlee}}, \bibinfo {author} {\bibfnamefont {A.~N.}\ \bibnamefont
  {Gorban}}, \ and\ \bibinfo {author} {\bibfnamefont {J.}~\bibnamefont
  {Levesley}},\ }\href {\doibase 10.1103/PhysRevE.74.037703} {\bibfield
  {journal} {\bibinfo  {journal} {Phys. Rev. E}\ }\textbf {\bibinfo {volume}
  {74}},\ \bibinfo {pages} {037703} (\bibinfo {year} {2006})}\BibitemShut
  {NoStop}%
\bibitem [{\citenamefont {Brownlee}\ \emph {et~al.}(2007)\citenamefont
  {Brownlee}, \citenamefont {Gorban},\ and\ \citenamefont
  {Levesley}}]{BROWNLEE_PRE_75_2007}%
  \BibitemOpen
  \bibfield  {author} {\bibinfo {author} {\bibfnamefont {R.~A.}\ \bibnamefont
  {Brownlee}}, \bibinfo {author} {\bibfnamefont {A.~N.}\ \bibnamefont
  {Gorban}}, \ and\ \bibinfo {author} {\bibfnamefont {J.}~\bibnamefont
  {Levesley}},\ }\href {\doibase 10.1103/PhysRevE.75.036711} {\bibfield
  {journal} {\bibinfo  {journal} {Phys. Rev. E}\ }\textbf {\bibinfo {volume}
  {75}},\ \bibinfo {pages} {036711} (\bibinfo {year} {2007})}\BibitemShut
  {NoStop}%
\bibitem [{\citenamefont {Brownlee}\ \emph {et~al.}(2008)\citenamefont
  {Brownlee}, \citenamefont {Gorban},\ and\ \citenamefont
  {Levesley}}]{BROWNLEE_PA_387_2008}%
  \BibitemOpen
  \bibfield  {author} {\bibinfo {author} {\bibfnamefont {R.}~\bibnamefont
  {Brownlee}}, \bibinfo {author} {\bibfnamefont {A.}~\bibnamefont {Gorban}}, \
  and\ \bibinfo {author} {\bibfnamefont {J.}~\bibnamefont {Levesley}},\ }\href
  {\doibase 10.1016/j.physa.2007.09.031} {\bibfield  {journal} {\bibinfo
  {journal} {Physica A}\ }\textbf {\bibinfo {volume} {387}},\ \bibinfo {pages}
  {385 } (\bibinfo {year} {2008})}\BibitemShut {NoStop}%
\bibitem [{\citenamefont {Gorban}\ and\ \citenamefont
  {Packwood}(2014)}]{GORBAN_PA_414_2014}%
  \BibitemOpen
  \bibfield  {author} {\bibinfo {author} {\bibfnamefont {A.}~\bibnamefont
  {Gorban}}\ and\ \bibinfo {author} {\bibfnamefont {D.}~\bibnamefont
  {Packwood}},\ }\href {\doibase 10.1016/j.physa.2014.07.052} {\bibfield
  {journal} {\bibinfo  {journal} {Physica A}\ }\textbf {\bibinfo {volume}
  {414}},\ \bibinfo {pages} {285 } (\bibinfo {year} {2014})}\BibitemShut
  {NoStop}%
\bibitem [{\citenamefont {d'Humi\`{e}res}(1992)}]{DHUMIERE_PAA_159_1992}%
  \BibitemOpen
  \bibfield  {author} {\bibinfo {author} {\bibfnamefont {D.}~\bibnamefont
  {d'Humi\`{e}res}},\ }\href {\doibase 10.2514/5.9781600866319.0450.0458}
  {\bibfield  {journal} {\bibinfo  {journal} {Prog. Astronaut. Aeronaut.}\
  }\textbf {\bibinfo {volume} {159}},\ \bibinfo {pages} {450} (\bibinfo {year}
  {1992})}\BibitemShut {NoStop}%
\bibitem [{\citenamefont {Lallemand}\ and\ \citenamefont
  {Luo}(2000)}]{LALLEMAND_PRE_61_2000}%
  \BibitemOpen
  \bibfield  {author} {\bibinfo {author} {\bibfnamefont {P.}~\bibnamefont
  {Lallemand}}\ and\ \bibinfo {author} {\bibfnamefont {L.-S.}\ \bibnamefont
  {Luo}},\ }\href {\doibase 10.1103/PhysRevE.61.6546} {\bibfield  {journal}
  {\bibinfo  {journal} {Phys. Rev. E}\ }\textbf {\bibinfo {volume} {61}},\
  \bibinfo {pages} {6546} (\bibinfo {year} {2000})}\BibitemShut {NoStop}%
\bibitem [{\citenamefont {Ginzburg}\ \emph
  {et~al.}(2008{\natexlab{a}})\citenamefont {Ginzburg}, \citenamefont
  {Verhaeghe},\ and\ \citenamefont {d'Humi\`{e}res}}]{GINZBURG_CCP_3_2008a}%
  \BibitemOpen
  \bibfield  {author} {\bibinfo {author} {\bibfnamefont {I.}~\bibnamefont
  {Ginzburg}}, \bibinfo {author} {\bibfnamefont {F.}~\bibnamefont {Verhaeghe}},
  \ and\ \bibinfo {author} {\bibfnamefont {D.}~\bibnamefont {d'Humi\`{e}res}},\
  }\href
  {https://www.researchgate.net/publication/281975432_Two-relaxation-time_Lattice_Boltzmann_scheme_about_parametrization_velocity_pressure_and_mixed_boundary_conditions}
  {\bibfield  {journal} {\bibinfo  {journal} {Commun. Comput. Phys.}\ }\textbf
  {\bibinfo {volume} {3}},\ \bibinfo {pages} {427} (\bibinfo {year}
  {2008}{\natexlab{a}})}\BibitemShut {NoStop}%
\bibitem [{\citenamefont {Ginzburg}\ \emph
  {et~al.}(2008{\natexlab{b}})\citenamefont {Ginzburg}, \citenamefont
  {Verhaeghe},\ and\ \citenamefont {d'Humi\`{e}res}}]{GINZBURG_CCP_3_2008b}%
  \BibitemOpen
  \bibfield  {author} {\bibinfo {author} {\bibfnamefont {I.}~\bibnamefont
  {Ginzburg}}, \bibinfo {author} {\bibfnamefont {F.}~\bibnamefont {Verhaeghe}},
  \ and\ \bibinfo {author} {\bibfnamefont {D.}~\bibnamefont {d'Humi\`{e}res}},\
  }\href
  {https://www.researchgate.net/publication/281975403_Study_of_Simple_Hydrodynamic_Solutions_with_the_Two-Relaxation-Times_Lattice_Boltzmann_Scheme}
  {\bibfield  {journal} {\bibinfo  {journal} {Commun. Comput. Phys.}\ }\textbf
  {\bibinfo {volume} {3}},\ \bibinfo {pages} {519} (\bibinfo {year}
  {2008}{\natexlab{b}})}\BibitemShut {NoStop}%
\bibitem [{\citenamefont {d'Humi\`{e}res}(2002)}]{DHUMIERES_TRS_360_2002}%
  \BibitemOpen
  \bibfield  {author} {\bibinfo {author} {\bibfnamefont {D.}~\bibnamefont
  {d'Humi\`{e}res}},\ }\href {\doibase 10.1098/rsta.2001.0955} {\bibfield
  {journal} {\bibinfo  {journal} {Philos. Trans. R . Soc. London, Ser. A}\
  }\textbf {\bibinfo {volume} {360}},\ \bibinfo {pages} {437} (\bibinfo {year}
  {2002})}\BibitemShut {NoStop}%
\bibitem [{\citenamefont {d'Humi\`{e}res}\ and\ \citenamefont
  {Ginzburg}(2009)}]{DHUMIERE_CMA_58_2009}%
  \BibitemOpen
  \bibfield  {author} {\bibinfo {author} {\bibfnamefont {D.}~\bibnamefont
  {d'Humi\`{e}res}}\ and\ \bibinfo {author} {\bibfnamefont {I.}~\bibnamefont
  {Ginzburg}},\ }\href {\doibase 10.1016/j.camwa.2009.02.008} {\bibfield
  {journal} {\bibinfo  {journal} {Comput. Math. Appl.}\ }\textbf {\bibinfo
  {volume} {58}},\ \bibinfo {pages} {823 } (\bibinfo {year}
  {2009})}\BibitemShut {NoStop}%
\bibitem [{\citenamefont {Ginzburg}\ \emph {et~al.}(2010)\citenamefont
  {Ginzburg}, \citenamefont {d'Humi\`eres},\ and\ \citenamefont
  {Kuzmin}}]{GINZBURG_JSP_139_2010}%
  \BibitemOpen
  \bibfield  {author} {\bibinfo {author} {\bibfnamefont {I.}~\bibnamefont
  {Ginzburg}}, \bibinfo {author} {\bibfnamefont {D.}~\bibnamefont
  {d'Humi\`eres}}, \ and\ \bibinfo {author} {\bibfnamefont {A.}~\bibnamefont
  {Kuzmin}},\ }\href {\doibase 10.1007/s10955-010-9969-9} {\bibfield  {journal}
  {\bibinfo  {journal} {J. Stat. Phys.}\ }\textbf {\bibinfo {volume} {139}},\
  \bibinfo {pages} {1090} (\bibinfo {year} {2010})}\BibitemShut {NoStop}%
\bibitem [{\citenamefont {Kuzmin}\ \emph {et~al.}(2011)\citenamefont {Kuzmin},
  \citenamefont {Ginzburg},\ and\ \citenamefont
  {Mohamad}}]{KUZMIN_CMWA_61_2011}%
  \BibitemOpen
  \bibfield  {author} {\bibinfo {author} {\bibfnamefont {A.}~\bibnamefont
  {Kuzmin}}, \bibinfo {author} {\bibfnamefont {I.}~\bibnamefont {Ginzburg}}, \
  and\ \bibinfo {author} {\bibfnamefont {A.}~\bibnamefont {Mohamad}},\ }\href
  {\doibase 10.1016/j.camwa.2010.07.036} {\bibfield  {journal} {\bibinfo
  {journal} {Comput. Math. Appl.}\ }\textbf {\bibinfo {volume} {61}},\ \bibinfo
  {pages} {3417 } (\bibinfo {year} {2011})}\BibitemShut {NoStop}%
\bibitem [{\citenamefont {Hao}\ and\ \citenamefont
  {Cheng}(2009)}]{HAO_JPS_186_2009}%
  \BibitemOpen
  \bibfield  {author} {\bibinfo {author} {\bibfnamefont {L.}~\bibnamefont
  {Hao}}\ and\ \bibinfo {author} {\bibfnamefont {P.}~\bibnamefont {Cheng}},\
  }\href {\doibase 10.1016/j.jpowsour.2008.09.086} {\bibfield  {journal}
  {\bibinfo  {journal} {J. Power Sources}\ }\textbf {\bibinfo {volume} {186}},\
  \bibinfo {pages} {104 } (\bibinfo {year} {2009})}\BibitemShut {NoStop}%
\bibitem [{\citenamefont {Xu}\ and\ \citenamefont
  {Sagaut}(2011)}]{XU_JCP_230_2011}%
  \BibitemOpen
  \bibfield  {author} {\bibinfo {author} {\bibfnamefont {H.}~\bibnamefont
  {Xu}}\ and\ \bibinfo {author} {\bibfnamefont {P.}~\bibnamefont {Sagaut}},\
  }\href {\doibase 10.1016/j.jcp.2011.03.040} {\bibfield  {journal} {\bibinfo
  {journal} {J. Comput. Phys.}\ }\textbf {\bibinfo {volume} {230}},\ \bibinfo
  {pages} {5353 } (\bibinfo {year} {2011})}\BibitemShut {NoStop}%
\bibitem [{\citenamefont {Wang}\ \emph {et~al.}(2013)\citenamefont {Wang},
  \citenamefont {Wang}, \citenamefont {Lallemand},\ and\ \citenamefont
  {Luo}}]{WANG_CMA_65_2013}%
  \BibitemOpen
  \bibfield  {author} {\bibinfo {author} {\bibfnamefont {J.}~\bibnamefont
  {Wang}}, \bibinfo {author} {\bibfnamefont {D.}~\bibnamefont {Wang}}, \bibinfo
  {author} {\bibfnamefont {P.}~\bibnamefont {Lallemand}}, \ and\ \bibinfo
  {author} {\bibfnamefont {L.-S.}\ \bibnamefont {Luo}},\ }\href {\doibase
  10.1016/j.camwa.2012.07.001} {\bibfield  {journal} {\bibinfo  {journal}
  {Comput. Math. Appl.}\ }\textbf {\bibinfo {volume} {65}},\ \bibinfo {pages}
  {262 } (\bibinfo {year} {2013})}\BibitemShut {NoStop}%
\bibitem [{\citenamefont {Servan-Camas}\ and\ \citenamefont
  {Tsai}(2009)}]{SERVANCAMAS_AWR_32_2009}%
  \BibitemOpen
  \bibfield  {author} {\bibinfo {author} {\bibfnamefont {B.}~\bibnamefont
  {Servan-Camas}}\ and\ \bibinfo {author} {\bibfnamefont {F.~T.-C.}\
  \bibnamefont {Tsai}},\ }\href {\doibase 10.1016/j.advwatres.2009.02.001}
  {\bibfield  {journal} {\bibinfo  {journal} {Adv. Water Resour.}\ }\textbf
  {\bibinfo {volume} {32}},\ \bibinfo {pages} {620 } (\bibinfo {year}
  {2009})}\BibitemShut {NoStop}%
\bibitem [{\citenamefont {Silva}\ and\ \citenamefont
  {Semiao}(2017)}]{SILVA_PRE_96_2017}%
  \BibitemOpen
  \bibfield  {author} {\bibinfo {author} {\bibfnamefont {G.}~\bibnamefont
  {Silva}}\ and\ \bibinfo {author} {\bibfnamefont {V.}~\bibnamefont {Semiao}},\
  }\href {\doibase 10.1103/PhysRevE.96.013311} {\bibfield  {journal} {\bibinfo
  {journal} {Phys. Rev. E}\ }\textbf {\bibinfo {volume} {96}},\ \bibinfo
  {pages} {013311} (\bibinfo {year} {2017})}\BibitemShut {NoStop}%
\bibitem [{\citenamefont {Geier}\ \emph {et~al.}(2006)\citenamefont {Geier},
  \citenamefont {Greiner},\ and\ \citenamefont {Korvink}}]{GEIER_PRE_73_2006}%
  \BibitemOpen
  \bibfield  {author} {\bibinfo {author} {\bibfnamefont {M.}~\bibnamefont
  {Geier}}, \bibinfo {author} {\bibfnamefont {A.}~\bibnamefont {Greiner}}, \
  and\ \bibinfo {author} {\bibfnamefont {J.~G.}\ \bibnamefont {Korvink}},\
  }\href {\doibase 10.1103/PhysRevE.73.066705} {\bibfield  {journal} {\bibinfo
  {journal} {Phys. Rev. E}\ }\textbf {\bibinfo {volume} {73}},\ \bibinfo
  {pages} {066705} (\bibinfo {year} {2006})}\BibitemShut {NoStop}%
\bibitem [{\citenamefont {Geier}\ \emph {et~al.}(2009)\citenamefont {Geier},
  \citenamefont {Greiner},\ and\ \citenamefont
  {Korvink}}]{GEIER_EPJST_171_2009}%
  \BibitemOpen
  \bibfield  {author} {\bibinfo {author} {\bibfnamefont {M.}~\bibnamefont
  {Geier}}, \bibinfo {author} {\bibfnamefont {A.}~\bibnamefont {Greiner}}, \
  and\ \bibinfo {author} {\bibfnamefont {J.~G.}\ \bibnamefont {Korvink}},\
  }\href {\doibase 10.1140/epjst/e2009-01011-1} {\bibfield  {journal} {\bibinfo
   {journal} {Eur. Phys. J. Spec. Top.}\ }\textbf {\bibinfo {volume} {171}},\
  \bibinfo {pages} {55} (\bibinfo {year} {2009})}\BibitemShut {NoStop}%
\bibitem [{\citenamefont {Geller}\ \emph {et~al.}(2013)\citenamefont {Geller},
  \citenamefont {Uphoff},\ and\ \citenamefont {Krafczyk}}]{GELLER_CMA_65_2013}%
  \BibitemOpen
  \bibfield  {author} {\bibinfo {author} {\bibfnamefont {S.}~\bibnamefont
  {Geller}}, \bibinfo {author} {\bibfnamefont {S.}~\bibnamefont {Uphoff}}, \
  and\ \bibinfo {author} {\bibfnamefont {M.}~\bibnamefont {Krafczyk}},\ }\href
  {\doibase 10.1016/j.camwa.2013.04.013} {\bibfield  {journal} {\bibinfo
  {journal} {Comput. Math. Appl.}\ }\textbf {\bibinfo {volume} {65}},\ \bibinfo
  {pages} {1956 } (\bibinfo {year} {2013})}\BibitemShut {NoStop}%
\bibitem [{\citenamefont {Lycett-Brown}\ and\ \citenamefont
  {Luo}(2014)}]{LYCETTBROWN_CMWA_67_2014}%
  \BibitemOpen
  \bibfield  {author} {\bibinfo {author} {\bibfnamefont {D.}~\bibnamefont
  {Lycett-Brown}}\ and\ \bibinfo {author} {\bibfnamefont {K.~H.}\ \bibnamefont
  {Luo}},\ }\href {\doibase 10.1016/j.camwa.2013.08.033} {\bibfield  {journal}
  {\bibinfo  {journal} {Comput. Math. Appl.}\ }\textbf {\bibinfo {volume}
  {67}},\ \bibinfo {pages} {350 } (\bibinfo {year} {2014})}\BibitemShut
  {NoStop}%
\bibitem [{\citenamefont {Dubois}\ \emph
  {et~al.}(2015{\natexlab{a}})\citenamefont {Dubois}, \citenamefont
  {F\'{e}vrier},\ and\ \citenamefont {Graille}}]{DUBOIS_CCP_17_2015}%
  \BibitemOpen
  \bibfield  {author} {\bibinfo {author} {\bibfnamefont {F.}~\bibnamefont
  {Dubois}}, \bibinfo {author} {\bibfnamefont {T.}~\bibnamefont {F\'{e}vrier}},
  \ and\ \bibinfo {author} {\bibfnamefont {B.}~\bibnamefont {Graille}},\ }\href
  {\doibase 10.4208/cicp.2014.m394} {\bibfield  {journal} {\bibinfo  {journal}
  {Commun. Comput. Phys.}\ }\textbf {\bibinfo {volume} {17}},\ \bibinfo {pages}
  {1088–1112} (\bibinfo {year} {2015}{\natexlab{a}})}\BibitemShut {NoStop}%
\bibitem [{\citenamefont {Ning}\ \emph {et~al.}(2016)\citenamefont {Ning},
  \citenamefont {Premnath},\ and\ \citenamefont {Patil}}]{NING_IJNMF_82_2016}%
  \BibitemOpen
  \bibfield  {author} {\bibinfo {author} {\bibfnamefont {Y.}~\bibnamefont
  {Ning}}, \bibinfo {author} {\bibfnamefont {K.~N.}\ \bibnamefont {Premnath}},
  \ and\ \bibinfo {author} {\bibfnamefont {D.~V.}\ \bibnamefont {Patil}},\
  }\href {\doibase 10.1002/fld.4208} {\bibfield  {journal} {\bibinfo  {journal}
  {Int. J. Numer. Meth. Fluids}\ }\textbf {\bibinfo {volume} {82}},\ \bibinfo
  {pages} {59} (\bibinfo {year} {2016})},\ \bibinfo {note}
  {fld.4208}\BibitemShut {NoStop}%
\bibitem [{\citenamefont {Fei}\ and\ \citenamefont
  {Luo}(2017)}]{FEI_PRE_96_2017}%
  \BibitemOpen
  \bibfield  {author} {\bibinfo {author} {\bibfnamefont {L.}~\bibnamefont
  {Fei}}\ and\ \bibinfo {author} {\bibfnamefont {K.~H.}\ \bibnamefont {Luo}},\
  }\href {\doibase 10.1103/PhysRevE.96.053307} {\bibfield  {journal} {\bibinfo
  {journal} {Phys. Rev. E}\ }\textbf {\bibinfo {volume} {96}},\ \bibinfo
  {pages} {053307} (\bibinfo {year} {2017})}\BibitemShut {NoStop}%
\bibitem [{\citenamefont {De~Rosis}(2017{\natexlab{a}})}]{DEROSIS_PRE_95_2017}%
  \BibitemOpen
  \bibfield  {author} {\bibinfo {author} {\bibfnamefont {A.}~\bibnamefont
  {De~Rosis}},\ }\href {\doibase 10.1103/PhysRevE.95.013310} {\bibfield
  {journal} {\bibinfo  {journal} {Phys. Rev. E}\ }\textbf {\bibinfo {volume}
  {95}},\ \bibinfo {pages} {013310} (\bibinfo {year}
  {2017}{\natexlab{a}})}\BibitemShut {NoStop}%
\bibitem [{\citenamefont {Fei}\ \emph {et~al.}(2018)\citenamefont {Fei},
  \citenamefont {Luo},\ and\ \citenamefont {Li}}]{FEI_PRE_97_2018}%
  \BibitemOpen
  \bibfield  {author} {\bibinfo {author} {\bibfnamefont {L.}~\bibnamefont
  {Fei}}, \bibinfo {author} {\bibfnamefont {K.~H.}\ \bibnamefont {Luo}}, \ and\
  \bibinfo {author} {\bibfnamefont {Q.}~\bibnamefont {Li}},\ }\href {\doibase
  10.1103/PhysRevE.97.053309} {\bibfield  {journal} {\bibinfo  {journal} {Phys.
  Rev. E}\ }\textbf {\bibinfo {volume} {97}},\ \bibinfo {pages} {053309}
  (\bibinfo {year} {2018})}\BibitemShut {NoStop}%
\bibitem [{\citenamefont {De~Rosis}\ and\ \citenamefont
  {Luo}(2019)}]{DEROSIS_PRE_99_2019}%
  \BibitemOpen
  \bibfield  {author} {\bibinfo {author} {\bibfnamefont {A.}~\bibnamefont
  {De~Rosis}}\ and\ \bibinfo {author} {\bibfnamefont {K.~H.}\ \bibnamefont
  {Luo}},\ }\href {\doibase 10.1103/PhysRevE.99.013301} {\bibfield  {journal}
  {\bibinfo  {journal} {Phys. Rev. E}\ }\textbf {\bibinfo {volume} {99}},\
  \bibinfo {pages} {013301} (\bibinfo {year} {2019})}\BibitemShut {NoStop}%
\bibitem [{\citenamefont {De~Rosis}\ \emph {et~al.}(2019)\citenamefont
  {De~Rosis}, \citenamefont {Huang},\ and\ \citenamefont
  {Coreixas}}]{DEROSIS_PRE_Submitted_2019}%
  \BibitemOpen
  \bibfield  {author} {\bibinfo {author} {\bibfnamefont {A.}~\bibnamefont
  {De~Rosis}}, \bibinfo {author} {\bibfnamefont {R.}~\bibnamefont {Huang}}, \
  and\ \bibinfo {author} {\bibfnamefont {C.}~\bibnamefont {Coreixas}},\
  }\href@noop {} {\bibfield  {journal} {\bibinfo  {journal} {submitted to Phys.
  Rev. E}\ } (\bibinfo {year} {2019})}\BibitemShut {NoStop}%
\bibitem [{\citenamefont {Lycett-Brown}\ and\ \citenamefont
  {Luo}(2016)}]{LYCETT_BROWN_PRE_94_2016}%
  \BibitemOpen
  \bibfield  {author} {\bibinfo {author} {\bibfnamefont {D.}~\bibnamefont
  {Lycett-Brown}}\ and\ \bibinfo {author} {\bibfnamefont {K.~H.}\ \bibnamefont
  {Luo}},\ }\href {\doibase 10.1103/PhysRevE.94.053313} {\bibfield  {journal}
  {\bibinfo  {journal} {Phys. Rev. E}\ }\textbf {\bibinfo {volume} {94}},\
  \bibinfo {pages} {053313} (\bibinfo {year} {2016})}\BibitemShut {NoStop}%
\bibitem [{\citenamefont {Dubois}\ \emph
  {et~al.}(2015{\natexlab{b}})\citenamefont {Dubois}, \citenamefont
  {F\'{e}vrier},\ and\ \citenamefont {Graille}}]{DUBOIS_CRM_343_2015}%
  \BibitemOpen
  \bibfield  {author} {\bibinfo {author} {\bibfnamefont {F.}~\bibnamefont
  {Dubois}}, \bibinfo {author} {\bibfnamefont {T.}~\bibnamefont {F\'{e}vrier}},
  \ and\ \bibinfo {author} {\bibfnamefont {B.}~\bibnamefont {Graille}},\ }\href
  {\doibase 10.1016/j.crme.2015.07.010} {\bibfield  {journal} {\bibinfo
  {journal} {C. R. M\'{e}canique}\ }\textbf {\bibinfo {volume} {343}},\
  \bibinfo {pages} {599 } (\bibinfo {year} {2015}{\natexlab{b}})}\BibitemShut
  {NoStop}%
\bibitem [{\citenamefont
  {De~Rosis}(2017{\natexlab{b}})}]{DEROSIS_CMAME_319_2017}%
  \BibitemOpen
  \bibfield  {author} {\bibinfo {author} {\bibfnamefont {A.}~\bibnamefont
  {De~Rosis}},\ }\href {\doibase 10.1016/j.cma.2017.03.001} {\bibfield
  {journal} {\bibinfo  {journal} {Comput. Methods in Appl. Mech. Eng.}\
  }\textbf {\bibinfo {volume} {319}},\ \bibinfo {pages} {379 } (\bibinfo {year}
  {2017}{\natexlab{b}})}\BibitemShut {NoStop}%
\bibitem [{\citenamefont {De~Rosis}\ \emph {et~al.}(2018)\citenamefont
  {De~Rosis}, \citenamefont {L\'{e}v\^{e}que},\ and\ \citenamefont
  {Chahine}}]{DEROSIS_JT_19_2018}%
  \BibitemOpen
  \bibfield  {author} {\bibinfo {author} {\bibfnamefont {A.}~\bibnamefont
  {De~Rosis}}, \bibinfo {author} {\bibfnamefont {E.}~\bibnamefont
  {L\'{e}v\^{e}que}}, \ and\ \bibinfo {author} {\bibfnamefont {R.}~\bibnamefont
  {Chahine}},\ }\href {\doibase 10.1080/14685248.2018.1461875} {\bibfield
  {journal} {\bibinfo  {journal} {J. Turb.}\ }\textbf {\bibinfo {volume}
  {19}},\ \bibinfo {pages} {446} (\bibinfo {year} {2018})}\BibitemShut
  {NoStop}%
\bibitem [{\citenamefont {Cook}(1951)}]{COOK_Bio_38_1951}%
  \BibitemOpen
  \bibfield  {author} {\bibinfo {author} {\bibfnamefont {M.}~\bibnamefont
  {Cook}},\ }\href {\doibase 10.2307/2332327} {\bibfield  {journal} {\bibinfo
  {journal} {Biometrika}\ }\textbf {\bibinfo {volume} {38}},\ \bibinfo {pages}
  {179} (\bibinfo {year} {1951})}\BibitemShut {NoStop}%
\bibitem [{\citenamefont {Seeger}\ and\ \citenamefont
  {Hoffmann}(2000)}]{SEEGER_CMT_12_2000}%
  \BibitemOpen
  \bibfield  {author} {\bibinfo {author} {\bibfnamefont {S.}~\bibnamefont
  {Seeger}}\ and\ \bibinfo {author} {\bibfnamefont {H.}~\bibnamefont
  {Hoffmann}},\ }\href {\doibase 10.1007/s001610050145} {\bibfield  {journal}
  {\bibinfo  {journal} {Contin. Mech. Thermodyn.}\ }\textbf {\bibinfo {volume}
  {12}},\ \bibinfo {pages} {403} (\bibinfo {year} {2000})}\BibitemShut
  {NoStop}%
\bibitem [{\citenamefont {Seeger}\ and\ \citenamefont
  {Hoffmann}(2002)}]{SEEGER_CMT_14_2002}%
  \BibitemOpen
  \bibfield  {author} {\bibinfo {author} {\bibfnamefont {S.}~\bibnamefont
  {Seeger}}\ and\ \bibinfo {author} {\bibfnamefont {K.~H.}\ \bibnamefont
  {Hoffmann}},\ }\href {\doibase 10.1007/s001610100067} {\bibfield  {journal}
  {\bibinfo  {journal} {Contin. Mech. Thermodyn.}\ }\textbf {\bibinfo {volume}
  {14}},\ \bibinfo {pages} {321} (\bibinfo {year} {2002})}\BibitemShut
  {NoStop}%
\bibitem [{\citenamefont {Seeger}(2003)}]{SEEGER_PhD_2003}%
  \BibitemOpen
  \bibfield  {author} {\bibinfo {author} {\bibfnamefont {S.}~\bibnamefont
  {Seeger}},\ }\emph {\bibinfo {title} {The Cumulant Method}},\ \href
  {http://nbn-resolving.de/urn:nbn:de:swb:ch1-200301201} {Ph.D. thesis},\
  \bibinfo  {school} {{TU}-{C}hemnitz} (\bibinfo {year} {2003})\BibitemShut
  {NoStop}%
\bibitem [{\citenamefont {Geier}\ \emph {et~al.}(2015)\citenamefont {Geier},
  \citenamefont {Sch\"{o}nherr}, \citenamefont {Pasquali},\ and\ \citenamefont
  {Krafczyk}}]{GEIER_CMA_70_2015}%
  \BibitemOpen
  \bibfield  {author} {\bibinfo {author} {\bibfnamefont {M.}~\bibnamefont
  {Geier}}, \bibinfo {author} {\bibfnamefont {M.}~\bibnamefont
  {Sch\"{o}nherr}}, \bibinfo {author} {\bibfnamefont {A.}~\bibnamefont
  {Pasquali}}, \ and\ \bibinfo {author} {\bibfnamefont {M.}~\bibnamefont
  {Krafczyk}},\ }\href {\doibase 10.1016/j.camwa.2015.05.001} {\bibfield
  {journal} {\bibinfo  {journal} {Comput. Math. Appl.}\ }\textbf {\bibinfo
  {volume} {70}},\ \bibinfo {pages} {507 } (\bibinfo {year}
  {2015})}\BibitemShut {NoStop}%
\bibitem [{\citenamefont {Junk}\ \emph {et~al.}(2005)\citenamefont {Junk},
  \citenamefont {Klar},\ and\ \citenamefont {Luo}}]{JUNK_JCP_210_2005}%
  \BibitemOpen
  \bibfield  {author} {\bibinfo {author} {\bibfnamefont {M.}~\bibnamefont
  {Junk}}, \bibinfo {author} {\bibfnamefont {A.}~\bibnamefont {Klar}}, \ and\
  \bibinfo {author} {\bibfnamefont {L.-S.}\ \bibnamefont {Luo}},\ }\href
  {\doibase 10.1016/j.jcp.2005.05.003} {\bibfield  {journal} {\bibinfo
  {journal} {J. Comput. Phys.}\ }\textbf {\bibinfo {volume} {210}},\ \bibinfo
  {pages} {676 } (\bibinfo {year} {2005})}\BibitemShut {NoStop}%
\bibitem [{\citenamefont {Kr\"{u}ger}\ \emph {et~al.}(2017)\citenamefont
  {Kr\"{u}ger}, \citenamefont {Kusumaatmaja}, \citenamefont {Kuzmin},
  \citenamefont {Shardt}, \citenamefont {Silva},\ and\ \citenamefont
  {Viggen}}]{KRUGER_Book_2017}%
  \BibitemOpen
  \bibfield  {author} {\bibinfo {author} {\bibfnamefont {T.}~\bibnamefont
  {Kr\"{u}ger}}, \bibinfo {author} {\bibfnamefont {H.}~\bibnamefont
  {Kusumaatmaja}}, \bibinfo {author} {\bibfnamefont {A.}~\bibnamefont
  {Kuzmin}}, \bibinfo {author} {\bibfnamefont {O.}~\bibnamefont {Shardt}},
  \bibinfo {author} {\bibfnamefont {G.}~\bibnamefont {Silva}}, \ and\ \bibinfo
  {author} {\bibfnamefont {E.~M.}\ \bibnamefont {Viggen}},\ }\href {\doibase
  10.1007/978-3-319-44649-3} {\emph {\bibinfo {title} {The Lattice {B}oltzmann
  Method: Principles and Practice}}}\ (\bibinfo  {publisher} {Springer
  International Publishing},\ \bibinfo {year} {2017})\BibitemShut {NoStop}%
\bibitem [{\citenamefont {Geier}\ \emph
  {et~al.}(2017{\natexlab{a}})\citenamefont {Geier}, \citenamefont {Pasquali},\
  and\ \citenamefont {Sch\"{o}nherr}}]{GEIER_JCP_348_2017a}%
  \BibitemOpen
  \bibfield  {author} {\bibinfo {author} {\bibfnamefont {M.}~\bibnamefont
  {Geier}}, \bibinfo {author} {\bibfnamefont {A.}~\bibnamefont {Pasquali}}, \
  and\ \bibinfo {author} {\bibfnamefont {M.}~\bibnamefont {Sch\"{o}nherr}},\
  }\href {\doibase 10.1016/j.jcp.2017.05.040} {\bibfield  {journal} {\bibinfo
  {journal} {J. Comput. Phys.}\ }\textbf {\bibinfo {volume} {348}},\ \bibinfo
  {pages} {862 } (\bibinfo {year} {2017}{\natexlab{a}})}\BibitemShut {NoStop}%
\bibitem [{\citenamefont {Geier}\ \emph
  {et~al.}(2017{\natexlab{b}})\citenamefont {Geier}, \citenamefont {Pasquali},\
  and\ \citenamefont {Sch\"{o}nherr}}]{GEIER_JCP_348_2017b}%
  \BibitemOpen
  \bibfield  {author} {\bibinfo {author} {\bibfnamefont {M.}~\bibnamefont
  {Geier}}, \bibinfo {author} {\bibfnamefont {A.}~\bibnamefont {Pasquali}}, \
  and\ \bibinfo {author} {\bibfnamefont {M.}~\bibnamefont {Sch\"{o}nherr}},\
  }\href {\doibase 10.1016/j.jcp.2017.07.004} {\bibfield  {journal} {\bibinfo
  {journal} {J. Comput. Phys.}\ }\textbf {\bibinfo {volume} {348}},\ \bibinfo
  {pages} {889 } (\bibinfo {year} {2017}{\natexlab{b}})}\BibitemShut {NoStop}%
\bibitem [{\citenamefont {Geier}\ and\ \citenamefont
  {Pasquali}(2018)}]{GEIER_CF_166_2018}%
  \BibitemOpen
  \bibfield  {author} {\bibinfo {author} {\bibfnamefont {M.}~\bibnamefont
  {Geier}}\ and\ \bibinfo {author} {\bibfnamefont {A.}~\bibnamefont
  {Pasquali}},\ }\href {\doibase 10.1016/j.compfluid.2018.01.015} {\bibfield
  {journal} {\bibinfo  {journal} {Comput. Fluids}\ }\textbf {\bibinfo {volume}
  {166}},\ \bibinfo {pages} {139 } (\bibinfo {year} {2018})}\BibitemShut
  {NoStop}%
\bibitem [{\citenamefont {Ladd}\ and\ \citenamefont
  {Verberg}(2001)}]{LADD_JSP_104_2001}%
  \BibitemOpen
  \bibfield  {author} {\bibinfo {author} {\bibfnamefont {A.~J.~C.}\
  \bibnamefont {Ladd}}\ and\ \bibinfo {author} {\bibfnamefont {R.}~\bibnamefont
  {Verberg}},\ }\href {\doibase 10.1023/A:1010414013942} {\bibfield  {journal}
  {\bibinfo  {journal} {J. Stat. Phys.}\ }\textbf {\bibinfo {volume} {104}},\
  \bibinfo {pages} {1191} (\bibinfo {year} {2001})}\BibitemShut {NoStop}%
\bibitem [{\citenamefont {Latt}\ and\ \citenamefont
  {Chopard}(2005)}]{LATT_ARXIV_2005}%
  \BibitemOpen
  \bibfield  {author} {\bibinfo {author} {\bibfnamefont {J.}~\bibnamefont
  {Latt}}\ and\ \bibinfo {author} {\bibfnamefont {B.}~\bibnamefont {Chopard}},\
  }\href@noop {} {\bibfield  {journal} {\bibinfo  {journal} {arXiv preprint
  physics/0506157}\ } (\bibinfo {year} {2005})},\ \Eprint
  {http://arxiv.org/abs/physics/0506157} {physics/0506157} \BibitemShut
  {NoStop}%
\bibitem [{\citenamefont {Latt}\ and\ \citenamefont
  {Chopard}(2006)}]{LATT_MCS_72_2006a}%
  \BibitemOpen
  \bibfield  {author} {\bibinfo {author} {\bibfnamefont {J.}~\bibnamefont
  {Latt}}\ and\ \bibinfo {author} {\bibfnamefont {B.}~\bibnamefont {Chopard}},\
  }\href {\doibase 10.1016/j.matcom.2006.05.017} {\bibfield  {journal}
  {\bibinfo  {journal} {Math. Comput. Simul.}\ }\textbf {\bibinfo {volume}
  {72}},\ \bibinfo {pages} {165} (\bibinfo {year} {2006})}\BibitemShut
  {NoStop}%
\bibitem [{\citenamefont {Chen}\ \emph {et~al.}(2006)\citenamefont {Chen},
  \citenamefont {Zhang}, \citenamefont {Staroselsky},\ and\ \citenamefont
  {Jhon}}]{CHEN_PA_362_2006}%
  \BibitemOpen
  \bibfield  {author} {\bibinfo {author} {\bibfnamefont {H.}~\bibnamefont
  {Chen}}, \bibinfo {author} {\bibfnamefont {R.}~\bibnamefont {Zhang}},
  \bibinfo {author} {\bibfnamefont {I.}~\bibnamefont {Staroselsky}}, \ and\
  \bibinfo {author} {\bibfnamefont {M.}~\bibnamefont {Jhon}},\ }\href {\doibase
  10.1016/j.physa.2005.09.008} {\bibfield  {journal} {\bibinfo  {journal}
  {Physica A}\ }\textbf {\bibinfo {volume} {362}},\ \bibinfo {pages} {125}
  (\bibinfo {year} {2006})}\BibitemShut {NoStop}%
\bibitem [{\citenamefont {Zhang}\ \emph {et~al.}(2006)\citenamefont {Zhang},
  \citenamefont {Shan},\ and\ \citenamefont {Chen}}]{ZHANG_PRE_74_2006}%
  \BibitemOpen
  \bibfield  {author} {\bibinfo {author} {\bibfnamefont {R.}~\bibnamefont
  {Zhang}}, \bibinfo {author} {\bibfnamefont {X.}~\bibnamefont {Shan}}, \ and\
  \bibinfo {author} {\bibfnamefont {H.}~\bibnamefont {Chen}},\ }\href {\doibase
  10.1103/PhysRevE.74.046703} {\bibfield  {journal} {\bibinfo  {journal} {Phys.
  Rev. E}\ }\textbf {\bibinfo {volume} {74}},\ \bibinfo {pages} {046703}
  (\bibinfo {year} {2006})}\BibitemShut {NoStop}%
\bibitem [{\citenamefont {{Malaspinas}}(2015)}]{MALASPINAS_ARXIV_2015}%
  \BibitemOpen
  \bibfield  {author} {\bibinfo {author} {\bibfnamefont {O.}~\bibnamefont
  {{Malaspinas}}},\ }\href {http://arxiv.org/pdf/1505.06900v1.pdf} {\bibfield
  {journal} {\bibinfo  {journal} {ArXiv e-prints}\ } (\bibinfo {year}
  {2015})},\ \Eprint {http://arxiv.org/abs/1505.06900} {arXiv:1505.06900
  [physics.flu-dyn]} \BibitemShut {NoStop}%
\bibitem [{\citenamefont {Coreixas}\ \emph {et~al.}(2017)\citenamefont
  {Coreixas}, \citenamefont {Wissocq}, \citenamefont {Puigt}, \citenamefont
  {Boussuge},\ and\ \citenamefont {Sagaut}}]{COREIXAS_PRE_96_2017}%
  \BibitemOpen
  \bibfield  {author} {\bibinfo {author} {\bibfnamefont {C.}~\bibnamefont
  {Coreixas}}, \bibinfo {author} {\bibfnamefont {G.}~\bibnamefont {Wissocq}},
  \bibinfo {author} {\bibfnamefont {G.}~\bibnamefont {Puigt}}, \bibinfo
  {author} {\bibfnamefont {J.-F.}\ \bibnamefont {Boussuge}}, \ and\ \bibinfo
  {author} {\bibfnamefont {P.}~\bibnamefont {Sagaut}},\ }\href {\doibase
  10.1103/PhysRevE.96.033306} {\bibfield  {journal} {\bibinfo  {journal} {Phys.
  Rev. E}\ }\textbf {\bibinfo {volume} {96}},\ \bibinfo {pages} {033306}
  (\bibinfo {year} {2017})}\BibitemShut {NoStop}%
\bibitem [{\citenamefont {Coreixas}(2018)}]{COREIXAS_PhD_2018}%
  \BibitemOpen
  \bibfield  {author} {\bibinfo {author} {\bibfnamefont {C.}~\bibnamefont
  {Coreixas}},\ }\emph {\bibinfo {title} {High-order extension of the recursive
  regularized lattice Boltzmann method}},\ \href
  {http://oatao.univ-toulouse.fr/19861/} {Ph.D. thesis},\ \bibinfo  {school}
  {INP Toulouse} (\bibinfo {year} {2018})\BibitemShut {NoStop}%
\bibitem [{\citenamefont {Brogi}\ \emph {et~al.}(2017)\citenamefont {Brogi},
  \citenamefont {Malaspinas}, \citenamefont {Chopard},\ and\ \citenamefont
  {Bonadonna}}]{BROGI_JASA_142_2017}%
  \BibitemOpen
  \bibfield  {author} {\bibinfo {author} {\bibfnamefont {F.}~\bibnamefont
  {Brogi}}, \bibinfo {author} {\bibfnamefont {O.}~\bibnamefont {Malaspinas}},
  \bibinfo {author} {\bibfnamefont {B.}~\bibnamefont {Chopard}}, \ and\
  \bibinfo {author} {\bibfnamefont {C.}~\bibnamefont {Bonadonna}},\ }\href
  {\doibase 10.1121/1.5006900} {\bibfield  {journal} {\bibinfo  {journal} {J.
  Acoust. Soc. Am}\ }\textbf {\bibinfo {volume} {142}},\ \bibinfo {pages}
  {2332} (\bibinfo {year} {2017})}\BibitemShut {NoStop}%
\bibitem [{\citenamefont {Mattila}\ \emph {et~al.}(2017)\citenamefont
  {Mattila}, \citenamefont {Philippi},\ and\ \citenamefont
  {Hegele~Jr.}}]{MATTILA_PF_29_2017}%
  \BibitemOpen
  \bibfield  {author} {\bibinfo {author} {\bibfnamefont {K.~K.}\ \bibnamefont
  {Mattila}}, \bibinfo {author} {\bibfnamefont {P.~C.}\ \bibnamefont
  {Philippi}}, \ and\ \bibinfo {author} {\bibfnamefont {L.~A.}\ \bibnamefont
  {Hegele~Jr.}},\ }\href {\doibase 10.1063/1.4981227} {\bibfield  {journal}
  {\bibinfo  {journal} {Phys. Fluids}\ }\textbf {\bibinfo {volume} {29}},\
  \bibinfo {pages} {046103} (\bibinfo {year} {2017})}\BibitemShut {NoStop}%
\bibitem [{\citenamefont {Niu}\ \emph {et~al.}(2007)\citenamefont {Niu},
  \citenamefont {Hyodo}, \citenamefont {Munekata},\ and\ \citenamefont
  {Suga}}]{NIU_PRE_76_2007}%
  \BibitemOpen
  \bibfield  {author} {\bibinfo {author} {\bibfnamefont {X.-D.}\ \bibnamefont
  {Niu}}, \bibinfo {author} {\bibfnamefont {S.-A.}\ \bibnamefont {Hyodo}},
  \bibinfo {author} {\bibfnamefont {T.}~\bibnamefont {Munekata}}, \ and\
  \bibinfo {author} {\bibfnamefont {K.}~\bibnamefont {Suga}},\ }\href {\doibase
  10.1103/PhysRevE.76.036711} {\bibfield  {journal} {\bibinfo  {journal} {Phys.
  Rev. E}\ }\textbf {\bibinfo {volume} {76}},\ \bibinfo {pages} {036711}
  (\bibinfo {year} {2007})}\BibitemShut {NoStop}%
\bibitem [{\citenamefont {Viggen}(2013)}]{VIGGEN_PRE_87_2013}%
  \BibitemOpen
  \bibfield  {author} {\bibinfo {author} {\bibfnamefont {E.~M.}\ \bibnamefont
  {Viggen}},\ }\href {\doibase 10.1103/PhysRevE.87.023306} {\bibfield
  {journal} {\bibinfo  {journal} {Phys. Rev. E}\ }\textbf {\bibinfo {volume}
  {87}},\ \bibinfo {pages} {023306} (\bibinfo {year} {2013})}\BibitemShut
  {NoStop}%
\bibitem [{\citenamefont {Wang}\ \emph {et~al.}(2015)\citenamefont {Wang},
  \citenamefont {Shi},\ and\ \citenamefont {Chai}}]{WANG_PRE_92_2015}%
  \BibitemOpen
  \bibfield  {author} {\bibinfo {author} {\bibfnamefont {L.}~\bibnamefont
  {Wang}}, \bibinfo {author} {\bibfnamefont {B.}~\bibnamefont {Shi}}, \ and\
  \bibinfo {author} {\bibfnamefont {Z.}~\bibnamefont {Chai}},\ }\href {\doibase
  10.1103/PhysRevE.92.043311} {\bibfield  {journal} {\bibinfo  {journal} {Phys.
  Rev. E}\ }\textbf {\bibinfo {volume} {92}},\ \bibinfo {pages} {043311}
  (\bibinfo {year} {2015})}\BibitemShut {NoStop}%
\bibitem [{\citenamefont {Ba}\ \emph {et~al.}(2018)\citenamefont {Ba},
  \citenamefont {Wang}, \citenamefont {Liu}, \citenamefont {Li},\ and\
  \citenamefont {He}}]{BA_PRE_97_2018}%
  \BibitemOpen
  \bibfield  {author} {\bibinfo {author} {\bibfnamefont {Y.}~\bibnamefont
  {Ba}}, \bibinfo {author} {\bibfnamefont {N.}~\bibnamefont {Wang}}, \bibinfo
  {author} {\bibfnamefont {H.}~\bibnamefont {Liu}}, \bibinfo {author}
  {\bibfnamefont {Q.}~\bibnamefont {Li}}, \ and\ \bibinfo {author}
  {\bibfnamefont {G.}~\bibnamefont {He}},\ }\href {\doibase
  10.1103/PhysRevE.97.033307} {\bibfield  {journal} {\bibinfo  {journal} {Phys.
  Rev. E}\ }\textbf {\bibinfo {volume} {97}},\ \bibinfo {pages} {033307}
  (\bibinfo {year} {2018})}\BibitemShut {NoStop}%
\bibitem [{\citenamefont {Montessori}\ \emph {et~al.}(2018)\citenamefont
  {Montessori}, \citenamefont {Lauricella}, \citenamefont {Rocca},
  \citenamefont {Succi}, \citenamefont {Stolovicki}, \citenamefont {Ziblat},\
  and\ \citenamefont {Weitz}}]{MONTESSORI_CF_167_2018}%
  \BibitemOpen
  \bibfield  {author} {\bibinfo {author} {\bibfnamefont {A.}~\bibnamefont
  {Montessori}}, \bibinfo {author} {\bibfnamefont {M.}~\bibnamefont
  {Lauricella}}, \bibinfo {author} {\bibfnamefont {M.~L.}\ \bibnamefont
  {Rocca}}, \bibinfo {author} {\bibfnamefont {S.}~\bibnamefont {Succi}},
  \bibinfo {author} {\bibfnamefont {E.}~\bibnamefont {Stolovicki}}, \bibinfo
  {author} {\bibfnamefont {R.}~\bibnamefont {Ziblat}}, \ and\ \bibinfo {author}
  {\bibfnamefont {D.}~\bibnamefont {Weitz}},\ }\href {\doibase
  10.1016/j.compfluid.2018.02.029} {\bibfield  {journal} {\bibinfo  {journal}
  {Comput. Fluids}\ }\textbf {\bibinfo {volume} {167}},\ \bibinfo {pages} {33 }
  (\bibinfo {year} {2018})}\BibitemShut {NoStop}%
\bibitem [{\citenamefont {Siebert}\ \emph {et~al.}(2008)\citenamefont
  {Siebert}, \citenamefont {Hegele},\ and\ \citenamefont
  {Philippi}}]{SIEBERT_PRE_77_2008}%
  \BibitemOpen
  \bibfield  {author} {\bibinfo {author} {\bibfnamefont {D.~N.}\ \bibnamefont
  {Siebert}}, \bibinfo {author} {\bibfnamefont {L.~A.}\ \bibnamefont {Hegele}},
  \ and\ \bibinfo {author} {\bibfnamefont {P.~C.}\ \bibnamefont {Philippi}},\
  }\href {\doibase 10.1103/PhysRevE.77.026707} {\bibfield  {journal} {\bibinfo
  {journal} {Phys. Rev. E}\ }\textbf {\bibinfo {volume} {77}},\ \bibinfo
  {pages} {026707} (\bibinfo {year} {2008})}\BibitemShut {NoStop}%
\bibitem [{\citenamefont {Adhikari}\ and\ \citenamefont
  {Succi}(2008)}]{ADHIKARI_PRE_78_2008}%
  \BibitemOpen
  \bibfield  {author} {\bibinfo {author} {\bibfnamefont {R.}~\bibnamefont
  {Adhikari}}\ and\ \bibinfo {author} {\bibfnamefont {S.}~\bibnamefont
  {Succi}},\ }\href {\doibase 10.1103/PhysRevE.78.066701} {\bibfield  {journal}
  {\bibinfo  {journal} {Phys. Rev. E}\ }\textbf {\bibinfo {volume} {78}},\
  \bibinfo {pages} {066701} (\bibinfo {year} {2008})}\BibitemShut {NoStop}%
\bibitem [{\citenamefont {Mari{\'e}}\ \emph {et~al.}(2009)\citenamefont
  {Mari{\'e}}, \citenamefont {Ricot},\ and\ \citenamefont
  {Sagaut}}]{MARIE_JCP_228_2009}%
  \BibitemOpen
  \bibfield  {author} {\bibinfo {author} {\bibfnamefont {S.}~\bibnamefont
  {Mari{\'e}}}, \bibinfo {author} {\bibfnamefont {D.}~\bibnamefont {Ricot}}, \
  and\ \bibinfo {author} {\bibfnamefont {P.}~\bibnamefont {Sagaut}},\ }\href
  {\doibase 10.1016/j.jcp.2008.10.021} {\bibfield  {journal} {\bibinfo
  {journal} {J. Comput. Phys.}\ }\textbf {\bibinfo {volume} {228}},\ \bibinfo
  {pages} {1056 } (\bibinfo {year} {2009})}\BibitemShut {NoStop}%
\bibitem [{\citenamefont {Hosseini}\ \emph {et~al.}(2017)\citenamefont
  {Hosseini}, \citenamefont {Darabiha}, \citenamefont {Th{\'e}venin},\ and\
  \citenamefont {Eshghinejadfard}}]{HOSSEINI_IJMPC_28_2017}%
  \BibitemOpen
  \bibfield  {author} {\bibinfo {author} {\bibfnamefont {S.~A.}\ \bibnamefont
  {Hosseini}}, \bibinfo {author} {\bibfnamefont {N.}~\bibnamefont {Darabiha}},
  \bibinfo {author} {\bibfnamefont {D.}~\bibnamefont {Th{\'e}venin}}, \ and\
  \bibinfo {author} {\bibfnamefont {A.}~\bibnamefont {Eshghinejadfard}},\
  }\href {\doibase 10.1142/S0129183117501418} {\bibfield  {journal} {\bibinfo
  {journal} {Int. J. Mod. Phys. C}\ }\textbf {\bibinfo {volume} {28}},\
  \bibinfo {pages} {1750141} (\bibinfo {year} {2017})}\BibitemShut {NoStop}%
\bibitem [{\citenamefont {Ch\'{a}vez-Modena}\ \emph {et~al.}(2018)\citenamefont
  {Ch\'{a}vez-Modena}, \citenamefont {Ferrer},\ and\ \citenamefont
  {Rubio}}]{CHAVEZMODENA_CF_172_2018}%
  \BibitemOpen
  \bibfield  {author} {\bibinfo {author} {\bibfnamefont {M.}~\bibnamefont
  {Ch\'{a}vez-Modena}}, \bibinfo {author} {\bibfnamefont {E.}~\bibnamefont
  {Ferrer}}, \ and\ \bibinfo {author} {\bibfnamefont {G.}~\bibnamefont
  {Rubio}},\ }\href {\doibase 10.1016/j.compfluid.2018.03.084} {\bibfield
  {journal} {\bibinfo  {journal} {Comput. Fluids}\ }\textbf {\bibinfo {volume}
  {172}},\ \bibinfo {pages} {397 } (\bibinfo {year} {2018})}\BibitemShut
  {NoStop}%
\bibitem [{\citenamefont {Wissocq}\ \emph {et~al.}(2019)\citenamefont
  {Wissocq}, \citenamefont {Sagaut},\ and\ \citenamefont
  {Boussuge}}]{WISSOCQ_JCP_380_2019}%
  \BibitemOpen
  \bibfield  {author} {\bibinfo {author} {\bibfnamefont {G.}~\bibnamefont
  {Wissocq}}, \bibinfo {author} {\bibfnamefont {P.}~\bibnamefont {Sagaut}}, \
  and\ \bibinfo {author} {\bibfnamefont {J.-F.}\ \bibnamefont {Boussuge}},\
  }\href {\doibase 10.1016/j.jcp.2018.12.015} {\bibfield  {journal} {\bibinfo
  {journal} {J. Comput. Phys.}\ }\textbf {\bibinfo {volume} {380}},\ \bibinfo
  {pages} {311 } (\bibinfo {year} {2019})}\BibitemShut {NoStop}%
\bibitem [{\citenamefont {Asinari}(2008)}]{ASINARI_PRE_78_2008}%
  \BibitemOpen
  \bibfield  {author} {\bibinfo {author} {\bibfnamefont {P.}~\bibnamefont
  {Asinari}},\ }\href {\doibase 10.1103/PhysRevE.78.016701} {\bibfield
  {journal} {\bibinfo  {journal} {Phys. Rev. E}\ }\textbf {\bibinfo {volume}
  {78}},\ \bibinfo {pages} {016701} (\bibinfo {year} {2008})}\BibitemShut
  {NoStop}%
\bibitem [{\citenamefont {Higuera}\ and\ \citenamefont
  {Jimenez}(1989)}]{HIGUERA_EPL_9_1989}%
  \BibitemOpen
  \bibfield  {author} {\bibinfo {author} {\bibfnamefont {F.}~\bibnamefont
  {Higuera}}\ and\ \bibinfo {author} {\bibfnamefont {J.}~\bibnamefont
  {Jimenez}},\ }\href {\doibase 10.1209/0295-5075/9/7/009} {\bibfield
  {journal} {\bibinfo  {journal} {Europhys. Lett.}\ }\textbf {\bibinfo {volume}
  {9}},\ \bibinfo {pages} {663} (\bibinfo {year} {1989})}\BibitemShut {NoStop}%
\bibitem [{\citenamefont {Benzi}\ \emph {et~al.}(1992)\citenamefont {Benzi},
  \citenamefont {Succi},\ and\ \citenamefont {Vergassola}}]{BENZI_PR_222_1992}%
  \BibitemOpen
  \bibfield  {author} {\bibinfo {author} {\bibfnamefont {R.}~\bibnamefont
  {Benzi}}, \bibinfo {author} {\bibfnamefont {S.}~\bibnamefont {Succi}}, \ and\
  \bibinfo {author} {\bibfnamefont {M.}~\bibnamefont {Vergassola}},\ }\href
  {\doibase 10.1016/0370-1573(92)90090-M} {\bibfield  {journal} {\bibinfo
  {journal} {Phys. Rep.}\ }\textbf {\bibinfo {volume} {222}},\ \bibinfo {pages}
  {145 } (\bibinfo {year} {1992})}\BibitemShut {NoStop}%
\bibitem [{SupMat()}]{SupMat}%
  \BibitemOpen
  SupMat,\ \href@noop {} {}\bibinfo {note} {See Supplemental Material at
  [\protect \url {COREIXAS_PRE_2019_Theo_SupMat.pdf}] for instructions
  regarding the implementation of most collision models considered in the
  present work.}\BibitemShut {Stop}%
\bibitem [{SupMatQ9()}]{SupMatQ9}%
  \BibitemOpen
  SupMatQ9,\ \href@noop {} {}\bibinfo {note} {See Supplemental Material for
  instructions regarding the implementation of D2Q9 collision models at
  [\protect \url {Comprehensive_Collision_Models_D2Q9.txt}].}\BibitemShut
  {Stop}%
\bibitem [{SupMatQ27()}]{SupMatQ27}%
  \BibitemOpen
  SupMatQ27,\ \href@noop {} {}\bibinfo {note} {See Supplemental Material for
  instructions regarding the implementation of D3Q27 collision models at
  [\protect \url {Comprehensive_Collision_Models_D3Q27.txt}].}\BibitemShut
  {Stop}%
\bibitem [{\citenamefont {Karlin}\ and\ \citenamefont
  {Asinari}(2010)}]{KARLIN_PA_389_2010}%
  \BibitemOpen
  \bibfield  {author} {\bibinfo {author} {\bibfnamefont {I.}~\bibnamefont
  {Karlin}}\ and\ \bibinfo {author} {\bibfnamefont {P.}~\bibnamefont
  {Asinari}},\ }\href {\doibase 10.1016/j.physa.2009.12.032} {\bibfield
  {journal} {\bibinfo  {journal} {Physica A}\ }\textbf {\bibinfo {volume}
  {389}},\ \bibinfo {pages} {1530 } (\bibinfo {year} {2010})}\BibitemShut
  {NoStop}%
\bibitem [{\citenamefont {Kendall}(1940)}]{KENDALL_AE_10_1940}%
  \BibitemOpen
  \bibfield  {author} {\bibinfo {author} {\bibfnamefont {M.}~\bibnamefont
  {Kendall}},\ }\href {\doibase 10.1111/j.1469-1809.1940.tb02261.x} {\bibfield
  {journal} {\bibinfo  {journal} {Ann. Eugen.}\ }\textbf {\bibinfo {volume}
  {10}},\ \bibinfo {pages} {392} (\bibinfo {year} {1940})}\BibitemShut
  {NoStop}%
\bibitem [{\citenamefont {Bell}(1934)}]{BELL_AM_1934}%
  \BibitemOpen
  \bibfield  {author} {\bibinfo {author} {\bibfnamefont {E.~T.}\ \bibnamefont
  {Bell}},\ }\href {\doibase 10.2307/1968431} {\bibfield  {journal} {\bibinfo
  {journal} {Ann. Math.}\ ,\ \bibinfo {pages} {258}} (\bibinfo {year}
  {1934})}\BibitemShut {NoStop}%
\bibitem [{\citenamefont {Withers}(1994)}]{WITHERS_1994}%
  \BibitemOpen
  \bibfield  {author} {\bibinfo {author} {\bibfnamefont {C.~S.}\ \bibnamefont
  {Withers}},\ }in\ \href
  {http://www.thebookshelf.auckland.ac.nz/docs/NZOperationalResearch/conferenceproceedings/1994-proceedings/ORSNZ-proceedings-1994-41.pdf}
  {\emph {\bibinfo {booktitle} {Proceedings of the 1994 New Zealand Statistical
  Association Conference, Massey University}}}\ (\bibinfo {year}
  {1994})\BibitemShut {NoStop}%
\bibitem [{\citenamefont {{B}hatnagar}\ \emph {et~al.}(1954)\citenamefont
  {{B}hatnagar}, \citenamefont {{G}ross},\ and\ \citenamefont
  {{K}rook}}]{BHATNAGAR_PR_94_1954}%
  \BibitemOpen
  \bibfield  {author} {\bibinfo {author} {\bibfnamefont {P.}~\bibnamefont
  {{B}hatnagar}}, \bibinfo {author} {\bibfnamefont {E.}~\bibnamefont
  {{G}ross}}, \ and\ \bibinfo {author} {\bibfnamefont {M.}~\bibnamefont
  {{K}rook}},\ }\href {\doibase 10.1103/PhysRev.94.511} {\bibfield  {journal}
  {\bibinfo  {journal} {Phys. Rev.}\ }\textbf {\bibinfo {volume} {94}},\
  \bibinfo {pages} {511} (\bibinfo {year} {1954})}\BibitemShut {NoStop}%
\bibitem [{\citenamefont {McCullagh}(2018)}]{MCCULLAGH_Book_2018}%
  \BibitemOpen
  \bibfield  {author} {\bibinfo {author} {\bibfnamefont {P.}~\bibnamefont
  {McCullagh}},\ }\href
  {https://www.crcpress.com/Tensor-Methods-in-Statistics-Monographs-on-Statistics-and-Applied-Probability/McCullagh/p/book/9781315898018}
  {\emph {\bibinfo {title} {Tensor Methods in Statistics: Monographs on
  Statistics and Applied Probability}}}\ (\bibinfo  {publisher} {Chapman and
  Hall/CRC},\ \bibinfo {year} {2018})\BibitemShut {NoStop}%
\bibitem [{\citenamefont {Kogan}(1965)}]{KOGAN_JAMM_29_1965}%
  \BibitemOpen
  \bibfield  {author} {\bibinfo {author} {\bibfnamefont {A.}~\bibnamefont
  {Kogan}},\ }\href {\doibase https://doi.org/10.1016/0021-8928(65)90157-7}
  {\bibfield  {journal} {\bibinfo  {journal} {J. Appl. Math. Mech.}\ }\textbf
  {\bibinfo {volume} {29}},\ \bibinfo {pages} {130 } (\bibinfo {year}
  {1965})}\BibitemShut {NoStop}%
\bibitem [{\citenamefont {Press\'e}\ \emph {et~al.}(2013)\citenamefont
  {Press\'e}, \citenamefont {Ghosh}, \citenamefont {Lee},\ and\ \citenamefont
  {Dill}}]{PRESSE_RMP_85_2013}%
  \BibitemOpen
  \bibfield  {author} {\bibinfo {author} {\bibfnamefont {S.}~\bibnamefont
  {Press\'e}}, \bibinfo {author} {\bibfnamefont {K.}~\bibnamefont {Ghosh}},
  \bibinfo {author} {\bibfnamefont {J.}~\bibnamefont {Lee}}, \ and\ \bibinfo
  {author} {\bibfnamefont {K.~A.}\ \bibnamefont {Dill}},\ }\href {\doibase
  10.1103/RevModPhys.85.1115} {\bibfield  {journal} {\bibinfo  {journal} {Rev.
  Mod. Phys.}\ }\textbf {\bibinfo {volume} {85}},\ \bibinfo {pages} {1115}
  (\bibinfo {year} {2013})}\BibitemShut {NoStop}%
\bibitem [{\citenamefont {Ansumali}\ \emph {et~al.}(2003)\citenamefont
  {Ansumali}, \citenamefont {Karlin},\ and\ \citenamefont
  {\"{O}ttinger}}]{ANSUMALI_EPL_63_2003}%
  \BibitemOpen
  \bibfield  {author} {\bibinfo {author} {\bibfnamefont {S.}~\bibnamefont
  {Ansumali}}, \bibinfo {author} {\bibfnamefont {I.~V.}\ \bibnamefont
  {Karlin}}, \ and\ \bibinfo {author} {\bibfnamefont {H.~C.}\ \bibnamefont
  {\"{O}ttinger}},\ }\href {\doibase 10.1209/epl/i2003-00496-6} {\bibfield
  {journal} {\bibinfo  {journal} {Europhys. Lett.}\ }\textbf {\bibinfo {volume}
  {63}},\ \bibinfo {pages} {798} (\bibinfo {year} {2003})}\BibitemShut
  {NoStop}%
\bibitem [{\citenamefont {Asinari}\ and\ \citenamefont
  {Karlin}(2009)}]{ASINARI_PRE_79_2009}%
  \BibitemOpen
  \bibfield  {author} {\bibinfo {author} {\bibfnamefont {P.}~\bibnamefont
  {Asinari}}\ and\ \bibinfo {author} {\bibfnamefont {I.~V.}\ \bibnamefont
  {Karlin}},\ }\href {\doibase 10.1103/PhysRevE.79.036703} {\bibfield
  {journal} {\bibinfo  {journal} {Phys. Rev. E}\ }\textbf {\bibinfo {volume}
  {79}},\ \bibinfo {pages} {036703} (\bibinfo {year} {2009})}\BibitemShut
  {NoStop}%
\bibitem [{\citenamefont {Frapolli}\ \emph {et~al.}(2015)\citenamefont
  {Frapolli}, \citenamefont {Chikatamarla},\ and\ \citenamefont
  {Karlin}}]{FRAPOLLI_PRE_92_2015}%
  \BibitemOpen
  \bibfield  {author} {\bibinfo {author} {\bibfnamefont {N.}~\bibnamefont
  {Frapolli}}, \bibinfo {author} {\bibfnamefont {S.~S.}\ \bibnamefont
  {Chikatamarla}}, \ and\ \bibinfo {author} {\bibfnamefont {I.~V.}\
  \bibnamefont {Karlin}},\ }\href {\doibase 10.1103/PhysRevE.92.061301}
  {\bibfield  {journal} {\bibinfo  {journal} {Phys. Rev. E}\ }\textbf {\bibinfo
  {volume} {92}},\ \bibinfo {pages} {061301} (\bibinfo {year}
  {2015})}\BibitemShut {NoStop}%
\bibitem [{\citenamefont {Atif}\ \emph {et~al.}(2018)\citenamefont {Atif},
  \citenamefont {Namburi},\ and\ \citenamefont {Ansumali}}]{ATIF_PRE_98_2018}%
  \BibitemOpen
  \bibfield  {author} {\bibinfo {author} {\bibfnamefont {M.}~\bibnamefont
  {Atif}}, \bibinfo {author} {\bibfnamefont {M.}~\bibnamefont {Namburi}}, \
  and\ \bibinfo {author} {\bibfnamefont {S.}~\bibnamefont {Ansumali}},\ }\href
  {\doibase 10.1103/PhysRevE.98.053311} {\bibfield  {journal} {\bibinfo
  {journal} {Phys. Rev. E}\ }\textbf {\bibinfo {volume} {98}},\ \bibinfo
  {pages} {053311} (\bibinfo {year} {2018})}\BibitemShut {NoStop}%
\bibitem [{\citenamefont {Qian}\ \emph {et~al.}(1992)\citenamefont {Qian},
  \citenamefont {D'Humières},\ and\ \citenamefont
  {Lallemand}}]{QIAN_EPL_17_1992}%
  \BibitemOpen
  \bibfield  {author} {\bibinfo {author} {\bibfnamefont {Y.~H.}\ \bibnamefont
  {Qian}}, \bibinfo {author} {\bibfnamefont {D.}~\bibnamefont {D'Humières}}, \
  and\ \bibinfo {author} {\bibfnamefont {P.}~\bibnamefont {Lallemand}},\ }\href
  {http://stacks.iop.org/0295-5075/17/i=6/a=001} {\bibfield  {journal}
  {\bibinfo  {journal} {Europhys. Lett.}\ }\textbf {\bibinfo {volume} {17}},\
  \bibinfo {pages} {479} (\bibinfo {year} {1992})}\BibitemShut {NoStop}%
\bibitem [{\citenamefont {Shan}\ \emph {et~al.}(2006)\citenamefont {Shan},
  \citenamefont {Yuan},\ and\ \citenamefont {Chen}}]{SHAN_JFM_550_2006}%
  \BibitemOpen
  \bibfield  {author} {\bibinfo {author} {\bibfnamefont {X.}~\bibnamefont
  {Shan}}, \bibinfo {author} {\bibfnamefont {X.-F.}\ \bibnamefont {Yuan}}, \
  and\ \bibinfo {author} {\bibfnamefont {H.}~\bibnamefont {Chen}},\ }\href
  {\doibase 10.1017/S0022112005008153} {\bibfield  {journal} {\bibinfo
  {journal} {J. Fluid Mech.}\ }\textbf {\bibinfo {volume} {550}},\ \bibinfo
  {pages} {413} (\bibinfo {year} {2006})}\BibitemShut {NoStop}%
\bibitem [{\citenamefont {Lee}\ \emph {et~al.}(2006)\citenamefont {Lee},
  \citenamefont {Lin},\ and\ \citenamefont {Chen}}]{LEE_JCP_215_2006}%
  \BibitemOpen
  \bibfield  {author} {\bibinfo {author} {\bibfnamefont {T.}~\bibnamefont
  {Lee}}, \bibinfo {author} {\bibfnamefont {C.-L.}\ \bibnamefont {Lin}}, \ and\
  \bibinfo {author} {\bibfnamefont {L.-D.}\ \bibnamefont {Chen}},\ }\href
  {\doibase 10.1016/j.jcp.2005.10.021} {\bibfield  {journal} {\bibinfo
  {journal} {J. Comput. Phys.}\ }\textbf {\bibinfo {volume} {215}},\ \bibinfo
  {pages} {133 } (\bibinfo {year} {2006})}\BibitemShut {NoStop}%
\bibitem [{\citenamefont {Surmas}\ \emph {et~al.}(2009)\citenamefont {Surmas},
  \citenamefont {Pico~Ortiz},\ and\ \citenamefont
  {Philippi}}]{SURMAS_EPJST_171_2009}%
  \BibitemOpen
  \bibfield  {author} {\bibinfo {author} {\bibfnamefont {R.}~\bibnamefont
  {Surmas}}, \bibinfo {author} {\bibfnamefont {C.~E.}\ \bibnamefont
  {Pico~Ortiz}}, \ and\ \bibinfo {author} {\bibfnamefont {P.~C.}\ \bibnamefont
  {Philippi}},\ }\href {\doibase 10.1140/epjst/e2009-01014-x} {\bibfield
  {journal} {\bibinfo  {journal} {Eur. Phys. J. Special Topics}\ }\textbf
  {\bibinfo {volume} {171}},\ \bibinfo {pages} {81} (\bibinfo {year}
  {2009})}\BibitemShut {NoStop}%
\bibitem [{\citenamefont {Guo}\ \emph {et~al.}(2011)\citenamefont {Guo},
  \citenamefont {Shi},\ and\ \citenamefont {Zheng}}]{GUO_TRS_369_2011}%
  \BibitemOpen
  \bibfield  {author} {\bibinfo {author} {\bibfnamefont {Z.}~\bibnamefont
  {Guo}}, \bibinfo {author} {\bibfnamefont {B.}~\bibnamefont {Shi}}, \ and\
  \bibinfo {author} {\bibfnamefont {C.}~\bibnamefont {Zheng}},\ }\href
  {\doibase 10.1098/rsta.2011.0023} {\bibfield  {journal} {\bibinfo  {journal}
  {Philos. Trans. R . Soc. London, Ser. A}\ }\textbf {\bibinfo {volume}
  {369}},\ \bibinfo {pages} {2283} (\bibinfo {year} {2011})}\BibitemShut
  {NoStop}%
\bibitem [{\citenamefont {Guo}\ \emph {et~al.}(2013)\citenamefont {Guo},
  \citenamefont {Xu},\ and\ \citenamefont {Wang}}]{GUO_PRE_88_2013}%
  \BibitemOpen
  \bibfield  {author} {\bibinfo {author} {\bibfnamefont {Z.}~\bibnamefont
  {Guo}}, \bibinfo {author} {\bibfnamefont {K.}~\bibnamefont {Xu}}, \ and\
  \bibinfo {author} {\bibfnamefont {R.}~\bibnamefont {Wang}},\ }\href {\doibase
  10.1103/PhysRevE.88.033305} {\bibfield  {journal} {\bibinfo  {journal} {Phys.
  Rev. E}\ }\textbf {\bibinfo {volume} {88}},\ \bibinfo {pages} {033305}
  (\bibinfo {year} {2013})}\BibitemShut {NoStop}%
\bibitem [{\citenamefont {Gendre}\ \emph {et~al.}(2017)\citenamefont {Gendre},
  \citenamefont {Ricot}, \citenamefont {Fritz},\ and\ \citenamefont
  {Sagaut}}]{GENDRE_PRE_96_2017}%
  \BibitemOpen
  \bibfield  {author} {\bibinfo {author} {\bibfnamefont {F.}~\bibnamefont
  {Gendre}}, \bibinfo {author} {\bibfnamefont {D.}~\bibnamefont {Ricot}},
  \bibinfo {author} {\bibfnamefont {G.}~\bibnamefont {Fritz}}, \ and\ \bibinfo
  {author} {\bibfnamefont {P.}~\bibnamefont {Sagaut}},\ }\href {\doibase
  10.1103/PhysRevE.96.023311} {\bibfield  {journal} {\bibinfo  {journal} {Phys.
  Rev. E}\ }\textbf {\bibinfo {volume} {96}},\ \bibinfo {pages} {023311}
  (\bibinfo {year} {2017})}\BibitemShut {NoStop}%
\bibitem [{\citenamefont {Suga}\ \emph {et~al.}(2015)\citenamefont {Suga},
  \citenamefont {Kuwata}, \citenamefont {Takashima},\ and\ \citenamefont
  {Chikasue}}]{SUGA_CMA_69_2015}%
  \BibitemOpen
  \bibfield  {author} {\bibinfo {author} {\bibfnamefont {K.}~\bibnamefont
  {Suga}}, \bibinfo {author} {\bibfnamefont {Y.}~\bibnamefont {Kuwata}},
  \bibinfo {author} {\bibfnamefont {K.}~\bibnamefont {Takashima}}, \ and\
  \bibinfo {author} {\bibfnamefont {R.}~\bibnamefont {Chikasue}},\ }\href
  {\doibase 10.1016/j.camwa.2015.01.010} {\bibfield  {journal} {\bibinfo
  {journal} {Comput. Math. Appl.}\ }\textbf {\bibinfo {volume} {69}},\ \bibinfo
  {pages} {518 } (\bibinfo {year} {2015})}\BibitemShut {NoStop}%
\bibitem [{\citenamefont {Fakhari}\ \emph {et~al.}(2017)\citenamefont
  {Fakhari}, \citenamefont {Bolster},\ and\ \citenamefont
  {Luo}}]{FAKHARI_JCP_341_2017}%
  \BibitemOpen
  \bibfield  {author} {\bibinfo {author} {\bibfnamefont {A.}~\bibnamefont
  {Fakhari}}, \bibinfo {author} {\bibfnamefont {D.}~\bibnamefont {Bolster}}, \
  and\ \bibinfo {author} {\bibfnamefont {L.-S.}\ \bibnamefont {Luo}},\ }\href
  {\doibase 10.1016/j.jcp.2017.03.062} {\bibfield  {journal} {\bibinfo
  {journal} {J. Comput. Phys.}\ }\textbf {\bibinfo {volume} {341}},\ \bibinfo
  {pages} {22 } (\bibinfo {year} {2017})}\BibitemShut {NoStop}%
\bibitem [{\citenamefont {Karlin}\ \emph
  {et~al.}(2011{\natexlab{a}})\citenamefont {Karlin}, \citenamefont
  {Lycett-Brown},\ and\ \citenamefont {Luo}}]{KARLIN_ARXIV_2011}%
  \BibitemOpen
  \bibfield  {author} {\bibinfo {author} {\bibfnamefont {I.~V.}\ \bibnamefont
  {Karlin}}, \bibinfo {author} {\bibfnamefont {D.}~\bibnamefont
  {Lycett-Brown}}, \ and\ \bibinfo {author} {\bibfnamefont {K.~H.}\
  \bibnamefont {Luo}},\ }\href {https://arxiv.org/abs/1107.3309} {\bibfield
  {journal} {\bibinfo  {journal} {arXiv preprint arXiv:1107.3309}\ } (\bibinfo
  {year} {2011}{\natexlab{a}})}\BibitemShut {NoStop}%
\bibitem [{\citenamefont {Ansumali}\ \emph {et~al.}(2007)\citenamefont
  {Ansumali}, \citenamefont {Arcidiacono}, \citenamefont {Chikatamarla},
  \citenamefont {Prasianakis}, \citenamefont {Gorban},\ and\ \citenamefont
  {Karlin}}]{ANSUMALI_EPJB_56_2007}%
  \BibitemOpen
  \bibfield  {author} {\bibinfo {author} {\bibfnamefont {S.}~\bibnamefont
  {Ansumali}}, \bibinfo {author} {\bibfnamefont {S.}~\bibnamefont
  {Arcidiacono}}, \bibinfo {author} {\bibfnamefont {S.~S.}\ \bibnamefont
  {Chikatamarla}}, \bibinfo {author} {\bibfnamefont {I.~N.}\ \bibnamefont
  {Prasianakis}}, \bibinfo {author} {\bibfnamefont {N.~A.}\ \bibnamefont
  {Gorban}}, \ and\ \bibinfo {author} {\bibfnamefont {V.~I.}\ \bibnamefont
  {Karlin}},\ }\href {\doibase 10.1140/epjb/e2007-00100-1} {\bibfield
  {journal} {\bibinfo  {journal} {Eur. Phys. J. B}\ }\textbf {\bibinfo {volume}
  {56}},\ \bibinfo {pages} {135} (\bibinfo {year} {2007})}\BibitemShut
  {NoStop}%
\bibitem [{\citenamefont {Huang}\ \emph {et~al.}(2018)\citenamefont {Huang},
  \citenamefont {Wu},\ and\ \citenamefont {Adams}}]{HUANG_PRE_97_2018}%
  \BibitemOpen
  \bibfield  {author} {\bibinfo {author} {\bibfnamefont {R.}~\bibnamefont
  {Huang}}, \bibinfo {author} {\bibfnamefont {H.}~\bibnamefont {Wu}}, \ and\
  \bibinfo {author} {\bibfnamefont {N.~A.}\ \bibnamefont {Adams}},\ }\href
  {\doibase 10.1103/PhysRevE.97.053308} {\bibfield  {journal} {\bibinfo
  {journal} {Phys. Rev. E}\ }\textbf {\bibinfo {volume} {97}},\ \bibinfo
  {pages} {053308} (\bibinfo {year} {2018})}\BibitemShut {NoStop}%
\bibitem [{\citenamefont {Shan}\ and\ \citenamefont
  {Chen}(2007)}]{SHAN_IJMPC_18_2007}%
  \BibitemOpen
  \bibfield  {author} {\bibinfo {author} {\bibfnamefont {X.}~\bibnamefont
  {Shan}}\ and\ \bibinfo {author} {\bibfnamefont {H.}~\bibnamefont {Chen}},\
  }\href {\doibase 10.1142/S0129183107010887} {\bibfield  {journal} {\bibinfo
  {journal} {Int. J. Mod. Phys. C}\ }\textbf {\bibinfo {volume} {18}},\
  \bibinfo {pages} {635} (\bibinfo {year} {2007})}\BibitemShut {NoStop}%
\bibitem [{\citenamefont {Chen}\ \emph {et~al.}(2014)\citenamefont {Chen},
  \citenamefont {Gopalakrishnan},\ and\ \citenamefont
  {Zhang}}]{CHEN_IJMPC_25_2014}%
  \BibitemOpen
  \bibfield  {author} {\bibinfo {author} {\bibfnamefont {H.}~\bibnamefont
  {Chen}}, \bibinfo {author} {\bibfnamefont {P.}~\bibnamefont
  {Gopalakrishnan}}, \ and\ \bibinfo {author} {\bibfnamefont {R.}~\bibnamefont
  {Zhang}},\ }\href {\doibase 10.1142/S0129183114500466} {\bibfield  {journal}
  {\bibinfo  {journal} {Int. J. Mod. Phys. C}\ }\textbf {\bibinfo {volume}
  {25}},\ \bibinfo {pages} {1450046} (\bibinfo {year} {2014})}\BibitemShut
  {NoStop}%
\bibitem [{\citenamefont {Chen}\ \emph {et~al.}(2015)\citenamefont {Chen},
  \citenamefont {Zhang},\ and\ \citenamefont
  {Gopalakrishnan}}]{CHEN_PATENT_Collision_2015}%
  \BibitemOpen
  \bibfield  {author} {\bibinfo {author} {\bibfnamefont {H.}~\bibnamefont
  {Chen}}, \bibinfo {author} {\bibfnamefont {R.}~\bibnamefont {Zhang}}, \ and\
  \bibinfo {author} {\bibfnamefont {P.}~\bibnamefont {Gopalakrishnan}},\ }\href
  {http://www.google.com/patents/CA2919062A1?cl=en} {\enquote {\bibinfo {title}
  {Lattice {B}oltzmann collision operators enforcing isotropy and {G}alilean
  invariance},}\ } (\bibinfo {year} {2015}),\ \bibinfo {note} {{CA} {P}atent
  {A}pp. {CA} 2,919,062}\BibitemShut {NoStop}%
\bibitem [{\citenamefont {Viggen}(2014)}]{VIGGEN_PhD_2014}%
  \BibitemOpen
  \bibfield  {author} {\bibinfo {author} {\bibfnamefont {E.~M.}\ \bibnamefont
  {Viggen}},\ }\emph {\bibinfo {title} {The lattice {B}oltzmann method:
  Fundamentals and acoustics}},\ \href
  {https://brage.bibsys.no/xmlui/handle/11250/2370925} {Ph.D. thesis},\
  \bibinfo  {school} {NTNU} (\bibinfo {year} {2014})\BibitemShut {NoStop}%
\bibitem [{\citenamefont {Chikatamarla}\ and\ \citenamefont
  {Karlin}(2009)}]{CHIKATAMARLA_PRE_79_2009}%
  \BibitemOpen
  \bibfield  {author} {\bibinfo {author} {\bibfnamefont {S.~S.}\ \bibnamefont
  {Chikatamarla}}\ and\ \bibinfo {author} {\bibfnamefont {I.~V.}\ \bibnamefont
  {Karlin}},\ }\href {\doibase 10.1103/PhysRevE.79.046701} {\bibfield
  {journal} {\bibinfo  {journal} {Phys. Rev. E}\ }\textbf {\bibinfo {volume}
  {79}},\ \bibinfo {pages} {046701} (\bibinfo {year} {2009})}\BibitemShut
  {NoStop}%
\bibitem [{\citenamefont {Frapolli}(2017)}]{FRAPOLLI_PhD_2017}%
  \BibitemOpen
  \bibfield  {author} {\bibinfo {author} {\bibfnamefont {N.}~\bibnamefont
  {Frapolli}},\ }\emph {\bibinfo {title} {Entropic lattice {B}oltzmann models
  for thermal and compressible flows}},\ \href
  {https://www.research-collection.ethz.ch/handle/20.500.11850/130664} {Ph.D.
  thesis},\ \bibinfo  {school} {ETH Zurich} (\bibinfo {year}
  {2017})\BibitemShut {NoStop}%
\bibitem [{\citenamefont {Grad}(1949)}]{GRADb_CPAM_2_1949}%
  \BibitemOpen
  \bibfield  {author} {\bibinfo {author} {\bibfnamefont {H.}~\bibnamefont
  {Grad}},\ }\href {\doibase 10.1002/cpa.3160020403} {\bibfield  {journal}
  {\bibinfo  {journal} {Commun. Pure Appl. Math.}\ }\textbf {\bibinfo {volume}
  {2}},\ \bibinfo {pages} {331} (\bibinfo {year} {1949})}\BibitemShut {NoStop}%
\bibitem [{\citenamefont {Shan}\ and\ \citenamefont
  {He}(1998)}]{SHAN_PRL_80_1998}%
  \BibitemOpen
  \bibfield  {author} {\bibinfo {author} {\bibfnamefont {X.}~\bibnamefont
  {Shan}}\ and\ \bibinfo {author} {\bibfnamefont {X.}~\bibnamefont {He}},\
  }\href {\doibase 10.1103/PhysRevLett.80.65} {\bibfield  {journal} {\bibinfo
  {journal} {Phys. Rev. Lett.}\ }\textbf {\bibinfo {volume} {80}},\ \bibinfo
  {pages} {65} (\bibinfo {year} {1998})}\BibitemShut {NoStop}%
\bibitem [{\citenamefont {Philippi}\ \emph {et~al.}(2006)\citenamefont
  {Philippi}, \citenamefont {Hegele}, \citenamefont {dos Santos},\ and\
  \citenamefont {Surmas}}]{PHILIPPI_PRE_73_2006}%
  \BibitemOpen
  \bibfield  {author} {\bibinfo {author} {\bibfnamefont {P.~C.}\ \bibnamefont
  {Philippi}}, \bibinfo {author} {\bibfnamefont {L.~A.}\ \bibnamefont
  {Hegele}}, \bibinfo {author} {\bibfnamefont {L.~O.~E.}\ \bibnamefont {dos
  Santos}}, \ and\ \bibinfo {author} {\bibfnamefont {R.}~\bibnamefont
  {Surmas}},\ }\href {\doibase 10.1103/PhysRevE.73.056702} {\bibfield
  {journal} {\bibinfo  {journal} {Phys. Rev. E}\ }\textbf {\bibinfo {volume}
  {73}},\ \bibinfo {pages} {056702} (\bibinfo {year} {2006})}\BibitemShut
  {NoStop}%
\bibitem [{\citenamefont {Karlin}\ \emph
  {et~al.}(2011{\natexlab{b}})\citenamefont {Karlin}, \citenamefont {Asinari},\
  and\ \citenamefont {Succi}}]{KARLIN_PTRSL_369_2011}%
  \BibitemOpen
  \bibfield  {author} {\bibinfo {author} {\bibfnamefont {I.}~\bibnamefont
  {Karlin}}, \bibinfo {author} {\bibfnamefont {P.}~\bibnamefont {Asinari}}, \
  and\ \bibinfo {author} {\bibfnamefont {S.}~\bibnamefont {Succi}},\ }\href
  {\doibase 10.1098/rsta.2011.0061} {\bibfield  {journal} {\bibinfo  {journal}
  {Philos. Trans. R . Soc. London, Ser. A}\ }\textbf {\bibinfo {volume}
  {369}},\ \bibinfo {pages} {2202} (\bibinfo {year}
  {2011}{\natexlab{b}})}\BibitemShut {NoStop}%
\bibitem [{\citenamefont {Dellar}(2002)}]{DELLAR_PRE_65_2002}%
  \BibitemOpen
  \bibfield  {author} {\bibinfo {author} {\bibfnamefont {P.~J.}\ \bibnamefont
  {Dellar}},\ }\href {\doibase 10.1103/PhysRevE.65.036309} {\bibfield
  {journal} {\bibinfo  {journal} {Phys. Rev. E}\ }\textbf {\bibinfo {volume}
  {65}},\ \bibinfo {pages} {036309} (\bibinfo {year} {2002})}\BibitemShut
  {NoStop}%
\bibitem [{\citenamefont {Dellar}(2014)}]{DELLAR_JCP_259_2014}%
  \BibitemOpen
  \bibfield  {author} {\bibinfo {author} {\bibfnamefont {P.~J.}\ \bibnamefont
  {Dellar}},\ }\href {\doibase 10.1016/j.jcp.2013.11.021} {\bibfield  {journal}
  {\bibinfo  {journal} {J. Comput. Phys.}\ }\textbf {\bibinfo {volume} {259}},\
  \bibinfo {pages} {270 } (\bibinfo {year} {2014})}\BibitemShut {NoStop}%
\bibitem [{\citenamefont {Premnath}\ and\ \citenamefont
  {Banerjee}(2011)}]{PREMNATH_JSP_143_2011}%
  \BibitemOpen
  \bibfield  {author} {\bibinfo {author} {\bibfnamefont {K.~N.}\ \bibnamefont
  {Premnath}}\ and\ \bibinfo {author} {\bibfnamefont {S.}~\bibnamefont
  {Banerjee}},\ }\href {\doibase 10.1007/s10955-011-0208-9} {\bibfield
  {journal} {\bibinfo  {journal} {J. Stat. Phys.}\ }\textbf {\bibinfo {volume}
  {143}},\ \bibinfo {pages} {747} (\bibinfo {year} {2011})}\BibitemShut
  {NoStop}%
\bibitem [{\citenamefont {Hirsch}(2007)}]{HIRSCH_Book_2007}%
  \BibitemOpen
  \bibfield  {author} {\bibinfo {author} {\bibfnamefont {C.}~\bibnamefont
  {Hirsch}},\ }\href
  {https://www.elsevier.com/books/numerical-computation-of-internal-and-external-flows-the-fundamentals-of-computational-fluid-dynamics/hirsch/978-0-7506-6594-0}
  {\emph {\bibinfo {title} {Numerical computation of internal and external
  flows: {T}he fundamentals of {C}omputational {F}luid {D}ynamics}}}\ (\bibinfo
   {publisher} {Elsevier},\ \bibinfo {year} {2007})\BibitemShut {NoStop}%
\bibitem [{\citenamefont {Huang}(1987)}]{HUANG_Book_2nd_1987}%
  \BibitemOpen
  \bibfield  {author} {\bibinfo {author} {\bibfnamefont {K.}~\bibnamefont
  {Huang}},\ }\href
  {http://gen.lib.rus.ec/book/index.php?md5=836019C9EC85D594F380D1D898B59713}
  {\emph {\bibinfo {title} {Statistical Mechanics}}},\ \bibinfo {edition}
  {2nd}\ ed.\ (\bibinfo  {publisher} {Wiley},\ \bibinfo {year}
  {1987})\BibitemShut {NoStop}%
\bibitem [{\citenamefont {Chapman}\ and\ \citenamefont
  {Cowling}(1970)}]{CHAPMAN_Book_3rd_1970}%
  \BibitemOpen
  \bibfield  {author} {\bibinfo {author} {\bibfnamefont {S.}~\bibnamefont
  {Chapman}}\ and\ \bibinfo {author} {\bibfnamefont {T.}~\bibnamefont
  {Cowling}},\ }\href
  {https://www.cambridge.org/ch/academic/subjects/mathematics/fluid-dynamics-and-solid-mechanics/mathematical-theory-non-uniform-gases-account-kinetic-theory-viscosity-thermal-conduction-and-diffusion-gases?format=PB&isbn=9780521408448}
  {\emph {\bibinfo {title} {The Mathematical Theory of Non-uniform Gases: An
  Account of the Kinetic Theory of Viscosity, Thermal Conduction and Diffusion
  in Gases}}}\ (\bibinfo  {publisher} {Cambridge University Press},\ \bibinfo
  {year} {1970})\BibitemShut {NoStop}%
\bibitem [{\citenamefont {Dubois}(2008)}]{DUBOIS_CMA_55_2008}%
  \BibitemOpen
  \bibfield  {author} {\bibinfo {author} {\bibfnamefont {F.}~\bibnamefont
  {Dubois}},\ }\href {\doibase 10.1016/j.camwa.2007.08.003} {\bibfield
  {journal} {\bibinfo  {journal} {Comput. Math. Appl.}\ }\textbf {\bibinfo
  {volume} {55}},\ \bibinfo {pages} {1441 } (\bibinfo {year}
  {2008})}\BibitemShut {NoStop}%
\bibitem [{\citenamefont {Otomo}\ \emph {et~al.}(2017)\citenamefont {Otomo},
  \citenamefont {Boghosian},\ and\ \citenamefont {Dubois}}]{OTOMO_PA_486_2017}%
  \BibitemOpen
  \bibfield  {author} {\bibinfo {author} {\bibfnamefont {H.}~\bibnamefont
  {Otomo}}, \bibinfo {author} {\bibfnamefont {B.~M.}\ \bibnamefont
  {Boghosian}}, \ and\ \bibinfo {author} {\bibfnamefont {F.}~\bibnamefont
  {Dubois}},\ }\href {\doibase https://doi.org/10.1016/j.physa.2017.06.010}
  {\bibfield  {journal} {\bibinfo  {journal} {Physica A}\ }\textbf {\bibinfo
  {volume} {486}},\ \bibinfo {pages} {1000 } (\bibinfo {year}
  {2017})}\BibitemShut {NoStop}%
\bibitem [{\citenamefont {H\'{a}zi}\ and\ \citenamefont
  {K\'{a}vr\'{a}n}(2006)}]{HAZI_JPA_39_2006}%
  \BibitemOpen
  \bibfield  {author} {\bibinfo {author} {\bibfnamefont {G.}~\bibnamefont
  {H\'{a}zi}}\ and\ \bibinfo {author} {\bibfnamefont {P.}~\bibnamefont
  {K\'{a}vr\'{a}n}},\ }\href {http://stacks.iop.org/0305-4470/39/i=12/a=019}
  {\bibfield  {journal} {\bibinfo  {journal} {J. Phys. A Math. Gen.}\ }\textbf
  {\bibinfo {volume} {39}},\ \bibinfo {pages} {3127} (\bibinfo {year}
  {2006})}\BibitemShut {NoStop}%
\bibitem [{\citenamefont {Dellar}(2013)}]{DELLAR_CMA_65_2013}%
  \BibitemOpen
  \bibfield  {author} {\bibinfo {author} {\bibfnamefont {P.~J.}\ \bibnamefont
  {Dellar}},\ }\href {\doibase 10.1016/j.camwa.2011.08.047} {\bibfield
  {journal} {\bibinfo  {journal} {Comput. Math. Appl.}\ }\textbf {\bibinfo
  {volume} {65}},\ \bibinfo {pages} {129 } (\bibinfo {year}
  {2013})}\BibitemShut {NoStop}%
\bibitem [{\citenamefont {Feng}\ \emph {et~al.}(2016)\citenamefont {Feng},
  \citenamefont {Sagaut},\ and\ \citenamefont {Tao}}]{FENG_CF_131_2016}%
  \BibitemOpen
  \bibfield  {author} {\bibinfo {author} {\bibfnamefont {Y.}~\bibnamefont
  {Feng}}, \bibinfo {author} {\bibfnamefont {P.}~\bibnamefont {Sagaut}}, \ and\
  \bibinfo {author} {\bibfnamefont {W.-Q.}\ \bibnamefont {Tao}},\ }\href
  {\doibase 10.1016/j.compfluid.2016.03.009} {\bibfield  {journal} {\bibinfo
  {journal} {Comput. Fluids}\ }\textbf {\bibinfo {volume} {131}},\ \bibinfo
  {pages} {45 } (\bibinfo {year} {2016})}\BibitemShut {NoStop}%
\bibitem [{\citenamefont {Jacob}\ \emph {et~al.}(2018)\citenamefont {Jacob},
  \citenamefont {Malaspinas},\ and\ \citenamefont {Sagaut}}]{JACOB_JT_0_2018}%
  \BibitemOpen
  \bibfield  {author} {\bibinfo {author} {\bibfnamefont {J.}~\bibnamefont
  {Jacob}}, \bibinfo {author} {\bibfnamefont {O.}~\bibnamefont {Malaspinas}}, \
  and\ \bibinfo {author} {\bibfnamefont {P.}~\bibnamefont {Sagaut}},\ }\href
  {\doibase 10.1080/14685248.2018.1540879} {\bibfield  {journal} {\bibinfo
  {journal} {J. Turb.}\ }\textbf {\bibinfo {volume} {0}},\ \bibinfo {pages} {1}
  (\bibinfo {year} {2018})}\BibitemShut {NoStop}%
\end{thebibliography}%

\end{document}